%% file: time_evolution_review.tex
\documentclass[3p,preprint,number,sort&compress]{elsarticle}

\usepackage{amsmath}
\usepackage{amssymb}
\usepackage{amsfonts,dsfont}
\usepackage[]{graphicx} %draft
\usepackage[fit]{truncate}
\usepackage{mathrsfs}
\usepackage[T1]{fontenc}
\usepackage{upgreek}
\usepackage{textcomp}
\usepackage{type1cm,type1ec}
\usepackage{xcolor,fancybox,colortbl}
\usepackage[american]{babel}
\usepackage{caption}
\usepackage{listings}
\usepackage{lineno}
\usepackage{algpseudocode}
\usepackage{algorithm}
\usepackage{multirow}
\usepackage{blkarray}
\usepackage{dsfont}
\usepackage{ifthen}
\usepackage{qcircuit}
\usepackage{verbatim}
\usepackage{booktabs}
\usepackage{url}
\usepackage{bookmark}
\usepackage{subcaption}
\usepackage{bm} % for bold font in math environment via \bm
\usepackage{bbm, bbold}
\usepackage{enumitem}
\usepackage{siunitx}
\usepackage{morefloats}
\usepackage{rotating}
\usepackage[customcolors]{hf-tikz} % want this to have colored blkarray rows and cols
\usepackage{xkeyval}
\usepackage{moreverb}
\usepackage{nicefrac}
\usepackage{grffile}
\usepackage{braket}
\usepackage[framemethod=TikZ]{mdframed}
\usepackage{hyperref}
\usepackage[nameinlink,capitalize]{cleveref}
\usepackage{calc}
\usepackage{pgf}
\usepackage{tikz}
\tikzset{>=stealth}
\usepackage{pgffor}
\usepackage{pgfplots}
\pgfplotsset{compat=1.8}
\usepackage{pgfplotstable}
\usepgfplotslibrary{groupplots}
\usepgfplotslibrary{fillbetween}
\usepgfplotslibrary{patchplots}
\usetikzlibrary
{
	calc,
	decorations,
	plotmarks,
	patterns,
	positioning,
	petri,
	arrows,
	intersections,
	decorations.markings,
	backgrounds,
	fit,
	matrix,
	graphs,
	shapes.geometric,
	decorations.pathreplacing, 
	decorations.pathmorphing,
	shapes.misc,
	shapes.multipart,
	shapes,
	through,
	tikzmark,
	arrows.meta
}
\tikzstyle{ghostA} = [text=red!70,thick, minimum size=2*(5pt-\pgflinewidth), inner sep=0pt, outer sep=0pt]
\tikzstyle{ghostB} = [text=blue!70,thick, minimum size=2*(5pt-\pgflinewidth), inner sep=0pt, outer sep=0pt]
\tikzstyle{siteA} = [regular polygon, regular polygon sides=3, shape border rotate= 30, draw=red!50,fill=red!20,thick,inner sep=0pt,minimum width=1.5em,font=\footnotesize]
\tikzstyle{siteB} = [regular polygon, regular polygon sides=3, shape border rotate= -30, draw=green!50,fill=green!20,thick,inner sep=0pt,minimum width=1.5em,font=\footnotesize]
\tikzstyle{op} = [regular polygon, regular polygon sides=4, draw=orange!50, fill=orange!20, thick, inner sep=0.2pt, minimum width=1.25em, minimum height=1.5em,font=\footnotesize]
\tikzstyle{site} = [circle,draw=blue!50,fill=blue!20,thick,inner sep=0.2pt,minimum width=1.25em,font=\footnotesize]
\tikzstyle{hiddensite} = [circle,draw=white!50,fill=white!20,thick,inner sep=0.2pt,minimum width=1.25em,font=\footnotesize]
\tikzstyle{nosite} = [circle,draw=white,fill=white,thick,inner sep=0.1pt,minimum width=1.5em]
\tikzstyle{ghost} = []
\tikzstyle{intersite} = [regular polygon, regular polygon sides=4, shape border rotate= 45, draw=black!50,fill=black!20,thick,inner sep=0pt,minimum width=1.5em]
\tikzstyle{ld} = [inner sep=1pt, font=\small]
\tikzset{->-/.style={decoration={
			markings,
			mark=at position .5 with {\arrow{>}}},postaction={decorate}}}
\tikzset{->-rc/.style={decoration={
			markings,
			mark=at position .5 with {\arrow{>}}},postaction={decorate},rounded corners}}

\tikzset{
  -|-/.style={
    to path={
      (\tikztostart) -| ($(\tikztostart)!#1!(\tikztotarget)$) |- (\tikztotarget)
      \tikztonodes
    }
  },
  -|-/.default=0.5,
  |-|/.style={
    to path={
      (\tikztostart) |- ($(\tikztostart)!#1!(\tikztotarget)$) -| (\tikztotarget)
      \tikztonodes
    }
  },
  |-|/.default=0.5,
}

\newboolean{buildtikzpics}
\setboolean{buildtikzpics}{true}
\setboolean{nopreprintline}{true}
\newif\ifrebuildtikz
\newif\ifChangeMode
\ChangeModetrue
\ChangeModefalse
\ifthenelse{\boolean{buildtikzpics}}
{
	\rebuildtikztrue
	\usetikzlibrary{external}
	\tikzexternalize[optimize=false,prefix=figures/autogen/]%,mode=list and make]
}
{
	\rebuildtikzfalse
}
\definecolor{colorA}{rgb} {0.58,0,0.8275}
\definecolor{colorB}{rgb} {0.11,0.663,0.51}
\definecolor{colorC}{rgb} {0.3373,0.7059,0.9137}
\definecolor{colorD}{rgb} {0.902,0.6235,0}
\definecolor{colorE}{rgb} {0.9451,0.902,0.3255}
\definecolor{lgrey}{gray}{.9}

\usepackage{ulem}
\usepackage{nicefrac}
\newcommand{\nodagger}[0]{{\phantom{\dagger}}}
\newcommand{\noprime}[0]{{\phantom{\prime}}}

\DeclareMathOperator*{\argmin}{arg\,min}

\newcommand{\I}{\ensuremath{\mathrm{i}\mkern1mu}}
\newcommand{\E}{\ensuremath{\mathrm{e}}}
\DeclareMathOperator{\Tr}{Tr}
\DeclareMathOperator{\Real}{Re}

\newcommand{\wii}{\ensuremath{W^{\mathrm{II}}} }
\newcommand{\wiii}{\ensuremath{W^{\mathrm{I,II}}} }

\newcommand{\syten}{\textsc{SyTen}}
\newcommand{\scipal}{\textsc{SciPal}}
\newcommand{\symmps}{\textsc{SymMPS}}
\newcommand{\overbar}[1]{\mkern 1.5mu\overline{\mkern-1.5mu#1\mkern-1.5mu}\mkern 1.5mu}
\newcommand{\legdist}{0.05}
\newcommand{\id}{\ensuremath{\mathbf{\hat 1}}}
\newcommand{\numexamples}{four}
\makeatletter
\newcommand{\vast}{\bBigg@{3}}
\makeatother

\begin{document}

%% Link using Sec.~X and Alg.~X
\crefname{algorithm}{Alg.}{Algs.}
\Crefname{algorithm}{Algorithm}{Algorithms}
\crefname{section}{Sec.}{Secs.}
\Crefname{section}{Section}{Sections}
\crefname{table}{Tab.}{Tabs.}

\title{Time-evolution methods for matrix-product states}

\author[1]{Sebastian Paeckel}
\author[1,2]{Thomas K\"ohler}
\author[3]{Andreas Swoboda}
\author[1]{Salvatore R. Manmana}
\author[3,4]{Ulrich Schollw\"ock}
\author[5,4]{Claudius Hubig\corref{cor1}}
\ead{claudius.hubig@mpq.mpg.de}

\cortext[cor1]{Corresponding author}
\address[1]{Institut f\"ur Theoretische Physik, Universit\"at G\"ottingen, 37077 G\"ottingen, Germany }
\address[2]{Department of Physics and Astronomy,  Uppsala University,  Box 516,  S-75120 Uppsala, Sweden}
\address[3]{Department of Physics, Arnold Sommerfeld Center for Theoretical Physics (ASC),\\
Fakult\"{a}t f\"{u}r Physik, Ludwig-Maximilians-Universit\"{a}t M\"{u}nchen, 80333 M\"{u}nchen, Germany}
\address[4]{Munich Center for Quantum Science and Technology (MCQST), 80799 M\"unchen, Germany}
\address[5]{Max-Planck-Institut f\"ur Quantenoptik, 85748 Garching, Germany}

\begin{abstract}
  Matrix-product states have become the de facto standard for the
  representation of one-dimensional quantum many body states. During
  the last few years, numerous new methods have been introduced to
  evaluate the time evolution of a matrix-product state.  Here, we
  will review and summarize the recent work on this topic as applied
  to finite quantum systems.  We will explain and compare the
  different methods available to construct a time-evolved
  matrix-product state, namely the time-evolving block decimation, the
  MPO \wiii method, the global Krylov method, the local Krylov method
  and the one- and two-site time-dependent variational principle. We
  will also apply these methods to \numexamples{} different
  representative examples of current problem settings in condensed
  matter physics.
\end{abstract}


\maketitle

\tableofcontents

\section{Introduction}
\label{sec:intro}
\input{content/intro.tex}

\section{Matrix-product states and operators}

\input{content/mps_mpo.tex}

\input{content/time-evolution-overview.tex}

\section{\label{sec:approx}Approximations to $\hat U(\delta)$}
\input{content/tebd.tex}

\clearpage

\input{content/zaletel_mpo.tex}

\input{content/krylov.tex}

\clearpage

\section{\label{sec:local}MPS-local methods}
\input{content/local-methods.tex}
\input{content/local-krylov.tex}

\input{content/tdvp.tex}

\clearpage

\input{content/tricks.tex}

\input{content/examples/intro.tex}
\input{content/examples/dsf.tex}
\input{content/examples/hubbard.tex}
\input{content/examples/2dmelting.tex}

\input{content/examples/otoc.tex}

\clearpage

\input{content/outlook.tex}

\section*{Acknowledgements}
We gratefully acknowledge insightful discussions with J.~Haegeman, I.~P.~McCulloch, E.~M.~Stoudenmire and F.~Verstraete.
We furthermore thank T.~Barthel, D.~Bauernfeind, P.~Emonts, J.~Haegeman, F.~Heidrich-Meisner, D.~Jansen, M.~Lubasch, I.~P.~McCulloch, P.~Moser, R.~M.~Noack, L.~Tagliacozzo and G.~Vidal for useful comments on an early draft of this manuscript.

Computational resources made available by the Department of Applied
Theoretical Physics, Clausthal Technical University, are gratefully
acknowledged.

\paragraph{Funding information}
C. H.~acknowledges funding through ERC Grant QUENOCOBA, ERC-2016-ADG
(Grant no. 742102).  T. K.~and S. R. M.~acknowledge funding from the
Deutsche Forschungsgemeinschaft (DFG, German Research Foundation)
through SFB/CRC1073 (project B03) -- 217133147.  S. R. M., S. P. and
U. S.~acknowledge funding from the DFG through Research Unit FOR 1807
(projects P1, P3 and P7) -- 207383564. C. H.~and U. S.~are supported by the DFG
under Germany's Excellence Strategy -- EXC-2111 -- 390814868.
T. K.~acknowledges funding through ERC Grant 1D-Engine, ERC-2017-STG (Grant No. 758935).

\end{document}

%% file: content/intro.tex
The dynamics of correlated quantum systems is a rich playground for
non-equilibrium physics.  The actual study, however, is a challenge
for experiments as well as for theory.  In the context of tensor
networks, numerical methods were introduced around 2004 and opened the
path to study the dynamical behavior of low-dimensional systems in a
controlled fashion.  In particular, the Trotterized time evolution of
matrix-product states (MPS) as produced by the density-matrix
renormalization group (TEBD, tDMRG, tMPS)\cite{white92:_densit,
  white93:_densit,
  vidal03:_effic_class_simul_sligh_entan_quant_comput,
  vidal04:_effic_simul_one_dimen_quant,
  white04:_real_time_evolut_using_densit, daley04:_time_hilber,
  zwolak04,
  verstraete04:_matrix_produc_densit_operat} made a breakthrough as it
combined powerful time-evolution schemes with the efficient truncation
of the exponentially-large Hilbert space intrinsic to quantum
many-body systems. Nowadays, DMRG methods formulated in the language
of matrix-product states\cite{rommer97:_class, dukelsky98:_equiv,
  mcculloch07:_from,
  schuch08:_entrop_scalin_simul_matrix_produc_states,
  oseledets11:_tensor_train_decom, schollwoeck11} (MPS) are the
de facto standard for the investigation of one-dimensional systems.
Here, we will focus on the time evolution of finite-dimensional quantum states along the
real and the imaginary time axis.
That is, we restrict ourselves to those methods which produce a
time-evolved MPS representation of a quantum state as opposed to
evaluating the time-dependence of a specific observable in some
subspace (as done by the Chebyshev\cite{RevModPhys.78.275, holzner11:_cheby,
  wolf14:_cheby, halimeh15:_cheby, xie18:_reort_cheby} and other
similar approaches) or methods which directly work in frequency space
(such as, for example, the correction vector\cite{kuehner99:_dynam, PhysRevB.66.045114, 10.1143/PTPS.176.143} approach).
Specifically, we will review and compare
in some detail the various MPS-based approaches which have been
developed in the past few years with sometimes large improvements over
the first generation of methods.

The main goal of these schemes is to combine the efficient truncation
of the Hilbert space with an accurate time-evolution
method. Naturally, time-evolution methods are dealt with regularly in
the mathematical literature (cf.~the review
\cite{moler03:_ninet_dubious_ways_comput_expon}), but the
applicability to MPS-based problem settings is often limited due to
the peculiarity of the MPS approach in approximating the Hilbert
space. Since we have quantum systems in mind, the underlying
differential equation is the time-dependent Schr\"odinger
equation (TDSE), whose solution can formally be recast to the task of
applying the time-evolution operator
\begin{equation}
  \hat U(\delta) = e^{-\I \delta \hat H} \quad
\end{equation}
with the time-independent Hamiltonian $\hat H$, time step size $\delta$ and
$\hbar \equiv 1$. It is therefore natural to ask for efficient MPS methods
for the time evolution of quantum states by combining suitable
approaches to matrix exponentials with the MPS formulation of quantum
(lattice) systems.  Indeed, most of this review article is concerned
precisely with this question and explores under which circumstances
which ansatz will be the most successful one in terms of accuracy and
efficiency.

In the context of quantum systems, first attempts to numerically
integrate the time-dependent Schr\"odinger equation beyond the reach
of exact diagonalization using sparse matrix
exponentials\cite{saad03:_iterat_method_spars_linear_system} are
reported in quantum chemistry\cite{park86:_unitar_lancz, kosloff88:_time}. When
attempting to do the same for MPS, two main approaches have to be
differentiated: the first evaluates or constructs the time-evolved
state $\Ket{\psi(t+\delta)} = \hat U(\delta)\Ket{\psi(t)}$ directly by
approximating the action of $\hat U(\delta)$ in a sufficiently small
subspace. In this approach, $\hat U(\delta)$ itself is never
calculated or approximated. The global Krylov
method\cite{garcia-ripoll06:_time_matrix_produc_states,
  dargel12:_lancz, wall12:_out} presented in \cref{sec:te:krylov}
simply translates the Lanczos formalism\cite{lanczos50} for unitary time evolution
to matrix-product states.  The local Krylov
approach\cite{schmitteckert04:_noneq, feiguin05:_time,
  manmana05:_time_quant_many_body_system, rodriguez06, wall12:_out,
  ronca17:_time_step_target_time_depen} (\cref{sec:local-krylov}) and
the time-dependent variational
principle\cite{haegeman11:_time_depen_variat_princ_quant_lattic,
  haegeman16:_unify} (\cref{sec:tdvp}) further adapt the Krylov method
to the special MPS structure and re-derive the TDSE in suitable local
subspaces.

Alternatively, one may ask for an approximation of $\hat U(\delta)$
which 
can be obtained efficiently and also can be applied efficiently
to the quantum state. One way to deal with such exponentials
is well-known in quantum field theory where it permits to set up path
integrals\cite{keller75:_feynm_integ}. It decomposes the
time-evolution operator for ``sufficiently small'' time steps using
the Suzuki-Trotter\cite{suzuki76:_gener_trott} decomposition. This
decomposition leads to conveniently smaller matrix exponentials. The
method can be directly applied in the context of matrix-product
states\cite{vidal04:_effic_simul_one_dimen_quant,
  white04:_real_time_evolut_using_densit, daley04:_time_hilber,
  zwolak04,
  verstraete04:_matrix_produc_densit_operat}, as explained in
\cref{sec:tebd}. Extending on it, we may ask for efficient
matrix-product operator (MPO) approximations of $\hat U(\delta)$ which
exploit the MPO structure directly\cite{zaletel15:_time} to allow
efficient exponentiation, cf.~\cref{sec:te:mpowii}.

A direct comparison of widely used approaches, including the recently
developed methods, is missing so far. Since those methods have
different strengths and weaknesses, it is difficult to say a priori
which one will be the most efficient for a given problem setting. We
will attempt to fill this gap by providing an overview and
comparison. In particular, we discuss in detail the Suzuki-Trotter decomposition and MPO \wiii
methods for constructing $\hat U(\delta)$ directly, and TDVP- and
Krylov-subspace-based approaches evaluating
$\hat U(\delta)\Ket{\psi}$. All methods are tested and compared by
treating \numexamples{} representative problems of dynamics in
correlated systems.  This includes the real- and imaginary time
evolution of chains, time-dependent correlators, critical and gapped
systems, and attempts at two-dimensional systems. To cross-check our
results, we use two independent implementations of the time-evolution
methods within the \scipal{}-\symmps\cite{kramer15:_scipal, paeckel:_symmps_toolk} and
\syten{}\cite{hubig:_syten_toolk,
  hubig17:_symmet_protec_tensor_networ} toolkits.

%%% Local Variables: 
%%% mode: latex
%%% TeX-master: "../time_evolution_review"
%%% End: 

%% file: content/mps_mpo.tex
The main problem in the numerical treatment of quantum many-body systems is the large Hilbert space which grows exponentially with system size\cite{schollwoeck11}.
This strongly restricts exact diagonalization approaches to small
system sizes; for spin systems one can reach $\sim 40-50$ lattice
sites\cite{wietek18:_sublat}.  A variety of numerical methods (e.g.,
quantum Monte Carlo\cite{sandvik91:_quant_monte_carlo}, dynamical
mean-field theory\cite{georges96:_dynam}, ...) have been developed to
overcome this restriction.
For one-dimensional systems, matrix-product state (MPS) methods are
extremely successful because they can efficiently represent
weakly-entangled states that obey the area law\cite{schollwoeck05, eisert10:_colloq,
  schollwoeck11}.
Here, we start by recapitulating general properties of the MPS ansatz class and its efficient numerical use, with the scaling costs of each application summarized in \cref{tab:scaling-costs} at the very end of the chapter. As in the rest of the review, we focus on finite matrix-product states, representing quantum states on finite-dimensional Hilbert spaces. While a few of the methods tested here translate also to infinite matrix-product states and a careful choice of boundary conditions\cite{phien12:_infin, binder15:_minim} alleviates finite-size effects, the review of time-evolution methods treating the thermodynamic limit is a topic on its own and is left for future review articles.
\subsection{Tensor notation}

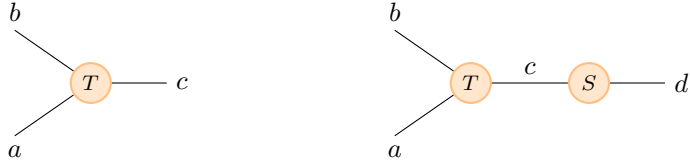
\begin{figure}
  \centering
  \tikzsetnextfilename{example-tensor}
  \begin{tikzpicture}
    \node[draw,circle,draw=orange!50, fill=orange!20, thick, inner sep=0.5pt,minimum width=1.5em,font=\footnotesize] (tensor) {$T$};
    \draw[] (tensor) -- +(-1,-0.7) node[below]{$a$};
    \draw[] (tensor) -- +(-1,+0.7) node[above]{$b$};
    \draw[] (tensor) -- +(+1,-0) node[right]{$c$};

    \node[draw,circle,draw=orange!50, fill=orange!20, thick, inner sep=0.5pt,minimum width=1.5em,font=\footnotesize] (t1) at (5,0) {$T$};
    \draw[] (t1) -- +(-1,-0.7) node[below]{$a$};
    \draw[] (t1) -- +(-1,+0.7) node[above]{$b$};

    \node[draw,circle,draw=orange!50, fill=orange!20, thick, inner sep=0.5pt,minimum width=1.5em,font=\footnotesize] (t2) [right=of t1] {$S$};
    \draw[] (t1) -- (t2) node[midway,above]{$c$};
    \draw[] (t2) -- +(+1,0) node[right]{$d$};
  \end{tikzpicture}
  \caption{\label{fig:example-tensor}Left: Graphical representation of
    the example tensor $T_{a,b,c}$ with three indices (legs) $a$, $b$
    and $c$. Right: Graphical representation of the tensor contraction
    $\sum_{c} T_{a,b,c} S_{c,d} = T \cdot S$.}
\end{figure}

The fundamental objects of MPS algorithms are tensors.
In our context, tensors are multi-dimensional collections of complex
or real numbers.
Each index of a tensor corresponds to one of its dimensions;
graphically, tensors are represented as shapes (circles, triangles,
squares) with one leg per tensor index
(cf.~\cref{fig:example-tensor}).
A tensor $T$ on e.g.~three associated vector spaces $A$, $B$ and $C$
has three indices $a=[1, \ldots, \mathrm{dim}(A)]$,
$b=[1, \ldots, \mathrm{dim}(B)]$ and $c=[1, \ldots,
\mathrm{dim}(C)]$.
The tensor then has scalar entries $T_{a,b,c}$.
We explicitly do not consider global
symmetries\footnote{The use of global symmetries is certainly crucial for efficient implementations, but at the same time orthogonal to the time-evolution methods discussed here. For introductions to the topic of global symmetries in MPS and more generically tensor networks, cf.~\cite{mcculloch02:_abelian, mcculloch02:_collec_phenom_stron_correl_elect_system, mcculloch07:_from, singh10:_tensor, singh11:_tensor_u, singh12:_tensor_su, weichselbaum12:_non, hubig17:_symmet_protec_tensor_networ, hubig18:_abelian}.} here and hence do not associate a
direction to our tensor legs (or, conversely, differentiate between
vector and dual spaces).
As such, the upstairs/downstairs location of
tensor indices is meaningless and
$T_{a,b,c} = T^{a}_{b,c} = T^{a,b,c}$.
By complex-conjugating every element of a tensor $T$, we obtain a new
tensor $\overbar{T}$ with elements
$\overbar{T}_{\overbar{a}, \overbar{b}, \overbar{c}} =
T_{a,b,c}^\star$.
We will use $\overbar{a}$ to denote indices of
conjugated tensors. Contracting $\overbar{T}$ and $T$ over the indices
$a$ and $b$, we write:
\begin{equation}
  \sum_{a, b} \overbar{T}_{a,b,\overbar{c}} T_{a,b,c} = X_{\overbar{c},c}
\end{equation}
which is equivalent to
\begin{equation}
  \sum_{\overbar{a},a,\overbar{b},b} \delta_{\overbar{a},a\vphantom{\overbar{b}}} \delta_{\overbar{b},b} \overbar{T}_{\overbar{a},\overbar{b},\overbar{c}} T_{a,b,c\vphantom{\overbar{b}}} \;.
\end{equation}
Given two tensors $A_{a,b,c}$ and $B_{b,d,c}$, the shorthand
$A \cdot B$ denotes the contraction over all shared indices:
\begin{equation}
  A \cdot B = \sum_{b,c} A_{a,b,c} B_{b,d,c} = Y_{a,d} \;.
\end{equation}
Tensor contractions can also be represented graphically by drawing
tensors with connected legs, cf.~\cref{fig:example-tensor}. Note that
we will, where possible without confusion, also use $a$, $b$, $c$,
\ldots to refer to the dimension $\mathrm{dim}(A)$ etc. respectively.

\subsection{Matrix-product states (MPS)}
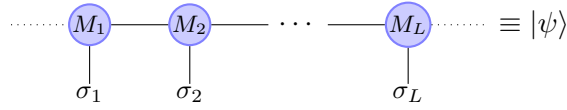
\begin{figure}
  \centering
  \tikzsetnextfilename{mps}
  %	\tikzset{external/export next=false}
  \begin{tikzpicture}
    \begin{scope}[node distance = 0.5 and 0.75]
      \node[site] (site1) {$M_1$};
      \node[ghost] (ghost0) [left=of site1] {};
      \node[site] (site2) [right=of site1] {$M_2$};
      \node[ghost] (dots) [right=of site2] {$\cdots$};
      \node[site] (siteL) [right=of dots] {$M_L$};
      
      \node (psi) [right=of siteL] {$\equiv \left|\psi\right\rangle$};
      
      \node[ld] (sigma1) [below=of site1] {$\sigma_{1}$};
      \node[ld] (sigma2) at (sigma1 -| site2) {$\sigma_{2}$};
      \node[ld] (sigmaL) at (sigma1 -| siteL) {$\sigma_{L}$};
      
      \draw[dotted] (ghost0) -- (site1);
      \draw[] (sigma1) -- (site1);
      \draw[] (site1) -- (site2);
      \draw[] (sigma2) -- (site2);
      \draw[] (site2) -- (dots);
      \draw[] (dots) -- (siteL);
      \draw[] (sigmaL) -- (siteL);
      \draw[dotted] (siteL) -- (psi);
    \end{scope}
  \end{tikzpicture}
  \caption{\label{fig:mps} Schematic of the tensor network of a matrix-product state (MPS).  Horizontal lines denote the internal indices
    with bond dimension $m$, whereas the vertical lines denote
    physical indices with dimension $\sigma$. Dotted lines to the left
    and right indicate the dummy indices $m_0$ and $m_L$.}
\end{figure}

Matrix-product states are efficient representations for
one-dimensional weakly-entangled quantum states.
The main idea is to represent the coefficient tensor
$c_{\sigma_1\ldots\sigma_L}$ of a general quantum state on a discrete
lattice
\begin{align}
	\label{eq:state}
	\lvert\psi\rangle = \sum_{\sigma_1, \ldots, \sigma_L} c_{\sigma_1\ldots\sigma_L} \lvert {\sigma_1 \cdots \sigma_L} \rangle
\end{align}
as a product of $L$ rank-3 tensors $M_i$ (cf.~\cref{fig:mps})
\begin{align}
	\label{eq:mps}
	\lvert \psi \rangle = \sum_{\substack{\sigma_1, \ldots, \sigma_L, \\ m_0, \ldots, m_L}} M^{\sigma_1}_{1; m_0, m_1} \cdots M^{\sigma_L}_{L;m_{L-1}, m_L} \lvert {\sigma_1 \cdots \sigma_L} \rangle \;,
\end{align}
where $m_0$ and $m_L$ are 1-dimensional dummy indices introduced for
consistency.
For a specific set of local states $\{ \sigma_1\ldots\sigma_L \}$, we
use a single matrix $M_j^{\sigma_j}$ per site. We hence evaluate a matrix
product to give a single entry $c_{\sigma_1\ldots\sigma_L}$, resulting in the name ``matrix-product'' states. We will use $\sigma_j$ to index the local
physical states on site $j$.
Depending on the bond dimensions $m_j$ of the tensors $M_j$ (also
called virtual or auxiliary dimension), different quantum states can
be represented exactly.
If we let $m_j$ grow exponentially by a factor of $\sigma$ per site
towards the center of the system, any quantum state can be written as
an MPS.
The entanglement between the part of an MPS to the left of bond
$(j,j+1)$ and to the right of this bond is bounded by $\log(m_j)$.
The required bond dimension $m = \mathrm{max}_j(m_j)$ to represent a
quantum state exactly hence grows exponentially with its entanglement
along left/right partitions.
On the other hand, lowly-entangled states require only a small
bond dimension, leading to an efficient representation of the quantum
state\cite{schollwoeck05, eisert10:_colloq, schollwoeck11}.

\subsection{Matrix-product operators (MPO)}

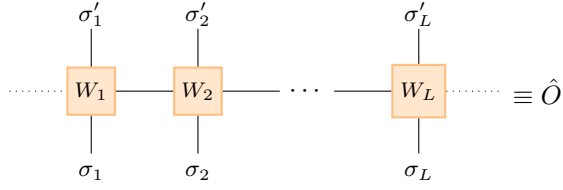
\begin{figure}
	\centering
%	\tikzset{external/export next=false}
	\tikzsetnextfilename{mpo}
	\begin{tikzpicture}
		\begin{scope}[node distance = 0.5 and 0.75]
			\node[op] (site1) {$W_1$};
			\node[ghost] (ghost0) [left=of site1] {};
			\node[op] (site2) [right=of site1] {$W_2$};
			\node[ghost] (dots) [right=of site2] {$\cdots$};
			\node[op] (siteL) [right=of dots] {$W_L$};
	
			\node (psi) [right=of siteL] {$\equiv \hat O$};
	
			\node[ld] (sigma1) [above=of site1] {$\sigma_{1}^\prime$};
			\node[ld] (sigma2) at (sigma1 -| site2) {$\sigma_{2}^\prime$};
			\node[ld] (sigmaL) at (sigma1 -| siteL) {$\sigma_{L}^\prime$};

			\node[ld] (sigma1p) [below=of site1] {$\sigma_{1}^\noprime$};
			\node[ld] (sigma2p) at (sigma1p -| site2) {$\sigma_{2}^\noprime$};
			\node[ld] (sigmaLp) at (sigma1p -| siteL) {$\sigma_{L}^\noprime$};
	
			\draw[dotted] (ghost0) -- (site1);
			\draw[] (site1) -- (site2);
			\draw[] (site2) -- (dots);
			\draw[] (dots) -- (siteL);
			\draw[dotted] (siteL) -- (psi);

			\draw[] (sigma1) -- (site1);
			\draw[] (sigma2) -- (site2);
			\draw[] (sigmaL) -- (siteL);

			\draw[] (sigma1p) -- (site1);
			\draw[] (sigma2p) -- (site2);
			\draw[] (sigmaLp) -- (siteL);
		\end{scope}
	\end{tikzpicture}
	\caption
	{
		\label{fig:mpo}
		Schematic of the tensor network of a matrix-product
                operator (MPO).  Horizontal lines denote the internal
                indices with bond dimension $w$, whereas the vertical
                lines denote physical indices with dimension
                $\sigma$.}
\end{figure}

By analogy to quantum states, we can express every operator as a matrix-product operator (MPO), i.e.,~a contraction of $L$ rank-4 tensors $W_j$:
\begin{align}
	\hat O 
	&=
		\sum_{\substack{\sigma_1^\noprime,\ldots,\sigma_L^\noprime,\\ \sigma_1^{\prime},\ldots,\sigma_L^{\prime}}}
			c_{\sigma_1^\noprime\ldots\sigma_L^\noprime, \sigma_1^{\prime}\ldots\sigma_L^{\prime}} 
			\ket{\sigma_1^\noprime\cdots\sigma_L^\noprime} 
			\bra{\sigma_1^{\prime}\cdots\sigma_L^{\prime}} 
\\
	&= 
		\sum_{\substack{\sigma_1^\noprime,\ldots,\sigma_L^\noprime,\\ \sigma_1^{\prime},\ldots,\sigma_L^{\prime}, \\ w_0,\ldots,w_L}}
			W^{\sigma_1^\noprime,\sigma_1^{\prime}}_{1; w_0, w_1}
			\ldots
			W^{\sigma_L^\noprime,\sigma_L^{\prime}}_{L; w_{L-1}, w_L} 
			\ket{\sigma_1^\noprime\cdots\sigma_L^\noprime}
			\bra{\sigma_1^{\prime}\cdots\sigma_L^{\prime}}
\;.
\end{align}
The only difference is that the tensor components $W_j$ are now rank-4
tensors to account for the domain and image Hilbert spaces. There are
different avenues to constructing matrix-product operators \cite{kin-lic16:_matrix_produc_operat_matrix_produc, hubig17:_gener, paeckel17:_autom}. When used to represent the Hamiltonian $\hat H$ or other operators which are sums of local terms, the construction can be understood by splitting the
system at bond $j$ (connecting sites $j$ and $j+1$). We then separate
terms of the operator that act only within their partition
$\hat{H}^{L}_{j-1},\hat{H}^{R}_{j+1}$ and those that connect the partitions
$\hat{h}^{L}_{j;a_j},\hat{h}^{R}_{j;a_j}$\footnote{This is actually what
  is called \textit{superblock Hamiltonian} in the standard
  DMRG\cite{dmrgbook}.}
\begin{equation}
	\hat{H} = \hat{H}^{L}_{j-1} \otimes \mathbf{\hat{1}}^{R}_{j} + \mathbf{\hat{1}}^{L}_{j}\otimes \hat{H}^{R}_{j+1} + \sum_{a_j=1}^{N_j}\hat{h}^{L}_{j;a_j} \otimes \hat{h}^{R}_{j;a_j} \; . \label{eq:mpo:splitting}
\end{equation}
Based on the tensor product structure there is an operator-valued matrix $\hat{W}_{j}$ which relates the partitioned representations between the bonds $j-1$ and $j$; e.g., for the right partition we have
\begin{equation}
	\left(
		\begin{array}{c}
			\hat{H}^{R}_{j-1} \\ \hat{h}^{R}_{j-1} \\ \mathbf{\hat{1}}^{R}_{j-1}
		\end{array}
	\right)
	=
	\tikzsetnextfilename{Zaletel_MPO_matrix_1}
%	\tikzset{external/export next=false}
	\begin{tikzpicture}
		[
			baseline=(MPO.center)
		]
		\begin{scope}
			\matrix (MPO) 
			[
				matrix of math nodes,
				left delimiter=(,
				right delimiter=),
				nodes in empty cells,
				nodes = 
				{
					minimum height=2em,
					anchor=center,
					text width=2em,
					text depth=.25ex,
					align=center,
					inner sep=0pt
				}
			]
			{
				\mathbf{\hat{1}}_{j} 	& \phantom{1} 	& \phantom{1}	& \phantom{1} \\
				\phantom{1}	     		& \phantom{1}	& \phantom{1} 	& \phantom{1} \\
				\phantom{1}	     		& \phantom{1}	& \phantom{1}	& \phantom{1} \\
				\phantom{1}				&		 	 	& \phantom{1} 	& \mathbf{\hat{1}}_{j} \\
			};
			\node [above=5pt of MPO-1-1] (IdL) {\scriptsize $1\phantom{_j}$};
			\node [above=5pt of MPO-1-4] (IdR) {\scriptsize $1\phantom{_j}$};
			\node at ($(IdL)!0.5!(IdR)$) {\scriptsize $N_{j}$};
			\node [left=15pt of MPO-1-1] (IdT) {\scriptsize $1\phantom{_j}$};
			\node [left=15pt of MPO-4-1] (IdB) {\scriptsize $1\phantom{_j}$};
			\node at ($(IdT)!0.5!(IdB)$)  {\scriptsize $N_{j-1}$};
			\draw [black,decorate,decoration={brace,mirror,raise=-2pt}] ([xshift=-15pt,yshift=-15pt]MPO-4-1.south west) -- node[below] {$\hat{W}_{j}$} ([xshift=15pt,yshift=-15pt]{MPO-4-4.south east});
		\end{scope}
		\begin{scope}[on background layer]
			\node [draw=blue!30, fill=blue!10, inner sep=-1pt, fit=(MPO-1-2)(MPO-1-3)] (cj) {};
			\node [draw=blue!30, fill=blue!10, inner sep=-1pt, fit=(MPO-1-4)] (dj) {};
			\node [draw=blue!30, fill=blue!10, inner sep=-1pt, fit=(MPO-2-2)(MPO-3-3)] (aj) {};
			\node [draw=blue!30, fill=blue!10, inner sep=-1pt, fit=(MPO-2-4)(MPO-3-4)] (bj) {};
		\end{scope}
		\begin{scope}
			\node [] (a) at (aj) {$\hat{A}_{j}$};
			\node [] (b) at (bj) {$\hat{B}_{j}$};
			\node [] (c) at (cj) {$\hat{C}_{j}$};
			\node [] (d) at (dj) {$\hat{D}_{j}$};
		\end{scope}
	\end{tikzpicture}
	\otimes
	\left(
		\begin{array}{c}
			\hat{H}^{R}_{j} \\ \hat{h}^{R}_{j} \\ \mathbf{\hat{1}}^{R}_{j}
		\end{array}
	\right) \; .\label{eq:mpo:recursion}
\end{equation}
The operator-valued matrices $\hat A$, $\hat B$, $\hat C$ and $\hat D$ then define the recursion relations to iteratively build the complete operator $\hat H$. This picture directly leads to a construction of MPOs based on
finite-state machines (FSM) \cite{crosswhite08:_finit, paeckel17:_autom}.

In analogy to matrix-product states with bonds $m_j$ and a maximal
bond dimension $m$, the bonds of matrix-product operators are labelled
by $w_j$ with the maximal bond dimension denoted by $w$.

\subsection{Canonical form}
\label{sec:mps_canonical}
Considering \cref{eq:mps}, we can clearly insert
resolutions of the identity $X X^{-1}$ in between any two MPS tensors
$M_j$ and $M_{j+1}$. Multiplying $X$ into $M_j$ and $X^{-1}$
into $M_{j+1}$ changes the numerical content of each tensor
while keeping the state invariant.
This gauge freedom can be exploited to increase the numerical stability of the algorithm and simplify many tensor contractions. Two possible choices to fix the gauge are enforcing the \emph{left} or \emph{right} normalization of the tensors $M_j$. We write $A_j$ for a left-normalized tensor and $B_j$ for a right-normalized tensor with the defining property that
\begin{align}
	\label{eq:normalized}
  \sum_{\sigma_j,m_{j-1}} {\bar{A}^{\sigma_j}}_{j; m_{j-1},\overbar{m}_{j}} A^{\sigma_j}_{j; m_{j-1},m_{j}} & = \mathbf{1}_{\overbar{m}_{j},m_{j}} \\
  \sum_{\sigma_j, m_{j}} B^{\sigma_j}_{j; m_{j-1}, m_{j}} {\bar B^{\sigma_j}}_{j; \overbar{m}_{j-1}, m_{j}} & = \mathbf{1}_{m_{j-1},\overbar{m}_{j-1}} \; .
\end{align}

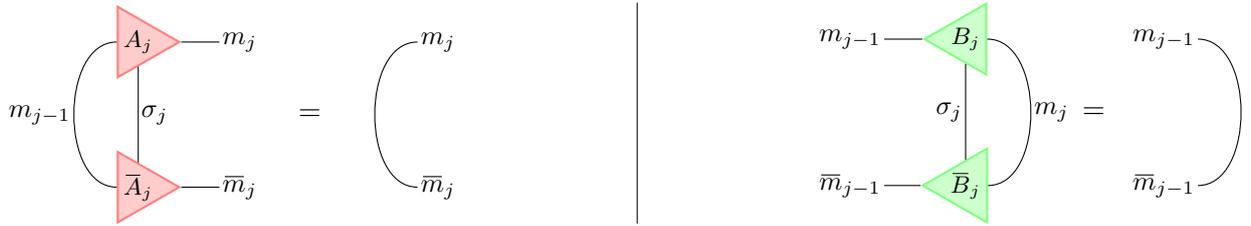
\begin{figure}
	\centering
%	\tikzset{external/export next=false}
	\tikzsetnextfilename{MPS_left_normalized}
	\begin{tikzpicture}
		\begin{scope}[node distance = 0.5]
			\node[siteA] (site1a) {$A_{j}$};
			\node[ld] (site2a) [right=of site1a] {$m_j$};

			\draw[] (site1a) -- (site2a);

			\node[ghost] (ghost1) [below = of site1a] {};

			\node[siteA] (site1b) [below = of ghost1] {$\overbar A_{j}$};
			\node[ld] (site2b) [right = of site1b] {$\overbar m_j$};

			\draw[] (site1b) -- (site2b);

			\draw[] (site1a) -- (site1b) node [ld,midway, right] (nonbendtext) {$\sigma_j$};
			\draw[] (site1a.west) to [bend right=90] node [ld,midway,left] (bendtext) {$m_{j-1}$} (site1b.west);

			\node[ghost] (ghost2) at ($(site2a)!0.5!(site2b)$) {};
			\node[ghost] (center) [right = of ghost2] {$=$};
			\node[ghost] (ghost3) [right = of center] {};
			\node[ghost] (ghost4) [right = of ghost3] {};

			\node[ld] (site4a) at (ghost4 |- site1a) {$m_j$};
			\node[ld] (site4b) at (ghost4 |- site1b) {$\overbar m_j$};
			\draw[] (site4a.west) to [bend right=90] (site4b.west);
		\end{scope}
	\end{tikzpicture}
	\hfill
	\vline
	\hfill
%	\tikzset{external/export next=false}
	\tikzsetnextfilename{MPS_right_normalized}
	\begin{tikzpicture}
		\begin{scope}[node distance = 0.5]
			\node[ld] (site1a) {$m_{j-1}$};
			\node[siteB] (site2a) [right=of site1a] {$B_{j}$};

			\draw[] (site1a) -- (site2a);

			\node[ghost] (ghost1) [below = of site2a] {};
			\node[siteB] (site2b) [below = of ghost1] {$\overbar B_{j}$};
			\node[ld] (site1b) at (site2b-|site1a) {$\overbar m_{j-1}$};

			\draw[] (site1b) -- (site2b);

			\draw[] (site2a) -- (site2b) node [ld,midway, left] (nonbendtext) {$\sigma_j$};
			\draw[] (site2a.east) to [bend left=90] node [ld,midway,right] (bendtext) {$m_{j}$} (site2b.east);

			\node[ghost] (ghost2) at ($(site2a)!0.5!(site2b)$) {};
			\node[ghost] (ghost3) [right = of ghost2] {};
			\node[ghost] (center) [right = of ghost3] {$=$};
			\node[ghost] (ghost4) [right = of center] {};

			\node[ld] (site4a) at (ghost4 |- site1a) {$m_{j-1}$};
			\node[ld] (site4b) at (ghost4 |- site1b) {$\overbar m_{j-1}$};
			\draw[] (site4a.east) to [bend left=90] (site4b.east);
		\end{scope}
	\end{tikzpicture}
	\caption
	{
		\label{fig:normalized}
		Left (Right) normalized tensor $A_{j}$ (red, right-pointing triangle) ($B_{j}$ (green, left-pointing triangle)) contracted with its adjoint resulting in an identity. 
	}
\end{figure}

\begin{figure} %left_right_normalized
	\centering
%	\tikzset{external/export next=false}
	\tikzsetnextfilename{left_right_normalized}
	\begin{tikzpicture}
		\begin{scope}[node distance = 0.5]
			\node[ghost] (site0)  {};
			\node[siteA,inner sep=-1pt, minimum width=1.5cm] (site1) [right = of site0] {$A_1$};
			\node[ghost] (site2) [right = of site1] {$\cdots$};
			\node[siteA,inner sep=-3pt, minimum width=1.5cm] (site3) [right = of site2] {$A_{j-1}$};
			\node[site,inner sep=0pt, minimum width=1cm]  (site4) [right = of site3] {$M_j$};
			\node[siteB,inner sep=-3pt, minimum width=1.5cm] (site5) [right = of site4] {$B_{j+1}$};
			\node[ghost] (site6) [right = of site5] {$\cdots$};
			\node[siteB,inner sep=-1pt, minimum width=1.5cm] (site7) [right = of site6] {$B_L$};
			\node[ghost] (site8) [right = of site7] {};

			\node[ld] (sigma1) [below=of site1] {$\sigma_{1}$};
			\node[ld] (sigma3) at (sigma1-| site3) {$\sigma_{j-1}$};
			\node[ld] (sigma4) at (sigma1-| site4) {$\sigma_{j}$};
			\node[ld] (sigma5) at (sigma1-| site5) {$\sigma_{j+1}$};
			\node[ld] (sigma7) at (sigma1-| site7) {$\sigma_{L}$};

			\draw[dotted] (site0) -- (site1);
			\draw[] (site1) -- (site2);
			\draw[] (site2) -- (site3);
			\draw[] (site3) -- (site4);
			\draw[] (site4) -- (site5);
			\draw[] (site5) -- (site6);
			\draw[] (site6) -- (site7);
			\draw[dotted] (site7) -- (site8);

			\draw[] (site1) -- (sigma1);
			\draw[] (site3) -- (sigma3);
			\draw[] (site4) -- (sigma4);
			\draw[] (site5) -- (sigma5);
			\draw[] (site7) -- (sigma7);
		\end{scope}
	\end{tikzpicture}
	\caption
	{
		\label{fig:left_right_normalized}
		MPS with active site $j$ and consequently left (right) normalized site tensor left (right) of site $j$ as defined in \cref{eq:normalized}.
	}
\end{figure}
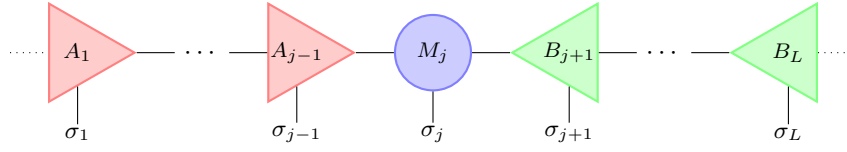

Graphically they are represented by red (left-normalized) or green (right-normalized) triangles (see \cref{fig:normalized}) where the orientation of the triangles also indicates the normalization.
Left-/Right-canonical MPS are now defined by requiring that they consist of left-/right-normalized tensors only.
Furthermore, a mixed-canonical MPS is defined by fixing a site $j$ with unnormalized tensor $M_j$ and demanding all site tensors to the left/right to be left-/right-normalized tensors (see \cref{fig:left_right_normalized}).
The unnormalized site is often called active site or orthogonality center.

\subsection{Normalizing an MPS}
\label{sec:mps_normalizing}
Given an unnormalized MPS with site tensors $M_j$, the first step is often to bring it into a left or right-canonical form.
This can be done by a series of QR decompositions: The tensor $M_j$ is
reshaped into a matrix $\tilde{M}$ with the left and physical tensor
legs forming the rows and the right tensor leg the columns of the
matrix, that is $\tilde{M}_{(\sigma_j m_{j-1}),m_j} = M_{j;m_{j-1}, m_j}^{\sigma_j}$. Applying a QR decomposition to this matrix, one obtains two
new matrices $Q_j$ and $R_{\underline{j}}$ (we label tensors living on
bonds $(j,j+1)$ with an underlined index ${}_{\underline{j}}$). The reshaping operation on $M_j$ is done in reverse on $Q$ to give the new left-normalized site tensor $A_j$ while the transfer tensor
$R_{\underline{j}}$ is multiplied into the next site tensor
$M_{j+1}$. Likewise reshaping the $M_j$ with right and physical legs
as rows and left tensor leg as columns results in a right-normalized
tensor $B_j$ (with the transfer tensor multiplied into the previous
site tensor $M_{j-1}$).

Starting on the left edge of the system and subsequently performing
left-normalizations on each MPS tensor results in a complete
left-normalized state. Equivalently, starting on the right edge of the
system and moving to the left results in a right-normalized state.

\subsection{Truncating an MPS}
\label{sec:truncation}
Operations on matrix-product states typically increase the bond
dimension of the state (e.g. MPO-MPS applications or the
addition of two MPS). Finding an optimal approximation to such a
quantum state for a smaller bond dimension is the purpose of this
section. This is of particular relevance to time-evolution methods, as
entanglement generically grows during real-time evolution and
time-evolved states hence per se already need a larger bond dimension
than e.g.~ground states. Hence finding good approximation methods is
crucial.

Let us consider a state $\ket{\psi}$ which is represented by an MPS
with an initial large bond dimension $m$.
We wish to find another state $\ket{\psi^\prime}$ with smaller bond dimension $m^\prime$ which approximates $\ket{\psi}$ well in the sense that it minimizes the Hilbert space distance
\begin{equation}
  \label{eq:truncation_minimize}
  \lVert \ket{\psi} - \ket{\psi^\prime} \rVert \;.
\end{equation}
The most direct way to proceed is to use a series of singular value decompositions to
successively truncate each bond of the MPS. On each individual bond,
the optimal choice is made, but this does not have to result in the
globally optimal state $\ket{\psi^\prime}$. It is also possible to
optimize each site tensor of $\ket{\psi^\prime}$ sequentially to
maximize the overlap between $\ket{\psi}$ and
$\ket{\psi^\prime}$. Multiple sweeps of this variational optimization
can be done to approach the global optimum as well as possible.
\subsubsection{Direct truncation via SVD}

\begin{figure}[!t]
	\begin{subfigure}{1.0\textwidth}
		\caption{\hfill \phantom.\label{fig:SVD_tensor}}
		\centering
		\tikzsetnextfilename{SVD_tensor}
%		\tikzset{external/export next=false}
		\begin{tikzpicture}
			\begin{scope}[node distance = 0.5 and 0.6]
				\node[ld]		(site0a)						{$m_{j-1}^\noprime$};
				\node[site]		(site1a)	[right=of site0a]	{};
				\node[ld]		(site2a)	[right=of site1a]	{$m_{j}^\noprime$};
				\node[ld]		(sigma1a)	[below=of site1a]	{$\sigma_{j}^\noprime$};
				\draw[]			(site0a)	--	(site1a);
				\draw[] 		(sigma1a)	--	(site1a);
				\draw[] 		(site1a)	--	(site2a);
				\node[ghost]	(betweensigma1andsite1a)	at ($(site1a)!0.5!(sigma1a)$)			{};
				\node[ghost]	(betweensigma2andsite2a)	at (betweensigma1andsite1a-|site2a)		{};
				\node[ghost]	(betweensigma0andsite0a)	at (betweensigma1andsite1a-|site0a)		{};
	
				\node[ghost]	(inter2ab)	[right=of site2a]	{};
				\node[ghost]	(inter2ab)					at (inter2ab|-betweensigma1andsite1a)	{};
				\node[ghost]	(inter1ab)	[above=of inter2ab]	{};
				\node[ghost]	(inter3ab)	[below=of inter2ab]	{};
				
				\node[ld]		(site0b)	[right=of inter1ab]	{$\sigma_{j}^\noprime, m_{j-1}^\noprime$};
				\node[site]		(site1b)	[right=of site0b]	{};
				\node[ld]		(site2b)	[right=of site1b]	{$m_{j}^\noprime$};
				\node[ghost]	(sigma1b)	[below=of site1b]	{};
				\draw[] 		($(site0b.east)+(0,\legdist)$)	--	($(site1b.west)+(0,\legdist)$);
				\draw[] 		($(site0b.east)+(0,-\legdist)$)	--	($(site1b.west)+(0,-\legdist)$);
				\draw[] 		(site1b) 	--	(site2b);
				\draw[->]		(betweensigma2andsite2a)		--	(site0b)	node [midway,above,sloped] (reshapec) {\small reshape};
	
				\node[ld]		(site0c)	[right=of inter3ab]	{$m_{j-1}^\noprime$};
				\node[site]		(site1c)	[right=of site0c]	{};
				\node[ld]		(site2c)	[right=of site1c]	{$\sigma_{j}^\noprime,m_{j}^\noprime$};	
				\draw[] 		($(site1c.east)+(0,\legdist)$)	--	($(site2c.west)+(0,\legdist)$);
				\draw[] 		($(site1c.east)+(0,-\legdist)$)	--	($(site2c.west)+(0,-\legdist)$);
				\draw[] 		(site0c)	--	(site1c);
				\draw[->]		(betweensigma2andsite2a)		--	(site0c)	node [midway,below,sloped] (reshapec) {\small reshape};
	
				\node[ghost]	(interbd)	[right=of site2c]	{};
				\node[ghost]	(interbd2)	[right=of interbd]	{};
				\node[ld]		(site0d)					at (interbd2|-site0b)					{$\sigma_{j}^\noprime, m_{j-1}^\noprime$};
				\node[siteA]	(site1d)	[right=of site0d]	{};
				\node[intersite](site2d)	[right=of site1d]	{};
				\node[siteB]	(site3d)	[right=of site2d]	{};
				\node[ld]		(site4d)	[right=of site3d]	{$m_{j}$};
				\draw[] 		($(site0d.east)+(0,\legdist)$)	--	($(site1d.west)+(0,\legdist)$);
				\draw[] 		($(site0d.east)+(0,-\legdist)$)	--	($(site1d.west)+(0,-\legdist)$);
				\draw[] 		(site1d)	--	(site2d)	node [ld,midway,above] {$m_j^\prime$};
				\draw[] 		(site2d)	--	(site3d)	node [ld,midway,above] {$s_j^\prime$};
				\draw[] 		(site3d)	--	(site4d);
				\draw[->]		(site2b)	--	(site0d)	node [midway,below,sloped] {\small SVD};
	
				\node[ghost]	(interce)	[right=of site2c]	{};
				\node[ghost]	(interce2)	[right=of interce]	{};
				\node[ld]		(site0e)					at (interce2|-site0c)					{$m_{j-1}^\noprime$};
				\node[siteA]	(site1e)	[right=of site0e]	{};
				\node[intersite](site2e)	[right=of site1e]	{};
				\node[siteB]	(site3e)	[right=of site2e]	{};
				\node[ld]		(site4e)	[right=of site3e]	{$\sigma_{j}^\noprime,m_{j}^\noprime$};
				\draw[] 		(site0e)	--	(site1e);
				\draw[]			(site1e)	--	(site2e)	node [ld,midway,above] {$s_j^\noprime$};
				\draw[]			(site2e)	--	(site3e)	node [ld,midway,above] {$m_{j-1}^\prime$};
				\draw[] 		($(site3e.east)+(0,\legdist)$)	--	($(site4e.west)+(0,\legdist)$);
				\draw[] 		($(site3e.east)+(0,-\legdist)$)	--	($(site4e.west)+(0,-\legdist)$);
				\draw[->]		(site2c)	--	(site0e)	node [midway,above,sloped] {\small SVD};
			\end{scope}
		\end{tikzpicture}
	\end{subfigure}
	\begin{subfigure}{1.0\textwidth}
		\caption{\hfill \phantom.\label{fig:DeSVD_tensor}}
		\centering
		\tikzsetnextfilename{DeSVD_tensor}
%		\tikzset{external/export next=false}
		\begin{tikzpicture}
			\begin{scope}[node distance = 0.5 and 0.6]
	
				\node[ld]		(site0d)								{$\sigma_{j},m_{j-1}$};
				\node[siteA]	(site1d)		[right=of site0d]		{};
				\node[intersite](site2d)		[right=of site1d]		{};
				\node[siteB]	(site3d)		[right=of site2d]		{};
				\node[siteB]	(site4d)		[right=of site3d]		{};
				\node[ld]		(sigma4d)		[below=of site4d]		{$\sigma_{j+1}^\noprime$};
				\node[ld]		(site5d)		[right=of site4d]		{$m_{j+1}^\noprime$};
				\node[ghost]	(sigma0d)		at (sigma4d-|site0d)	{};
				\draw[] 		($(site0d.east)+(0,\legdist)$)	--	($(site1d.west)+(0,\legdist)$);
				\draw[] 		($(site0d.east)+(0,-\legdist)$)	--	($(site1d.west)+(0,-\legdist)$);
				\draw[] (site1d) -- (site2d) node [ld,midway,above] {$m_j^\prime$};
				\draw[] (site2d) -- (site3d) node [ld,midway,above] {$s_j^\prime$};
				\draw[] (site3d) -- (site4d) node [ld,midway,above] {$m_{j}^\noprime$};
				\draw[] (site4d) -- (sigma4d);
				\draw[] (site4d) -- (site5d);
	
				\node[ghost]	(interdf)		[right=of site5d]		{};
				\node[ghost]	(interdf2)		[right=of interdf]		{};
				\node[ld]		(site0f)		[right=of interdf2]		{$m_{j-1}^\noprime$};
				\node[siteA]	(site1f)		[right=of site0f]		{};
				\node[ld]		(sigma1f)		at (sigma4d-|site1f)	{$\sigma_{j}^\noprime$};
				\node[site]		(site2f)		[right=of site1f]		{};
				\node[ld]		(sigma2f)		at (sigma4d-|site2f)	{$\sigma_{j+1}^\noprime$};
				\node[ld]		(site3f)		[right=of site2f]		{$m_{j+1}^\noprime$};
				\draw[] (site0f) -- (site1f);
				\draw[] (site1f) -- (site2f) node [ld,midway,above] {$m_j^\prime$};
				\draw[] (site1f) -- (sigma1f);
				\draw[] (site2f) -- (sigma2f);
				\draw[] (site2f) -- (site3f);
				\draw[->]	(site5d) -- (site0f) node [midway,above,sloped] {\small reshape} node [midway,below,sloped] {\small recombine};
	
				\node[ld]		(site0e)		[below=of sigma0d]	{$m_{j-2}^\noprime$};
				\node[siteA]	(site1e)		[right=of site0e]	{};
				\node[ld]		(sigma1e)		[below=of site1e]	{$\sigma_{j-1}^\noprime$};
				\node[siteA]	(site2e)		at (site1e-|site2d)	{};
				\node[intersite](site3e)		[right=of site2e]	{};
				\node[siteB]	(site4e)		[right=of site3e]	{};
				\node[ld]		(site5e)		[right=of site4e]	{$\sigma_{j}^\noprime,m_{j}^\noprime$};
				\draw[] (site0e) -- (site1e);
				\draw[] (site1e) -- (sigma1e);
				\draw[] (site1e) -- (site2e) node [ld,midway,above] {$m_{j-1}^\noprime$};
				\draw[] (site2e) -- (site3e) node [ld,midway,above] {$s_j^\noprime$};
				\draw[] (site3e) -- (site4e) node [ld,midway,above] {$m_{j-1}^\prime$};
				\draw[] ($(site4e.east)+(0,\legdist)$)	--	($(site5e.west)+(0,\legdist)$);
				\draw[] ($(site4e.east)+(0,-\legdist)$)	--	($(site5e.west)+(0,-\legdist)$);
	
				\node[ghost]	(intereg)		[right=of site5e]		{};
				\node[ghost]	(intereg2)		[right=of intereg]		{};
				\node[ld]		(site0g)		[right=of intereg2]		{$m_{j-2}^\noprime$};
				\node[site]		(site1g)		[right=of site0g]		{};
				\node[ld]		(sigma1g)		at (sigma1e-|site1g)	{$\sigma_{j-1}^\noprime$};
				\node[siteB]	(site2g)		[right=of site1g]		{};
				\node[ld]		(sigma2g)		at (sigma1e-|site2g)	{$\sigma_{j}^\noprime$};
				\node[ld]		(site3g)		[right=of site2g]		{$m_{j}^\noprime$};
				\draw[] (site0g) -- (site1g);
				\draw[] (site1g) -- (site2g) node [ld,midway,above] {$m_{j-1}^\prime$};
				\draw[] (site1g) -- (sigma1g);
				\draw[] (site2g) -- (sigma2g);
				\draw[] (site2g) -- (site3g);
				\draw[->]	(site5e) -- (site0g) node [midway,above,sloped] {\small recombine} node [midway,below,sloped] {\small reshape};
			\end{scope}
		\end{tikzpicture}
	\end{subfigure}
	\caption
	{
		(\subref{fig:SVD_tensor}) The singular value decomposition of a rank-three tensor $M$ into $U S V$ within a truncation sweep to the right (top) or to the left (bottom).
		(\subref{fig:DeSVD_tensor}) Assignment of the result of the singular value decomposition (\subref{fig:SVD_tensor}) into the new left (right) normalized rank three tensor $A_j$ ($B_j$) and the new active site $M_{j+1}$ ($M_{j-1}$) at the top (bottom).
              }
\end{figure}
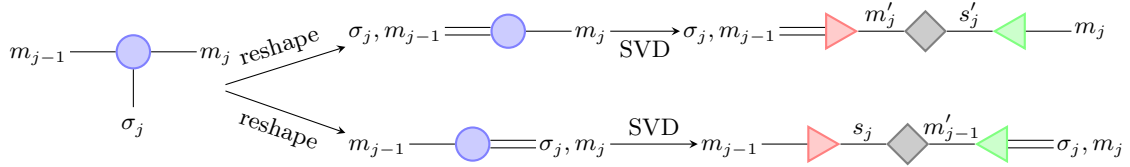
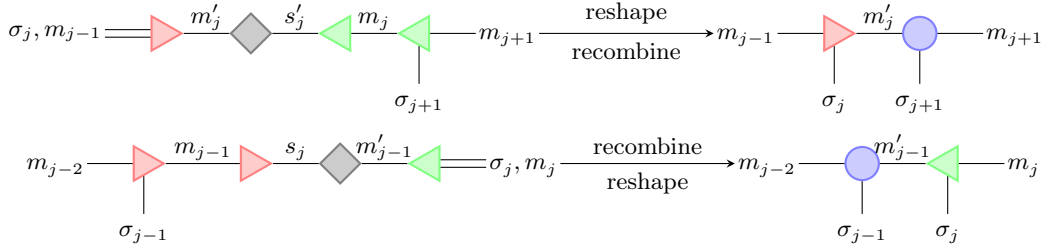

Consider a cut of the MPS on one bond $(j,j+1)$ into a left and right half
such that the state $\ket{\psi}$ can be represented in effective left
and right basis sets as
\begin{align}
	\label{eq:generaldecomposition}
	\ket{\psi} = \sum_{l,r=1}^{m_j} \Psi_{l,r} \ket{l}_L \otimes \ket{r}_R 
\;.
\end{align}
The coefficient tensor $\Psi_{l,r}$ then occupies the bond between sites
$j$ and $j+1$, such rank-2 tensors are called bond tensors in the following.
For orthonormal left and right basis sets as realized in an MPS with
orthogonality center on bond $(j,j+1)$, we can use a singular value
decomposition (SVD, cf.~Ref.~\cite{schollwoeck11} Sec.~4.5) of the
$\Psi$ tensor to obtain the approximation which is optimal under the
condition that all other site tensors are kept fixed.
That is, we decompose
\begin{equation}
  \Psi^\nodagger_{l,r} = \sum_{s=1}^m U^\nodagger_{l,s} S^\nodagger_{s,s} V^\dagger_{s,r}
\end{equation}
with $U$ and $V^\dagger$ left-/right-unitary and $S$ diagonal and non-negative. An
approximation is then given by
\begin{equation}
  \Psi^\prime_{l,r} = \sum_{s=1}^{m^\prime \leq m} U^\prime_{l,s} S^\prime_{s,s} V^{\prime\dagger}_{s,r}
\end{equation}
where we only keep the $m^\prime$ largest singular values and also
correspondingly truncate the rows and columns of $U$ and $V^\dagger$.
The error of this approximation of $\ket{\psi}$ is given by
\begin{equation}
  \epsilon \approx \sqrt{\sum_{k=m^\prime+1}^m S_{k,k}^2}
\end{equation}
where the argument to the square root is also known as the discarded weight. In practice, one selects $m^\prime$ such that some target discarded weight (e.g.~$10^{-10}$, corresponding to error $\approx 10^{-5} \cdot L$) is obtained.
Instead of working on the bond tensor $\Psi_{l,r}$ we can also apply the
decomposition directly to the MPS tensor $M_j$ by a suitable reshaping
(cf.~\cref{fig:SVD_tensor,fig:DeSVD_tensor}). Starting e.g.~on the
left end of a right-normalized MPS, we can sweep left-to-right and
sequentially truncate every bond, each time obtaining the locally
optimal state.

Due to this sweeping through the system, the truncation of site $L-1$
becomes dependent on the truncation of site $1$ but not vice versa.
In the case of small truncations the error resulting from this
asymmetry is small and can be ignored.
If truncation errors are large, however, this asymmetry leads to a
series of optimal approximations which together do not constitute a
globally optimal approximation, as each of the singular value
decompositions always only produces the optimal approximation on a
particular bond.
To increase accuracy in this case, a subsequent variational
optimization of the state may be necessary.

\subsubsection{Variational truncation}

\begin{figure}
	\centering
	\tikzsetnextfilename{iterative_truncation}
%	\tikzset{external/export next=false}
	\begin{tikzpicture}
		\begin{scope}[node distance = 0.5]
			\node[ghost]	(site1)						{$\langle \phi \rvert$};
			\node[ghost]	(site2)	[right = of site1]	{$\cdots$};
			\node[siteA]	(site3)	[right = of site2]	{};
			\node[nosite]	(site4)	[right = of site3]	{};
			\node[siteB]	(site5)	[right = of site4]	{};
			\node[ghost]	(site6)	[right = of site5]	{$\cdots$};

			\node[ghost]	(sigma1)	[above=of site1]	{$\lvert \phi \rangle$};
			\node[ghost]	(sigma2)	at (sigma1-| site2)	{$\cdots$};
			\node[siteA]	(sigma3)	at (sigma1-| site3)	{};
			\node[site]		(sigma4)	at (sigma1-| site4)	{};
			\node[siteB]	(sigma5)	at (sigma1-| site5)	{};
			\node[ghost]	(sigma6)	at (sigma1-| site6)	{$\cdots$};

			\draw[]	(site2) -- (site3);
			\draw[]	(site3) -- (site4);
			\draw[]	(site4) -- (site5);
			\draw[]	(site5) -- (site6);

			\draw[]	(site3) -- (sigma3);
			\draw[]	(site4) -- (sigma4);
			\draw[]	(site5) -- (sigma5);

			\draw[]	(sigma2) -- (sigma3);
			\draw[]	(sigma3) -- (sigma4);
			\draw[]	(sigma4) -- (sigma5);
			\draw[]	(sigma5) -- (sigma6);

			\node[draw,fill opacity=.2,draw opacity=.6,gray,thick,rounded corners,fill=gray!60,fit=(sigma2) (sigma3) (site2) (site3),label=below:{$\delta_{m_{j-1}, \overbar m_{j-1}}$}] {};

			\node[draw,fill opacity=.2,draw opacity=.6,gray,thick,rounded corners,fill=gray!60,fit=(sigma5) (sigma5) (site6) (site6),label=below:{$\delta_{m_{j}, \overbar m_{j}}$}] {};

			\node[ghost]	(site3b)		[right = of site6]				{};
			\node[ghost]	(site3a)		[right = of site3b]				{};
			\node[ghost]	(sigma3a)		at (sigma2-| site6)				{};
			\node[ghost]	(inter)			at ($(site3a)!0.5!(sigma3a)$)	{$=$}; 
			
			\node[ghost]	(site2)			at (site3a|-site2)				{$\cdots$};
			\node[siteA]	(site3)			[right = of site2]				{};
			\node[nosite]	(site4)			[right = of site3]				{};
			\node[siteB]	(site5)			[right = of site4]				{};
			\node[ghost]	(site6)			[right = of site5]				{$\cdots$};
			\node[ghost]	(site7)			[right = of site6]				{$\langle \phi \rvert$};

			\node[ghost]	(sigma2)		at (sigma2-| site2)				{$\cdots$};
			\node[site]		(sigma3)		at (sigma2-| site3)				{};
			\node[site]		(sigma4)		at (sigma2-| site4)				{};
			\node[site]		(sigma5)		at (sigma2-| site5)				{};
			\node[ghost]	(sigma6)		at (sigma2-| site6)				{$\cdots$};
			\node[ghost]	(sigma7)		at (sigma2-| site7)				{$\lvert \psi \rangle$};

			\draw[]	(site2) -- (site3);
			\draw[]	(site3) -- (site4);
			\draw[]	(site4) -- (site5);
			\draw[]	(site5) -- (site6);

			\draw[]	(site3) -- (sigma3);
			\draw[]	(site4) -- (sigma4);
			\draw[]	(site5) -- (sigma5);

			\draw[]	(sigma2) -- (sigma3);
			\draw[]	(sigma3) -- (sigma4);
			\draw[]	(sigma4) -- (sigma5);
			\draw[]	(sigma5) -- (sigma6);

			\node[draw,fill opacity=.2,draw opacity=.6,red,thick,rounded corners,fill=red!60,fit=(sigma2) (sigma3) (site2) (site3),label=below:{$\tilde L_{j-1; m_{j-1}}^{\overbar m_{j-1}}$}] {};

			\node[draw,fill opacity=.2,draw opacity=.6,green,thick,rounded corners,fill=green!60,fit=(sigma5) (sigma5) (site6) (site6),label=below:{$\tilde R_{j+1; m_{j}}^{\overbar m_{j}}$}] {};

		\end{scope}
	\end{tikzpicture}
	\caption
	{
		\label{fig:iterative_truncation} 
		Iterative variational truncation considering the truncated state to be in a mixed canonical form.
		The left hand side can then be reduced to the new optimal site tensor $M^\prime_j$.
		The right hand side, which needs to be considered completely, can nevertheless be calculate iteratively via the bond tensors $\tilde{L}$ and $\tilde{R}$.
	}
\end{figure}
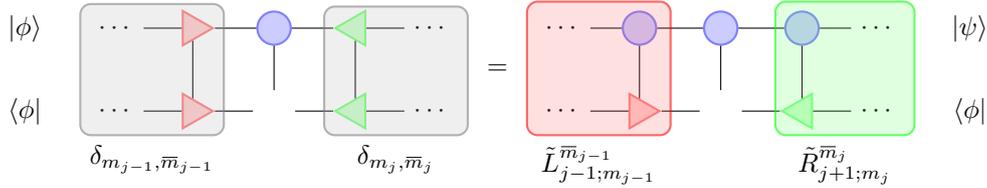

To overcome the problems of the
direct truncation by SVD, an iterative sweeping mechanism is often
employed. By sweeping multiple times through the system and finding on each
site the locally optimal tensor, it is more likely that one
obtains the globally optimal MPS. 
We start from an initial guess state $\lvert \phi \rangle$ with
tensors $M^\prime_j$ and a chosen bond dimension $m^\prime$ and
variationally minimize the distance to the un-truncated state
$\lvert \psi \rangle$ with tensors $M_j$.
We must stress that the convergence of the variational optimization
algorithm strongly depends on the inital guess state.
Being unlucky and starting from an unsuitable initial state, the
variational optimization may take a long time to converge or even may
get completely stuck in a locally optimal but globally suboptimal
state.
Typically, a good choice for the initial state is produced by the direct
truncation using the SVD.
Once we have an initial state, the distance to minimize is given by
\begin{align}
	\label{eq:mps_distance}
	\lVert \lvert \psi \rangle - \lvert \phi \rangle \rVert^2 = 
	\langle \psi \vert \psi \rangle - \langle \psi \vert \phi \rangle - \langle \phi \vert \psi \rangle + \langle \phi \vert \phi \rangle \;.
\end{align}
We now keep all tensors but $M_j^\prime$ fixed and hence only minimize
the single site tensor $M_j^\prime$, differentiating with respect to
$\overbar M_j^\prime$.
Because this tensor only occurs in the second half of \cref{eq:mps_distance}, the new (optimized) $M_j^\prime$ can be obtained via
\begin{align}
	\frac{\partial}{\partial \overbar{M_j}^{\prime}} \left( \langle \phi \vert \phi \rangle - \langle \phi \vert \psi \rangle \right) = 0 \;.
\end{align}
Let us now consider that the truncated state is in a mixed canonical form with the active site at position $j$.
Then the new tensor is given by
\begin{align}
	M_{j; m_{j-1}^\noprime, m_j^\noprime}^{\prime \sigma_j^\noprime} =
		\sum_{\overbar m_{j-1}^\noprime} 
			\tilde L_{j-1; m_{j-1}^\noprime}^{\overbar m_{j-1}^\noprime} 
			\left( 
				\sum_{\overbar m_{j}^\noprime} 
					\tilde R_{j+1; m_{j}^\noprime}^{\overbar m_{j}^\noprime}
					M_{j; m_{j-1}^\noprime, m_{j}^\noprime}^{\sigma_j^\noprime}
			\right) \;,
\end{align}
in which the left (right) parts of the tensor network (\cref{fig:iterative_truncation}, right hand side) are contracted into $\tilde L$ ($\tilde R$),
\begin{align}
	\label{eq:mps_trunc_bound_tensors}
	\tilde L_{j-1; m_{j-1}^\noprime}^{\overbar m_{j-1}} &= 
		\sum_{\substack{\sigma_{j-1}^\noprime,\\ \overbar m_{j-2}, \\ m_{j-2}}} 
			\overbar A_{j-1; \overbar m_{j-2}^\noprime, \overbar m_{j-1}^\noprime}^{\prime \sigma_{j-1}^\noprime}
			\left( 
				\cdots 
				\left( 
					\sum_{\substack{\sigma_1^\noprime,\\ \overbar m_0,\\ m_0}}
						\overbar A_{1; \overbar m_0, \overbar m_1}^{\prime \sigma_1^\noprime}
						M_{1; m_0, m_1}^{\sigma_1}
				\right) 
				\cdots 
			\right) 
			M_{j-1; m_{j-2}^\noprime, m_{j-1}^\noprime}^{\sigma_{j-1}^\noprime}
	\\
	\tilde R_{j+1;m_{j}}^{\overbar m_{j}} &= 
		\sum_{\substack{\sigma_{j+1},\\ \overbar m_{j+1},\\m_{j+1}}}
			\overbar B_{j+1;\overbar m_j^\noprime, \overbar m_{j+1}^\noprime}^{\prime \sigma_{j+1}^\noprime}
			\left( 
				\cdots 
				\left( 
					\sum_{\substack{\sigma_{L},\\ \overbar m_{L},\\m_{L}}}
						\overbar B_{L; \overbar m_{L-1}^\noprime, \overbar m_L^\noprime}^{\prime \sigma_L^\noprime}
						M_{L; m_{L-1}^\noprime,  m_L^\noprime}^{\sigma_L^\noprime}
				\right)
				\cdots 
			\right) 
			M_{j+1; m_j^\noprime, m_{j+1}^\noprime}^{\sigma_{j+1}^\noprime}
	\;.
\end{align}
It is never necessary to calculate the complete contraction of the
boundary tensors \cref{eq:mps_trunc_bound_tensors}, because the next
tensor in sweep direction was already calculated in the sweep before
and the other tensor is obtained by reusing the one from the previous
sweep step.

In practice, one should consider a two-site variational
optimization in which neighboring site tensors $M^\prime_j$,
$M^\prime_{j+1}$ are optimized at the same time. This allows for
flexibility in the bond dimension and distribution of quantum number
sectors on the bond $m_j$.  For convenience we will demonstrate the
necessity of a two-site update to permit for an increased bond
dimension.  Consider the reshaped updated tensor
\begin{equation}
  \Theta_{\sigma_{j+1} m_{j+1}}^{\sigma_j m_{j-1}} = M^{\prime \sigma_j}_{j; m_{j-1}, m_j} \cdot M^{\prime \sigma_{j+1}}_{j+1; m_j, m_{j+1}} \;.
\end{equation}
Any matrix factorization $\Theta = X\cdot Y$ may then create an index
with intermediate bond dimension which is bounded by
$\min\{m_{j-1} \sigma_j, \sigma_{j+1} m_{j+1}\}$ and potentially larger
than $m_j$, which is possible iff at least two MPS sites are
considered at the same time.

\subsection{\label{sec:mps:finitetemp}Finite temperatures}

MPS by default can only represent pure quantum states. To 
describe the mixed states encountered at finite temperatures, one therefore
requires an additional ingredient. Two possible choices exist within
the MPS framework: minimally entangled typical thermal
states\cite{stoudenmire10:_minim} and thermofield doublet
states\cite{fano57:_descr_states_quant_mechan_densit,FeigWhite05:Finite_temp_DMRG_u_enlarged_HS,verstraete04:_matrix_produc_densit_operat}. 
In the following, we will concentrate on thermofield doublet states or ``purifications'' of a mixed state; a detailed comparison between both methods can be found in~\cite{binder15:_minim}. 
The purification works by doubling the Hilbert space
$\mathcal{H} \to \mathcal{H}_p \otimes \mathcal{H}_a$ where $a$
denotes the auxiliary or ancilla degrees of freedom.
The density matrix we wish to represent is just the trace over the
auxiliary space of our purified quantum state
$\ket{\psi} \in \mathcal{H}_p \otimes \mathcal{H}_a$:
\begin{equation}
  \hat \rho = \Tr_{a} \ket{\psi} \bra{\psi} \quad.
\end{equation}
At infinite temperature ($\beta = 0$), $\hat \rho$ is the identity matrix.
Hence, an infinite-temperature quantum state in the grand-canonical ensemble
can be constructed simply as a product state of maximally
entangled states of physical and auxiliary sites.
To construct a \emph{canonical} infinite-temperature state, it is best
to start with a complete vacuum state. In the Hubbard
model, for instance, with creators on physical (auxiliary) sites denoted by
$\hat c^\dagger_{j,\sigma\vphantom{()}}$ ($\hat c^\dagger_{a(j),\sigma}$), one applies
the operator
\begin{align}
	\hat C^\dagger_{\mathrm{tot}} = \sum_{j=1}^{L} \hat c^\dagger_{j,\uparrow\vphantom{()}} \hat c^\dagger_{a(j),\uparrow} + \hat c^\dagger_{j,\downarrow\vphantom{()}} \hat c^\dagger_{a(j),\downarrow}
\end{align}
repeatedly to create equal superpositions of particles until the
desired filling is reached\cite{barthel16:_matrix}. To obtain a
finite-temperature state at $\beta$, we time-evolve the infinite-temperature
initial state along the imaginary axis over a range $\beta/2$. The density matrix is then
\begin{equation}
  \hat \rho \propto \Tr_{a}\{ e^{-(\beta/2) \hat H} \ket{\psi} \bra{\psi} e^{-(\beta/2) \hat H} \}  \propto e^{-\beta \hat H}\quad
\end{equation}
up to normalization which is taken care of by keeping the purified
state $\ket{\psi}$ normalized at all times. For a more detailed
discussion, cf.~Ref.~\cite{schollwoeck11} Sec.~7.2.

The dynamics of this density operator is described by the von-Neumann equation. In the case of the purification, this is identical to the Liouville equation which has two copies of the Hamiltonian acting on both the physical and the auxiliary space\cite{barnett87:_liouv}, but with a negative sign in the auxiliary degrees of freedom.
Formally, this corresponds to a time evolution, which is forward in time on the physical degrees of freedom, but backward in time on the auxiliary space. 
Despite this formal resemblance, as one traces out the auxiliary degrees of freedom, a gauge freedom exists, which allows to apply \emph{any} unitary operator $\hat V_a$ acting only on the auxiliary space to the state $\ket{\psi}$.
It is clear that this leaves the density matrix $\hat \rho = \Tr_{a} \ket{\psi} \bra{\psi} = \Tr^\nodagger_{a} \hat V_a^\nodagger \ket{\psi} \bra{\psi} \hat V^\dagger_a$ invariant.
The simplest choice then is to obtain the real-time evolution by time-evolving the purified state with the Hamiltonian on the physical space, and leave the auxiliary degrees untouched.

Even more, the question arises, if there are unitaries acting on the auxiliary space, which may reduce the entanglement of the purified state $\ket{\psi}$  - finding such an optimized scheme would be extremely important in practice.
Multiple choices for such unitary operators have been worked out, whose effect on computational efficiency depends on the system at hand.
For example, during a real-time evolution of the physical system under the Hamiltonian $\hat H$, an easy and often helpful\cite{karrasch12:_finit_temper_dynam_densit_matrix, karrasch13:_reduc} choice is to evolve the auxiliary system under the Hamiltonian $-\hat H$, which formally resembles to the solution of the Liouville equation for the purified density operator as mentioned above.  
Finding further schemes is a topic of present research, and for details and more elaborate approaches we refer to the literature, e.g.~Refs.~\cite{barthel13:_precis, hauschild18:_findin}.
Furthermore, also cf.~\cref{sec:tricks:pip} later for a detailled discussion on how specifically one should purify a density matrix when aiming to evaluate a series of time-dependent observables.

\subsection{Application of an MPO to an
  MPS} \label{sec:mpo-mps-application} One of the most important
operations within the framework of matrix-product states is the
application of an MPO to an MPS.
In theory this can be done by a straightforward tensor
multiplication of the corresponding site tensors of the MPS and MPO.
In practice this is not the method of choice as in most applications
the resulting target state $\hat{O}\ket{\psi}$ has a much higher bond
dimension $m^{\prime}=m \cdot w$, which, however, is mostly not
required to represent the target state efficiently.
It is therefore helpful to look at different approaches --
nevertheless, for pedagogical reasons we will begin with the direct
application before we turn to more elaborate application schemes.

\subsubsection{Direct application} \label{sec:mpo-mps-application:direct}
The direct application of an MPO to an MPS is obtained by regrouping the contractions such that the target MPS tensor can be obtained as a tensor product of the individual site tensors
\begin{align}
	\hat O \lvert \psi \rangle 
	=& \sum_{\substack{\sigma_1^\noprime,\ldots,\sigma_L^\noprime,\\ \sigma_1^{\prime},\ldots,\sigma_L^{\prime}, \\ \sigma_1^{\prime\prime},\ldots,\sigma_L^{\prime\prime}}}
	\sum_{\substack{m_0^\noprime, \ldots, m_L^\noprime, \\ w_0^\noprime, \ldots, w_L^\noprime}}
	\left(
		W_{1; w_0^\noprime, w_1^\noprime}^{\sigma_1^\noprime\sigma_1^{\prime}} 
		\cdots 
		W_{L; w_{L-1}^\noprime, w_L^\noprime}^{\sigma_L^\noprime\sigma_L^{\prime}}
	\right)
	\lvert \sigma_1^\noprime\ldots\sigma_L^\noprime \rangle \langle \sigma_1^{\prime}\ldots\sigma_L^{\prime} \lvert \nonumber\\
	\phantom{=}& \phantom{\sum_{\substack{\sigma_1^\noprime,\ldots,\sigma_L^\noprime,\\ \sigma_1^{\prime},\ldots,\sigma_L^{\prime}, \\ \sigma_1^{\prime\prime},\ldots,\sigma_L^{\prime\prime}}} 	\sum_{\substack{m_0^\noprime, \ldots, m_L^\noprime, \\ w_0^\noprime, \ldots, w_L^\noprime}} }
	\quad\times
	\left(
		M_{1; m_0^\noprime, m_1^\noprime}^{\sigma_1^{\prime\prime}}
		\cdots
		M_{L; m_{L-1}^\noprime, m_L^\noprime}^{\sigma_L^{\prime\prime}}
	\right)
	\lvert \sigma_1^{\prime\prime}\ldots\sigma_L^{\prime\prime} \rangle\\
	=&  \sum_{\substack{\sigma_1^\noprime\ldots\sigma_L^\noprime, \\ m_0^\noprime, \ldots, m_L^\noprime, \\ w_0^\noprime, \ldots, w_L^\noprime}}
	M_{1; (w_0^\noprime m_0^\noprime), (w_1^\noprime m_1^\noprime)}^{\prime \sigma_1^\noprime}\cdots M_{L; (w_{L-1}^\noprime m_{L-1}^\noprime), (w_L^\noprime m_L^\noprime)}^{\prime \sigma_L^\noprime} \lvert\sigma_1^\noprime\ldots\sigma_L^\noprime\rangle = \lvert\phi\rangle \,,
\end{align}
with the tensors $M^{\prime}_j$ given by
\begin{align}
	\label{eq:mpompsdirectresult}
	M_{j; (w_{j-1}^\noprime m_{j-1}^\noprime), (w_j^\noprime m_j^\noprime)}^{\prime \sigma_j^\noprime} =
	\sum_{\sigma_j^{\prime}}
		W_{j; w_{j-1},w_j}^{\sigma_j^\noprime\sigma_j^{\prime}}
		M_{j; m_{j-1}, m_j}^{\sigma_j^{\prime}} \;.
\end{align}

The resulting state $\lvert\phi\rangle$ is therefore again an MPS, but with a larger dimension $m^\prime = m \cdot w$ with $w$ being the matrix dimension of the MPO site tensors $W_j$.
Repeated application of an operator onto a state in this manner hence quickly increases the dimension of the state and truncation becomes necessary with computational costs scaling as $m^3 w^3 d$ per site if we do an SVD truncation or $m^{\prime 2} m w d$ per site for the variational compression.

\subsubsection{Variational application} \label{sec:mpo-mps-application:variational}

\begin{figure}
	\centering
	\tikzsetnextfilename{variational_mpo_mps}
%	\tikzset{external/export next=false}
	\begin{tikzpicture}
		\begin{scope}[node distance = 0.5]
			\node[ghost]	(site1)						{$\langle \phi \rvert$};
			\node[ghost]	(site2)	[right = of site1]	{$\cdots$};
			\node[siteA]	(site3)	[right = of site2]	{};
			\node[nosite]	(site4)	[right = of site3]	{};
			\node[siteB]	(site5)	[right = of site4]	{};
			\node[ghost]	(site6)	[right = of site5]	{$\cdots$};
			
			\node[ghost]	(op1)	[above = of site3]	{};
			\node[nosite]	(site7)	at (op1-|site4)		{};
			
			\node[siteA]	(sigma3)	[above = of op1]	{};
			\node[ghost]	(sigma2)	at (site2|-sigma3)	{$\cdots$};
			\node[site]	(sigma4)	at (sigma2-| site4)	{};
			\node[siteB]	(sigma5)	at (sigma2-| site5)	{};
			\node[ghost]	(sigma6)	at (sigma2-| site6)	{$\cdots$};
			\node[ghost]	(sigma1)	at (sigma2-| site1)	{$\lvert \phi \rangle$};

			\draw[]	(site2) -- (site3);
			\draw[]	(site3) -- (site4);
			\draw[]	(site4) -- (site5);
			\draw[]	(site5) -- (site6);

			\draw[]	(site3) -- (sigma3)	node [midway,left]	{};
			\draw[]	(site7) -- (sigma4);
			\draw[]	(site5) -- (sigma5)	node [midway,right]	{};

			\draw[]	(sigma2) -- (sigma3);
			\draw[]	(sigma3) -- (sigma4);
			\draw[]	(sigma4) -- (sigma5);
			\draw[]	(sigma5) -- (sigma6);

			\node[draw,fill opacity=.2,draw opacity=.6,gray,thick,rounded corners,fill=gray!60,fit=(sigma2) (sigma3) (site2) (site3),label=below:{$\delta_{m_{j-1}^\noprime, \overbar m_{j-1}^\prime}$}] {};

			\node[draw,fill opacity=.2,draw opacity=.6,gray,thick,rounded corners,fill=gray!60,fit=(sigma5) (sigma5) (site6) (site6),label=below:{$\delta_{m_{j}^\noprime, \overbar m_{j}^\prime}$}] {};

			\node[ghost]	(site3b)		[right = of site6]				{};
			\node[ghost]	(site3a)		[right = of site3b]				{};
			\node[ghost]	(sigma3a)		at (sigma2-| site6)				{};
			\node[ghost]	(inter)			at ($(site3a)!0.5!(sigma3a)$)	{$=$}; 
			
			\node[ghost]	(site2)			at (site3a|-site2)				{$\cdots$};
			\node[siteA]	(site3)			[right = of site2]				{};
			\node[nosite]	(site4)			[right = of site3]				{};
			\node[siteB]	(site5)			[right = of site4]				{};
			\node[ghost]	(site6)			[right = of site5]				{$\cdots$};
			\node[ghost]	(site7)			[right = of site6]				{$\langle \phi \lvert$};

			\node[ghost]	(op2)			at (op1-|site2)					{$\cdots$};
			\node[op]	(op3)			at (op1-|site3)					{};
			\node[op]	(op4)			at (op1-|site4)					{};
			\node[op]	(op5)			at (op1-|site5)					{};
			\node[ghost]	(op6)			at (op1-|site6)					{$\cdots$};
			\node[ghost]	(op7)		at (op1-|site7)					{$\hat O$};
			
			\node[ghost]	(sigma2)		at (sigma2-| site2)				{$\cdots$};
			\node[site]	(sigma3)		at (sigma2-| site3)				{};
			\node[site]	(sigma4)		at (sigma2-| site4)				{};
			\node[site]	(sigma5)		at (sigma2-| site5)				{};
			\node[ghost]	(sigma6)		at (sigma2-| site6)				{$\cdots$};
			\node[ghost]	(sigma7)	at (sigma2-| site7)				{$\lvert \psi \rangle$};

			\draw[]	(site2) -- (site3);
			\draw[]	(site3) -- (site4);
			\draw[]	(site4) -- (site5);
			\draw[]	(site5) -- (site6);
			
			\draw[] (op2) -- (op3);
			\draw[] (op3) -- (op4);
			\draw[] (op4) -- (op5);
			\draw[] (op5) -- (op6);

			\draw[]	(site3) -- (op3)	node [midway,left]	{};
			\draw[]	(op3) -- (sigma3)	node [midway,left]	{};
			\draw[]	(site4) -- (op4);
			\draw[] (op4) -- (sigma4);
			\draw[]	(site5) -- (op5)	node [midway,right]	{};
			\draw[]	(op5) -- (sigma5)	node [midway,right]	{};

			\draw[]	(sigma2) -- (sigma3);
			\draw[]	(sigma3) -- (sigma4);
			\draw[]	(sigma4) -- (sigma5);
			\draw[]	(sigma5) -- (sigma6);

			\node[draw,fill opacity=.2,draw opacity=.6,red,thick,rounded corners,fill=red!60,fit=(sigma2) (sigma3) (site2) (site3),label=below:{$L_{j-1; m_{j-1}^\noprime}^{\overbar m_{j-1}^\prime, w^{\noprime}_{j-1}}$}] {};

			\node[draw,fill opacity=.2,draw opacity=.6,green,thick,rounded corners,fill=green!60,fit=(sigma5) (sigma5) (site6) (site6),label=below:{$R_{j+1; m_{j}^\noprime}^{\overbar m_{j}^\prime, w^{\noprime}_{j}}$}] {};

		\end{scope}
	\end{tikzpicture}
	\caption
	{
		\label{fig:variational_mpo_mps} 
		The variational application of an MPO to an MPS considering the truncated guess state $\lvert \phi \rangle$ to be in a mixed canonical form.
		The left hand side can then be reduced to the active side $j$ that we want to optimize.
		The right hand side, which needs to be evaluated completely, can nevertheless be calculated iteratively via the bond tensors $L_{j-1}$ and $R_{j+1}$.
	}
\end{figure}
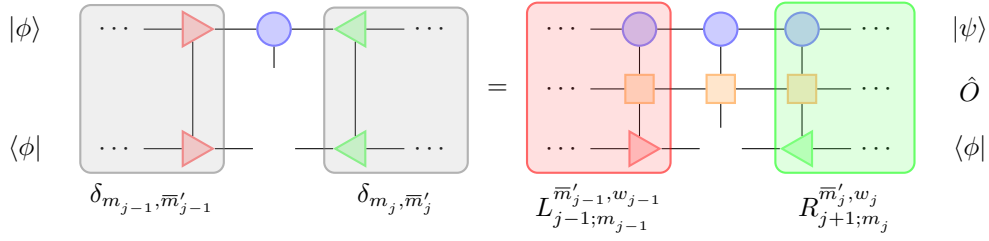
In the spirit of variationally compressing a state towards a target
state we can try to apply the same considerations to the application
of an MPO to a source state $\ket{\psi}$ with the subsequent
compression of the state in one optimization step.
Therefore we seek to minimize the distance between a guess state
$\lvert \phi \rangle$ with tensors $M^\prime_j$ and bond dimension
$m^\prime$ and the source state with the MPO applied to it, which we
denote as $\lvert \hat{O}\psi \rangle$,
\begin{gather}
	\lvert \phi \rangle = \min_{\lvert \phi \rangle}\left\Vert \ket{\phi} - \ket{\hat{O}\psi} \right\Vert^{2} \label{eq:mps_mpo_distance} =  \min_{\ket{\phi}} \left( \braket{\phi|\phi} - \braket{\phi|\hat{O}\psi} - \braket{\hat{O}\psi|\phi} + \braket{\hat{O}\psi|\hat{O}\psi} \right)\\
	\Rightarrow \frac{\partial}{\partial \overbar M_{j}^{\prime}} \left( \braket{\phi|\phi} - \braket{\phi|\hat{O}\psi} \right) \stackrel{!}{=} 0
\end{gather}
for all guess site tensors $\overbar M_{j}^{\prime}$.
If we keep the current guess state in a mixed canonical form, the above set of coupled equations again reduces to a local update scheme for the target tensors
\begin{align}
	M_{j; \overbar{m}_{j-1}^\prime, \overbar{m}_{j}^\prime}^{\prime \sigma_{j}^\noprime}
	&= \sum_{m^{\noprime}_{j-1},w^{\noprime}_{j-1}}
		L_{j-1; m^{\noprime}_{j-1}}^{\overbar m^{\prime}_{j-1}, w^{\noprime}_{j-1}}
		\left(
			\sum_{\sigma^{\prime}_{j}, m^{\noprime}_{j}, w^{\noprime}_{j}}
				W_{j; w^{\noprime}_{j-1}, w^{\noprime}_{j}}^{\sigma^{\noprime}_{j}\sigma^{\prime}_{j}}
				M_{j; m^{\noprime}_{j-1},m^{\noprime}_{j}}^{\sigma^{\prime}_{j}} 
				R_{j+1; m^{\noprime}_{j}}^{\overbar m^{\prime}_{j}, w^{\noprime}_{j}}
		\right) \;,
\end{align}
where the boundary tensors $L_{j-1},R_{j+1}$ can be built recursively by sweeping through the system and evaluating the contractions
\begin{align}
	L_{j-1; m^{\noprime}_{j-1}}^{ \overbar m^{\prime}_{j-1}, w^{\noprime}_{j-1}} &= 
		\sum_{\substack{\sigma^{\noprime}_{j-1}, \sigma^{\prime}_{j-1}, \\ \overbar{m}^\prime_{j-2}, m_{j-2}^\noprime, w_{j-2}^\noprime}}
			\overbar A_{j-1; \overbar m_{j-2}^\prime, \overbar m_{j-1}^\prime}^{\prime \sigma^{\noprime}_{j-1}} 
			L_{j-2; m^{\noprime}_{j-2}}^{\overbar m^{\prime}_{j-2}, w^{\noprime}_{j-2}}
			W_{j-1; w_{j-2}^\noprime, w_{j-1}^\noprime}^{\sigma^{\noprime}_{j-1}\sigma^{\prime}_{j-1}}
			M_{j-1; m_{j-2}^\noprime, m_{j-1}^\noprime}^{\sigma^{\prime}_{j-1}} \label{eq:left-mps-mpo-mps}
		\\
	R_{j+1; m^{\noprime}_{j}}^{\overbar m^{\prime}_{j}, w^{\noprime}_{j}} &=
		\sum_{\substack{\sigma^{\noprime}_{j+1}, \sigma^{\prime}_{j+1}, \\ w_{j+1}^\noprime, \overbar{m}_{j+1}^\prime,m_{j+1}^\noprime}}
			W_{j+1; w_{j}^\noprime, w_{j+1}^\noprime}^{\sigma^{\noprime}_{j+1}\sigma^{\prime}_{j+1}} 
			M_{j+1; m_j^\noprime, m_{j+1}^\noprime}^{\sigma^{\prime}_{j+1}} 
			R_{j+2; m_{j+1}^\noprime}^{\overbar m_{j+1}^\prime, w_{j+1}^\noprime} 
			\overbar B_{j+1; \overbar m_{j}^\prime, \overbar m_{j+1}^\prime}^{\prime \sigma^{\noprime}_{j+1}} \;. \label{eq:right-mps-mpo-mps}
\end{align}
The overall onsite contractions are depicted in
\cref{fig:variational_mpo_mps}. Care must be taken to use the
(typically) optimal contraction order
$\left(\left( L_{j-1} \cdot A_j\right) \cdot W_j\right) \cdot
\bar{A}_j$.

\subsubsection{The zip-up method} \label{sec:mpo-mps-application:zip-up}
\begin{figure}
	\centering
	\tikzsetnextfilename{mpo_zipup}
%	\tikzset{external/export next=false}
	\begin{tikzpicture}
		\begin{scope}
		[
			node distance = 0.75 and 1.05
		]
			\node[ghost]	(pos0)						{};
			\node[ghost]	(pos1)	[right = of pos0]	{};
			\node[ghost]	(pos2)	[right = of pos1]	{};
			\node[ghost]	(pos3)	[right = of pos2]	{};
			\node[ghost]	(pos4)	[right = of pos3]	{};
			\node[ghost]	(op0)	[above = of pos0]	{};
			\node[op]		(op1)	at (op0 -| pos1)	{};
			\node[op]		(op2)	at (op0 -| pos2)	{};
			\node[op]		(op3)	at (op0 -| pos3)	{};
			\node[ghost]		(op4)	at (op0 -| pos4)	{};
			\node[ghost]	(site0)	[above = of op0]	{};
			\node[siteB]		(site1)	at (site0 -| pos1)	{};
			\node[siteB]		(site2)	at (site0 -| pos2)	{};
			\node[siteB]		(site3)	at (site0 -| pos3)	{};
			\node[ghost]		(site4)	at (site0 -| pos4)	{};
			\draw[dotted]	(site0) -- node[ld, above] {$m_0^\noprime$} (site1) ;
			\draw[]	(site1) -- node[ld, above] {$m_1^\noprime$} (site2);
			\draw[]	(site2) -- node[ld, above] {$m_2^\noprime$} (site3);
			\draw[]	(site3) -- node[ld, above] {$m_3^\noprime$} (site4);
			\draw[]	(site1) -- node[ld, right] {$\sigma_1^\prime$} (op1);
			\draw[]	(site2) -- node[ld, right] {$\sigma_2^\prime$} (op2);
			\draw[]	(site3) -- node[ld, right] {$\sigma_3^\prime$} (op3);
			\draw[dotted]	(op0) -- node[ld, above] {$w_0^\noprime$} (op1);
			\draw[]	(op1) -- node[ld, above] {$w_1^\noprime$} (op2);
			\draw[]	(op2) -- node[ld, above] {$w_2^\noprime$} (op3);
			\draw[]	(op3) -- node[ld, above] {$w_3^\noprime$} (op4);
			\draw[]	(op1) -- node[ld, right] {$\sigma_1^\noprime$} (pos1);
			\draw[]	(op2) -- node[ld, right] {$\sigma_2^\noprime$} (pos2);
			\draw[]	(op3) -- node[ld, right] {$\sigma_3^\noprime$} (pos3);
			\node[draw,fill opacity=.2,inner sep=4pt,draw opacity=.6,blue,thick,rounded corners,fill=blue!60,fit=(site1) (op1) (site1) (op1),label={[shift={(0.0,0.0)}]{$M_{1; w_0^\noprime m_0^\noprime, w_1^\noprime m_1^\noprime}^{\prime \sigma_1^\noprime}$}}] {};
			\node[ghost]	(pos0b)	[right = of pos4]	{};
			\node[ghost]	(pos1b)	[right = of pos0b]		{};
			\node[ghost]	(pos1b1)	[right = of pos1b]	{};
			\node[ghost]	(pos1b2)	[right = of pos1b1]	{};
			\node[ghost]	(pos2b)	[right = of pos1b2]		{};
			\node[ghost]	(pos3b)	[right = of pos2b]		{};
			\node[ghost]	(pos4b)	[right = of pos3b]		{};
			\node[ghost]	(op0b)	[above = of pos0b]	{};
			\node[nosite]	(op1b)	at (op0b -| pos1b)	{};
			\node[nosite]	(op1b2)	at (op0b -| pos1b2)	{};
			\node[op]		(op2b)	at (op0b -| pos2b)	{};
			\node[op]		(op3b)	at (op0b -| pos3b)	{};
			\node[ghost]		(op4b)	at (op0b -| pos4b)	{};
			\node[ghost]	(site0b)	[above = of op0b]	{};
			\node[siteA]	(site1b)	at (site0b -| pos1b)	{};
			\node[intersite]	(site1b1)	at (site0b -| pos1b1)	{};
			\node[siteB]	(site1b2)	at (site0b -| pos1b2)	{};
			\node[siteB]		(site2b)	at (site0b -| pos2b)	{};
			\node[siteB]		(site3b)	at (site0b -| pos3b)	{};
			\node[ghost]		(site4b)	at (site0b -| pos4b)	{};
			\draw[dotted]	(site0b) -- node[ld, above,xshift=-0.5em] {$w_0^\noprime m_0^\noprime$} (site1b) ;
			\draw[]	(site1b) -- node[ld, above] {$s_1^\noprime$} (site1b1);
			\draw[]	(site1b1) -- node[ld, above] (s1pb) {$s_1^\prime$} (site1b2);
			\draw[]	($(site1b2.east)+(0,0.1)$) to[-|-=0.75] node[ld, above,yshift=0.1em] {$m_1^\noprime$} ($(site2b.west)$);
			\draw[]	(site2b) -- node[ld, above] {$m_2^\noprime$} (site3b);
			\draw[]	(site3b) -- node[ld, above] {$m_3^\noprime$} (site4b);
			\draw[]	(site1b) -- node[ld, right] {$\sigma_1^\noprime$} (pos1b);
			\draw[]	(site2b) -- node[ld, right] {$\sigma_2^\prime$} (op2b);
			\draw[]	(site3b) -- node[ld, right] {$\sigma_3^\prime$} (op3b);
			\draw[]	($(site1b2.east)+(0,-0.1)$) to[-|-=0.4] node[ld, above,xshift=0.6em] {$w_1^\noprime$} (op2b);
			\draw[]	(op2b) -- node[ld, above] {$w_2^\noprime$} (op3b);
			\draw[]	(op3b) -- node[ld, above] {$w_3^\noprime$} (op4b);
			\draw[]	(op2b) -- node[ld, right] {$\sigma_2^\noprime$} (pos2b);
			\draw[]	(op3b) -- node[ld, right] {$\sigma_3^\noprime$} (pos3b);
			\node
			[
				draw,
				fill opacity=.2,
				inner sep=7.5pt,
				draw opacity=.6,
				yellow!80!red,
				thick,
				rounded corners,
				fill=yellow!60!red,
				fit=(site1b) (s1pb) (site1b2) (site1b1),
				label=
					{[shift={(0.0,0.0)}]
						{
							$\text{SVD}(M_{1; w_0^\noprime m_0^\noprime, w_1^\noprime m_1^\noprime}^{\prime \sigma_1^\noprime})$
						}
					}
			] {};
			\node
			[
				draw,
				fill opacity=.2,
				inner sep=4.5pt,
				draw opacity=.6,
				blue,
				thick,
				rounded corners,
				fill=blue!60,
				fit=(site1b1) (s1pb) (op1b2) (op2b),
				label=
					{[shift={(-0.0,-2.9)}]
						{
							$M_{2; s_1^\noprime, w_2^\noprime m_2^\noprime}^{\prime \sigma_2^\noprime}$
						}
					}
			] {};
			\node[] at ($(op4)!0.5!(site0b)$) {$\rightarrow$};
		\end{scope}	
	\end{tikzpicture}
	\caption
	{
		\label{fig:mpo_mps_zipup}
		The essential steps of the zip-up method proposed in Ref.~\cite{stoudenmire10:_minim}.
		(left) Initial tensor network that shall be contracted, consisting of a right-canonical MPS and an MPO that only slightly destroys the canonical form.
		The first step is to interpret the combination of the first MPS tensor and the first MPO tensor as a new tensor $M_{1; w_0^{\protect\noprime} m_0^{\protect\noprime}, w_1^{\protect\noprime} m_1^{\protect\noprime}}^{\prime \sigma_1^{\protect\noprime}}$, which is slightly non-normalized and therefore framed in blue.
		(right) The next step is to apply an SVD (including a relaxed truncation) on the tensor $M_{1}^{\prime}$ and finally build the next tensor $M_{2}^{\prime}$ leaving the contracted, left-normalized MPS tensor $A_{1; m_0^{\protect\noprime} w_0^{\protect\noprime}, s_1^{\protect\noprime}}^{\prime\sigma_1^{\protect\noprime}}$ on the left side.
	}
\end{figure}
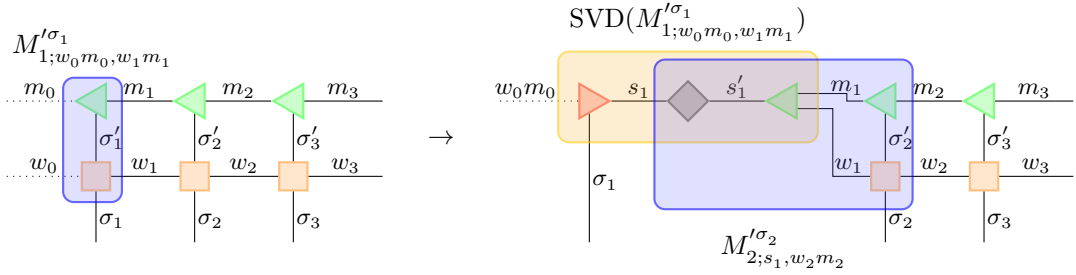

An alternative to the direct application of an MPO to an MPS
is the zip-up method described in Ref.~\cite{stoudenmire10:_minim}.
The central assumption is that the operator, such as a time-evolution operator, only slightly destroys the canonical form of the MPS.
Hence, a modest truncation is already possible during the contraction
process without too much loss of information because there is only a
small loss of orthogonality in the used left and right basis sets.

The first step is similar to \cref{eq:mpompsdirectresult}, but we work with a right-normalized initial tensor $B$,
\begin{align}
	M_{1; w_0^{\noprime} m_0^{\noprime}, w_1^{\noprime} m_1^{\noprime}}^{\prime \sigma_1^{\noprime}} = \sum_{\sigma_1^\prime}  W_{1; w_0^{\noprime}, w_1^{\noprime}}^{\sigma_1^\noprime \sigma_1^{\prime}} B_{1; m_0^{\noprime}, m_1^{\noprime}}^{\sigma_1^\prime} \;.
\end{align}
Note that $m_0^\noprime$ and $w_0^\noprime$ are dummy indices and can
therefore easily be fused into a new dummy index.
Applying the SVD with a relaxed truncation criterion we obtain the
left-normalized tensor $A_{1}$, with a single right index
$m^{\prime}_{1}$
\begin{align}
	\text{SVD}(M_{1; w_0^\noprime m_0^\noprime, w_1^\noprime m_1^\noprime}^{\prime \sigma_1^\noprime}) \approx
		\sum_{s_1^\noprime, s_1^\prime} 
			A_{1; m_0^\noprime w_0^\noprime, s_1^\noprime}^{\prime \sigma_1^\noprime} 
			S_{\underline 1; s_1^\noprime, s_1^\prime}^\prime 
			V_{\underline 1; s_1^\prime, w_1^\noprime m_1^\noprime}^\prime \;.
\end{align}
In the next step, the remaining parts of the result of the SVD are
incorporated as before, now also including the MPO tensor, to
obtain the next slightly unnormalized tensor
\begin{align}
	M_{2; s_1^\noprime, w_2^\noprime m_2^\noprime}^{\prime \sigma_2^\noprime} = 
	\sum_{\sigma_2^\prime, s_1^\prime, m_1^\noprime, w_1^\noprime} 
		S_{\underline 1; s_1^\noprime, s_1^\prime}^\prime 
		V_{\underline 1; s_1^\prime, w_1^\noprime m_1^\noprime}^\prime
		W_{2; w_1^\noprime, w_2^\noprime}^{\sigma_2^\noprime \sigma_2^\prime}
		B_{2; m_1^\noprime, m_2^\noprime}^{\sigma_2^\prime} \;.
\end{align}
This procedure is depicted in \cref{fig:mpo_mps_zipup}. It is repeated
until the right end of the system is reached and hence the complete
operator is applied.
For the relaxed truncation scheme a maximal growth factor of the MPS
bond dimension $m^{\prime}=2m$ and a truncated weight of
$\nicefrac{1}{10}$ of the target weight turns out to be a suitable
choice.
The contraction then has complexity
$2 m^3 w \sigma + 2 m^2 w^2 \sigma^2$. The main cost is in the SVD of
a $2m\sigma \times mw$ matrix at cost $\mathcal{O}(m^3 \sigma w)$,
i.e.~linear in $w$.
The MPS is in left canonical form now. 
A subsequent compression as described in \cref{sec:truncation} should be applied to obtain the resulting MPS with the target bond dimension.

\paragraph{Our experience} The zip-up method
(Sec.~\ref{sec:mpo-mps-application:zip-up}) is typically sufficiently
accurate and fast. In some cases, it is worthwhile to follow up on it with
some additional sweeps of the variational optimization
(Sec.~\ref{sec:mpo-mps-application:variational}) to increase
accuracy.
In particular if the bond dimension of the state is already large
and the operator close to the identity (as would be the case for a
small time step), variational optimization will converge quickly and
accurately.
It hence may be useful to also consider the MPS bond dimension when dynamically selecting the best application method.
Conversely, when using the zip-up method, repeated truncation sweeps according to a defined singular value threshold should be used to obtain the most efficient representation of the state.
Further interesting possibilities may be the truncation based on the
left/right density matrix\cite{stoudenmire18:_densit} or by optimizing
the local density matrix\cite{white18:_quant} which aims at preserving
local observables.

\begin{table}
  \centering
  \caption{\label{tab:scaling-costs}Approximate cost of various MPS and MPO operations. MPS
    bond dimensions are $m$ and -- if two MPS are used -- $n$, MPO
    bond dimensions are $w$ and the local dimension is
    $\sigma$. Different operations (matrix multiplication, QR and
    singular value decompositions) are given separately due to the
    very different cost of each. The overall factor $L$ of
    the system size is not given explicitly, each method scales
    linearly in $L$. Particularly expensive terms are highlighted in bold.}
  \begin{tabular}{lc|c|c}
    \hspace{1cm} & Matrix multiplications & QR & SVD \\
    \hline\hline
    \multicolumn{4}{l}{Overlap of two MPS} \\
              & $m n^2 \sigma + m^2 n \sigma$ & none & none \\ % just the two tensor contractions
    
    \hline
    \multicolumn{4}{l}{Expectation value of MPO between two MPS} \\
              & $mn^2 \sigma w + m n \sigma^2 w^2 + m^2 n \sigma w$ & none & none \\
    \hline
    \multicolumn{4}{l}{Variational truncation of MPS from $m$ to $n$; single-site; one optimization sweep} \\
              & $2 m n^2 \sigma + \boldsymbol{2 m^2 n \sigma} + n^3 \sigma$ & $n^3 \sigma$ & none \\ % two tensor contractions to build L^\prime, two to build the new tensor, QR to orthonormalize (+ transfer tensor in mm)
    \hline
    \multicolumn{4}{l}{Variational truncation of MPS from $m$ to $n$; two-site; one optimization sweep} \\
              & $2 m n^2 \sigma + \boldsymbol{2 m^2 n \sigma} + m n^2 \sigma^2 + n^3 \sigma$ & none & $\boldsymbol{n^3 \sigma^3} $ \\ % two tensor contractions to build L^\prime, three to build the new tensor, SVD to orthonormalize
    \hline
    \multicolumn{4}{l}{Direct truncation of MPS from $m$ to $n$ with SVD (incl. initial normalization)} \\
              & $m^3 + \boldsymbol{m^3 \sigma} + m n^2 + m^2 n \sigma$ & $ \boldsymbol{m^3 \sigma} $ & $ \boldsymbol{\mathrm{min}\{m^2 n \sigma,m n^2 \sigma^2\}} $ \\ % matrix multiplication from transfer tensors (QR and SVD), QR from initial normalization, SVD from truncation with already-truncated left leg
    \hline
    \multicolumn{4}{l}{Direct MPO-MPS application and subsequent direct truncation with SVD to bond dimension $n$} \\
              & $m^2 \sigma^2 w^2 + m^3 w^3 + \boldsymbol{m^3 \sigma w^3} + m n^2 w + m^2 n \sigma w^2$ & $\boldsymbol{m^3 \sigma w^3}$ & $\boldsymbol{m n^2 \sigma^2 w }$ \\ % copied from direct truncation with m -> mw and two initial terms in mm added
    \hline
    \multicolumn{4}{p{0.95\textwidth}}{Zip-up MPO-MPS application with target bond dimension $n$ and intermediate bond dimension $2n$ (incl. initial normalization; second line is the final truncation sweep from $2n$ to $n$)} \\
              & $m^3 + m^3 \sigma + \boldsymbol{2m^2 n \sigma w} + 2m n \sigma^2 w^2 + 4 m n^2 w$ & $m^3 \sigma$ & $\boldsymbol{4 \sigma^2 n^2 m w}$  \\ % first two MM and QR from initial normalisation; second two mm from local tensor contraction; fifth mm from first transfer tensor; sixth and seventh mm and second SVD from final truncation sweep
              & $ + 4n^3 \sigma + 2 n^3$         &  &  $+ 4 n^3 \sigma$ \\
    \hline
    \multicolumn{4}{l}{Variational MPO-MPS application with target bond dimension $n$; single-site; one optimization sweep} \\
              & $2 mn^2 \sigma w + 2 m n \sigma^2 w^2 + \boldsymbol{2 m^2 n \sigma w} + n^3 \sigma$ & $n^3 \sigma$ & none \\ % expectation value plus new tensor plus plus QR
    \hline
    \multicolumn{4}{l}{Variational MPO-MPS application with target bond dimension $n$; two-site; one optimization sweep} \\
              & $mn^2 \sigma w + 3 m n \sigma^2 w^2 + \boldsymbol{3 m^2 n \sigma w} + mn^2 \sigma^2 w + n^3 + n^3 \sigma$ & none & $\boldsymbol{n^3 \sigma^3}$ \\
    \hline

  \end{tabular}
\end{table}

\begin{figure}
  \centering
  \tikzsetnextfilename{expectation_value}
  % \tikzset{external/export next=false}
  \begin{tikzpicture}
    \begin{scope}[node distance = 0.5]
      \node[site] (x1) {};
      \node[op]   (o1) [below = of x1] {};
      \node[site] (y1)  [below = of o1] {};
      \draw (x1) -- (o1);
      \draw (o1) -- (y1);
      \foreach \start in {1,...,11}
      {
        \pgfmathtruncatemacro{\thisnode}{1 + \start};
        \pgfmathtruncatemacro{\lastnode}{\start};
        \node[site] (x\thisnode) [right = of x\lastnode] {};
        \node[op]   (o\thisnode) [below = of x\thisnode] {};
        \node[site] (y\thisnode) [below = of o\thisnode] {};
        \draw (x\lastnode) -- (x\thisnode);
        \draw (o\lastnode) -- (o\thisnode);
        \draw (y\lastnode) -- (y\thisnode);
        \draw (x\thisnode) -- (o\thisnode);
        \draw (o\thisnode) -- (y\thisnode);
      }

      \draw [decorate, decoration={brace,amplitude=3pt}] ($(x12.east)+(1,0.4)$) -- +(0,-0.8) node [black,midway,right,xshift=3pt] {\footnotesize $\Ket{\psi}$};
      \draw [decorate, decoration={brace,amplitude=3pt}] ($(o12.east)+(1.05,0.4)$) -- +(0,-0.8) node [black,midway,right,xshift=3pt] {\footnotesize $\hat O$};
      \draw [decorate, decoration={brace,amplitude=3pt}] ($(y12.east)+(1,0.4)$) -- +(0,-0.8) node [black,midway,right,xshift=3pt] {\footnotesize $\Bra{\phi}$};

      \node[draw, fill, opacity=0.05, inner sep=3pt, draw opacity=0.6, black, thick, rounded corners, fill=black, fit=(x1) (o1) (y1)] (s1) {};

      \foreach \start in {1,...,5}
      {
        \pgfmathtruncatemacro{\thisnode}{1 + \start};
        \pgfmathtruncatemacro{\lastnode}{\start};
        \node[draw, fill, opacity=0.05, inner sep=3pt, draw opacity=0.6, black, thick, rounded corners, fill=black, fit=(s\lastnode) (x\thisnode) (o\thisnode) (y\thisnode)] (s\thisnode) {};
      }

      \node[draw, fill, opacity=0.05, inner sep=3pt, draw opacity=0.6, black, thick, rounded corners, fill=black, fit=(x12) (o12) (y12)] (s12) {};
      \foreach \start in {1,...,5}
      {
        \pgfmathtruncatemacro{\thisnode}{12 - \start};
        \pgfmathtruncatemacro{\lastnode}{12 - \start + 1};
        \node[draw, fill, opacity=0.05, inner sep=3pt, draw opacity=0.6, black, thick, rounded corners, fill=black, fit=(s\lastnode) (x\thisnode) (o\thisnode) (y\thisnode)] (s\thisnode) {};
      }

      \node[draw, fill, opacity=0.05, inner sep=3pt, draw opacity=0.6, black, thick, rounded corners, fill=black, fit=(s6) (s7)] (sall) {};

      % \node[draw, fill, opacity=0.1, inner sep=3pt, draw opacity=0.6, black, thick, rounded corners, fill=black, fit=(s2) (x3) (o3) (y3)] (s3) {};
      % \node[draw, fill, opacity=0.1, inner sep=3pt, draw opacity=0.6, black, thick, rounded corners, fill=black, fit=(s3) (x4) (o4) (y4)] (s4) {};
      % \node[draw, fill, opacity=0.1, inner sep=3pt, draw opacity=0.6, black, thick, rounded corners, fill=black, fit=(s4) (x5) (o5) (y5)] (s5) {};
    \end{scope}
  \end{tikzpicture}
  \caption{\label{fig:expectation_value} The expectation value
    $\Bra{\phi} \hat O \Ket{\psi}$ of an operator between two
    (possibly different) states represented as MPO and MPS
    respectively. The optimal contraction order is sideways, e.g.~from
    left to right, as indicated by the shading. To add an additional
    column, the optimal evaluation order is
    $L_{j-1}^{\protect\nodagger} \cdot M_j^{\protect\nodagger} \cdot W_j^{\protect\nodagger} \cdot M_j^\dagger$. An easy option
    for two-fold parallelization is the concurrent evaluation of
    $L_{j-1}$ and $R_j$ from left and right respectively. Here and in all following diagrams, we leave off the dummy left/right indices indicated by dotted lines earlier.}
\end{figure}
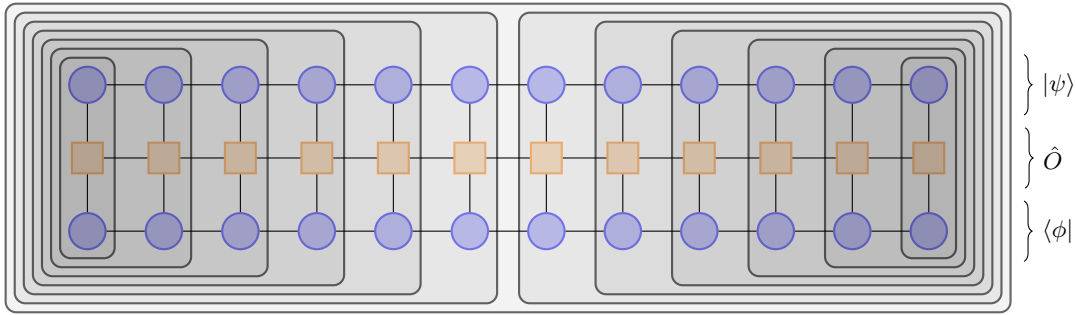

\subsection{\label{sec:expectation-values}Expectation values}

In standard dense numerical linear algebra, to evaluate the
expectation value $\Bra{\phi} \hat O \Ket{\psi}$, it is necessary to
evaluate $\hat O \Ket{\psi}$ and subsequently the overlap between
$\Bra{\phi}$ and $\hat O \Ket{\psi}$. As we have seen in the previous
section, MPO-MPS products are relatively costly to evaluate. Luckily,
the tensor network representing $\Bra{\phi} \hat O \Ket{\psi}$
(cf.~\cref{fig:expectation_value}) allows many different contraction
orders. Contracting it from left-to-right (using iteratively updated
tensors $L_j$ as defined in \cref{eq:left-mps-mpo-mps}) or
right-to-left (using $R_j$ as defined in \cref{eq:right-mps-mpo-mps}
respectively) leads to an asymptotic contraction cost of
$O\left(L \left(m^3 w \sigma + m^2 w^2 \sigma^2\right)\right)$. The
optimal contraction sequence is also indicated by the shading in
\cref{fig:expectation_value}.

%%% Local Variables: 
%%% mode: latex
%%% TeX-master: "../time_evolution_review"
%%% End: 

%% file: content/time-evolution-overview.tex
\section{\label{sec:te}Overview of time-evolution methods}

In the following sections, we will discuss in detail five different
time-evolution methods for MPS which are currently in use to solve the
time-dependent Schr\"odinger equation (TDSE). Each of them has
different strengths and weaknesses, requiring a careful consideration
of each individual problem to determine the most promising approach.
In \cref{sec:examples} we then examine some prototypical examples in
an attempt to provide some guidance.

The first two methods (\cref{sec:approx}) approximate the
time-evolution operator $\hat U(\delta) = e^{-\I \delta \hat H}$, which is
then repeatedly applied to the MPS $\ket{\psi(t)}$ to yield
time-evolved states $\ket{\psi(t+\delta)}$, $\ket{\psi(t+2\delta)}$
etc. The \emph{time-evolving block decimation,} also known as the
Trotter method and abbreviated here as TEBD, was developed around 2004
both for MPS\cite{vidal04:_effic_simul_one_dimen_quant,
  zwolak04, verstraete04:_matrix_produc_densit_operat} and the original
classical
DMRG\cite{white04:_real_time_evolut_using_densit,daley04:_time_hilber}.
It uses a Trotter-Suzuki decomposition of $\hat U(\delta)$, which is
in particular applicable to short-ranged Hamiltonians.
The time evolution is unitary up to the inherent truncation error, but
the energy is typically not conserved.
The main variants are using a second-order decomposition (TEBD2 in the
following) and a fourth-order decomposition (TEBD4) to minimise the
error due to the finite time step. While TEBD2 is essentially always
useful, as it produces only two- to three times as many exponential
terms as a first-order decomposition, TEBD4 produces four to five
times as many exponentials as TEBD2. Depending on the desired error,
this may or may not be worthwhile.
In contrast, the \emph{MPO \wiii} method was introduced only
recently\cite{zaletel15:_time} and relies on a different decomposition
scheme for the time-evolution operator $\hat U(\delta)$ particularly
suited to construct an efficient representation as a matrix-product
operator.
It can directly deal with long-range interactions and
generally generates smaller MPOs than TEBD.
Its primary downside is that the evolution is not exactly unitary.
The time step error of both TEBD and the MPO \wiii method is larger
than in the other methods described below.

Compared to these methods, algorithms based on Krylov
subspaces\cite{paige:_approx_krylov,
  barrett94:_templ_solut_linear_system} directly approximate the
action of $\hat U(\delta)$ onto $\ket{\psi}$, i.e.~produce a state
$\ket{\psi(t+\delta)}$ without explicitly constructing
$\hat U(\delta)$ in the full Hilbert space while preserving the
unitarity of the time evolution.
The main advantage lies in a very precise evaluation of
$\Ket{\psi(t+\delta)}$ with a very small inherent finite time-step
error\cite{Hochbruck1997}.

The \emph{global Krylov} algorithm (\cref{sec:te:krylov}) is related
to the time-dependent Lanczos method in exact diagonalization
approaches to time evolution\cite{park86:_unitar_lancz, kosloff88:_time,
  saad03:_iterat_method_spars_linear_system} and has only
partially\cite{garcia-ripoll06:_time_matrix_produc_states,dargel12:_lancz,
  wall12:_out} been adapted to the MPS setting.  Interestingly,
evaluation of observables on a very fine time grid
(e.g.~$\delta = 10^{-5}$) is possible, which would be prohibitively
expensive using any of the time-stepping methods.  The downside of
this global Krylov method is its need to represent potentially highly
entangled Krylov vectors as matrix-product states.

These highly-entangled global Krylov vectors do not need to be
represented if instead of working globally, one works in the local
reduced basis (\cref{sec:local}). From a DMRG perspective, this can be
seen as a variant of the time-step targeting
method\cite{feiguin05:_time, manmana05:_time_quant_many_body_system,
  rodriguez06,ronca17:_time_step_target_time_depen}.
Its primary objective is to solve the TDSE locally on each pair of
lattice sites to construct an optimal MPS for the time-evolved state.
Conversely, this local method can no longer evaluate observables
precisely at arbitrarily-small time steps $\delta' \ll \delta$ (as no
complete MPS is available at such intermediate times) but works much
like TEBD2 as a time-stepper, producing a single final state
$|\psi(t+\delta)\rangle$.  In contrast to TEBD and the MPO \wii it
allows for the treatment of arbitrary Hamiltonians. An uncontrolled
projection error may, however, lead to incorrect results as the MPS-projected
Hamiltonian cannot represent the actual physical Hamiltonian well if
the MPS used for the projection is very small. A further development
of this approach is the \emph{time-dependent variational
  principle}\cite{haegeman11:_time_depen_variat_princ_quant_lattic,haegeman16:_unify}
(TDVP). The TDVP can be considered an approach to remedy the dominant
errors in the local Krylov approach by a thorough theoretical analysis
leading to an optimized evolution method. Its implementation is nearly
identical to the local Krylov method, but the different derivation of
the local equations leads to smaller errors because it arrives at a
closed solution of a series of coupled equations.  In particular,
using the two-site variant of TDVP (2TDVP), we know that
nearest-neighbor Hamiltonians do not incur a projection error which is
often the primary error source in the local methods.  The single-site
variant (1TDVP) has a larger projection error and also always works at
constant MPS bond dimension but violates neither unitarity nor energy
conservation during the time evolution.

We complement this part of the review by presenting a subjective selection of useful tricks
which are applicable to most of the time-evolution methods and which
can help in the treatment of complicated systems. We will discuss in
some detail: (i) how to combine the time evolution in the Heisenberg
and the Schr\"odinger picture, respectively, to reach longer times; (ii)
how to select time steps to increase the accuracy of the integrator;
(iii) how removing low-lying eigenstates and the application of
linear prediction helps in calculating Green's functions; (iv) how to
specifically select the optimal choice of purification schemes for
finite-temperature calculations; and (v) briefly summarize the local
basis optimization technique to treat systems with many bosonic
degrees of freedom.

%%% Local Variables: 
%%% mode: latex
%%% TeX-master: "../time_evolution_review"
%%% End: 

%% file: content/tebd.tex
The following two subsections will summarize two popular choices to
approximate the time-evolution operator $\hat U(\delta)$, which, when
applied to an MPS $\ket{\psi(t)}$ gives a new MPS
$\ket{\psi(t+\delta)}$ at time $t + \delta$. Future developments to
construct such $\hat U(\delta)$ in an easier and more generic manner
are conceivable.

\subsection{\label{sec:tebd}Time-evolving block decimation (TEBD) or Trotter decomposition}

At its heart, this method relies on a Trotter-Suzuki\cite{suzuki76:_gener_trott} decomposition and
subsequent approximation of the time-evolution operator
$\hat U^\textrm{exact}(\delta)$. It was developed around 2004 both in
the (then new) context of
MPS\cite{vidal04:_effic_simul_one_dimen_quant,
  verstraete04:_matrix_produc_densit_operat} and the classical
system-environment DMRG\cite{white04:_real_time_evolut_using_densit,
  daley04:_time_hilber} and remained the most popular
time-evolution method for both settings in the following years.\footnote{In these papers, TEBD is introduced without relying on MPO/MPS arithmetic but instead by applying quantum gates to individual state tensors. Here, we will employ MPOs to simplify notation and implementation.}

As an illustrative example let us first consider the nearest-neighbor
Heisenberg chain. Its Hamiltonian is given by
\begin{align}
  \hat H & = \sum_{j} \hat h_{j,j+1} \\
  \hat h^{\vphantom{y}}_{j,j+1} & = \hat s^{x\vphantom{y}}_j \hat s^{x\vphantom{y}}_{j+1} + \hat s^y_j \hat s^y_{j+1} + \hat s^{z\vphantom{y}}_j \hat s^{z\vphantom{y}}_{j+1} \;.
\end{align}
The exact evolution is then given by
\begin{align}
  \hat U^\mathrm{exact}(\delta) = e^{-\I \delta \hat H} \;.
\end{align}
If we split $\hat H$ into two summands $\hat H_{\mathrm{even}}$ and
$\hat H_{\mathrm{odd}}$ as
\begin{align}
  \hat H_{\textrm{even}} & = \sum_{j \textrm{ even}} \hat h_{j,j+1} \\
  \hat H_{\textrm{odd}} & = \sum_{j \textrm{ odd}} \hat h_{j,j+1} \\
  \hat H & = \hat H_{\mathrm{even}} + \hat H_{\mathrm{odd}}\;,
\end{align}
we can use the Baker-Campbell-Hausdorff formula to approximate
\begin{align}
  \hat U^\textrm{exact}(\delta) & = e^{-\I \delta \hat H} \\
                              & = e^{-\I \delta \hat H_{\textrm{even}}} e^{-\I \delta \hat H_{\mathrm{odd}}} e^{-\I \delta^2 \left[ \hat H_{\textrm{even}}, \hat H_{\textrm{odd}} \right]} \\ 
                              & \approx e^{-\I \delta \hat H_{\textrm{even}}} e^{-\I \delta \hat H_{\textrm{odd}}} \\
                              & \equiv \hat U^{\textrm{TEBD1}}(\delta) \;.
\end{align}
Now, both $e^{-\I \delta \hat H_{\mathrm{even}}}$ and
$e^{-\I \delta \hat H_{\mathrm{odd}}}$ are easy to evaluate: All the
summands $\hat h_{j,j+1}$ commute with each other so it suffices to
exponentiate them individually:
\begin{equation}
  e^{-\I \delta \hat H_{\textrm{even}}} = e^{-\I \delta \sum_{j \textrm{ even}} \hat h_{j,j+1}} = \prod_{j \textrm{ even}} e^{-\I \delta \hat h_{j,j+1}} \;.
\end{equation}
Writing $e^{-\I \delta \hat h_{j,j+1}}$ into an MPO, we have identity
tensors with bond dimension $w=1$ everywhere except on sites $j$ and
$j+1$. Splitting the two-site gate into two individual tensors (e.g.~with an SVD) results in a bond between those sites which has at most dimension
$\sigma^2$. Hence writing $e^{-\I \delta \hat H_{\mathrm{even}}}$ as
an MPO is also efficient, it has alternating bond dimensions
$\sigma^2$ and $1$. Accordingly,
$e^{-\I \delta \hat H_{\mathrm{odd}}}$ written as an MPO has
alternating bond dimensions $1$ and $\sigma^2$ and their product
$e^{-\I \delta \hat H_{\mathrm{even}}} e^{-\I \delta \hat
  H_{\mathrm{odd}}}$
can be written with bond dimensions $\sigma^2$ throughout.

If we keep the decomposition of $\hat H$ into $\hat H_{\mathrm{even}}$
and $\hat H_{\mathrm{odd}}$ fixed and hence the commutator
$\left[ \hat H_\mathrm{even}, \hat H_\mathrm{odd} \right]$ constant,
the error per time step $\delta$ between
$\hat U^{\mathrm{TEBD1}}(\delta)$ and
$\hat U^{\mathrm{exact}}(\delta)$ is of second order:
\begin{equation}
  \hat U^\mathrm{exact}(\delta) = \hat U^{\mathrm{TEBD1}}(\delta) + O(\delta^2) \;
\end{equation}
because we can approximate
\begin{equation}
  e^{-\I \delta^2 \left[ \hat H_1, \hat H_2 \right]} \approx \id -\I \delta^2 \left[ \hat H_1, \hat H_2 \right] \;.
\end{equation}
Considering a longer interval $T$, which we divide into
$\nicefrac{T}{\delta}$ smaller intervals, at which we apply
$\hat U^{\mathrm{TEBD1}}(\delta)$, our errors accumulate to
$\nicefrac{T}{\delta} \, O(\delta^2) = O(\delta)$ per time interval
$T$.  The error is hence of first order in $\delta$ which gives
$\hat U^{\mathrm{TEBD1}}(\delta)$ its name as first-order TEBD.  By
symmetrizing the decomposition, it is straightforward to construct
\begin{equation}
  \hat U^{\mathrm{TEBD2}}(\delta) \equiv e^{-\I \nicefrac{\delta}{2} \hat H_{\mathrm{even}}} e^{-\I \delta \hat H_{\mathrm{odd}}}e^{-\I \nicefrac{\delta}{2} \hat H_{\mathrm{even}}} \;,
\end{equation}
which has a third-order error per step
\begin{equation}
  \hat U^\mathrm{exact}(\delta) = \hat U^{\mathrm{TEBD2}}(\delta) + O(\delta^3)
\end{equation}
and hence a second-order error per time interval. Similarly, we can
construct a fourth-order TEBD evolution operator as
\begin{equation}
  \hat U^{\mathrm{TEBD4}}(\delta) \equiv \hat U^{\mathrm{TEBD2}}(\delta_1) \hat U^{\mathrm{TEBD2}}(\delta_1) \hat U^{\mathrm{TEBD2}}(\delta_2) \hat U^{\mathrm{TEBD2}}(\delta_1) \hat U^{\mathrm{TEBD2}}(\delta_1)
\end{equation}
with
\begin{align}
  \delta_1 & \equiv \frac{1}{4 - 4^{\nicefrac{1}{3}}} \delta \\
  \delta_2 & \equiv \left(1 - 4 \delta_1\right) \delta \;.
\end{align}
We then have
\begin{equation}
\hat U^\mathrm{exact}(\delta) = \hat U^{\mathrm{TEBD4}}(\delta) + O(\delta^5)
\end{equation}
and hence a fourth-order error per time step.

\begin{figure}
  \centering
  \tikzsetnextfilename{tebd_layers}
%  \tikzset{external/export next=false}
  \begin{tikzpicture}
    \begin{scope}[node distance = 0.5]
      \node[site] (s1) {};
      \node[site] (s2) [right = of s1] {};
      \node[site] (s3) [right = of s2] {};
      \node[site] (s4) [right = of s3] {};
      \node[site] (s5) [right = of s4] {};
      \node[site] (s6) [right = of s5] {};
      \node[site] (s7) [right = of s6] {};
      \node[site] (s8) [right = of s7] {};
      \node[site] (s9) [right = of s8] {};
      \node[site] (sA) [right = of s9] {};
      \node[site] (sB) [right = of sA] {};
      \node[site] (sC) [right = of sB] {};

      \draw [decorate,decoration={brace,amplitude=3pt}] ($(sC.east)+(0.17,+0.4)$) -- +(0,-0.8) node [black,midway,right,xshift=3pt] {\footnotesize $\Ket{\psi}$};

      \draw[] (s1) -- (s2);
      \draw[] (s2) -- (s3);
      \draw[] (s3) -- (s4);
      \draw[] (s4) -- (s5);
      \draw[] (s5) -- (s6);
      \draw[] (s6) -- (s7);
      \draw[] (s7) -- (s8);
      \draw[] (s8) -- (s9);
      \draw[] (s9) -- (sA);
      \draw[] (sA) -- (sB);
      \draw[] (sB) -- (sC);

% layer 1
      \node[op] (u11) [below = of s1] {};
      \node[op] (u12) [below = of s2] {};
      \node[op] (u13) [below = of s3] {};
      \node[op] (u14) [below = of s4] {};
      \node[op] (u15) [below = of s5] {};
      \node[op] (u16) [below = of s6] {};
      \node[op] (u17) [below = of s7] {};
      \node[op] (u18) [below = of s8] {};
      \node[op] (u19) [below = of s9] {};
      \node[op] (u1A) [below = of sA] {};
      \node[op] (u1B) [below = of sB] {};
      \node[op] (u1C) [below = of sC] {};

      \draw [decorate,decoration={brace,amplitude=3pt}] ($(u1C.east)+(0.2,+0.4)$) -- +(0,-0.8) node [black,midway,right,yshift=0.1em,xshift=3pt] {\footnotesize $\hat U_1(\nicefrac{\delta}{2}) = e^{-\I \nicefrac{\delta}{2} \hat H_1}$};

      \draw[line width=0.5mm] (u11) -- (u12);
      \draw[line width=0.5mm] (u12) -- (u13);
      \draw[line width=0.25mm, dotted] (u13) -- (u14);
      \draw[line width=0.5mm] (u14) -- (u15);
      \draw[line width=0.5mm] (u15) -- (u16);
      \draw[line width=0.25mm, dotted] (u16) -- (u17);
      \draw[line width=0.5mm] (u17) -- (u18);
      \draw[line width=0.5mm] (u18) -- (u19);
      \draw[line width=0.25mm, dotted] (u19) -- (u1A);
      \draw[line width=0.5mm] (u1A) -- (u1B);
      \draw[line width=0.5mm] (u1B) -- (u1C);

      \draw[] (s1) -- (u11);
      \draw[] (s2) -- (u12);
      \draw[] (s3) -- (u13);
      \draw[] (s4) -- (u14);
      \draw[] (s5) -- (u15);
      \draw[] (s6) -- (u16);
      \draw[] (s7) -- (u17);
      \draw[] (s8) -- (u18);
      \draw[] (s9) -- (u19);
      \draw[] (sA) -- (u1A);
      \draw[] (sB) -- (u1B);
      \draw[] (sC) -- (u1C);

% layer 2
      \node[op] (u21) [below = of u11] {};
      \node[op] (u22) [below = of u12] {};
      \node[op] (u23) [below = of u13] {};
      \node[op] (u24) [below = of u14] {};
      \node[op] (u25) [below = of u15] {};
      \node[op] (u26) [below = of u16] {};
      \node[op] (u27) [below = of u17] {};
      \node[op] (u28) [below = of u18] {};
      \node[op] (u29) [below = of u19] {};
      \node[op] (u2A) [below = of u1A] {};
      \node[op] (u2B) [below = of u1B] {};
      \node[op] (u2C) [below = of u1C] {};

      \draw [decorate,decoration={brace,amplitude=3pt}] ($(u2C.east)+(0.2,+0.4)$) -- +(0,-0.8) node [black,midway,right,yshift=0.1em,xshift=3pt] {\footnotesize $\hat U_2(\nicefrac{\delta}{2}) = e^{-\I \nicefrac{\delta}{2} \hat H_2}$};

      \draw[line width=0.25mm, dotted] (u21) -- (u22);
      \draw[line width=0.5mm]    (u22) -- (u23);
      \draw[line width=0.5mm]    (u23) -- (u24);
      \draw[line width=0.25mm, dotted] (u24) -- (u25);
      \draw[line width=0.5mm]    (u25) -- (u26);
      \draw[line width=0.5mm]    (u26) -- (u27);
      \draw[line width=0.25mm, dotted] (u27) -- (u28);
      \draw[line width=0.5mm]    (u28) -- (u29);
      \draw[line width=0.5mm]    (u29) -- (u2A);
      \draw[line width=0.25mm, dotted] (u2A) -- (u2B);
      \draw[line width=0.25mm, dotted] (u2B) -- (u2C);

      \draw[] (u11) -- (u21);
      \draw[] (u12) -- (u22);
      \draw[] (u13) -- (u23);
      \draw[] (u14) -- (u24);
      \draw[] (u15) -- (u25);
      \draw[] (u16) -- (u26);
      \draw[] (u17) -- (u27);
      \draw[] (u18) -- (u28);
      \draw[] (u19) -- (u29);
      \draw[] (u1A) -- (u2A);
      \draw[] (u1B) -- (u2B);
      \draw[] (u1C) -- (u2C);

% layer 3
      \node[op] (u31) [below = of u21] {};
      \node[op] (u32) [below = of u22] {};
      \node[op] (u33) [below = of u23] {};
      \node[op] (u34) [below = of u24] {};
      \node[op] (u35) [below = of u25] {};
      \node[op] (u36) [below = of u26] {};
      \node[op] (u37) [below = of u27] {};
      \node[op] (u38) [below = of u28] {};
      \node[op] (u39) [below = of u29] {};
      \node[op] (u3A) [below = of u2A] {};
      \node[op] (u3B) [below = of u2B] {};
      \node[op] (u3C) [below = of u2C] {};

      \draw [decorate,decoration={brace,amplitude=3pt}] ($(u3C.east)+(0.2,+0.4)$) -- +(0,-0.8) node [black,midway,right,yshift=0.1em,xshift=3pt] {\footnotesize $\hat U_3(\delta) = e^{-\I \delta \hat H_3}$};

      \draw[line width=0.25mm, dotted] (u31) -- (u32);
      \draw[line width=0.25mm, dotted]    (u32) -- (u33);
      \draw[line width=0.5mm]    (u33) -- (u34);
      \draw[line width=0.5mm] (u34) -- (u35);
      \draw[line width=0.25mm, dotted]    (u35) -- (u36);
      \draw[line width=0.5mm]    (u36) -- (u37);
      \draw[line width=0.5mm] (u37) -- (u38);
      \draw[line width=0.25mm, dotted]    (u38) -- (u39);
      \draw[line width=0.5mm]    (u39) -- (u3A);
      \draw[line width=0.5mm] (u3A) -- (u3B);
      \draw[line width=0.25mm, dotted] (u3B) -- (u3C);

      \draw[] (u21) -- (u31);
      \draw[] (u22) -- (u32);
      \draw[] (u23) -- (u33);
      \draw[] (u24) -- (u34);
      \draw[] (u25) -- (u35);
      \draw[] (u26) -- (u36);
      \draw[] (u27) -- (u37);
      \draw[] (u28) -- (u38);
      \draw[] (u29) -- (u39);
      \draw[] (u2A) -- (u3A);
      \draw[] (u2B) -- (u3B);
      \draw[] (u2C) -- (u3C);
% layer 4
      \node[op] (u41) [below = of u31] {};
      \node[op] (u42) [below = of u32] {};
      \node[op] (u43) [below = of u33] {};
      \node[op] (u44) [below = of u34] {};
      \node[op] (u45) [below = of u35] {};
      \node[op] (u46) [below = of u36] {};
      \node[op] (u47) [below = of u37] {};
      \node[op] (u48) [below = of u38] {};
      \node[op] (u49) [below = of u39] {};
      \node[op] (u4A) [below = of u3A] {};
      \node[op] (u4B) [below = of u3B] {};
      \node[op] (u4C) [below = of u3C] {};

      \draw [decorate,decoration={brace,amplitude=3pt}] ($(u4C.east)+(0.2,+0.4)$) -- +(0,-0.8) node [black,midway,right,yshift=0.1em,xshift=3pt] {\footnotesize $\hat U_2(\nicefrac{\delta}{2}) = e^{-\I \nicefrac{\delta}{2} \hat H_2}$};

      \draw[line width=0.25mm, dotted] (u41) -- (u42);
      \draw[line width=0.5mm]    (u42) -- (u43);
      \draw[line width=0.5mm]    (u43) -- (u44);
      \draw[line width=0.25mm, dotted] (u44) -- (u45);
      \draw[line width=0.5mm]    (u45) -- (u46);
      \draw[line width=0.5mm]    (u46) -- (u47);
      \draw[line width=0.25mm, dotted] (u47) -- (u48);
      \draw[line width=0.5mm]    (u48) -- (u49);
      \draw[line width=0.5mm]    (u49) -- (u4A);
      \draw[line width=0.25mm, dotted] (u4A) -- (u4B);
      \draw[line width=0.25mm, dotted] (u4B) -- (u4C);

      \draw[] (u31) -- (u41);
      \draw[] (u32) -- (u42);
      \draw[] (u33) -- (u43);
      \draw[] (u34) -- (u44);
      \draw[] (u35) -- (u45);
      \draw[] (u36) -- (u46);
      \draw[] (u37) -- (u47);
      \draw[] (u38) -- (u48);
      \draw[] (u39) -- (u49);
      \draw[] (u3A) -- (u4A);
      \draw[] (u3B) -- (u4B);
      \draw[] (u3C) -- (u4C);

% layer 5
      \node[op] (u51) [below = of u41] {};
      \node[op] (u52) [below = of u42] {};
      \node[op] (u53) [below = of u43] {};
      \node[op] (u54) [below = of u44] {};
      \node[op] (u55) [below = of u45] {};
      \node[op] (u56) [below = of u46] {};
      \node[op] (u57) [below = of u47] {};
      \node[op] (u58) [below = of u48] {};
      \node[op] (u59) [below = of u49] {};
      \node[op] (u5A) [below = of u4A] {};
      \node[op] (u5B) [below = of u4B] {};
      \node[op] (u5C) [below = of u4C] {};

      \draw [decorate,decoration={brace,amplitude=3pt}] ($(u5C.east)+(0.2,+0.4)$) -- +(0,-0.8) node [black,midway,right,yshift=0.1em,xshift=3pt] {\footnotesize $\hat U_1(\nicefrac{\delta}{2}) = e^{-\I \nicefrac{\delta}{2} \hat H_1}$};

      \draw[line width=0.5mm] (u51) -- (u52);
      \draw[line width=0.5mm]    (u52) -- (u53);
      \draw[line width=0.25mm, dotted]    (u53) -- (u54);
      \draw[line width=0.5mm] (u54) -- (u55);
      \draw[line width=0.5mm]    (u55) -- (u56);
      \draw[line width=0.25mm, dotted]    (u56) -- (u57);
      \draw[line width=0.5mm] (u57) -- (u58);
      \draw[line width=0.5mm]    (u58) -- (u59);
      \draw[line width=0.25mm, dotted]    (u59) -- (u5A);
      \draw[line width=0.5mm] (u5A) -- (u5B);
      \draw[line width=0.5mm] (u5B) -- (u5C);

      \draw[] (u41) -- (u51);
      \draw[] (u42) -- (u52);
      \draw[] (u43) -- (u53);
      \draw[] (u44) -- (u54);
      \draw[] (u45) -- (u55);
      \draw[] (u46) -- (u56);
      \draw[] (u47) -- (u57);
      \draw[] (u48) -- (u58);
      \draw[] (u49) -- (u59);
      \draw[] (u4A) -- (u5A);
      \draw[] (u4B) -- (u5B);
      \draw[] (u4C) -- (u5C);

      \node[draw,fill opacity=0.1, inner sep=5pt, draw opacity=0.6, black, thick, rounded corners, fill=black!80, fit=(u51) (u52) (u53), label={[shift={(-0.5cm,-1.4cm)}]{\footnotesize $e^{-\I \nicefrac{\delta}{2} \left( \hat s_1 \cdot \left( \hat s_2 \times \hat s_3 \right) + \mathrm{h.c.}\right)}$}}] {};

      \node[draw,fill opacity=0.1, inner sep=5pt, draw opacity=0.6, black, thick, rounded corners, fill=black!80, fit=(u57) (u58) (u59), label={[shift={(+0.5cm,-1.4cm)}]{\footnotesize $e^{-\I \nicefrac{\delta}{2} \left( \hat s_7 \cdot \left( \hat s_8 \times \hat s_9 \right) + \mathrm{h.c.}\right)}$}}] {};

    \end{scope}
  \end{tikzpicture}
  \caption{\label{fig:tebd}Exemplary structure of a tensor network
    representing a time-evolved state
    $\hat U^\mathrm{TEBD2}(\delta)\Ket{\psi}$ of 12 sites. The state
    is written as an MPS $\Ket{\psi}$ at the top and evolved using
    TEBD2 under the Hamiltonian
    $\hat H = \sum_{j=1}^{10} \hat s_j \cdot \left( \hat s_{j+1}
      \times \hat s_{j+2} \right) + \mathrm{h.c.}$.  The three-site
    action of the Hamiltonian necessitates the decomposition into
    three sums $\hat H_{1,2,3}$. To achieve a second-order error, the
    evolution is symmetrized by applying the operators
    $\hat U_{1,2,3}$ in reverse order after each half-step. The
    individual time-evolution operators written as MPOs have two thick
    bonds alternating with a 1-dimensional dummy bond.}
\end{figure}
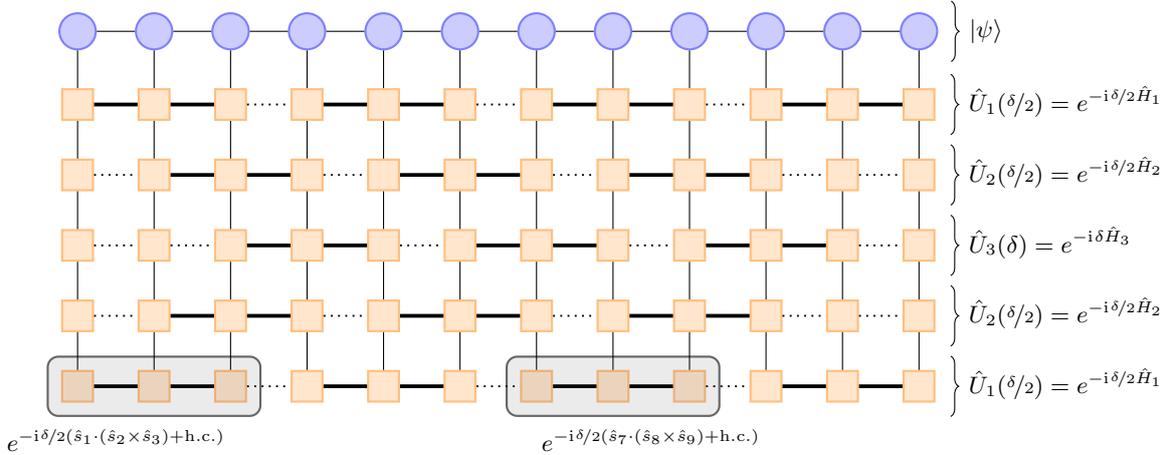

This approach can be generalized to more complicated divisions of the
full Hamiltonian $\hat H$ into $N_H$ internally-commuting parts $\hat H_1$,
$\hat H_2$, $\ldots$:
\begin{equation}
  \hat H = \sum_{\alpha=1}^{N_H} \hat H_\alpha \;,
\end{equation}
where each $\hat H_\alpha$ is a sum
\begin{equation}
  \hat H_\alpha = \sum_{k=1}^{N_\alpha} \hat h_{\alpha}^k \label{eq:tebd:split_H}
\end{equation}
such that $\hat h_{\alpha}^k$ can be diagonalized efficiently
and $[ \hat h_{\alpha}^k, \hat h_{\alpha}^l ] = 0$. The
local terms $\hat h_{\alpha}^k$ do not need to be limited to two sites,
but it must still be possible to evaluate their exponential
efficiently. A first-order TEBD time evolution operator
$\hat U^\mathrm{TEBD1}(\delta)$ can then be written as
\begin{align}
  \hat U^\mathrm{TEBD1}(\delta) & = e^{-\I \delta \hat H_1} e^{-\I \delta \hat H_2} e^{-\I \delta \hat H_3} \cdots e^{-\I \delta \hat H_{N_H}}\\
                                & = \prod_{\alpha=1}^{N_H} e^{-\I \delta \hat H_\alpha} \;.
\end{align}
The error per time interval is still of first order as before. The
second-order decomposition is likewise obtained as a symmetrization of
the $\hat U^\mathrm{TEBD1}(\delta)$:
\begin{align}
  \hat U^\mathrm{TEBD2}(\delta) & = e^{-\I \nicefrac{\delta}{2} \hat H_1} e^{-\I \nicefrac{\delta}{2} \hat H_2} \cdots e^{-\I \nicefrac{\delta}{2} \hat H_{N_H-1}} e^{-\I \delta \hat H_{N_H}} e^{-\I \nicefrac{\delta}{2} \hat H_{N_H-1}} \cdots e^{-\I \nicefrac{\delta}{2} \hat H_2} e^{-\I \nicefrac{\delta}{2} \hat H_1} \\
                                & = \prod_{\alpha=1}^{N_H} e^{-\I \nicefrac{\delta}{2} \hat H_\alpha} \prod_{\alpha=N_H}^1 e^{-\I \nicefrac{\delta}{2} \hat H_\alpha} \;.
\end{align}
It has the same second-order error per time step as above, an
exemplary tensor network for the case of a three-site interaction is
given in \cref{fig:tebd}. A fourth-order decomposition could be
constructed, but the number of terms required grows very quickly with
the number of summands $N_H$ in the Hamiltonian, making the
fourth-order decomposition less attractive. Recently,
Ref.~\cite{barthel19:_optim_lie_trott_suzuk} investigated optimized decomposition schemes beyond the naive approaches presented here. These schemes have the potential to further reduce errors and the number of exponentials required, but still need to be tested in practice. 

\subsubsection{\label{sec:tebd:errors}Errors}

TEBD suffers from two errors, both of which are controllable and can
be estimated fairly straightforwardly. The first is the time step
error of order $O(\delta^2)$ per time step for first-order TEBD and
$O(\delta^3)$ per time step for second-order TEBD. With a fixed total
interval $T$ divided into $N = T / \delta$ steps, the error over the
entire interval is $O(\delta)$ and $O(\delta^2)$ respectively for
first and second order TEBD. This time step-inherent error does not
violate unitarity of a real-time evolution as each of the constructed
operators $e^{-\I \delta \hat H_\alpha}$ is unitary.  However,
the energy and other conserved quantities are not
necessarily constant if the time-step error is large. The primary
option to reduce this discretization error is to chose a smaller
time-step size $\delta$ or a higher-order decomposition like TEBD4. It
is not always optimal to use TEBD4, as higher-order decompositions
typically result in more MPOs to apply to the state sequentially to do
a single time step. For example, if a Trotter error of $10^{-2}$ per
unit time is desired, 1TEBD needs a step size $\delta = 0.01$ and
accordingly 100 steps to reach time $t=1$. For the same error, TEBD2
needs a step size $\delta = 0.1$ and hence $10$ steps to reach $t=1$
while TEBD4 could work with a step size of $\sqrt{0.1} \approx 0.3$
and hence approximately three steps. However, where TEBD1 (e.g.) needs
three exponentials
$e^{-\I \delta H_1} e^{-\I \delta H_2} e^{-\I \delta H_3}$ , TEBD2 has
to apply five exponentials
$e^{-\I \nicefrac{\delta}{2} H_1} e^{-\I \nicefrac{\delta}{2} H_2}
e^{-\I \delta H_3} e^{-\I \nicefrac{\delta}{2} H_2} e^{-\I
  \nicefrac{\delta}{2} H_1}$
(or four, if we combine $e^{-\I \nicefrac{\delta}{2} H_1}$ of
neighboring steps) and TEBD4 has to work with 21 (20 when combining
neighboring steps) exponentials. That is, while TEBD4 only needs three
times fewer steps than TEBD2 for a final error of $10^{-2}$, each
individual step needs approximately five times as much work. In
comparison, if we target an error of $10^{-8}$, TEBD1 needs $10^8$
steps, TEBD2 needs $10^4$ steps and TEBD4 only needs $10^2$ steps --
for doing five times as much work per step, we reduce the number of
steps by a factor of 100 when going from TEBD2 to TEBD4.

As an alternative to such
higher-order decompositions, one may combine a large-scale Trotter
decomposition with a small-scale Krylov-based approach to make the
\emph{commutator} of the Hamiltonian summands $\hat H_\alpha$
smaller\cite{hashizume18:_dynam_phase_trans_two_dimen}.

The second source of errors is due to the mandatory truncation of the
time-evolved state to a realistic bond dimension. This truncation
error can also be controlled and measured by the discarded weight
during the truncation as usual. It affects both the unitarity of the
evolution and the conserved quantities. By gradually increasing the
MPS bond dimension, it is straightforward to estimate convergence of
the calculation with respect to this error.

\subsubsection{\label{sec:tebd:details}Long-range interactions}

Long-range terms in the Hamiltonian can be treated by introducing swap
gates\cite{stoudenmire10:_minim} in order to exponentiate each
individual contribution. Each swap gate exchanges the physical indices of two nearest neighbors, hence moving the interacting sites next to each other. The evolution term
$e^{-\mathrm{i}\delta \hat h_{\alpha}^k}$ is then sandwiched between
two series of swap gates which first bring the sites onto which
$\hat h_{\alpha}^k$ acts next to each other and then, after the evolution,
move them back to their original places. A smart ordering of the
individual evolution terms often allows to cancel two series of swap
gates from different $\hat h_{\alpha}^k$ and $\hat h_{\alpha}^l$
against each other. Nearly the entire overhead of such swap gates can
be removed in this way.

\subsubsection{Algorithm}

The two steps of the TEBD method (constructing the evolution operator
and repeatedly applying it to the state) are described in
\cref{alg:tebd} at the example of the second-order TEBD2 method.

To construct the time-evolution operator, one first needs to split the
Hamiltonian into internally-commuting parts $\hat H_\alpha$
(cf.~\cref{eq:tebd:split_H}). Each summand $\hat h_{\alpha}^k$ of each
part is then individually exponentiated as
$\mathrm{exp}\left(-\mathrm{i}\delta \hat h_{\alpha}^k\right)$. If the
summand is non-local, swap gates are introduced before and after the
exponentiation such that the exponentiation works on nearest-neighbor
sites. One then obtains a number of MPO component tensors (one for
each site between the leftmost and rightmost sites on which
$\hat h_{\alpha}^k$ acts). These tensors are placed within an MPO spanning
the entire system to eventually yield $\hat U_\alpha$. Once all $\hat U_\alpha$
are constructed (in a specific, selected ordering), one has a
representation of the first-order decomposition of $\hat U$.  Going
from this first-order decomposition to a second-order decomposition is
very straightforward, one only has to first replace $\delta$ by
$\delta/2$ in the initial exponentiation and second append the reversed
list of MPOs to itself, i.e.~store a list
$\{ \hat U_1, \hat U_2, \hat U_3, \hat U_3, \hat U_2, \hat U_1 \}$.
This new $\hat U$ is then symmetric, each half moves time forward by
$\delta/2$ for a total step of size $\delta$.

To then time-evolve the state, one only has to apply $\hat U$
repeatedly to it, each time advancing the state by a step $\delta$ in
time. In practice, we found it useful to write the time-evolved state
$|\psi(t)\rangle$ to disk after every step. In this way, the time
evolution can proceed as quickly as possible and evaluations (e.g. for
expectation values, overlaps with other states etc.) may proceed in
parallel by using the generated state files, possibly even on separate
computers.

An interesting avenue to
parallelization\cite{urbanek16:_paral_bose_hubbar} in TEBD can be used
if instead of working in the MPO framework all $n$-site gates are kept
separate. For very small time steps, gates should not affect singular
value spectra on far-away bonds. It should hence be possible to apply
many gates independently and also truncate many bonds independently
(using a slightly different gauge not discussed here). While not
extensively utilized in the literature yet, such real-space
parallelization promises much larger speed-ups than local (e.g. BLAS)
parallelization options if the associated errors are kept sufficiently
small.

\paragraph{Splitting or combining $\hat U(\delta)$}
It may provide an additional computational advantage to combine
multiple MPOs representing $\hat U_\alpha(\delta)$ into one larger MPO
representing $\hat U(\delta)$.
For example in the
case of the nearest-neighbor Heisenberg model, one arrives naively at
two MPOs $\hat U_{\mathrm{even}}$ and $\hat U_{\mathrm{odd}}$, each
exponentials of $\hat H_{\mathrm{even}}$ and $\hat H_{\mathrm{odd}}$
respectively. These MPOs have alternating bond dimensions
$1 - 4 - 1 - 4 - \ldots$ and $4 - 1 - 4 - 1 - \ldots$.  Multiplying
both MPOs together, one obtains a single new MPO with bond dimension
$4$ throughout the system. Applying the new MPO only requires a single
sweep over the system, whereas applying the two original MPOs requires
two sequential sweeps. While the computational effort to leading order
is identical, the former option is much more cache friendly, requiring
access of each individual site tensor from RAM (or even a hard disk
drive) only once instead of twice.

Similarly, if the Hamiltonian needs to be split into multiple terms
$\hat H_\alpha$, it is potentially helpful to first multiply the MPOs
for $\hat U_\alpha$ and then apply the resulting $\hat U$ to the MPS.
In general, applying many $\hat U_\alpha$ with very small bond
dimensions leads to a large overhead, whereas $\hat U_\alpha$ with
very large bond dimensions are more costly to multiply with states.
The optimal trade-off depends on the problem at hand.  As a rule of
thumb, one should multiply two time-evolution operators if this does
not increase the maximal bond dimension (as in the example above).  If
the bond dimension of the time-evolution operators is very small, it
might also make sense to combine them until a comparatively large bond
dimension (e.g.  $w \approx 10$) is obtained which reduces the
overhead associated to many small matrix multiplications.

\begin{algorithm}[h!]
  \caption{\label{alg:tebd}The second-order TEBD2 method. The main
    input is an analytic expression for the Hamiltonian $\hat H$
    necessary to split it into its constitutent internally-commuting
    parts. Additionally, one needs to select an application method
    \textit{apply} to evaluate MPO-MPS products (including truncation
    as required). To run, one first calls \textsc{Prepare-$U(\delta)$}
    and then repeatedly \textsc{Timestep} to generate the time-evolved
    states.}
  \begin{algorithmic}[1]
    \Procedure{Prepare-$U_\alpha(\delta)$}{Analytic expression for $\hat H_\alpha$, step size $\delta$}
    \For{\textbf{each} summand $\hat h_{\alpha}^k$ of $\hat H_\alpha$}
      \State $\hat u_\alpha^k \gets \Call{exp}{-\mathrm{i}\delta \hat h_{\alpha}^k}$ \Comment{If $\hat h_{\alpha}^k$ is long-range, swap gates are used around this line}
      \State Split $\hat u_{\alpha}^k$ into MPO component tensors by SVD or QR
      \State Place those tensors on appropriate sites of MPO $\hat U_\alpha$
    \EndFor
    \State \textbf{return} $\hat U_\alpha$
    \EndProcedure
    \Procedure{Prepare-$U(\delta)$}{Analytic expression for $\hat H$, step size $\delta$}
    \State Split $\hat H$ into internally-commuting parts $\{ \hat H_1, \ldots, \hat H_{N_H} \}$ \Comment{typically done by hand}
    \State $\hat U \gets \{ \Call{Prepare-$U_\alpha(\delta/2)$}{\hat H_1}, \ldots, \Call{Prepare-$U_\alpha(\delta/2)$}{\hat H_{N_H}} \}$
    \State $\hat U \gets \hat U + \Call{reverse}{\hat U}$ \Comment{Construct second-order Trotter decomposition, hence $\delta/2$}
    \State Multiply pairs of constituents $\hat U_\alpha, \hat U_{\alpha+1}$ of $\hat U$ if possible within bond dimensions
    \State \textbf{return} $\hat U$
    \EndProcedure
    \Procedure{Timestep}{List of MPOs $\hat U$, initial state $|\psi\rangle$, application method \textit{apply}}
    \For{\textbf{each} term $\hat U_\alpha$ in $\hat U$}
    \State $|\psi\rangle \gets \Call{apply}{\hat U_\alpha, |\psi\rangle}$
    \EndFor
    \State \textbf{return } $|\psi\rangle$
    \EndProcedure
  \end{algorithmic}
\end{algorithm}

%%% Local Variables: 
%%% mode: latex
%%% TeX-master: "../time_evolution_review"
%%% End: 

%% file: content/zaletel_mpo.tex
\subsection{\label{sec:te:mpowii}The MPO \wiii method}
In Ref.~\cite{zaletel15:_time} Zaletel et al. proposed a generalization of the Euler approximation of the operator exponential $e^{-\mathrm i \delta \hat{H}}$, which can be implemented efficiently using MPOs.
In this scheme, the error per site is independent of the system size.
Furthermore, the construction is capable of dealing with long-ranged interaction terms making it suitable for two-dimensional systems, too.
In this section, we will sketch the derivation of the MPO representations $\wiii$ and discuss how to construct these operators as well as the numerical errors and the stability of the overall method.

\subsubsection{Motivation and construction}
The general idea is to exploit the intrinsic factorization of MPO representations of operators.
We consider operators which have a local structure in the sense that they are given by a sum of terms $\hat{H}_{j}$ acting on a subset of the lattice, starting at site $j$: $\hat{H} = \sum_{j}\hat{H}_{j}$.
The Euler approximation of the operator exponential is $e^{-\mathrm i \delta \hat{H}} = 1 - \mathrm i \delta\sum_{j}\hat{H}_{j} + \mathcal{O}(\delta^2)$ and since there are $\sim L^2$ contributions from all possible combinations of local terms with finite support the error \textit{per site} is $\sim L\delta^{2}$; hence the approximation becomes more and more unstable with increasing system size.
Ref.~\cite{zaletel15:_time} introduced the following local version of the Euler stepper
\begin{align}
	e^{-\mathrm i \delta \hat{H}} &= 1 -\mathrm i \delta \sum_{j}\hat{H}_{j} - \delta^{2}{\sum_{j<k}}^{\prime}\hat{H}_{j}\hat{H}_{k} + \mathcal{O}(L\delta^2) + \mathcal{O}(\delta^3)\\ &\approx \prod_{j}(1 - \mathrm i \delta\hat{H}_{j})  \equiv \hat{U}^{\mathrm I}(\delta) \; ,
\end{align}
where the primed sum indicates that the local operator terms $\hat{H}_{j},\hat{H}_{k}$ do not act on a common subset of the lattice, i.e., they do not overlap.
Even though the error of $\hat U^{\mathrm{I}}$ is still of order
$\delta^2$ in the step size there are only $\mathcal{O}(L)$
contributions which are missed, namely those combinations of local
terms with overlapping support.
Hence, the overall error is bounded by $\mathcal{O}(L\delta^2)$, and
thus the error per site is constant in the system size.

Recall now the decomposition of an MPO into a left, right and local part
\begin{equation}
	\hat{H} = \hat{H}^{L}_{j-1} \otimes \mathbf{\hat{1}}^{R}_{j} + \mathbf{\hat{1}}^{L}_{j}\otimes \hat{H}^{R}_{j+1} + \sum_{a_j=1}^{N_j}\hat{h}^{L}_{j;a_j} \otimes \hat{h}^{R}_{j;a_j} \; . \label{eq:mpowii:splitting}
\end{equation}
with $N_j$ interaction terms crossing bond $j$, or equivalently:
\begin{equation}
	\left(
		\begin{array}{c}
			\hat{H}^{R}_{j-1} \\ \hat{h}^{R}_{j-1} \\ \mathbf{\hat{1}}^{R}_{j-1}
		\end{array}
	\right)
	=
	\tikzsetnextfilename{Zaletel_MPO_matrix_2}
%	\tikzset{external/export next=false}
	\begin{tikzpicture}
		[
			baseline=(MPO.center)
		]
		\begin{scope}
			\matrix (MPO) 
			[
				matrix of math nodes,
				left delimiter=(,
				right delimiter=),
				nodes in empty cells,
				nodes = 
				{
					minimum height=2em,
					anchor=center,
					text width=2em,
					text depth=.25ex,
					align=center,
					inner sep=0pt
				}
			]
			{
				\mathbf{\hat{1}}_{j} 	& \phantom{1} 	& \phantom{1}	& \phantom{1} \\
				\phantom{1}	     		& \phantom{1}	& \phantom{1} 	& \phantom{1} \\
				\phantom{1}	     		& \phantom{1}	& \phantom{1}	& \phantom{1} \\
				\phantom{1}				&		 	 	& \phantom{1} 	& \mathbf{\hat{1}}_{j} \\
			};
			\node [above=5pt of MPO-1-1] (IdL) {\scriptsize $1\phantom{_j}$};
			\node [above=5pt of MPO-1-4] (IdR) {\scriptsize $1\phantom{_j}$};
			\node at ($(IdL)!0.5!(IdR)$) {\scriptsize $N_{j}$};
			\node [left=15pt of MPO-1-1] (IdT) {\scriptsize $1\phantom{_j}$};
			\node [left=15pt of MPO-4-1] (IdB) {\scriptsize $1\phantom{_j}$};
			\node at ($(IdT)!0.5!(IdB)$)  {\scriptsize $N_{j-1}$};
			\draw [black,decorate,decoration={brace,mirror,raise=-2pt}] ([xshift=-15pt,yshift=-15pt]MPO-4-1.south west) -- node[below] {$\hat{W}_{j}$} ([xshift=15pt,yshift=-15pt]{MPO-4-4.south east});
		\end{scope}
		\begin{scope}[on background layer]
			\node [draw=blue!30, fill=blue!10, inner sep=-1pt, fit=(MPO-1-2)(MPO-1-3)] (cj) {};
			\node [draw=blue!30, fill=blue!10, inner sep=-1pt, fit=(MPO-1-4)] (dj) {};
			\node [draw=blue!30, fill=blue!10, inner sep=-1pt, fit=(MPO-2-2)(MPO-3-3)] (aj) {};
			\node [draw=blue!30, fill=blue!10, inner sep=-1pt, fit=(MPO-2-4)(MPO-3-4)] (bj) {};
		\end{scope}
		\begin{scope}
			\node [] (a) at (aj) {$\hat{A}_{j}$};
			\node [] (b) at (bj) {$\hat{B}_{j}$};
			\node [] (c) at (cj) {$\hat{C}_{j}$};
			\node [] (d) at (dj) {$\hat{D}_{j}$};
		\end{scope}
	\end{tikzpicture}
	\otimes
	\left(
		\begin{array}{c}
			\hat{H}^{R}_{j} \\ \hat{h}^{R}_{j} \\ \mathbf{\hat{1}}^{R}_{j}
		\end{array}
	\right) \; .\label{eq:mpowii:recursion}
\end{equation}
The operator-valued matrices $\hat{A}_{j},\hat{B}_{j},\hat{C}_{j},\hat{D}_{j}$ specify the local structure of the interactions described by $\hat{H}$ and the set of all matrices $\hat{W}_{1}\ldots\hat{W}_{L}$ define the MPO representation of $\hat{H}$.
The MPO representation $W^{\mathrm I}$ of $\hat U^{\mathrm{I}}$ is given by
\begin{equation}
	\hat{W}^{\mathrm I}_{j} = 
	\tikzsetnextfilename{WI_MPO_Matrix}
%	\tikzset{external/export next=false}
	\begin{tikzpicture}
		[
			baseline = (W_I.center)
		]
		\begin{scope}
			\matrix (W_I)
			[
				matrix of math nodes,
				left delimiter=(,
				right delimiter=),
				nodes in empty cells,
				nodes = 
				{
					minimum height=3em,
					anchor=center,
					text width=3em,
					text depth=.25ex,
					align=center,
					inner sep=0pt
				}
			]
			{
				\scriptstyle \mathbf{\hat{1}}_{j} + \delta\hat{D}_{j} 	& \phantom{1} & \phantom{1}	\\
				\phantom{1}												& \phantom{1} & \phantom{1}	\\
				\phantom{1}		 										& \phantom{1} & \phantom{1}	\\ 
			};
		\end{scope}
		\begin{scope}[on background layer]
			\node [draw=blue!30, fill=blue!10, inner sep=-1pt, fit=(W_I-1-1)] {};
			\node [draw=blue!30, fill=blue!10, inner sep=-1pt, fit=(W_I-1-2)(W_I-1-3)] (cj) {};
			\node [draw=blue!30, fill=blue!10, inner sep=-1pt, fit=(W_I-2-1)(W_I-3-1)] (bj) {};
			\node [draw=blue!30, fill=blue!10, inner sep=-1pt, fit=(W_I-2-2)(W_I-3-3)] (aj) {};
		\end{scope}
		\begin{scope}
			\node [] (c) at (cj) {$\sqrt{\delta}\hat{C}_{j}$};
			\node [] (b) at (bj) {$\sqrt{\delta}\hat{B}_{j}$};
			\node [] (a) at (aj) {$\hat{A}_{j}$};
		\end{scope}
	\end{tikzpicture} 
	\; .
\end{equation}
In this case the representation of $\hat{H}$ as finite state machine is particularly useful since it permits to directly deduce the MPO representation for $W^{\mathrm I}$.
The MPO bond dimension of $W^{\mathrm I}$ is $w -1$ with $w$ the bond dimension of $\hat{H}$.
Hence, this MPO can be applied numerically very efficiently.

However, the restriction of $W^{\mathrm I}$ to treat only non-overlapping local operator terms $\hat{H}_{j}$ is very strong and fails to even reproduce the correct time evolution of a purely on-site Hamiltonian. For example, $W^{\mathrm I}$ for $\hat{H}=\sum_{j}\hat{s}^{z}_{j}$ generates only operator strings $\hat{s}^{z}_{k}\cdots\hat{s}^{z}_{k+n}$ but not $\hat s^z_k \hat s^z_k$.
An improvement is to permit operator strings that overlap on one site:
\begin{equation}
	\hat{U}^{\mathrm{II}}(\delta) = 1 - \mathrm i \delta\sum_{j}\hat{H}_{j} - \frac{\delta^2}{2}{\sum_{j,k}}^{\prime\prime}\hat{H}_{j}\hat{H}_{k} +\cdots  \; ,
\end{equation}
where the double-primed sum only excludes terms $\hat{H}_{j},\hat{H}_{k}$ overlapping at more than one site, sharing a bond.
For instance, consider the expansion of the time-evolution operator for the $S=1/2$-Heisenberg chain: expressions of the form $\left(\hat{s}^{+}_{j}\hat{s}^{-}_{j+1}\right)\left(\hat{s}^{z}_{j+1}\right)$ are kept, while those of the form $\left(\hat{s}^{+}_{j}\hat{s}^{-}_{j+1}\right)\left(\hat{s}^{z}_{j}\hat{s}^{z}_{j+1}\right)$ are discarded.
Hence, the error is again of order $\delta^2$ but contributions with arbitrary powers of single-site terms are treated exactly.
There is no closed general MPO representation for $\hat{U}^{\mathrm{II}}$ but we can give an approximation which has an error $\mathcal{O}(\delta^3)$ and hence does not affect the second-order approximation of $\hat{U}^{\mathrm{II}}$.
In the following we will first explicitly demonstrate how to numerically construct the MPO representation $W^{\mathrm{II}}$ from the block-triangular structure  of the MPO representation of $\hat{H}$ and subsequently motivate the used formalism.
As for $\hat U^{\mathrm{I}}$, the MPO representation of $\hat U^{\mathrm{II}}$ is of the form
\begin{equation}
	\hat{W}^{\mathrm{II}}_{j} = 
	\left(
		\begin{array}{cc}
			\hat{W}^{\mathrm{II}}_{D_j} & \hat{W}^{\mathrm{II}}_{C_j} \\
			\hat{W}^{\mathrm{II}}_{B_j} & \hat{W}^{\mathrm{II}}_{A_j}
		\end{array}
	\right) \; .
\end{equation}
In order to construct the matrices $W^{\mathrm{II}}_{\left\{A_j,B_j,C_j,D_j\right\}}$ we employ transition amplitudes between hard-core bosonic states (cf.~\cref{sec:te:mpowii::derivation}) following the notation in Ref.~\cite{zaletel15:_time}.
Let $\mathcal{H}_{2,a_j}$ denote the $a_j$-th hard-core bosonic Hilbert space. $N_j$ of these spaces form $\mathcal{H}_{c}=\bigotimes_{a_j=1}^{N_j}\mathcal{H}_{2,a_j}$ and we work in the joint Hilbert space $\mathcal{H}_{c}\otimes\mathcal{H}_{\bar{c}}$ which spans $2 N_j$ individual hard-core bosonic Hilbert spaces. The ladder operators acting on $\mathcal{H}_{c}$ and $\mathcal{H}_{\overbar{c}}$ are $\hat{c}^{\dagger}_{a_j},\hat{c}^{\nodagger}_{a_j}, \hat{\bar{c}}^{\dagger}_{\bar{a}_j}$ and $\hat{\bar{c}}^{\nodagger}_{\bar{a}_j}$ respectively.

The generator of the matrix elements is a map on the joint bosonic and physical Hilbert spaces $\mathcal{H}_{c}\otimes\mathcal{H}_{\bar{c}}\otimes\mathcal{H}_{\mathrm{phys}}$ and given by the operator-valued exponentials 
\begin{equation}
	\hat{\Phi}_{j;a_j,\bar{a}_j} = e^{\hat{F}_{j;a_j,\bar{a}_j}} = e^{ \hat{c}^{\dagger}_{a_j}\hat{\bar{c}}^{\dagger}_{\bar{a}_j}\hat{A}^{\nodagger}_{j;a_j,\bar{a}_j} + \sqrt{\delta}(\hat{c}^{\dagger}_{a_j}\hat{B}^{\nodagger}_{j;a_j} + \hat{\bar{c}}^{\dagger}_{\bar{a}_j}\hat{C}^{\nodagger}_{j;\bar{a}_j}) + \delta\hat{D}^\nodagger_j}\; .
\end{equation}
Denoting the combined bosonic vacuum state by $\ket{0}\otimes\ket{\bar{0}} \equiv \ket{0,\bar{0}}$ the following transition amplitudes determine the operator-valued entries of the MPO representation $\hat{W}^{\mathrm{II}}_j$:
\begin{align}
	\hat{W}^{\mathrm{II}}_{A_j;a_j,\bar{a}_j} 
		&= 
			\bra{0,\bar{0}}
			\hat{c}_{a_j}
			\hat{\bar{c}}_{\bar{a}_j} 
			\hat{\Phi}_{j;a_j,\bar{a}_j} 
			\ket{0,\bar{0}} 
			= 
				\bra{0}
				\hat{c}_{a_j}
				\hat{\bar c}_{\bar{a}_j}
				e^{ 
					\hat{c}^{\dagger}_{a_j}
					\hat{\bar{c}}^{\dagger}_{\bar{a}_j}
					\hat{A}^\nodagger_{j;a_j,\bar{a}_j} 
					+ 
					\sqrt{\delta}
					(
						\hat{c}^{\dagger}_{a_j}
						\hat{B}^\nodagger_{j;a_j}
						+
						\hat{\bar{c}}^{\dagger}_{\bar{a}_j}
						\hat{C}^\nodagger_{j;\bar{a}_j}
					)
					+
					\delta\hat{D}^\nodagger_j
				}
				\ket{0} 
				\label{eq:WII:matrix-element-A}
\\
	\hat{W}^{\mathrm{II}}_{B_j;a_j} 
		&= 
			\bra{0,\bar{0}}
			\hat{c}_{a_j} 
			\hat{\Phi}_{j;a_j,\bar{a}_j} 
			\ket{0,\bar{0}} 
			= 
				\bra{0}
				\hat{c}_{a_j} 
				e^{
					\sqrt{\delta}
					\hat{c}^{\dagger}_{a_j}
					\hat{B}^\nodagger_{j;a_j}
					+ 
					\delta\hat{D}^\nodagger_j
				}
				\ket{0}
				\label{eq:WII:matrix-element-B}
\\
	\hat{W}^{\mathrm{II}}_{C_j;\bar{a}_j} 
		&= 
			\bra{0,\bar{0}}
			\hat{\bar{c}}_{\bar{a}_j} 
			\hat{\Phi}_{j;a_j,\bar{a}_j} 
			\ket{0,\bar{0}} 
			= 
				\bra{\bar{0}}
				\hat{\bar{c}}_{\bar{a}_j} 
				e^{ 
					\sqrt{\delta}
					\hat{\bar{c}}^{\dagger}_{\bar{a}_j}
					\hat{C}^\nodagger_{j;\bar{a}_j}
					+ 
					\delta\hat{D}^\nodagger_j
				} 
				\ket{\bar{0}} 
				\label{eq:WII:matrix-element-C}
\\
	\hat{W}^{\mathrm{II}}_{D_j} 
		&= 
			\bra{0,\bar{0}}
			\hat{\Phi}_{j;a_j,\bar{a}_j}
			\ket{0,\bar{0}} 
			= 
				e^{ 
					\delta\hat{D}^\nodagger_j
				} 
\; . \label{eq:WII:matrix-element-D}
\end{align}
For clarity, in the following we will explicitly demonstrate the calculation of the elements of $W^{\mathrm{II}}_{A_j;a_j,\bar{a}_j}$.
Explicitly, we use
\begin{align}
  \ket{0}^{\nodagger}_{a_j} &=\left(\begin{array}{c} 1 \\ 0 \end{array}\right) \in \mathcal{H}_{2,a_j}, & 
  \hat{c}^{\nodagger}_{a_j} &= \left(\begin{array}{cc} 0 & 1 \\ 0 & 0 \end{array} \right), & 
  \hat{c}^{\dagger}_{a_j} &= \left(\begin{array}{cc} 0 & 0 \\ 1 & 0 \end{array} \right), & 
  \mathbf{\hat{1}}^{\nodagger}_{a_j} &= \left(\begin{array}{cc} 1 & 0 \\ 0 & 1 \end{array} \right)
\end{align}
to represent the hard-core bosonic states and operators.
Since $\hat{F}_{j;a_j,\bar{a}_j}$ only contains creation operators for the modes $a_j,\bar{a}_j$ the vacuum expectation value over all $2N_{j}$ bosonic modes can be simplified to the relevant modes, i.e., we can replace $\ket{0,\bar{0}} \rightarrow \ket{0_{a_j},\bar{0}_{\bar{a}_j}}$.
Therefore, the generator is obtained by exponentiating the matrix
\begin{align}
	\hat{F}^{\nodagger}_{j;a_j,\bar{a}_j} &= 
	\hat{c}^{\dagger}_{a_j}\otimes\hat{\bar{c}}^{\dagger}_{\bar{a}_j} \hat{A}^{\nodagger}_{j;a_j,\bar{a}_j} + \sqrt{\delta}\left(\hat{c}^{\dagger}_{a_j}\otimes\mathbf{\hat{1}}^{\nodagger}_{\bar{a}_j}\hat{B}^{\nodagger}_{j;a_j} + \mathbf{\hat{1}}^{\nodagger}_{a_j}\otimes \hat{\bar{c}}^{\dagger}_{\bar{a}_j} \hat{C}^{\nodagger}_{j;\bar{a}_j} \right) + \mathbf{\hat{1}}^{\nodagger}_{a_j}\otimes\mathbf{\hat{1}}^{\nodagger}_{\bar{a}_j} \delta\hat{D}^\nodagger_j \\
	&=
	\left(
		\begin{array}{cccc}
			\delta\hat{D}^\nodagger_j & 0 & 0 & 0 \\
			\sqrt{\delta}\hat{C}_{j;\bar{a}_j} & \delta\hat{D}^\nodagger_j & 0 & 0 \\
			\sqrt{\delta}\hat{B}_{j;a_j} & 0 & \delta\hat{D}^\nodagger_j & 0 \\
			\hat{A}_{j;a_j,\bar{a}_j} & \sqrt{\delta}\hat{B}_{j;a_j} & \sqrt{\delta}\hat{C}_{j;\bar{a}_j} & \delta\hat{D}^\nodagger_j
		\end{array}
	\right) \; ,
\end{align}
which has been truncated to the relevant bosonic Hilbert spaces $\mathcal{H}_{2,a_j},\mathcal{H}_{2,\bar{a}_j}$. %, too.
The entries of $\hat{W}^{\mathrm{II}}_{A_j;a_j,\bar{a}_j}$ are then obtained by evaluating the power series of the matrix exponential and calculating the vacuum expectation value
\begin{align}
	\hat{W}^{\mathrm{II}}_{A_j;a_j,\bar{a}_j} &= \bra{0,\bar{0}}\hat{c}_{a_j}\hat{\bar{c}}_{\bar{a}_j} \hat{\Phi}_{j;a_j,\bar{a}_j} \ket{0,\bar{0}} = \bra{1_{a_j},\bar{1}_{\bar{a}_j}} \hat{\Phi}_{j;a_j,\bar{a}_j} \ket{0,\bar{0}} \notag \\
	&= \left(\begin{array}{cccc} 0 & 0 & 0 & 1 \end{array}\right) 
	\exp\left\{
	\left(
		\begin{array}{cccc}
			\delta\hat{D}^\nodagger_j & 0 & 0 & 0 \\
			\sqrt{\delta}\hat{C}_{j;\bar{a}_j} & \delta\hat{D}^\nodagger_j & 0 & 0 \\
			\sqrt{\delta}\hat{B}_{j;a_j} & 0 & \delta\hat{D}^\nodagger_j & 0 \\
			\hat{A}_{j;a_j,\bar{a}_j} & \sqrt{\delta}\hat{B}_{j;a_j} & \sqrt{\delta}\hat{C}_{j;\bar{a}_j} & \delta\hat{D}^\nodagger_j
		\end{array}
	\right)\right\}
	\left(\begin{array}{c} 1 \\ 0 \\ 0 \\ 0 \end{array}\right)\; .
\end{align}
A compact notation for all matrix elements of \wii can be obtained if we let the annihilation operators $\hat{c}_{a_j},\hat{c}_{\overbar{a}_j}$ in \cref{eq:WII:matrix-element-A,eq:WII:matrix-element-B,eq:WII:matrix-element-C,eq:WII:matrix-element-D} act on the bra $\bra{0,\overbar{0}}$ and define the formal symbol $\mathcal{S}_j\in\left\{A_j,B_j,C_j,D_j \right\}$
\begin{align}
	\hat{W}^{\mathrm{II}}_{\mathcal{S}_j;a_j,\overbar{a}_j} 
	&=
	\left( \begin{array}{cccc} \delta_{\mathcal{S}_j,D_j} & \delta_{\mathcal{S}_j,C_j} & \delta_{\mathcal{S}_j,B_j} & \delta_{\mathcal{S}_j,A_j} \end{array}\right)
	\exp\left\{
	\left(
		\begin{array}{cccc}
			\delta\hat{D}^\nodagger_j & 0 & 0 & 0 \\
			\sqrt{\delta}\hat{C}_{j;\bar{a}_j} & \delta\hat{D}^\nodagger_j & 0 & 0 \\
			\sqrt{\delta}\hat{B}_{j;a_j} & 0 & \delta\hat{D}^\nodagger_j & 0 \\
			\hat{A}_{j;a_j,\bar{a}_j} & \sqrt{\delta}\hat{B}_{j;a_j} & \sqrt{\delta}\hat{C}_{j;\bar{a}_j} & \delta\hat{D}^\nodagger_j
		\end{array}
	\right)\right\}
	\left(\begin{array}{c} 1 \\ 0 \\ 0 \\ 0 \end{array}\right)\; .
\end{align}
Hence, the $N^{2}_j$ exponentials $\exp\left\{\hat{F}_{j;a_j,\overbar{a}_j}\right\}$ already contain all relevant information to construct the stepper \wii and in particular there is no need to calculate different exponentials as suggested by \cref{eq:WII:matrix-element-A} -- \cref{eq:WII:matrix-element-D}.

\subsubsection{\label{sec:te:mpowii::derivation}Detailed derivation of the $\wii$ representation}
We will present the construction of $\wii$ in more detail, following Ref.~\cite{zaletel15:_time}, in this section.
Before digging into the details, let us briefly sketch the derivation:
The goal is to find an MPO representation of the time stepper $\hat{U}^{\mathrm{II}}(\delta)$ capturing interaction terms in the series expansion as described above.
The corresponding MPO site tensors $\wii$ are obtained by making use of the MPO recursion \cref{eq:mpowii:recursion} to factorize the exponential $e^{-\mathrm i \delta \hat{H}}$.
The factorization itself is performed exploiting complex Gaussian integrals via the introduction of auxiliary fields $\phi_{j},\overbar{\phi}_{j}$ on each bond which are integrated over.
The resulting MPO is of the form
\begin{align}
	W^{\mathrm{II}} &= \int D[\phi_{1},\overbar{\phi}_{1}] U_{\overbar{\phi}_{1}}e^{-\overbar{\phi}_{1}\phi_{1}}\int D[\phi_{2},\overbar{\phi}_{2}]U_{\phi_{1},\overbar{\phi}_{2}}e^{-\overbar{\phi}_{2}\phi_{2}}U_{\phi_{2},\overbar{\phi}_{3}}\cdots \;,
\end{align}
that is, the bond indices $\phi_{j},\overbar{\phi}_{j}$ are continuous degrees of freedom.
In a final step, these bond indices are discretized using bosonic coherent state path integrals yielding the desired expression for the MPO site tensors.
We begin by employing the bond representation of $\hat{H}$ (compare \cref{eq:mpowii:splitting}) and write the operator as a formal scalar product over auxiliary degrees of freedom $a_j = 1\ldots N_j$, where $N_j$ again is the number of interaction terms crossing the bond $j$,
\begin{align}
	\hat{H} &= \hat{H}^{L}_{j-1}\otimes \mathbf{\hat{1}}^{R}_{j} + \mathbf{\hat{1}}^{L}_{j}\otimes \hat{H}^{R}_{j+1} + 
	\left(\begin{array}{ccc}\hat{h}^{L}_{j;1} & \ldots & \hat{h}^{L}_{j;N_j} \end{array}\right)
	\left(\begin{array}{c}\hat{h}^{R}_{j;1} \\ \vdots \\ \hat{h}^{R}_{j;N_j} \end{array} \right) \\
	&\equiv 
	\hat{H}^{L}_{j-1}\otimes \mathbf{\hat{1}}^{R}_{j} + \mathbf{\hat{1}}^{L}_{j}\otimes \hat{H}^{R}_{j+1} + \hat{J}^{t}_{j} \hat{\overbar J}^{\phantom{t}}_{j}
\end{align}
and the product between the entries of the operator-valued ``vectors'' $\hat{J}_{j}$ and $\hat{\overbar J}_{j}$ are tensor products between operators acting on partitioned Hilbert spaces to the left and right of the bond $j$, respectively.
Next, we introduce complex vector fields $\phi_{j;a_j}$ and their complex conjugate $\overbar \phi_{j;a_j}$ and define a mapping from the auxiliary indices into the full Hilbert space
\begin{align}
  \hat{J}^{t}_{j}\cdot \overbar{\phi}^{\phantom{t}}_{j}: \quad \left(\mathcal{H}^{L}_{j} \rightarrow \mathcal{H}^{L}_{j}\right)^{\otimes N_j} \times \mathbb{C}^{N_j} & \longrightarrow \left( \mathcal{H} \to \mathcal{H}\right) \notag \\
  \left(\hat{J}_{j} , \overbar{\phi}_{j}\right) & \longmapsto \hat{J}^{t}_{j}\cdot \overbar{\phi}^{\phantom{t}}_{j} := \hat{h}^{L}_{j;1}\overbar{\phi}^{\phantom{L}}_{j;1}\mathbf{\hat 1}^{R}_{j} + \cdots + \hat{h}^{L}_{j;N_j} \overbar{\phi}^{\vphantom{R}}_{j;N_j}\mathbf{\hat 1}^{R}_{j} \\
  \phi_{j}\cdot \hat{\overbar J}_{j}: \quad \mathbb{C}^{N_j} \times \left(\mathcal{H}^{R}_{j} \rightarrow \mathcal{H}^{R}_{j}\right)^{\otimes N_j} &\longrightarrow  \left( \mathcal{H} \to \mathcal{H}\right) \notag\\
  \left(\phi_{j}, \hat{\overbar J}_{j}\right) & \longmapsto \phi_{j}\cdot \hat{\overbar J}_{j} := \mathbf{\hat 1}^{L}_{j}\phi^{\vphantom{R}}_{j;1}\hat{h}^{R}_{j;1} + \cdots + \mathbf{\hat 1}^{L}_{j}\phi^{\vphantom{R}}_{j;N_j}\hat{h}^{R}_{j;N_j} \;,
\end{align}
where we will suppress the identities acting on the left and right partitions of the Hilbert space in the following.
We then use complex Gaussian integrals to decouple each left and right interaction term
\begin{align}
	e^{\hat{J}_{j;a_j} \hat{\overbar J}_{j;a_j}} &= \frac{1}{\pi}\int d\phi_{j;a_j}d\overbar{\phi}_{j;a_j} e^{-\overbar{\phi}_{j;a_j}\cdot\phi_{j;a_j} + \hat{J}_{j;a_j}\cdot\overbar{\phi}_{j;a_j} + \phi_{j;a_j}\cdot\hat{\overbar J}_{j;a_j} } \;.
\end{align}
Furthermore, we factorize the operator exponential $e^{-\mathrm i\delta \hat{H}}$ (setting $\tau = -\mathrm i \delta$)
\begin{align}
	e^{\tau\hat{H}} &= e^{\tau \left( \hat{H}^{L}_{j-1}\otimes \mathbf{\hat{1}}^{R}_{j} + \mathbf{\hat{1}}^{L}_{j}\otimes \hat{H}^{R}_{j+1} + \hat{J}^{t}_{j} \hat{\overbar J}^{\phantom{t}}_{j} \right)} \notag\\
	&= e^{\tau \hat{H}^{L}_{j-1}\otimes \mathbf{\hat{1}}^{R}_{j}} e^{\tau \mathbf{\hat{1}}^{L}_{j}\otimes \hat{H}^{R}_{j+1}} e^{\tau \hat{J}^{t}_{j} \hat{\overbar J}^{\phantom{t}}_{j}} + \mathcal{O}(\tau^2) \label{eq:te:wii:Trotterization}
\end{align}
where the error occurs at second order in $\tau$ if the operators $\hat{J}_{j},\hat{\overbar J}_{j},\hat{H}^{L}_{j-1},\hat{H}^{R}_{j+1}$ do not commute.\footnote{The commutators need to be evaluated on the full Hilbert space by completing the partitioned operators $\hat{J}_{j},\hat{\overbar J}_{j},\hat{H}^{L}_{j-1},\hat{H}^{R}_{j+1}$ which is achieved by taking the appropriate tensor products with $\mathbf{\hat 1}^{L}_{j}$ and $\mathbf{\hat 1}^{R}_{j}$.}
Exploiting $N_j$ complex Gaussian integrals and defining $D\left[\phi_{j},\overbar{\phi}_{j} \right] = \prod_{a_j} \frac{d\phi_{j;a_j}}{\sqrt{\pi}}\frac{d\overbar{\phi}_{j;a_j}}{\sqrt{\pi}}$ we obtain
\begin{align}
	e^{\tau\hat{H}} 
	&= 
	\int D\left[\phi^{\vphantom{R}}_{j},\overbar{\phi}^{\vphantom{R}}_{j} \right] e^{\tau\hat{H}^{L}_{j-1}\otimes \mathbf{\hat{1}}^{R}_{j}} e^{\tau\mathbf{\hat{1}}^{L}_{j}\otimes \hat{H}^{R}_{j+1}} e^{-\overbar{\phi}^{\vphantom{R}}_{j}\cdot\phi^{\vphantom{R}}_{j} + \sqrt{\tau}\hat{J}^{\vphantom{R}}_{j}\cdot\overbar{\phi}^{\vphantom{R}}_{j} + \sqrt{\tau}\phi^{\vphantom{R}}_{j}\cdot\hat{\overbar J}^{\vphantom{R}}_{j}} + \mathcal{O}(\tau^{2})\notag \\
	&=
	\int D\left[\phi_{j},\overbar{\phi}_{j} \right] e^{\tau\hat{H}^{L}_{j-1} \cdot \mathbf{\hat{1}}^{R}_{j} + \sqrt{\tau}\hat{J}_{j}\cdot\overbar{\phi}_{j}} e^{-\overbar{\phi}_{j}\cdot\phi_{j}} e^{\tau\mathbf{\hat{1}}^{L}_{j} \cdot \hat{H}^{R}_{j+1} + \sqrt{\tau}\phi_{j}\cdot\hat{\overbar J}_{j}} + \mathcal{O}(\tau^{2}) \; .
\end{align}
Here we used that the identities $\mathbf{\hat 1}^{L}_{j}$ and $\mathbf{\hat 1}^{R}_{j}$ can also be interpreted as maps from an auxiliary index into the full Hilbert space.
Now the MPO recursion \cref{eq:mpowii:recursion} can be applied to, e.g., shift the active bond $j\rightarrow j+1$ in the third exponential
\begin{align}
	e^{\tau \mathbf{\hat{1}}^{L}_{j}\cdot \hat{H}^{R}_{j+1} + \sqrt{\tau}\phi^{\vphantom{R}}_{j}\hat{\overbar J}^{\vphantom{R}}_{j}} 
	&= 
	e^{ \tau \mathbf{\hat 1}^{\vphantom{R}}_{j}\cdot \hat{H}^{R}_{j+1} + \sqrt{\tau}\phi^{\vphantom{R}}_{j}\cdot\hat{A}^{\vphantom{R}}_{j}\hat{\overbar{J}}^{\vphantom{R}}_{j+1} + \sqrt{\tau}\phi^{\vphantom{R}}_{j}\cdot\hat{B}^{\vphantom{R}}_{j}\mathbf{\hat 1}^{R}_{j+1} + \tau \hat{C}^{\vphantom{R}}_{j}\hat{\overbar{J}}^{\vphantom{R}}_{j+1} + \tau \hat{D}^{\vphantom{R}}_{j}\cdot \mathbf{\hat 1}^{R}_{j+1}}\; .
\end{align}
From now on we replace the formal dot product $\hat{D}^{\vphantom{R}}_{j}\cdot \mathbf{\hat 1}^{R}_{j+1}$ by its action on the full Hilbert space which reduces for the case of $\hat{D}^{\vphantom{R}}_{j}$ to on-site terms only.
The same holds for the dot product $\phi^{\vphantom{R}}_{j}\cdot\hat{B}^{\vphantom{R}}_{j}\mathbf{\hat 1}^{R}_{j+1} \equiv \phi^{\vphantom{R}}_{j}\cdot\hat{B}^{\vphantom{R}}_{j}$ which only connects sites to the left of $j$.
In the exponent the summands $\sqrt{\tau}\phi^{\vphantom{R}}_{j}\cdot\hat{B}^{\vphantom{R}}_{j}\cdot\mathbf{\hat 1}^{R}_{j+1} ,\tau \hat{D}^{\vphantom{R}}_{j}\cdot \mathbf{\hat 1}^{R}_{j+1}$ and $\tau \mathbf{\hat 1}_{j}\cdot \hat{H}^{R}_{j+1}$ already act separately on site $j$ and the right partition of the Hilbert space $\mathcal{H}^{R}_{j+1}$.
In order to separate the remaining summands, too, we introduce another set of auxiliary fields $\phi^{\vphantom{R}}_{j+1},\overbar{\phi}^{\vphantom{R}}_{j+1}$ by insertion of more complex Gaussian integrals
\begin{gather}
	e^{\sqrt{\tau}\phi_{j}\cdot\hat{A}_{j}\hat{\overbar{J}}_{j+1} + \tau \hat{C}_{j}\cdot\hat{\overbar{J}}_{j+1}}
	= 
	\int D\left[\phi_{j+1},\overbar{\phi}_{j+1} \right] e^{-\overbar{\phi}_{j+1}\phi_{j+1}} e^{\phi_{j}\cdot\hat{A}_{j}\cdot\overbar{\phi}_{j+1} + \sqrt{\tau}\hat{C}_{j}\cdot\overbar{\phi}_{j+1}+\sqrt{\tau}\phi_{j+1}\cdot \hat{\overbar{J}}_{j+1}} \notag \\
	\Rightarrow 
	e^{\tau \mathbf{\hat{1}}^{L}_{j}\cdot \hat{H}^{R}_{j+1} + \sqrt{\tau}\phi^{\vphantom{R}}_{j}\cdot\hat{\overbar J}^{\vphantom{R}}_{j}} = \int D\left[\phi_{j+1},\overbar{\phi}_{j+1} \right] U_{\phi_{j},\overbar{\phi}_{j+1}}e^{-\overbar{\phi}^{\vphantom{R}}_{j+1}\phi^{\vphantom{R}}_{j+1}} e^{\tau\mathbf{\hat 1}^{L}_{j+1}\cdot\hat{H}^{R}_{j+2} + \sqrt{\tau}\phi^{\vphantom{R}}_{j+1}\cdot\hat{\overbar{J}}^{\vphantom{R}}_{j+1}}
\end{gather}
with $U_{\phi_{j},\overbar{\phi}_{j+1}} = e^{\phi_{j}\cdot\hat{A}_{j}\cdot\overbar{\phi}_{j+1} + \sqrt{\tau}\phi_{j}\cdot\hat{B}_{j} + \sqrt{\tau}\hat{C}_{j}\cdot\overbar{\phi}_{j+1} + \tau\hat{D}_{j}}$.
The last equation defines a recursion 
\begin{align}
	e^{\tau \mathbf{\hat{1}}^{L}_{j}\cdot \hat{H}^{R}_{j+1} + \sqrt{\tau}\phi_{j}\cdot\hat{\overbar J}_{j}} &= \mathcal{F}_{j,j+1}\left[e^{\tau \mathbf{\hat{1}}^{L}_{j+1}\cdot \hat{H}^{R}_{j+2} + \sqrt{\tau}\phi_{j+1}\cdot\hat{\overbar J}_{j+1}}\right]
\end{align}
generated by the integration of $U_{\phi_{j},\overbar{\phi}_{j+1}}e^{-\overbar{\phi}_{j+1}\phi_{j+1}}$ over the continuous bond degrees of freedom $\phi_{j},\overbar{\phi}_{j+1}$.
Iterating through the whole system we finally obtain
\begin{align}
	e^{\tau\hat{H}} &= \int D[\phi_{1},\overbar{\phi}_{1}] U_{\overbar{\phi}_{1}}e^{-\overbar{\phi}_{1}\phi_{1}}\int D[\phi_{2},\overbar{\phi}_{2}]U_{\phi_{1},\overbar{\phi}_{2}}e^{-\overbar{\phi}_{2}\phi_{2}}U_{\phi_{2},\overbar{\phi}_{3}}\cdots + \mathcal{O}(\tau^2),
\end{align}
which is a first-order approximation in $\tau$ of $e^{\tau\hat{H}}$ in terms of an MPO with continuous bond labels $\phi_{j} = \left(\phi_{1},\cdots ,\phi_{N_j} \right)$.
In a final step the continuous bond labels $\phi_{a_j}$ are transformed into discrete ones.
We note that for a set of $N_j$ bosonic ladder operators $\hat{c}^{\nodagger}_{a_j},\hat{c}^{\dagger}_{a_j}$ we can rewrite the tensors at site $j$ as expectation values for coherent states 
\begin{align}
	\ket{\phi_{a_j}} &= \sum_{n}\frac{\phi_{a_j}^{n}}{n!} \left(\hat{c}^{\dagger}_{a_j}\right)^{n} \ket{0} \;,
\end{align}
which have the properties
\begin{align}
	 \hat{c}^{\nodagger}_{a_j}\ket{\phi^{\nodagger}_{a_j}} = \phi^{\nodagger}_{a_j}\ket{\phi^{\nodagger}_{a_j}}, \quad
	\bra{\phi^{\nodagger}_{a_j}}\hat{c}^{\dagger}_{a_j} = \bra{\phi^{\nodagger}_{a_j}} \overbar{\phi}^{\nodagger}_{a_j}, \quad \braket{\phi^{\nodagger}_{a_j}|\phi^{\nodagger}_{a_j}} = e^{\overbar{\phi}_{a_j} \phi_{a_j}}\; .
\end{align}
For instance, we may consider the case when $N_j \equiv 1$.
Then a single bond label $\phi_{j;a_j}\equiv z_{j}$ is sufficient and (abbreviating $\hat{c}^{\nodagger}_{a_j}\equiv \hat{c}^{\nodagger}_{j}$, $\hat{c}^{\dagger}_{a_j}\equiv \hat{c}^{\dagger}_{j}$) we have at the bond $(j,j+1)$
\begin{align}
	e^{z^{\nodagger}_{j} \hat{A}^{\nodagger}_{j}\overbar{z}^{\nodagger}_{j+1} + \sqrt{\tau}z^{\nodagger}_{j} \hat{B}^{\nodagger}_{j} + \sqrt{\tau}\hat{C}^{\nodagger}_{j}\overbar{z}^{\nodagger}_{j+1} + \tau \hat{D}^{\nodagger}_{j}}
	&= 
	\braket{ \overbar{z}_{j}, z_{j+1} | e^{\hat{c}^{\dagger}_{j} \hat{A}^{\nodagger}_{j} \hat{c}^{\dagger}_{j+1} + \sqrt{\tau}\hat{c}^{\dagger}_{j}\hat{B}^{\nodagger}_{j} + \sqrt{\tau}\hat{C}^{\nodagger}_{j}\hat{c}^{\dagger}_{j+1} + \tau \hat{D}^{\nodagger}_{j}} | 0, 0} \notag \\
	&=
	\sum_{n_{j},n_{j+1}}
	\underbrace{\braket{ n_{j}, \overbar{n}_{j+1} | e^{\hat{c}^{\dagger}_{j} \hat{A}^{\nodagger}_{j} \hat{c}^{\dagger}_{j+1} + \sqrt{\tau}\hat{c}^{\dagger}_{j}\hat{B}^{\nodagger}_{j} + \sqrt{\tau}\hat{C}^{\nodagger}_{j}\hat{c}^{\dagger}_{j+1} + \tau \hat{D}^{\nodagger}_{j}} | 0, 0}}_{\equiv \hat{U}_{n_{j},\overbar{n}_{j+1}}} \frac{ z_{j}^{n_{j}} \overbar{z}_{j+1}^{\overbar{n}_{j+1}}}{\sqrt{n_{j}!\, \overbar{n}_{j+1}!}} \; .
\end{align}
In the last equality we expanded the coherent states $\ket{z_{j}},\ket{z_{j+1}}$ in the bosonic occupations number basis $\ket{n_{j}},\ket{\overbar{n}_{j+1}}$.
Now, we turn to the integral over one pair of bond labels $z_{j},\overbar{z}_{j}$, considering only those factors $\frac{z^{k_{j}}_{j}}{\sqrt{k_j!}}$ which contribute to the integration,
\begin{align}
	\int\frac{dz_{j}d\overbar{z}_{j}}{\pi}\hat{U}_{z_{j-1},\overbar{z}_{j}}e^{-\overbar{z}_{j}z_{j}}\hat{U}_{z_{j},\overbar{z}_{j+1}}
	&=
	\sum_{n_{j},\overbar{n}_{j}}\int \frac{dz_{j}d\overbar{z}_{j}}{\pi} \hat{U}_{n_{j-1}, \overbar{n}_{j}} e^{-\overbar{z}_{j}z_{j}} \hat{U}_{n_{j},\overbar{n}_{j+1}} \frac{\overbar{z}^{\overbar{n}_{j}}_{j} z^{n_{j}}_{j}}{\sqrt{\overbar{n}_{j}!\,n_{j}!}} \notag \\
	&= \sum_{n_{j},\overbar{n}_{j}} \hat{U}_{n_{j-1},\overbar{n}_{j}}\hat{U}_{n_{j},\overbar{n}_{j+1}} \int_{0}^{\infty} d\rho_{j} \frac{\rho^{\overbar{n}_{j} + n_{j}+1}_{j}}{\sqrt{\overbar{n}_{j}!\,n_{j}!}} e^{-\rho_{j}^{2}}\int_{0}^{2\pi}\frac{d \varphi_{j}}{\pi}e^{\mathrm i \varphi_{j}(n_{j}-\overbar{n}_{j})} \notag \\
	&=
	\sum_{n_{j}} \hat{U}_{n_{j-1},n_{j}}\hat{U}_{n_{j},\overbar{n}_{j+1}}  \frac{2}{n_{j}!} \underbrace{\int_{0}^{\infty} d\rho \rho^{2n_{j}+1}e^{-\rho^{2}}}_{=\frac{1}{2}\Gamma(n_{j})} \notag \\
	&=
	\sum_{n_{j}} \hat{U}_{n_{j-1},n_{j}}\hat{U}_{n_{j},\overbar{n}_{j+1}} \; .
\end{align}
We parametrized the complex integration in polar coordinates in the second identity and exploited the representation of the $\Gamma$ function in the last line.
For the general case $N_j>1$ we define the boson occupation number $\mathbf{n}_j = (n_{1}\cdots n_{N_j})$ for each bond.
Since we can always reorder the $\phi_{a_j}$, we can perform the same integrals for each continuous bond label so that we finally find
\begin{align}
	\hat{U}(\delta) &= \sum_{\mathbf{n}_{1}\cdots \mathbf{n}_{L}} \hat{U}_{\mathbf{\overbar n}_{1}} \hat{U}_{\mathbf{n}_{1},\mathbf{\overbar n}_{2}} \cdots + \mathcal{O}(\delta^2),
\end{align}
yielding the desired MPO representation.
A pedestrian way to obtain a compact MPO representation from this very formal derivation is to analyze $\hat{U}_{\mathbf{n}_{j-1},\mathbf{\overbar n}_{j}}$ when truncating the bosonic Hilbert spaces to a maximal boson occupation $b$, i.e.,  $n_{j;a_j} \in \left\{0,\ldots, b\right\}$.
We define bosonic field operators $\hat{\varphi}^{\dagger}_j = \left(\hat{c}^{\dagger}_{1} \cdots \hat{c}^{\dagger}_{L} \right)$ and denote the maximal occupation number by an upper index $(b)$.
Then, for each $b$ we have operator-valued bosonic matrix elements
\begin{align}
	\hat{U}^{(b)}_{j; \mathbf{n}_{j},\mathbf{\overbar n}_{j+1}} &= \braket{\mathbf{n}_{j},\mathbf{\overbar n}_{j+1}| \underbrace{e^{\hat{\varphi}^{\dagger}_{j} \cdot \hat{A}^{\nodagger}_{j} \cdot \hat{\overbar \varphi}^{\dagger}_{j+1} + \sqrt{\tau}\hat{\varphi}^{\dagger}_{j}\cdot\hat{B}^{\nodagger}_{j} + \sqrt{\tau}\hat{C}^{\nodagger}_{j} \cdot \hat{\overbar \varphi}^{\dagger}_{j+1} + \tau \hat{D}^{\nodagger}_{j}}}_{\equiv \hat{\Phi}_{j}} | \mathbf{0},\mathbf{\overbar 0}}
\end{align}
of maps $\hat{\Phi}_{j}$ on the joint Hilbert spaces of $b$ bosons $\mathcal{H}_{b;a_j}$ for every $a_j\in\left\{1,\ldots,N_j\right\}$ at each bond and the physical Hilbert space $\mathcal{H}_{\mathrm{phys}}$ at each site
\begin{align}
	\hat{\Phi}_{j}: \bigotimes_{a_j}\mathcal{H}_{b;a_j} \otimes \bigotimes_{\overbar{a}_j}\mathcal{H}_{\overbar{b}; \overbar{a}_j}\otimes \mathcal{H}_{\mathrm{phys}} &\longrightarrow \bigotimes_{a_j}\mathcal{H}_{b;a_j} \otimes \bigotimes_{\overbar{a}_j}\mathcal{H}_{\overbar{b}; \overbar{a}_j}\otimes \mathcal{H}_{\mathrm{phys}}\; .
\end{align}
Let us consider $b\equiv0$, then the matrix elements collapse to pure on-site terms $\hat{U}^{(0)}_{j;\mathbf{0},\mathbf{\overbar 0}} = \braket{\mathbf{0},\mathbf{\overbar 0}| \hat{\Phi}_{j} | \mathbf{0},\mathbf{\overbar 0}} = e^{\tau \hat{D}_{j}}$.
The next order $b \equiv 1$ additionally collects contributions from all interactions crossing either the bond $(j-1,j)$ or $(j,j+1)$
\begin{equation}
	\hat{U}^{(1)}_{j;\mathbf{0},\mathbf{\overbar 0}} = \braket{\mathbf{0},\mathbf{\overbar 0}| \hat{\Phi}_{j} |\mathbf{0},\mathbf{\overbar 0}}, \quad
	\hat{U}^{(1)}_{j;\mathbf{0},\mathbf{\overbar 1}} = \braket{\mathbf{0},\mathbf{\overbar 1}| \hat{\Phi}_{j} |\mathbf{0},\mathbf{\overbar 0}}, \quad
	\hat{U}^{(1)}_{j;\mathbf{1},\mathbf{\overbar 0}} = \braket{\mathbf{1},\mathbf{\overbar 0}| \hat{\Phi}_{j} |\mathbf{0},\mathbf{\overbar 0}}, \quad
	\hat{U}^{(1)}_{j;\mathbf{1},\mathbf{\overbar 1}} = \braket{\mathbf{1},\mathbf{\overbar 1}| \hat{\Phi}_{j} |\mathbf{0},\mathbf{\overbar 0}} \; ,
\end{equation}
which can be arranged as an operator-valued matrix
\begin{align}
	\hat{U}^{(1)}_{j}
	&=
	\left(
		\begin{array}{cc}
			\braket{\mathbf{0},\mathbf{\overbar 0}| \hat{\Phi}_{j} |\mathbf{0},\mathbf{\overbar 0}} & \braket{\mathbf{0},\mathbf{\overbar 1}| \hat{\Phi}_{j} |\mathbf{0},\mathbf{\overbar 0}} \\
			\braket{\mathbf{1},\mathbf{\overbar 0}| \hat{\Phi}_{j} |\mathbf{0},\mathbf{\overbar 0}} & \braket{\mathbf{1},\mathbf{\overbar 1}| \hat{\Phi}_{j} |\mathbf{0},\mathbf{\overbar 0}}
		\end{array}
	\right)
	\equiv
	\left(
		\begin{array}{cc}
			\hat{W}^{\mathrm{II}}_{D_j} & \hat{W}^{\mathrm{II}}_{C_j} \\
			\hat{W}^{\mathrm{II}}_{B_j} & \hat{W}^{\mathrm{II}}_{A_j}
		\end{array}
	\right)
	\equiv \hat{W}^{\mathrm{II}}_{j}
\end{align}
and yields the suggested form of $W^\mathrm{II}$.
The lower right matrix element $\hat{W}^{\mathrm{II}}_{A_j}$ contains two bosons.
When formally contracting $\hat{U}^{(1)}_{j}$ with the neighboring matrix $\hat{U}^{(1)}_{j+1}$, these bosons connect, for instance, all local operators from $\hat{A}_{j}$ to local operators from $\hat{B}_{j+1}$.
Following this procedure, we find that in this way truncating the MPO approximation to $b=1$ we keep all local operator strings that overlap at most at one site and hence the error is $\mathcal{O}(\tau^2)$.

\subsubsection{\label{sec:te:wii:errors}Errors}
The MPO \wii approximation to the time-evolution operator $\hat{U}(\delta)$ primarly exhibits an error $\mathcal{O}(\delta^2)$ due to the truncation of the auxiliary degrees to hard-core bosons with maximal occupation $b\equiv 1$.
The Trotter error created by \cref{eq:te:wii:Trotterization} was shown in Ref.~\cite{zaletel15:_time} to be $\mathcal{O}(\tau^3)$ so that it is subleading compared to the auxiliary boson field truncation error.
However, the \wii MPO representation is not invariant under the particular choice of the decomposition of the Hamiltonian into local terms $\hat{H} = \sum_{j}\hat{H}_{j}$.
This degree of freedom can be used to reduce the truncation error and is discussed in detail in Ref.~\cite{zaletel15:_time}.
We want to point out that the method of complex time steps discussed in \cref{sec:te:tricks:timestepchoice} can transform a first-order stepper with error $\mathcal{O}(\delta^2)$ into a second-order stepper with error $\mathcal{O}(\delta^3)$.
This is particularly suitable for the \wii MPO representation since the \wii bond dimension $w=N_j+1$ is comparably small also for long-ranged Hamiltonians.
An improved stepper is hence available without too much numerical effort and should be used (it is used in this form in the experiments run later).
In turn, this construction in general destroys the unitarity of the stepper.
Furthermore, as already discussed in \cref{sec:tebd:errors} for TEBD, the choice of the MPO application method can significantly improve (or reduce) the accuracy of the time stepper.

%% file: content/krylov.tex
\section{\label{sec:te:krylov}The global Krylov method}

\begin{algorithm}[tb]
  \caption{\label{alg:krylov}The Krylov method. The main input is the
    Hamiltonian, the initial state, and the (possibly complex) time
    step.  Additionally, a procedure \textsc{ApplyAndOrthonormalize}
    is needed, which in turn requires the operator-state product and
    the orthogonalization of states.  For details on a variational
    approach to this, see \cref{sec:apply_ortho}. The function
    \textsc{ComputeEffectiveH} only needs to update the new elements
    of $T_{j+1}$ compared to $T_j$.}
  \begin{algorithmic}[1]
  \Procedure{Orthonormalize}{$\ket{w}$, $\{\ket{v_0},\cdots,\ket{v_k}\}$}
  \State $\ket{w'} \gets \ket{w} - \sum_{\alpha=0}^{k}\braket{w|v_\alpha}\ket{v_\alpha}$ \Comment{or variational orthonormalization for MPS}
  \If{$\|\ket{w'}\| < \varepsilon$}
  \State \Return invariant subspace found \Comment{evolution exact using just $\{\ket{v_0},\cdots,\ket{v_k}\}$}
  \EndIf
  \State \Return{$\frac{\ket{w'}}{\|\ket{w'}\|}$}
  \EndProcedure
    \Procedure{ApplyAndOrthonormalize}{$\hat{H}$, $\{\ket{v_k},\cdots\ket{v_0}\}$}
    \State $\ket{w^\prime_k} \gets \hat{H}\ket{v_k}$
    \State \Return{\Call{Orthonormalize}{$\ket{w_k'}, \{\ket{v_0},\cdots\ket{v_k}\}$}}
    \EndProcedure
    \Procedure{Timestep}{$\hat{H}$, $\ket{\psi(t)}$, $\delta$}
    \State $\ket{v_0} \gets \ket{\psi(t)}/ \|\ket{\psi(t)}\|$
    \For{$j\gets 1\ldots$}
    \State $|v_j\rangle \gets \Call{ApplyAndOrthonormalize}{\hat H, \{\ket{v_0},\cdots\ket{v_{j-1}}\}}$
    \State $T_{j+1}\gets\Call{ComputeEffectiveH}{T_j, \hat H,  \{\ket{v_0},\cdots\ket{v_{j}}\}}$ \Comment{$\left(T_{j+1}\right)_{k,l}=\braket{v_k|\hat{H}|v_l}$}
    \State $c_{j+1} \gets \E^{-\I\delta T_{j+1}}\,e_{j+1}^1$
    \If{$c_{j+1}$ converged}
    \State\textbf{break}
    \EndIf
    \EndFor
    \State \textbf{return } $\|\ket{\psi(t)}\| \sum_{i=0}^j c_{j+1}^i \ket{v_i}$
    \EndProcedure
  \end{algorithmic}
\end{algorithm}

Krylov subspace methods\cite{park86:_unitar_lancz, kosloff88:_time,
  saad03:_iterat_method_spars_linear_system,
  garcia-ripoll06:_time_matrix_produc_states, dargel12:_lancz,
  wall12:_out} (e.g.~the Lanczos method\cite{lanczos50,barrett94:_templ_solut_linear_system}) are well-known iterative techniques from the field of
numerical linear algebra.
In their application to time-dependent problems, one approximates
the action of $\hat U^\mathrm{exact}(\delta)$ onto the quantum state
$\ket{\psi(t)}$ directly, resulting in a time-evolved state
$\ket{\psi(t+\delta)}$.  It does \emph{not} provide access to the
exponential $e^{-\I \delta \hat H}$ in the standard physical basis.
The most straight-forward approach is to ignore the special structure
of the MPS/MPO representation and directly implement the iterative
procedure, as detailed below.  This is what we call the \textit{global
  Krylov method}.  In contrast, a variant exploiting the structure of
the MPS ansatz will be discussed in \cref{sec:local-krylov}.  We first
introduce the Krylov method independent of the specific representation
(dense vectors as in exact diagonalization, MPS, tree tensor networks
etc.) used.  This algorithm is also used as the local integrator for
the local Krylov method of \cref{sec:local-krylov} and the TDVP in
\cref{sec:tdvp}. Subsequently, we will discuss particular caveats when
applying the method globally to matrix-product states.

The Krylov subspace $\mathcal{K}_N$ of a Hamiltonian $\hat H$ and
initial state $\ket{\psi}$ is defined as the span of vectors
$\{ \ket{\psi}, \hat H \ket{\psi}, \ldots, \hat H^{N-1} \ket{\psi}
\}$.  This space is spanned by the Krylov vectors $\ket{v_0}$,
$\ket{v_1}$, $\ldots$, $\ket{v_{N-1}}$ such that the first Krylov
vector $|v_0\rangle$ is set to $\ket{\psi}$ normalized to have norm
$1$, and the subsequent Krylov vectors $\ket{v_i}$ are constructed by
applying $\hat H$ to $\ket{v_{i-1}}$ and orthonormalizing against all
previous Krylov vectors equivalently to the Lanczos algorithm.
In exact arithmetics with Hermitian $\hat H$, this way to construct a Krylov subspace reduces to orthogonalizing against the previous two vectors $\ket{v_{i-1}}$ and
$\ket{v_{i-2}}$, which is equivalent to the Lanczos algorithm \cite{barrett94:_templ_solut_linear_system}. 
However, due to round-off errors intrinsic to a numerical implementation, orthogonality of the Krylov vectors is usually lost.
If the precision required of each solution is low, one may abstain from avoiding this problem and simply work in a very small subspace.
However, due to the accumulation of errors during a time evolution and the calculation of spectral or time-dependent properties, it is necessary to cure this problem.
Hence, one typically needs to explicitly orthogonalize each new Krylov vector against
\emph{all} previous Krylov vectors.%
\footnote{Alternatively, one may only orthogonalize against the previous two vectors and achieve complete orthogonality by a subsequent basis transformation as detailed in Ref.~\cite{dargel12:_lancz}.}
The method then proceeds by
searching for the element of $\mathcal{K}_N$ which approximates the
result of the exact evolution most closely:
\begin{equation}
  \hat U^\mathrm{exact}(\delta)\ket{\psi(t)} \approx \argmin_{\ket{u} \in \mathcal{K}_N}\| \ket{u} - \hat U^\mathrm{exact}(\delta)\ket{\psi(t)} \|\equiv \ket{\psi_N(t + \delta)} \;.
\end{equation}
To do so, we define the projector onto $\mathcal{K}_N$
\begin{align} % \Bra and \Ket on the left matrix look very strange, hence the manual scaling with \vast, which is a bigger \Bigg
  \hat P_N & = \sum_{i=0}^{N-1} \ket{v_{i}}\langle v_i | \\
           & = \begin{pmatrix} \vast|\begin{matrix} \vdots \\ v_0 \\ \vdots \end{matrix}\vast\rangle & \vast|\begin{matrix} \vdots \\ v_{1} \\ \vdots \end{matrix}\vast\rangle & \cdots & \vast|\begin{matrix} \vdots \\ v_{N-1} \\ \vdots \end{matrix}\vast\rangle \end{pmatrix} \cdot \begin{pmatrix} \bra{\cdots v_{0\hspace{0.5cm}\hbox{}} \cdots} \\ \bra{\cdots v_{1\hspace{0.5cm}\hbox{}} \cdots} \\ \vdots\\ \bra{\cdots v_{N-1} \cdots} \end{pmatrix} \equiv V_N^\dagger V^\nodagger_N
\end{align}
where we have introduced matrices $V_N^\nodagger$ and $V_N^\dagger$ to represent the maps from the Hilbert space onto the Krylov space and vice versa. 
The solution to the above minimization problem is given by
\begin{equation}
  \ket{\psi_N(t + \delta)} = \hat P_N^\dagger \hat U^\mathrm{exact}(\delta) \hat P_N \ket{\psi(t)} \;.
\end{equation}
Note that for $N = {\rm dim} \mathcal{H} \equiv N_{\mathcal{H}}$ this is exact. 
Expanding the projectors and writing down the formal Taylor series for
$\hat U^\mathrm{exact}(\delta)$, we find:
\begin{align}
  |\psi_{N_{\mathcal{H}}}(t + \delta)\rangle & = \sum_{i=0}^{N_{\mathcal{H}}-1} \ket{v_{i}}\langle v_i | e^{-\I \delta \hat H} \sum_{i^\prime=0}^{N_{\mathcal{H}}-1} \ket{v_{i^\prime}}\braket{v_{i^\prime} \vert \psi(t)} \\
                             & = \sum_{i=0}^{N_{\mathcal{H}}-1} \ket{v_{i}}\langle v_i | \sum_{n=0}^\infty \frac{\left(-\I \delta\right)^n}{n!} \hat H^n \sum_{i^\prime=0}^{N_{\mathcal{H}}-1} \ket{v_{i^\prime}}\braket{v_{i^\prime} \vert \psi(t)} \\
                             & = V^\dagger_{N_{\mathcal{H}}} \sum_{n=0}^\infty \frac{\left(-\I \delta\right)^n}{n!} \underbrace{V^\nodagger_{N_{\mathcal{H}}} \hat H^n V^\dagger_{N_{\mathcal{H}}}}_{\equiv \left(T_{N_{\mathcal{H}}}\right)^n} V^\nodagger_{N_{\mathcal{H}}} \ket{\psi(t)} \\
                             & \approx V^\dagger_N \sum_{n=0}^\infty \frac{\left(-\I \delta\right)^n}{n!} \left(T_N\right)^n V^\nodagger_N \ket{\psi(t)} = V^\dagger_N e^{-\I \delta T_N} V^\nodagger_N \ket{\psi(t)} \label{eq:krylov:final} \;,
\end{align}
where $N \ll N_{{\mathcal{H}}}$ and $T_N$ is the Krylov-space representation of $\hat H$ with
coefficients
\begin{equation}
  \left(T_N \right)_{i,i^\prime} = \bra{v_i}\hat H\ket{v_{i^\prime}} \;.
\end{equation}
The Krylov approximation is introduced in \cref{eq:krylov:final}. %we assume that 
%In the following we will always set $\Ket{v_0} = ||\Ket{\psi(t)}|| \Ket{\psi(t)}$. 
Note that for $n > N-1$
\begin{equation}
	V_{N_{\mathcal{H}}}^\nodagger \hat H^n V_{N_{\mathcal{H}}}^\dagger V_{N_{\mathcal{H}}}\Ket{\psi(t)} \neq \left(T_N\right)^n V_N\Ket{\psi(t)}\;.
\end{equation}
This implies that the error in the Taylor-expansion of the time-
evolution operator is of order $\nicefrac{\delta^n}{n!}$, indicating
that already a few iterations suffice to give a very small error in
the integrator.  If $V_N^\dagger V_N^\nodagger$ were a proper
identity, we could insert it in between any power of $\hat H$ and
obtain exactly $V_N^\nodagger \hat H^n V_N^\dagger = T_N^n$.  However,
our Krylov subspace is much smaller than the true Hilbert space and
the projection induced by $V_N^\dagger V_N^\nodagger$ is hence very
large, $V_N^\dagger V_N^\nodagger \neq \mathbf{1}$. But due to the
special structure of this Krylov space,
$V_N^\dagger T_N^n V_N^\nodagger\Ket{\psi(t)}$ converges much more
quickly to $\hat H^n\Ket{\psi(t)}$ than expected if we, e.g.,~selected
random vectors in the full Hilbert space to construct our subspace.

In exact arithmetic and with Hermitian $\hat H$, $T_N$ would be
tridiagonal, i.e., $\left(T_N\right)_{i,i^\prime} = 0$ if $|i-i^\prime| > 1$.  While
in practice this is not necessarily true, we found that it improves
numerical stability and accuracy of the results to \emph{assume} $T_N$
to be tridiagonal and only evaluate those elements while forcibly
setting all other elements to zero. Returning to our equation
\cref{eq:krylov:final}, $V^\nodagger_N \ket{\psi(t)}$ is the
Krylov-space vector
\begin{equation}
  (\| \ket{\psi(t)} \|, 0, 0, \ldots)^T
\end{equation}
as all other Krylov vectors are orthogonal to $\ket{v_0}$ and
$\ket{v_0}$ is the normalized version of $\ket{\psi(t)}$.  $T_N$ can
be exponentiated efficiently using standard diagonalization routines
from \textsc{Lapack}, as it is only of size $N \times N$.  With
$T_N = Q_N^\dagger D_N Q_N$ this yields
\begin{equation}
  e^{-\I \delta T_N} = Q^\dagger_N e^{-\I \delta D_N} Q_N^\nodagger \;.
\end{equation}
For a given number of Krylov vectors and step size $\delta$, we hence obtain
\begin{align}
  \ket{\psi_N(t + \delta)} & = \| \ket{\psi(t)}\| V^\dagger_N  Q^\dagger_N e^{-\I \delta D_N} Q_N^\nodagger e^1_N \\
                           & = \| \ket{\psi(t)}\| V^\dagger_N c^\nodagger_N
\end{align}
with the coefficient vector
$c_N = Q^\dagger_N e^{-\I \delta D_N} Q_N^\nodagger e^1_N$ and $e^1_N$
the $N$-dimensional unit vector $(1, 0, 0, \ldots)^T$.  For typical
problems as presented in the example section, the number of Krylov
vectors used by us was between $3$ and $10$.  The algorithmic
procedure is summarized in \cref{alg:krylov}.

\paragraph{Evaluation of expectation values
  $\braket{\hat O(t+\delta)}$}
It is not actually necessary to construct the time-evolved state
$\ket{\psi_N(t+\delta)}$ to evaluate $\braket{\hat O(t+\delta)}$ for
arbitrary $\delta$. Instead, evaluating $\bra{v_i} \hat O \ket{v_{i^\prime}}$
for all $i,i^\prime=[0, \ldots, N-1]$ and constructing only the coefficient
vector $c_N$ is sufficient to evaluate the observable. One can hence
potentially avoid adding up the Krylov vectors and constructing the
time-evolved state $\Ket{\psi(t+\delta)}$. Indeed
(cf.~\cref{sec:krylov_dynamic_step} later), by evaluating the
expectation values $\bra{v_i} \hat O \ket{v_{i^\prime}}$, it becomes possible
to calculate the value of the observable at \emph{any} time
$\delta^\prime < \delta$ by only operating on small $N \times N$
matrices. Hence very small time steps unattainable by the other
time-stepping methods (e.g.~$\delta^\prime = 10^{-5}$) become
available.

\subsection{\label{sec:krylov:errors}Errors}

Elaborate bounds are available to estimate the error
incurred during the Krylov approximation\cite{Hochbruck1997}.
Unfortunately, these bounds are proven under the
assumption of exact arithmetic and hence do not necessarily apply in
the context of matrix-product states. The main take-away, which is
confirmed in practice, is that the Krylov error is in $O( \delta^N )$
as long as
$\sqrt{W \delta} \leq N$
where $W$ is the spectral range, which, in turn, is roughly of the
same order of magnitude as the system size. For a typical system of
$L=100$ sites with $\delta = 0.1$, this condition is fulfilled as soon
as $N \approx 3$; more precise bounds are available in exact arithmetic.\footnote{In exact arithmetic for a spectral width
  $W$ of $\hat H$, the error is smaller than
  $10 e^{-\nicefrac{N^2}{(5 W \delta)}}$ if $N$ is between
  $\sqrt{4 W \delta}$ and $2 W \delta$ and the error is smaller than
  $\left(\nicefrac{10}{W \delta}\right) e^{-W \delta} \left(
    \nicefrac{e W \delta}{N} \right)^N$ if $N > 2 W \delta$
  \cite{Hochbruck1997}. It is unknown how this translates to the case
  of inexact arithmetic.}

Hence, there are two approaches to measure the convergence of the Krylov
space: (i) The bottom-rightmost entry of the effective matrix
$T_n$ measures the weight scattered out of the Krylov space by
application of the Hamiltonian and typically decays
exponentially; (ii) the full Hilbert-space 2-norm distance
between two sequential iterations is cheaply available through the
coefficients of Krylov vectors produced in the two iterations. In our
experience, this second measure makes for an excellent convergence
criterion.

In addition to the inherent Krylov error which can often be made
extremely small ($O(10^{-10})$ or smaller), the Krylov method of
course also suffers from the standard MPS truncation error -- this
error, too, can be measured precisely (via the discarded weight) and
be made very small. As such, both errors of the global Krylov method
can be made extremely small at \emph{finite time-step size}, albeit at
relatively large numerical cost. The method hence in particular
excels if used to evaluate states very precisely, e.g.,~to measure the
errors of other methods on short time scales.

\subsection{Application to matrix-product states}
Up to this point, there was no need to narrow the description down to
a specific representation, which serves as a proof for the versatility
of the Krylov method. In our practical calculations, however, we wish
to use MPS to represent the time-evolved quantum states and
intermediate Krylov vectors and an MPO to represent the Hamiltonian
$\hat H$, which requires a few minor adaptations for efficiency and
accuracy. Note that in stark contrast to the TEBD and MPO \wiii
method, \emph{only} an MPO representation of $\hat H$ and no analytical
or other decomposition is required.

First and foremost, the most obvious improvement is in the calculation
of the last entry of the effective Krylov matrix $T_N$. In exact or
dense arithmetic, the evaluation of
$\bra{v_{N-1}} \hat H \ket{v_{N-1}}$ requires computing the
matrix-vector product $\hat H \ket{v_{N-1}}$. This is not the case in
the MPS approach: Indeed, evaluating the expectation value
$\bra{v_{N-1}} \hat H \ket{v_{N-1}}$ is much cheaper than calculating
the MPS representing $\hat H \ket{v_{N-1}}$,
cf.~\cref{sec:expectation-values}. As such, to generate a
$N \times N$-dimensional effective Krylov matrix $T_N$, one only needs
to evaluate $N-1$ MPO-MPS products and avoids the MPO-MPS product for
the most costly application on the last Krylov vector. In our
experience, the bond dimension of every additional Krylov vector grows
superlinearly, making this optimization very worthwhile.

To construct the time-evolved state $\ket{\psi(t+\delta)}$, it is
necessary to sum $N$ Krylov vectors together. Various methods to do so exist, in our implementation, we sequentially add two vectors (resulting in a new MPS with bond dimension $2m$) which are then truncated back down to the target bond dimension $m$. In $N-1$ steps, all $N$ Krylov vectors are summed together at cost $O(N (2m)^3)$. One could follow-up on this procedure with some sweeps of variational optimization or alternatively directly variationally optimize, but this does not appear to be necessary for our application.

\subsection{Loss of orthogonality} \label{sec:apply_ortho}

A typical problem of Krylov-subspace methods is the possible loss of
orthogonality of the basis vectors due to the finite-precision
arithmetic of floating point operations.  This matter becomes substantially
more pressing in the matrix-product algebra, as truncation is crucial
to keeping computations feasible.  If many Krylov vectors are desired,
truncation errors affecting the orthogonality of the basis vectors do
not simply add to the overall error (see above), but may quickly
degrade the overall quality of the Krylov space, leading to a poor
result.  In this case, it is necessary to check for orthogonality in
the basis and eventually re-orthogonalize the basis vectors
successively.  However, if one uses a simple Gram-Schmidt procedure to
orthogonalize vectors by successive additions of MPS, new truncation
errors are introduced during this procedure, which will quite often
entail the same problem.

In our experience, it has proven fruitful to orthogonalize the new
Krylov states variationally against all other Krylov states.  This is
essentially variational compression of a state under the additional
constraints of having zero overlap with all previous Krylov states,
see \cref{sec:truncation}.  The additional constraints can be
incorporated with the method of Lagrange multipliers: For each
constraint (orthogonal vector $\ket{v_i}$), introduce the
respective Lagrange multiplier $\beta_i$.  To minimize
$\|\hat{H}\ket{v_i}-\ket{v_{i+1}}\|^2$ under the constraints
$\{\braket{v_{i+1}|v_{i^\prime}} = 0\}_{0\leq i^\prime \leq i}$, we actually minimize
\begin{alignat}{2}
\left\|\hat{H}\,\ket{v_i}-\ket{v_{i+1}}\right\|^2 + \sum_{i^\prime}\,\beta_{i^\prime}\,\braket{v_{i+1}|v_{i^\prime}} \label{eq:krylov:ortho}
\end{alignat}
with regard to $\ket{v_i}$ and the $\beta_{i^\prime}$.  As with variational
compression, this can also be solved by iteratively solving the local
one- or two-site problems (without explicitly evaluating $\braket{\hat H^2}$). Care should be taken to ensure local orthogonality by using the pseudo-inverse of the Gram matrix as explained in Ref.~\cite{wall12:_out}. Using a two-site approach entails an
additional truncation step after each local optimization step and
implies again a loss of orthogonality. However, the two-site approach
converges much better than the single-site approach towards the global
optimum. In practice, we hence first do a few sweeps using the
two-site optimization (or, similarly, a single-site optimization with
subspace expansion\cite{hubig15:_stric_dmrg}) and follow up with a few
sweeps of fully single-site optimization without expansion and hence
also without truncation. The resulting state is then exactly
orthogonal to all previous states. Note that when initially starting
the optimization \cref{eq:krylov:ortho}, the available vector space on
the first few sites is likely to be very small
(e.g.,~$\sigma^2\cdot (m_2 = \sigma^2)$) and the orthogonalization
hence overconstrained. To avoid this problem, one should add the
constraints one-by-one during subsequent sweeps.

This variational orthogonalization can either be used as a separate orthogonalization step after the MPO-MPS application (using any of the algorithms presented in \cref{sec:mpo-mps-application}) or it can be combined with the variational operator application (cf.~\cref{sec:mpo-mps-application:variational}).
Whether it is better to first do the MPO-MPS application using the zip-up method and then variationally orthogonalize the result or to do both steps at once depends on the system at hand: in particular with long-range interactions, the variational approach may require more sweeps to converge whereas short-range interactions are dealt with very efficiently there.

\subsection{Dynamic step sizing} \label{sec:krylov_dynamic_step}

Dynamic step sizing is one of the most interesting and powerful
features of this method and can be used in several ways.  The idea is
that a Krylov subspace, which was computed for some time step
$\delta$, can be recycled for some other step length.  It is possible
to distinguish two cases: interpolation and extrapolation.

\subsubsection{Interpolation}

In some applications, the time evolution needs to be
performed on a very fine grid in time.  The time-stepping methods
would involve a single step for each point of the grid, which can
quickly turn cumbersome or even impossible.  On the other hand, if we have a Krylov
subspace that we used to perform a large time step, it can be re-used
to compute any intermediate smaller time step at the same or higher
accuracy.  This immediately follows from the construction of the
Krylov space and the convergence criteria/assumptions made above.  As
the diagonalization of the effective Hamiltonian is already known, all
we need to do is exponentiate the diagonal times the new time step,
map back into the Krylov basis to get the coefficient vector, and
compute the new MPS as a superposition of Krylov vectors.  If one is
only interested in expectation values of an observable $\hat{O}$, it
is advantageous to compute its projection into the Krylov space via
$\left(O_{N}\right)_{i,i^\prime}=\braket{\phi_i|\hat{O}|\phi_{i^\prime}}$ with complexity
$\sim\mathcal{O}(n^2)$.  With the coefficient vector $c_N$, the
desired expectation value can be computed as
$c_N^\dagger O_N^\nodagger c^\nodagger_N$, entirely skipping the more
expensive superposition of Krylov states.

\subsubsection{Extrapolation\label{sec:krylov_dynamic_step:extra}} Albeit trickier to implement, extrapolation can
significantly improve performance when used as a kind of automatic
adaptive step sizing scheme.  The idea is as follows: Given a Krylov
space, it is also often possible to recycle it for larger step sizes,
by only adding a small number of additional Krylov vectors (or none at
all).  It follows that the optimal Krylov subspace dimension minimizes
the ratio of the time needed to compute its basis and the number of
steps that it can be used for. As crude approximations of these
quantities, we assume that the cost of any new Krylov vector grows
exponentially, i.e.~the ratio of the costs of successive vectors is
fixed. Furthermore, we also assume that any new Krylov vector allows
us as many additional time steps as the previous Krylov vector. We then continuously monitor the time needed to construct a new Krylov vector and the number of steps we are able to take with it. Once a decision has to be taken whether to extend the Krylov space or rebuild it from scratch, we use those values as estimates for our decision. In
practice, this heuristic has proven to be quite reliable.

%%% Local Variables: 
%%% mode: latex
%%% TeX-master: "../time_evolution_review"
%%% End: 

%% file: content/local-methods.tex
\begin{figure}
  \centering
%	\tikzset{external/export next=false}
	\tikzsetnextfilename{krylov_effective_state}
	\begin{tikzpicture}
		\begin{scope}[node distance = 0.5 and 0.4]
			\node[ghost]	(0s0)						{$\overbar{\psi}^{L}_{j-1}$, $\overbar{\psi}^{R}_{j+1}$};
			\node[site]		(0s1)	[right = of 0s0]	{};
			\node[ghost]	(0s2)	[right = of 0s1]	{$\cdots$};
			\node[site]		(0s3)	[right = of 0s2]	{};
			\node[nosite]	(0s4)	[right = of 0s3]	{};
			\node[site]		(0s5)	[right = of 0s4]	{};
			\node[ghost]	(0s6)	[right = of 0s5]	{$\cdots$};
			\node[site]		(0s7)	[right = of 0s6]	{};
			
			\node[site]		(1o1)	[above = of 0s1]	{};			
			\node[ghost]	(1o0)	at (0s0|-1o1)		{\(\phantom{\text{$\overbar{\psi}^{L}_{j-1}$, $\overbar{\psi}^{R}_{j+1}$}}\)};		
			\node at (1o0) {$\ket{\psi}$};	
			\node[ghost]	(1o2)	at (0s2|-1o1)		{$\cdots$};			
			\node[site]		(1o3)	at (0s3|-1o1)		{};			
			\node[site]		(1o4)	at (0s4|-1o1)		{};			
			\node[site]		(1o5)	at (0s5|-1o1)		{};			
			\node[ghost]	(1o6)	at (0s6|-1o1)		{$\cdots$};			
			\node[site]		(1o7)	at (0s7|-1o1)		{};			

			\draw (0s1) -- (0s2);
			\draw (0s2) -- (0s3);
			\draw (0s3) -- (0s4);
			\draw (0s4) -- (0s5);
			\draw (0s5) -- (0s6);
			\draw (0s6) -- (0s7);

			\draw (1o1) -- (1o2);
			\draw (1o2) -- (1o3);
			\draw (1o3) -- (1o4);
			\draw (1o4) -- (1o5);
			\draw (1o5) -- (1o6);
			\draw (1o6) -- (1o7);

			\draw (0s1) -- (1o1);
			\draw (0s3) -- (1o3);
			\draw (0s4) -- (1o4);
			\draw (0s5) -- (1o5);
			\draw (0s7) -- (1o7);

			\node[draw,fill opacity=.2,draw opacity=.6,gray,thick,rounded corners,fill=gray!60,fit= (0s1) (0s3),label=below:{$\overbar{\psi}^{L}_{j-1}$}] {};

			\node[draw,fill opacity=.2,draw opacity=.6,gray,thick,rounded corners,fill=gray!60,fit= (0s5) (0s7),label=below:{$\overbar{\psi}^{R}_{j+1}$}] {};
		\end{scope}
	\end{tikzpicture}
  \caption{\label{fig:krylov_effective_state} The effective state
    $|\psi^{\mathrm{eff}}_j\rangle$ is obtained by projecting the MPS
    with itself. In case of a mixed-orthogonal MPS with orthogonality
    center on site $j$,
    $|\psi^{\mathrm{eff}}_j\rangle$ is simply the local site tensor
    $M_j$ and the left and right projectors are identity matrices.}
\end{figure}
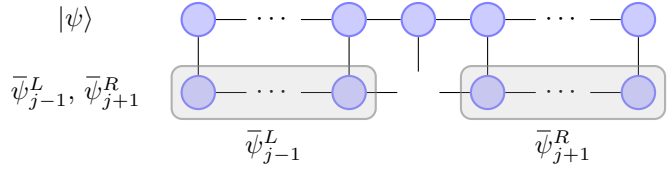

\begin{figure}
  \centering
%	\tikzset{external/export next=false}
	\tikzsetnextfilename{krylov_effective_operator}
	\begin{tikzpicture}
		\begin{scope}[node distance = 0.5 and 0.4]
			\node[ghost]	(0s0)						{$\overbar{\psi}^{L}_{j-1}$, $\overbar{\psi}^{R}_{j+1}$};
			\node[site]		(0s1)	[right = of 0s0]	{};
			\node[ghost]	(0s2)	[right = of 0s1]	{$\cdots$};
			\node[site]		(0s3)	[right = of 0s2]	{};
			\node[nosite]	(0s4)	[right = of 0s3]	{};
			\node[site]		(0s5)	[right = of 0s4]	{};
			\node[ghost]	(0s6)	[right = of 0s5]	{$\cdots$};
			\node[site]		(0s7)	[right = of 0s6]	{};
			
			\node[op]		(1o1)	[above = of 0s1]	{};			
			\node[ghost]	(1o0)	at (0s0|-1o1)		{\(\phantom{\text{$\overbar{\psi}^{L}_{j-1}$, $\overbar{\psi}^{R}_{j+1}$}}\)};		
			\node at (1o0) {$\hat H$};	
			\node[ghost]	(1o2)	at (0s2|-1o1)		{$\cdots$};			
			\node[op]		(1o3)	at (0s3|-1o1)		{};			
			\node[op]		(1o4)	at (0s4|-1o1)		{};			
			\node[op]		(1o5)	at (0s5|-1o1)		{};			
			\node[ghost]	(1o6)	at (0s6|-1o1)		{$\cdots$};			
			\node[op]		(1o7)	at (0s7|-1o1)		{};			

			\node[site]		(2s1)	[above = of 1o1]	{};			
			\node[ghost]	(2s0)	at (0s0|-2s1)		{${\psi}^{L}_{j-1}$, ${\psi}^{R}_{j+1}$};			
			\node[ghost]	(2s2)	at (0s2|-2s1)		{$\cdots$};			
			\node[site]		(2s3)	at (0s3|-2s1)		{};			
			\node[nosite]	(2s4)	at (0s4|-2s1)		{};			
			\node[site]		(2s5)	at (0s5|-2s1)		{};			
			\node[ghost]	(2s6)	at (0s6|-2s1)		{$\cdots$};			
			\node[site]		(2s7)	at (0s7|-2s1)		{};		

			\draw (0s1) -- (0s2);
			\draw (0s2) -- (0s3);
			\draw (0s3) -- (0s4);
			\draw (0s4) -- (0s5);
			\draw (0s5) -- (0s6);
			\draw (0s6) -- (0s7);

			\draw (1o1) -- (1o2);
			\draw (1o2) -- (1o3);
			\draw (1o3) -- (1o4);
			\draw (1o4) -- (1o5);
			\draw (1o5) -- (1o6);
			\draw (1o6) -- (1o7);

			\draw (2s1) -- (2s2);
			\draw (2s2) -- (2s3);
			\draw (2s3) -- (2s4);
			\draw (2s4) -- (2s5);
			\draw (2s5) -- (2s6);
			\draw (2s6) -- (2s7);

			\draw (0s1) -- (1o1);
			\draw (0s3) -- (1o3);
			\draw (0s4) -- (1o4);
			\draw (0s5) -- (1o5);
			\draw (0s7) -- (1o7);

			\draw (1o1) -- (2s1);
			\draw (1o3) -- (2s3);
			\draw (1o4) -- (2s4);
			\draw (1o5) -- (2s5);
			\draw (1o7) -- (2s7);

			\node[draw,fill opacity=.2,draw opacity=.6,gray,thick,rounded corners,fill=gray!60,fit= (0s1) (0s3),label=below:{$\overbar{\psi}^{L}_{j-1}$}] {};

			\node[draw,fill opacity=.2,draw opacity=.6,gray,thick,rounded corners,fill=gray!60,fit= (0s5) (0s7),label=below:{$\overbar{\psi}^{R}_{j+1}$}] {};

			\node[draw,fill opacity=.2,draw opacity=.6,gray,thick,rounded corners,fill=gray!60,fit= (2s1) (2s3),label=above:{$\psi^{L}_{j-1}$}] {};

			\node[draw,fill opacity=.2,draw opacity=.6,gray,thick,rounded corners,fill=gray!60,fit= (2s5) (2s7),label=above:{$\psi^{R}_{j+1}$}] {};
		\end{scope}
	\end{tikzpicture}
  \caption{\label{fig:krylov_effective_operator}The effective
    Hamiltonian $\hat H^{\mathrm{eff}}_j$ obtained by projecting the
    MPO for $\hat H$ using $\psi^{L}_{j-1}$, $\psi^{R}_{j+1}$, $\overbar{\psi}^{L}_{j-1}$,
    $\overbar{\psi}^{R}_{j+1}$. Note that this tensor is never explicitly constructed!
    Only its action on the state tensor $M_j$ is evaluated.}
\end{figure}
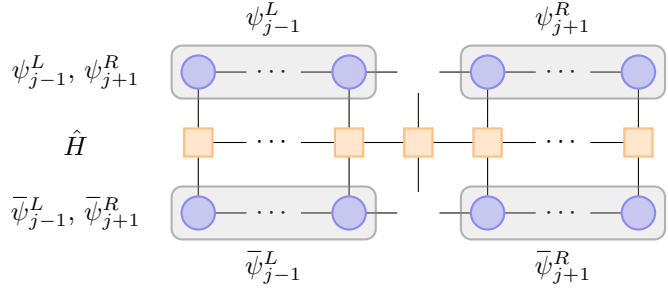

The global Krylov method of the previous section uses generic features
of Krylov subspaces to compute the time evolution of an MPS. However,
working ``globally'' and only using MPS to represent the Krylov
vectors comes with the major drawback that those vectors are typically
much more entangled than the actual time-evolved state
$\Ket{\psi(t+\delta)}$. To circumvent this problem, it may be
desirable to work in the local basis of reduced sites.\footnote{From a
  practical point of view, this is also motivated by the DMRG
  formulation which \emph{always} works in local spaces and hence
  makes it very difficult to formulate a global Krylov method.} There
are two different ways to obtain such a local basis, one being via an
attempt to directly Lie-Trotter decompose the Hamiltonian into local
subspaces\cite{feiguin05:_time,
  manmana05:_time_quant_many_body_system,rodriguez06,garcia-ripoll06:_time_matrix_produc_states,
  ronca17:_time_step_target_time_depen}, the other by enforcing a
constraint that the evolution lies within the MPS
manifold\cite{haegeman11:_time_depen_variat_princ_quant_lattic,
  haegeman16:_unify}. Both have been used in the literature, but their
precise derivation requires some work. In the following, we will first
present a very heuristic understanding of this approach, followed by a
detailed derivation of the first approach, which leads to the local
Krylov method presented in \cref{sec:local-krylov}, and subsequently
the second approach, which results in the time-dependent variational
principle, TDVP, presented in \cref{sec:tdvp}.

We start with an MPS $|\psi\rangle$, a Hamiltonian $\hat H$ and a site
$j$. The tensors $\psi^{L}_{j-1} \equiv (M_1, \ldots, M_{j-1})$ and
$\psi^{R}_{j+1} \equiv (M_{j+1}, \ldots, M_L)$ then constitute a left and
right map, respectively, from the joint Hilbert space on sites $1$
through $j-1$ onto the bond space $m_{j-1}$ and from the joint Hilbert
space on sites $j+1$ through $L$ onto the bond space $m_j$.

By sandwiching the Hamiltonian between two copies of $\psi^{L}_{j-1}$ and
$\psi^{R}_{j+1}$, we can transform it to act on the effective single-site
space on site $j$ (of dimension $\sigma_j \times m_{j-1} \times m_j$).
Let us call this effective single-site Hamiltonian
$\hat H_j^{\mathrm{eff}}$,
cf.~\cref{fig:krylov_effective_operator}. Likewise, we can take the
MPS $\ket{\psi}$ and act on it with a single copy of $\bar{\psi}^{L}_{j-1}$
and $\bar{\psi}^{R}_{j+1}$ to obtain the effective single-site tensor
$\tilde{M}_j$ representing the effective state
$\ket{\psi_j^{\mathrm{eff}}}$,
cf.~\cref{fig:krylov_effective_state}. If we move the orthogonality
center of the MPS to site $j$ first, the transformation of the state
is an identity operation, as the $\bar{A}_1$ of the transformation operator
cancels with the $A_1$ in the state etc. This is desirable for many
reasons, primarily numerical stability and avoiding a generalized
eigenvalue problem in favor of a standard eigenvalue problem.

Instead of mapping onto the space of a single site, we can also map
onto the space of two sites. This results in an effective state
$|\psi_{(j,j+1)}^{\mathrm{eff}}\rangle$ given by the product of two
site tensors $M_j \cdot M_{j+1}$ over their shared bond and likewise
an effective Hamiltonian $\hat H_{(j,j+1)}^{\mathrm{eff}}$. On the
other hand, if we map onto the space of a single bond between sites
$j$ and $j+1$ (without a physical index), we obtain the center matrix\footnote{As before, we will use
underlined indices ${}_{\underline{j}}$ to refer to bond tensors
between sites $j$ and $j+1$ whereas ${}_j$ (without the underline)
refers to the site tensor or effective Hamiltonian tensors on site
$j$.}
$C_{\underline{j}}$ to represent the effective state
$|\psi_{\underline{j}}^{\mathrm{eff}}\rangle$ and the effective center
site Hamiltonian $\hat H_{\underline{j}}^{\mathrm{eff}}$. 

In any of these smaller spaces -- typically in the basis of left- and
right-orthogonal environments as produced naturally during left-to-right and right-to-left sweeps -- it is very easy to solve the
time-dependent Schr\"odinger equation exactly from time $t$ to time
$t+\delta$ using a local Krylov procedure or any other exponential
solver\cite{moler03:_ninet_dubious_ways_comput_expon}. This results then in a new effective state
$|\psi_{j}^{\mathrm{eff}}(t+\delta)\rangle$. Unfortunately, the new
state will be represented in a relatively bad basis of the
environments $\psi^L_{\vphantom{j}}$ and $\psi^R_{\vphantom{j}}$ which are optimized to represent
the original state at time $t$. To alleviate this issue, after
evolving the site tensor $M_j$ forwards by $\delta$, one now wishes to
obtain a global state $\ket{\psi(t)}$ at the original time but in the
new left auxiliary basis. If this is achieved, we may iteratively
update the basis transformation tensors to be suitable for the new
state $\ket{\psi(t+\delta)}$.

This is where the local Krylov method and the TDVP diverge and proceed
differently.\footnote{An interesting intermediate between the local Krylov method based on time-step targetting and the TDVP as introduced in Refs.~\cite{haegeman11:_time_depen_variat_princ_quant_lattic, haegeman16:_unify} was presented in Refs.~\cite{koffel12:_entan_entrop_long_range_ising, hauke13:_spread_correl_long_range_inter_quant_system} where the time-step targetting was combined with a variational principle to solve the local TDSE.} However, regardless of how one obtains this old state in
the new basis, by sweeping once through the system, all site tensors
are incrementally optimized for a representation of the next time step
while the global state is kept at the original time. Upon reaching the
far end of the system, we keep the local state tensor at the new time
and hence obtain a global state at the new time $t+\delta$ written
using near-optimal basis transformation tensors.

Instead of solving the forward problem a single site at a time, it is
also possible to consider the effective Hamiltonian and state of two
neighboring sites. In this two-site scheme, one obtains a
forward-evolved tensor $M_{(j,j+1)}$ which needs to be split up using
a SVD, with then the single-site tensor on site $j+1$
backwards-evolved to keep the state at the original time. The
advantage is the variable bond dimension which allows a successive
adaptation of the MPS bond dimension to the entanglement growth in the system. The
downside is that an additional truncation error is introduced. In
practice, it appears that a hybrid scheme which first uses the
two-site method to increase the bond dimension to the usable maximum
and then switches to the single-site variant may fare best, at least
for some observables\cite{goto18:_perfor}.

Depending on the particular way in which we obtain the local problems,
solving the system of local Schr\"odinger equations exactly (using a
small direct solver each time) is possible. The result is that both
the energy and the norm (where applicable) of the state are conserved
exactly. This conservation may extend to other global observables\cite{leviatan17:_quant_matrix_produc_states, goto18:_perfor}. In
practice, the TDVP achieves this goal.

Finally, one may turn the first-order integrator of sweeping once
through the system into a second-order integrator by sweeping once
left-to-right and then right-to-left through the system, each time
with a reduced time step $\delta/2$.

%%% Local Variables: 
%%% mode: latex
%%% TeX-master: "../time_evolution_review"
%%% End: 

%% file: content/local-krylov.tex
\subsection{\label{sec:local-krylov}The local Krylov method}

Deriving a Lie-Trotter integration scheme while working in the local
reduced spaces is in fact equivalent to translating the time-step
targeting
DMRG\cite{feiguin05:_time,ronca17:_time_step_target_time_depen,manmana05:_time_quant_many_body_system,rodriguez06,garcia-ripoll06:_time_matrix_produc_states}
into the MPS framework.
Crucially, the MPS framework makes it possible to precisely analyze
the errors made, something which would be very difficult -- and to our
knowledge has not been done -- in the standard environment-system DMRG
picture.
To integrate the local time-dependent Schr\"odinger equations
resulting from the Lie-Trotter decomposition, we will use a
Krylov-based
approach\cite{manmana05:_time_quant_many_body_system,rodriguez06,garcia-ripoll06:_time_matrix_produc_states}.
This approach has the advantage of a very precise solution of the
local equations and a large degree of similarity with both
ground-state search DMRG and the time-dependent variational principle
(cf.~\cref{sec:tdvp}).
Alternatively, Runge-Kutta integrators have also been used extensively
with only minor changes to the overall method\cite{feiguin05:_time,ronca17:_time_step_target_time_depen}.
In particular, the error analysis presented here is also valid for the Runge-Kutta integrator, though one of course also has to include the additional time-step error of this integrator.

We begin by looking at the Lie-Trotter decomposition of the time-dependent Schr\"odinger equation
\begin{align}
        -\mathrm i\frac{d}{dt}\ket{\psi} &= \hat{H}\ket{\psi} \equiv \sum_{\nu}\hat{H}_{\nu}\ket{\psi}\; .
\end{align}
The goal is to find a decomposition scheme $\hat H = \sum_{\nu} \hat H_{\nu}$ such that we can integrate each summand separately by taking advantage of the MPS representation of the state vector $\ket{\psi}$.
Therefore we define orthogonal projectors $\hat{P}^{L,\ket{\psi}}_{j}$
and $\hat{P}^{R,\ket{\psi}}_{j}$ acting on the physical degrees of
freedom in a partition of the Hilbert space.
For that purpose we introduce bipartitions $\mathcal{H}=\mathcal{H}^{L}_{j}\otimes\mathcal{H}^{R}_{j+1}$ where $\mathcal{H}^{L}_{j} = \mathcal{H}_{1}\otimes\cdots\otimes\mathcal{H}_{j}$ and $\mathcal{H}^{R}_{j+1} = \mathcal{H}_{j+1}\otimes\cdots\otimes\mathcal{H}_{L}$ and declare
%.
\begin{align}
        \hat{P}^{L,\ket{\psi}}_{j} 
			&: 
				\mathcal{H}^{L}_{j} \otimes \mathcal{H}^{R}_{j+1} 
				\longrightarrow 
				\mathcal{H}^{L}_{j} \otimes \mathcal{H}^{R}_{j+1} \notag
\\
        \hat{P}^{L,\ket{\psi}}_{j} 
			&= 
				\sum_{
					\substack{
						\sigma_{1}, \ldots, \sigma_{j}, \\ 
						\bar{\sigma}_{1}, \ldots, \bar{\sigma}_{j}, \\
						m^{\noprime}_{j}
					}
				} 
					\underbrace{
						A_{1} \cdots A_{j}
					}_{
						\equiv \psi^{L}_{j;m_j}
					}
					\underbrace{
						\bar{A}_{j} \cdots \bar{A}_{1}
					}_{
						\equiv \bar{\psi}^{L}_{j;m_{j}}
					}
					\ket{\sigma_{1} \cdots \sigma_{j}}
					\bra{\bar\sigma_{1} \cdots \bar\sigma_{j}} 
					\otimes 
					\mathbf{\hat{1}}^{R}_{j+1}
\\
        \hat{P}^{R,\ket{\psi}}_{j} 
			&: 
				\mathcal{H}^{L}_{j-1} \otimes \mathcal{H}^{R}_{j} 
				\longrightarrow 
				\mathcal{H}^{L}_{j-1} \otimes \mathcal{H}^{R}_{j} \notag
\\
        \hat{P}^{R,\ket{\psi}}_{j} 
			&= 
				\mathbf{\hat{1}}^{L}_{j-1} 
				\otimes 
				\sum_{
					\substack{
						\sigma_{j}, \ldots, \sigma_{L}, \\
						\bar{\sigma}_{j}, \ldots, \bar{\sigma}_{L}, \\
						m^{\noprime}_{j-1}
					}
				}
					\underbrace{
						\overbar{B}_{L} \cdots \overbar{B}_{j}
					}_{
						\equiv \bar{\psi}^{R}_{j;m_{j-1}}
					}
					\underbrace{
						B_{j} \cdots B_{L}
					}_{
						\equiv \psi^{R}_{j;m_{j-1}}
					}
					\ket{\sigma_{j} \cdots \sigma_{L}}
					\bra{\bar\sigma_{j} \cdots \bar\sigma_{L}}
\end{align}
with mappings
$\psi^{L}_{j;m_{j}}$ from a part of the
physical Hilbert space into the bond space $m_j$.
By construction, these operators fulfill $\left(\hat{P}^{L/R,\ket{\psi}}_{j}\right)^2 = \hat{P}^{L/R,\ket{\psi}}_{j}$ and $\left(\hat{P}^{L/R,\ket{\psi}}_{j}\right)^{\dagger} = \hat{P}^{L/R,\ket{\psi}}_{j}$, i.e., they are projectors.
They are explicitly constructed from left-/right orthogonalized MPS site tensors.
\begin{figure}
  \centering
  \tikzsetnextfilename{tdvp_projectors}
   \tikzset{external/export next=false}
  \begin{tikzpicture}
    \begin{scope}[node distance=0.7]
      \node[siteA] (site1dag) {$\bar{A}_1$};
      \node[ghost] (site0dag) [left =of site1dag]{};
      \node[ghost] (sitem1dag) [left =of site0dag]{};
      \node[siteA] (site2dag) [right=of site1dag] {$\bar{A}_2$};
      \node[siteA] (site3dag) [right=of site2dag] {$\bar{A}_3$};

      \node[ghost] (site4dag) [right=of site3dag] {};

      \node[siteB] (site5dag) [right=of site4dag] {$\overbar{B}_5$};
      \node[siteB] (site6dag) [right=of site5dag] {$\overbar{B}_6$};
      \node[ghost] (site7dag) [right =of site6dag]{};
      \node[ghost] (site8dag) [right =of site7dag]{};

      \draw [decorate,decoration={brace,amplitude=3pt}] ($(site8dag.west)+(0.2,+0.5)$) -- +(0,-1) node [black,midway,right,xshift=3pt] {\footnotesize $\overbar{\psi}^R_5$};

      \draw [decorate,decoration={brace,amplitude=3pt,mirror}] ($(sitem1dag.east)+(-0.2,+0.5)$) -- +(0,-1) node [black,midway,left,xshift=-3pt] {\footnotesize $\overbar{\psi}^L_3$};

      \node[siteA] (site1) [below=of site1dag] {${A}_1$};
      \node[ghost] (site0) at (site1-|site0dag) {};
      \node[ghost] (sitem1) at (site1-|sitem1dag) {};
      \node[siteA] (site2) at (site1-|site2dag) {${A}_2$};
      \node[siteA] (site3) at (site1-|site3dag) {${A}_3$};

      \node[ghost] (site4) at (site1-|site4dag) {};

      \node[siteB] (site5) at (site1-|site5dag) {${B}_5$};
      \node[siteB] (site6) at (site1-|site6dag) {${B}_6$};
      \node[ghost] (site7) at (site1-|site7dag) {};
      \node[ghost] (site8) at (site1-|site8dag) {};

      \node[ghost,ld] (sigma1) [below=of site1] {$\sigma_1$};
      \node[ghost,ld] (sigma2) at (sigma1-|site2) {$\sigma_2$};
      \node[ghost,ld] (sigma3) at (sigma1-|site3) {$\sigma_3$};
      \node[ghost,ld] (sigma4) at (sigma1-|site4) {$\sigma_4$};
      \node[ghost,ld] (sigma5) at (sigma1-|site5) {$\sigma_5$};
      \node[ghost,ld] (sigma6) at (sigma1-|site6) {$\sigma_6$};

      \node[ghost,ld] (sigma1dag) [above=of site1dag] {$\bar{\sigma}_1$};
      \node[ghost,ld] (sigma2dag) at (sigma1dag-|site2dag) {$\bar{\sigma}_2$};
      \node[ghost,ld] (sigma3dag) at (sigma1dag-|site3dag) {$\bar{\sigma}_3$};
      \node[ghost,ld] (sigma4dag) at (sigma1dag-|site4dag) {$\bar{\sigma}_4$};
      \node[ghost,ld] (sigma5dag) at (sigma1dag-|site5dag) {$\bar{\sigma}_5$};
      \node[ghost,ld] (sigma6dag) at (sigma1dag-|site6dag) {$\bar{\sigma}_6$};

      \draw [decorate,decoration={brace,amplitude=3pt}] ($(site8.west)+(0.2,+0.5)$) -- +(0,-1) node [black,midway,right,xshift=3pt] {\footnotesize $\psi^R_5$};

      \draw [decorate,decoration={brace,amplitude=3pt,mirror}] ($(sitem1.east)+(-0.2,+0.5)$) -- +(0,-1) node [black,midway,left,xshift=-3pt] {\footnotesize $\psi^L_3$};

      \draw [] (site1.south) -| (sigma1.north);
      \draw [] (site2.south) -| (sigma2.north);
      \draw [] (site3.south) -| (sigma3.north);
      \draw [] (site5.south) -| (sigma5.north);
      \draw [] (site6.south) -| (sigma6.north);

      \draw [] (site1dag.north) -| (sigma1dag.south);
      \draw [] (site2dag.north) -| (sigma2dag.south);
      \draw [] (site3dag.north) -| (sigma3dag.south);
      \draw [] (site5dag.north) -| (sigma5dag.south);
      \draw [] (site6dag.north) -| (sigma6dag.south);

      \draw [] (site1) -- (site2);
      \draw [] (site2) -- (site3);
      \draw [] (site1dag) -- (site2dag);
      \draw [] (site2dag) -- (site3dag);

      \draw [] (site3.east) -| ($(site3dag.east)!0.3!(site4dag)$) -| (site3dag.east);

      \draw [] (site5) -- (site6);
      \draw [] (site5dag) -- (site6dag);

      \draw [] (site5.west) -| ($(site5dag.west)!0.3!(site4dag)$) -| (site5dag.west);

      \draw [] (sigma4) -- (sigma4dag);

  	  \node at ($(site0dag)!0.5!(site0)$) (site0center) {};
  	  \node at ($(site1dag)!0.5!(site1)$) (site1center) {};
	  \node at ($(site0center)!0.15!(site1center)$) {$\hat{P}^{L,\ket{\psi}}_3$};
	  \node[draw,fill opacity=.2,draw opacity=.6,inner sep=1em,gray,thick,rounded corners,fill=gray!60,fit= (site1dag) (site3dag) (site1) (site0center)] (PL) {};

  	  \node at ($(site7dag)!0.5!(site7)$) (site7center) {};
  	  \node at ($(site6dag)!0.5!(site6)$) (site6center) {};
	  \node at ($(site6center)!0.85!(site7center)$) {$\hat{P}^{R,\ket{\psi}}_5$};
	  \node[draw,fill opacity=.2,draw opacity=.6,inner sep=1em,gray,thick,rounded corners,fill=gray!60,fit= (site5dag) (site6dag) (site5) (site7center)] (PR) {};

    \end{scope}
  \end{tikzpicture}
  \caption{\label{fig:tdvp_projectors} Projector
    $\hat P_{3}^{L, |\psi\rangle} \otimes \mathbf{\hat 1}_4 \otimes
    \hat P_{5}^{R,|\psi\rangle}$ as defined in
    \cref{eq:local-krylov:projector} and also used in the TDVP
    projector \cref{eq:tdvp:projector}, here at the example of a
    six-site system. The other terms in \cref{eq:tdvp:projector} are
    constructed correspondingly.}
\end{figure}
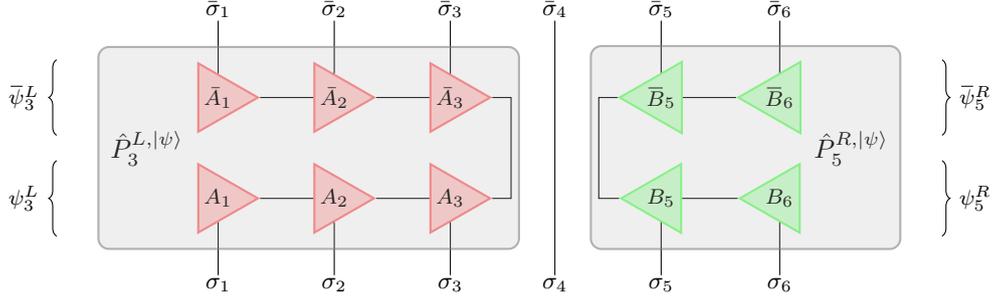
The action of such projectors onto an MPS representation of $\ket{\psi}$ in canonical form with
orthogonality center at site $j+1$ $(j-1)$ is then given
via
\begin{align}
        \hat{P}^{L,\ket{\psi}}_{j}\ket{\psi}
	        &= 
				\sum_{
					\substack{
						\sigma^{\noprime}_{1}, \ldots, \sigma^{\noprime}_{j},\\
						\sigma^{\prime}_{1}, \ldots, \sigma^{\prime}_{j},\\
						\bar{\sigma}^{\prime}_{1}, \ldots, \bar{\sigma}^{\prime}_{j},\\
						m^{\noprime}_{1}, \ldots, m^{\noprime}_{j}, \bar{m}^{\prime}_{j}
					}
				}
					\psi^{L; \sigma^{\prime}_{1}, \ldots, \sigma^{\prime}_{j}}_{j;m^{\prime}_{j}} 
					\underbrace{
						\bar{\psi}^{L;\bar{\sigma}^{\prime}_{1}, \ldots, \bar{\sigma}^{\prime}_{j}}_{j;\bar{m}^{\prime}_j}
						A^{\sigma^{\noprime}_{1}}_{1;m^{\noprime}_{1}}
						\cdots
						A^{\sigma^{\noprime}_{j}}_{j;m^{\noprime}_{j-1}, m^{\noprime}_{j}}
					}_{
						= 
						\delta_{\bar{m}^{\prime}_{j}, m^{\noprime}_j}
						\delta_{\bar{\sigma}^{\prime}_{1}, \sigma^{\noprime}_1}
						\cdots\,
						\delta_{\bar{\sigma}^{\prime}_{j}, \sigma^{\noprime}_j}
					}
				\sum_{
					\sigma^\noprime_{j+1}, \ldots, \sigma^\noprime_L} 
					M_{j+1} B_{j+2} \cdots B_{L} 
					\ket{\sigma^{\noprime}_{1} \cdots \sigma^{\noprime}_{L}} \notag
\\
	        &= 
				\sum_{\sigma^{\noprime}_{1}, \ldots, \sigma^{\noprime}_{L}}
					A_{1} \cdots A_{j} M_{j+1} B_{j+2} \cdots B_L
					\ket{\sigma^{\noprime}_{1} \cdots \sigma^{\noprime}_{L}} = \ket{\psi} 
\\
		\hat{P}^{R,\ket{\psi}}_{j}\ket{\psi}
	        &= 
				\sum_{\sigma^\noprime_1, \ldots, \sigma^\noprime_{j-1}} 
					A_{1} \cdots A_{j-2} M_{j-1} 
				\sum_{
					\substack{
						\sigma^{\noprime}_{j}, \ldots, \sigma^{\noprime}_{L},\\ 
						\sigma^{\prime}_{j}, \ldots, \sigma^{\prime}_{L},\\
						\bar{\sigma}^{\prime}_{j}, \ldots, \bar{\sigma}^{\prime}_{L},\\
						m^{\noprime}_{j}, \ldots, m^{\noprime}_{L}, \bar{m}^{\prime}_{j}
					}
				}
					\underbrace{
						B^{\sigma^{\noprime}_{j}}_{j;m^{\noprime}_{j-1}, m^{\noprime}_{j}}
						\cdots
						B^{\sigma^{\noprime}_{L}}_{L;m^{\noprime}_{L-1}}
						\bar{\psi}^{R;\bar{\sigma}^{\prime}_{j}, \ldots, \bar{\sigma}^{\prime}_{L}}_{j;\bar{m}^{\prime}_{j-1}}
					}_{
						\delta_{m^{\noprime}_{j},\bar{m}^{\prime}_{j}}
						\delta_{\bar{\sigma}^{\prime}_{j}, \sigma^{\noprime}_j}
						\cdots\,
						\delta_{\bar{\sigma}^{\prime}_{L}, \sigma^{\noprime}_L}
					} 
					\psi^{R;\sigma^{\prime}_{j}, \ldots, \sigma^{\prime}_{L}}_{j;m^{\prime}_{j-1}}
					\ket{\sigma^{\noprime}_{1} \cdots \sigma^{\noprime}_{L}} \notag 
\\
	        &= 
				\sum_{\sigma^{\noprime}_{1}\ldots\sigma^{\noprime}_{L}} 
					A_{1} \cdots A_{j-2} M_{j-1} B_{j} \cdots B_{L} 
					\ket{\sigma^{\noprime}_{1} \cdots \sigma^{\noprime}_{L}} = \ket{\psi} \;.
\end{align}
That is, if the state $\ket{\psi}$ has orthogonality center to the right (left) of the target index $j$, the projectors constructed from it act as an identity on their Hilbert space partition.
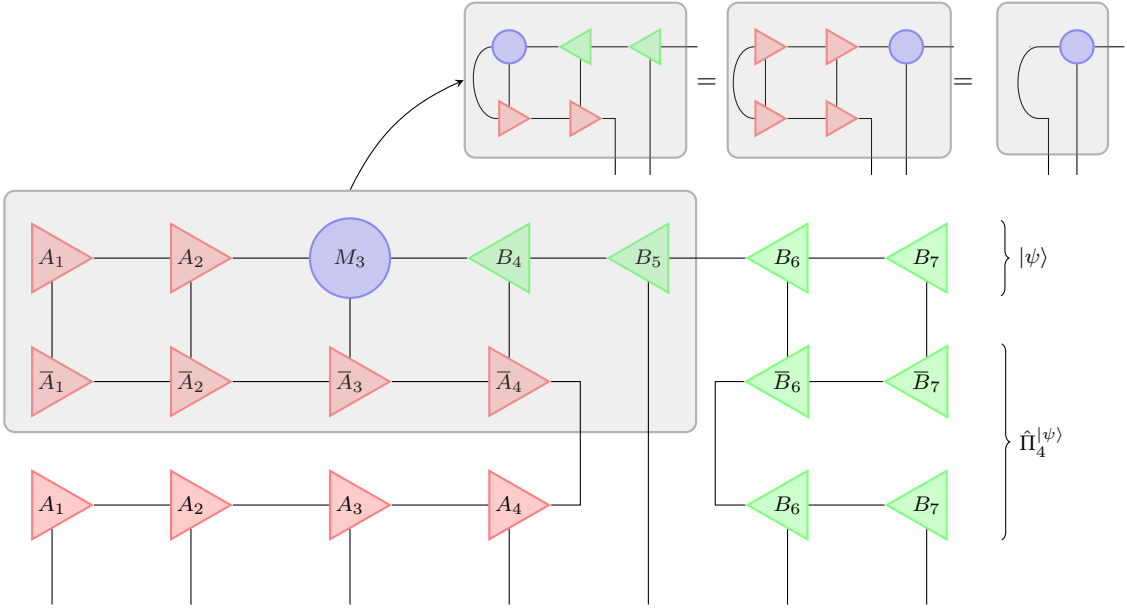
\begin{figure}[t!]
    \centering
    \tikzsetnextfilename{fig_local_krylov_projector_action}
%     \tikzset{external/export next=false}
    \begin{tikzpicture}
      \begin{scope}[node distance=1.0]
        \node[siteA] (site1) {$A_{1}$};
        \node[siteA] (site2) [right=of site1] {$A_{2}$};
        \node[site, minimum height=3em] (site3) [right=of site2] {$M_{3}$};
        \node[siteB] (site4) [right=of site3] {$B_{4}$};
        \node[siteB] (site5anchor) [right=of site4] {$B_{5}$};
        \node[siteB] (site6) [right=of site5anchor] {$B_{6}$};
        \node[siteB] (site7) [right=of site6] {$B_{7}$};

        \draw [decorate,decoration={brace,amplitude=3pt}] ($(site7.east)+(0.7,+0.5)$) -- +(0,-1) node [black,midway,right,xshift=3pt] {\footnotesize $\Ket{\psi}$};

        \node[siteA] (sitePdag1) [below=of site1] {$\overbar{A}_{1}$};
        \node[siteA] (sitePdag2) at (site2|-sitePdag1) {$\overbar{A}_{2}$};
        \node[siteA] (sitePdag3) at (site3|-sitePdag2) {$\overbar{A}_{3}$};
        \node[siteA] (sitePdag4) at (site4|-sitePdag1) {$\overbar{A}_{4}$};
        \node[ghost] (sitePdag5) at (site5anchor|-sitePdag1) {};
        \node[siteB] (sitePdag6) at (site6|-sitePdag1) {$\overbar{B}_{6}$};
        \node[siteB] (sitePdag7) at (site7|-sitePdag1) {$\overbar{B}_{7}$};

        \node[siteA] (siteP1) [below=of sitePdag1] {$A_{1}$};
        \node[siteA] (siteP2) at (sitePdag2|-siteP1) {$A_{2}$};
        \node[siteA] (siteP3) at (sitePdag3|-siteP2) {$A_{3}$};
        \node[siteA] (siteP4) at (sitePdag4|-siteP1) {$A_{4}$};
        \node[ghost] (siteP5) at (sitePdag5|-siteP1) {};
        \node[siteB] (siteP6) at (sitePdag6|-siteP1) {$B_{6}$};
        \node[siteB] (siteP7) at (sitePdag7|-siteP1) {$B_{7}$};

        \draw [decorate,decoration={brace,amplitude=3pt}] ($(sitePdag7.east)+(0.7,+0.5)$) -- +(0,-2.6) node [black,midway,right,xshift=3pt] {\footnotesize $\hat \Pi_{4}^{\ket{\psi}}$};

        \node[draw,fill opacity=.2,draw opacity=.6,gray,inner sep=1em,thick,rounded corners,fill=gray!60,fit=(site1) (site3) (site4) (site5anchor) (sitePdag3) (sitePdag5)] (graybox1) {};

        \node[ghost] (sitedag1) [below=of siteP1] {};
        \node[ghost] (sitedag2) at (siteP2|-sitedag1) {};
        \node[ghost] (sitedag3) at (siteP3|-sitedag1) {};
        \node[ghost] (sitedag4) at (siteP4|-sitedag1) {};
        \node[ghost] (sitedag5) at (siteP5|-sitedag1) {};
        \node[ghost] (sitedag6) at (siteP6|-sitedag1) {};
        \node[ghost] (sitedag7) at (siteP7|-sitedag1) {};
        \draw [] (site1) -- (site2);
        \draw [] (site2) -- (site3);
        \draw [] (site3) -- (site4);
        \draw [] (site4) -- (site5anchor);
        \draw [] (site5anchor) -- (site6);
        \draw [] (site6) -- (site7);

        \draw [] (site1.south) -| (sitePdag1.north);
        \draw [] (site2.south) -| (sitePdag2.north);
        \draw [] (site3.south) -| (sitePdag3.north);
        \draw [] (site4.south) -| (sitePdag4.north);
        \draw [] (site6.south) -| (sitePdag6.north);
        \draw [] (site7.south) -| (sitePdag7.north);

        \draw[] (sitePdag1) -- (sitePdag2);
        \draw[] (sitePdag2) -- (sitePdag3);
        \draw[] (sitePdag3) -- (sitePdag4);
        \draw[] (sitePdag6) -- (sitePdag7);

        \draw[] (siteP1) -- (siteP2);
        \draw[] (siteP2) -- (siteP3);
        \draw[] (siteP3) -- (siteP4);
        \draw[] (siteP6) -- (siteP7);

        \draw [] (sitePdag4.east) -| ($(siteP4.east)!0.3!(siteP5)$) -| (siteP4.east);
        \draw [] (sitePdag6.west) |- ($(sitePdag6.west)!0.3!(sitePdag5)$) |- (siteP6.west);

        \draw [] (sitedag1.north) -| (siteP1.south);
        \draw [] (sitedag2.north) -| (siteP2.south);
        \draw [] (sitedag3.north) -| (siteP3.south);
        \draw [] (sitedag4.north) -| (siteP4.south);
        \draw [] (sitedag6.north) -| (siteP6.south);
        \draw [] (sitedag7.north) -| (siteP7.south);

        \draw [] (site5anchor.south) -| (sitedag5.north);
	    \begin{scope}[node distance=1.0 and 0.5]
			\node[ghost, above=of site4] (tsite3ghost) {}; 
			\node[site, above=of tsite3ghost] (tsite3) {}; 
			\node[siteA] at ($(tsite3ghost)!0.3!(tsite3)$) (tsite3dag) {}; 

			\node[siteA, right=of tsite3dag] (tsite4dag) {};
			\node[siteB] at (tsite4dag|-tsite3) (tsite4) {}; 
			
			\node[ghost, right=of tsite4dag] (tsite5dag) {};
			\node[siteB] at (tsite5dag|-tsite4) (tsite5) {}; 
			\node[ghost, below = of tsite5dag] (tsite5dagdag) {};
			\node[ghost] at ($(tsite5dag)!0.5!(tsite5dagdag)$) (tsite5dagdag) {};
			
			\draw (tsite3) -- (tsite3dag);
			\draw (tsite4) -- (tsite4dag);
			\draw (tsite5) -- (tsite5dagdag.south);
			\node[ghost] at (tsite5dagdag-|tsite4) (tsite4dagdag) {};
			\node[ghost] at ($(tsite4dagdag)!0.5!(tsite5dagdag)$) (tsite4p5dagdag) {};
			\draw (tsite4dag) -| (tsite4p5dagdag.south);
%			\draw (tsite4dag) -- (tsite4dagdag.south);
%			\node[ghost] at (tsite5dagdag-|tsite3) (tsite3dagdag) {};
%			\draw (tsite3dag) -- (tsite3dagdag.south);
			\draw (tsite3) -- (tsite4);
			\draw (tsite4) -- (tsite5);
			\draw (tsite3dag) -- (tsite4dag);
			
			\node[ghost, right=of tsite5dag] (tsite6dag) {};
			\node[ghost] at (tsite6dag|-tsite5) (tsite6) {};
			\node[ghost] at ($(tsite6dag)!0.5!(tsite6)$) (tsite6mid) {$=$};

			\draw (tsite5) -- (tsite6);

			\draw (tsite3) to [bend right=90] (tsite3dag);
        \node[draw,fill opacity=.2,draw opacity=.6,gray,inner sep=1em,thick,rounded corners,fill=gray!60,fit=(tsite3) (tsite4) (tsite5dag) (tsite3dag)] (graybox2) {};

			\node[siteA, right=of tsite6] (tsite7) {}; 
			\node[siteA] at (tsite7|-tsite6dag) (tsite7dag) {}; 

			\node[siteA, right=of tsite7dag] (tsite8dag) {};
			\node[siteA] at (tsite8dag|-tsite3) (tsite8) {}; 
			
			\node[ghost, right=of tsite8dag] (tsite9dag) {};
			\node[hiddensite] at (tsite9dag|-tsite8) (tnosite9) {}; 
			\node[site] at (tsite9dag|-tsite8) (tsite9) {}; 
			\node[ghost, below = of tsite9dag] (tsite9dagdag) {};
			\node[ghost] at ($(tsite9dag)!0.5!(tsite9dagdag)$) (tsite9dagdag) {};
		
			\draw (tsite7) -- (tsite7dag);
			\draw (tsite8) -- (tsite8dag);
			\draw (tsite9) -- (tsite9dagdag.south);
			\node[ghost] at (tsite9dagdag-|tsite8) (tsite8dagdag) {};
			\node[ghost] at ($(tsite8dagdag)!0.5!(tsite9dagdag)$) (tsite8p5dagdag) {};
			\draw (tsite8dag) -| (tsite8p5dagdag.south);
%			\node[ghost] at (tsite9dagdag-|tsite8) (tsite8dagdag) {};
%			\draw (tsite8dag) -- (tsite8dagdag.south);
%			\node[ghost] at (tsite9dagdag-|tsite7) (tsite7dagdag) {};
%			\draw (tsite7dag) -- (tsite7dagdag.south);
			\draw (tsite7) -- (tsite8);
			\draw (tsite8) -- (tsite9);
			\draw (tsite7dag) -- (tsite8dag);
			
			\node[ghost, right=of tsite9dag] (tsite10dag) {};
			\node[ghost] at (tsite10dag|-tsite9) (tsite10) {};
			\node[ghost] at ($(tsite10dag)!0.5!(tsite10)$) (tsite10mid) {$=$};

			\draw (tsite9) -- (tsite10);

			\draw (tsite7) to [bend right=90] (tsite7dag);

        \node[draw,fill opacity=.2,draw opacity=.6,gray,inner sep=1em,thick,rounded corners,fill=gray!60,fit=(tsite7) (tnosite9) (tsite8) (tsite9dag) (tsite7dag)] {};

			\node[ghost, right=of tsite10dag] (tsite11dag) {};
			\node[ghost] at (tsite11dag|-tsite10) (tsite11) {}; 
			\node[ghost, below = of tsite11dag] (tsite11dagdag) {};
		
			\node[ghost, right=of tsite11dag] (tsite12dag) {};
			\node[site] at (tsite12dag|-tsite11) (tsite12) {}; 
			\node[ghost, below = of tsite12dag] (tsite12dagdag) {};
			\node[ghost] at ($(tsite12dag)!0.5!(tsite12dagdag)$) (tsite12dagdag) {};
			\node[ghost, right=of tsite12dag] (tsite13dag) {};
			\node[ghost] at (tsite13dag|-tsite12) (tsite13) {};

			\node[ghost] at (tsite12dagdag-|tsite11) (tsite11dagdag) {};
			\node[ghost] at ($(tsite11dagdag)!0.5!(tsite12dagdag)$) (tsite11p5dagdag) {};
			\node[ghost] at (tsite11p5dagdag|-tsite11dag) (tsite11p5dag) {};
			\node[ghost] at (tsite11p5dagdag|-tsite11) (tsite11p5) {};
			\draw (tsite12) -- (tsite11p5.west);
			\draw (tsite11p5) to [bend right=90] (tsite11p5dag);
			\draw (tsite11p5dag.west) -| (tsite11p5dagdag.south);

			\draw (tsite11) -- (tsite11);
			\draw (tsite12) -- (tsite13);
			\draw (tsite12) -- (tsite12dagdag.south);

        \node[draw,fill opacity=.2,draw opacity=.6,gray,inner xsep=0.5em,inner ysep=1em,thick,rounded corners,fill=gray!60,fit= (tsite11) (tsite12) (tsite12dag)] {};

		\end{scope}
		\draw[->] (graybox1.north) to [bend left = 20] (graybox2.west);
      \end{scope}
    \end{tikzpicture}
  \caption{\label{fig:local-krylov:projector-action}
    Shifting the center of orthogonality from $3\rightarrow 5$ under the action of $\hat \Pi^{\ket{\psi}}_{5} \ket{\psi} = \hat{P}^{L,\ket{\psi}}_{4}\otimes \mathbf{\hat{1}}_{5} \otimes \hat{P}^{R,\ket{\psi}}_{6} \ket{ \psi }$. The orthogonality center is implicitly shifted to site 5 (without changing the state content, c.f. gray boxes) which gives identities on sites 1 through 4 and sites 6 and 7 and a new orthogonality center tensor $M_5$. The completely contracted upper two rows then define the new tensor $M_5$.}
\end{figure}
Next we define the projector on the reduced site-space at site $j$
(cf.~\cref{fig:tdvp_projectors}) via
\begin{align}
        \hat{\Pi}^{\ket{\psi}}_{j} \equiv \hat{P}^{L,\ket{\psi}}_{j-1}\otimes \mathbf{\hat 1}_{j} \otimes \hat{P}^{R,\ket{\psi}}_{j+1} \;. \label{eq:local-krylov:projector}
\end{align}
The action of such a projector $\hat{\Pi}^{\ket{\psi}}_{j}$ on a state $\ket{\psi}$ is to shift the orthogonality center of $\ket{\psi}$ to the site $j$ which can be shown by applying the manipulation depicted in \cref{fig:local-krylov:projector-action} on to $\ket{\psi}$ repeatedly.
Therein gauge invariance is employed so that the action of the projector is trivial on the site tensors $A_{k}/B_{k}, k \neq j$ redering $j$ the center of orthogonality.
Thus, the quantum state $\ket{\psi}$ remains unchanged under the action of $\hat{\Pi}^{\ket{\psi}}_{j}$ and therefore
\begin{gather}
  \braket{\phi|\hat{\Pi}^{\ket{\psi}}_{j}|\psi}  = \braket{\phi | \psi} \label{eq:local-krylov:projector-action}\\
  \Rightarrow \hat{\Pi}^{\ket{\psi}}_{j} \ket{\phi} = \left(\frac{\ket{\psi^{\perp}}\bra{\psi^{\perp}}}{\braket{\psi^{\perp}|\psi^{\perp}}} + \frac{\ket{\psi}\bra{\psi}}{\braket{\psi|\psi}} \right)\hat{\Pi}^{\ket{\psi}}_{j}\ket{\phi} = \frac{\braket{\psi^{\perp}|\hat{\Pi}^{\ket{\psi}}_{j}|\phi} }{\braket{\psi^{\perp}|\psi^{\perp}}}\ket{\psi^{\perp}} + \frac{\braket{\psi|\hat{\Pi}^{\ket{\psi}}_{j}|\phi} }{\braket{\psi|\psi}}\ket{\psi} = \frac{\braket{\psi|\phi}}{\braket{\psi|\psi}}\ket{\psi}
\end{gather}
It is instructive to think of $\hat{\Pi}^{\ket{\psi}}_{j}$ as an operator acting on both the physical and gauge degrees of freedom of the MPS representation.
In the physical system it acts as a projector on the physical indices
$\sigma_{j}$ of the source state $\ket{\psi}$.
In the gauge degrees of freedom of the MPS representation,
$\hat{\Pi}^{\ket{\psi}}_{j}\ket{\phi}$ fixes the orthogonality center
to site $j$.
As the physical content of the state is independent of the location of its orthogonality center, we must have
\begin{align}
  \braket{\psi|\psi} &= \left(\bra{\psi} \hat \Pi_j^{\ket{\psi}}\right) \left( \hat \Pi_i^{\ket{\psi}} \ket{\psi} \right)
\end{align}
as also immediately follows from \cref{eq:local-krylov:projector-action}.
Now, we can reformulate the action of the Hamiltonian by decomposing it into representations $\hat{H}^{\hat \Pi^{\ket{\psi}}_{j}}$ acting only onto reduced site-spaces:
\begin{align}
  \hat{H}
  &\approx
  \frac{1}{L\Vert \ket{\psi}\Vert}\sum_j \hat \Pi^{\ket{\psi}}_j \hat H \hat \Pi^{\ket{\psi}}_j \equiv \frac{1}{L\Vert \ket{\psi}\Vert}\sum_{j} \hat{H}^{\hat \Pi^{\ket{\psi}}_{j}} \label{eq:local-krylov:approx-lie-decomp}\\
  \hat{H}\ket{\psi} &\approx \frac{1}{L\Vert \ket{\psi}\Vert}\sum_{j}\hat{\Pi}^{\ket{\psi}}_{j}\hat{H} \hat \Pi^{\ket{\psi}}_j\ket{\psi} \\
  & =
  \frac{1}{L\Vert \ket{\psi}\Vert}\sum_{j}\hat{H}^{\hat \Pi^{\ket{\psi}}_{j}}\ket{\psi}
\end{align}
which indeed yields a Lie-Trotter decomposition of the time-dependent Schr\"odinger equation
\begin{align}
        -\mathrm i\frac{d}{dt}\ket{\psi(t)} &= \frac{1}{L\Vert \ket{\psi}\Vert}\sum_{j}\hat{H}^{\hat \Pi^{\ket{\psi}}_{j}}\ket{\psi(t)}\;.\label{eq:local-krylov:decomposition}
\end{align}
Typically, this decomposition is not exact, it depends for instance on the size of
the chosen basis, but given a sufficiently large MPS bond dimension,
the error made will be small.
Having obtained the Lie-Trotter decomposition, we proceed by formulating a recursive integration scheme which is suitable for the particular structure of the MPS representation.
For this purpose note that a first-order approximation to the evolved state can be obtained by solving each problem
$-\mathrm i\frac{d}{dt}\ket{\psi(t)} = \hat{H}^{\hat \Pi^{\ket{\psi}}_{j}}\ket{\psi(t)}$
independently.\footnote{The prefactors $\frac{1}{L\Vert \ket{\psi} \Vert}$ are absorbed into the normalization of the site tensor that is currently evolved.}
Let $\ket{\psi^{\hat \Pi^{\ket{\psi}}_{j}}(t)}$ be the solution of the $j$-th problem.
An approximation to the overall time-evolved state from $t\rightarrow t+\delta$ can then be obtained by sequentially solving the initial value problems (setting
$\ket{\psi^{\hat \Pi^{\ket{\psi}}_{0}}(t)} \equiv \ket{\psi(t)}$)
\begin{align}
  -\mathrm i \frac{d}{dt} \ket{\psi^{\hat \Pi^{\ket{\psi}}_{j}}(t)} &= \hat{H}^{\hat \Pi^{\ket{\psi}}_{j}}\ket{\psi^{\hat \Pi^{\ket{\psi}}_{j}}(t)} \nonumber \\
  \textrm{and} \quad \ket{\psi^{\hat \Pi^{\ket{\psi}}_{j}}(t)} & = \ket{\psi^{\hat \Pi^{\ket{\psi}}_{j-1}}(t+\delta)} \label{eq:lie_trotter_approximation}
\end{align}
and identifying $\ket{\psi(t+\delta)} \equiv \ket{\psi^{\hat \Pi^{\ket{\psi}}_{L}}(t+\delta)}$ with the approximated time evolved state.
Comparing the formal Taylor expansions of the exactly integrated state $\ket{\psi(t+\delta)}_{\mathrm{exact}}$ with the approximation one readily finds
\begin{align}
        \ket{\psi(t+\delta)}_{\mathrm{exact}}
	        &= 
				\ket{\psi(t)} 
				- 
				\mathrm i \delta\hat{H}\ket{\psi(t)} 
				-
				\frac{\delta^2}{2}\hat{H}^{2}\ket{\psi(t)} 
				+ 
				\cdots 
\\
        \ket{\psi^{\hat \Pi^{\ket{\psi}}_{j}}(t+\delta)}
	        &= 
				\ket{\psi^{\hat \Pi^{\ket{\psi}}_{j}}(t)} 
				+
				\delta\frac{d}{dt}\ket{\psi^{\hat \Pi^{\ket{\psi}}_{j}}(t)} 
				+ 
				\frac{\delta^2}{2}\frac{d^2}{dt^2}\ket{\psi^{\hat \Pi^{\ket{\psi}}_{j}}(t)} 
				+
				\cdots \notag
\\
	        &= 
				\ket{\psi^{\hat \Pi^{\ket{\psi}}_{j-1}}(t+\delta)} 
				- 
				\mathrm i \delta \hat{H}^{\hat \Pi^{\ket{\psi}}_{j}} \ket{\psi^{\hat \Pi^{\ket{\psi}}_{j}}(t)} 
				-
				\frac{\delta^2}{2}\hat{H}^{\hat \Pi^{\ket{\psi}}_{j}} \hat{H}^{\hat \Pi^{\ket{\psi}}_{j}}\ket{\psi^{\hat \Pi^{\ket{\psi}}_{j}}(t)} 
				+
				\cdots 
\\
        \Rightarrow \ket{\psi(t+\delta)}
	        &= 
				\ket{\psi(t)} 
				-
				\mathrm i \delta\hat{H}\ket{\psi(t)} 
				-
				\frac{\delta^2}{2}\sum_{i\leq j}\hat{H}^{\hat \Pi^{\ket{\psi}}_{j}}\hat{H}^{\hat \Pi^{\ket{\psi}}_{i}}\ket{\psi(t)} 
				\cdots 
\label{eq:local-krylov:taylor-comparison}
\end{align}
where we have already applied \cref{eq:lie_trotter_approximation} to the expansion of the partially time-evolved intermediate state $\ket{\psi^{\hat \Pi^{\ket{\psi}}_j}(t+\delta)}$ around $\delta = 0$.
Replacing $\ket{\psi^{\hat \Pi^{\ket{\psi}}_{j-1}}(t+\delta)}$ by its Taylor expansion around $\delta=0$ generates a recursion for the partially time-evolved states.
Applying this recursion $L$ times and using the initial condition $\ket{\psi^{\hat \Pi^{\ket{\psi}}_0}(t)} = \ket{\psi(t)}$ yields the last equation.
As expected, $\ket{\psi^{\hat \Pi^{\ket{\psi}}_L} (t+\delta)}$ coincides with \cref{eq:local-krylov:decomposition} up to the first order of the expansion.
The first error occurs at second-order terms in the expansion and is given by
\begin{align}
        \ket{\psi(t+\delta)}_{\mathrm{exact}} - \ket{\psi(t+\delta)} = \frac{\delta^2}{2}\sum_{i<j}\left[\hat{H}^{\hat \Pi^{\ket{\psi}}_{i}}, \hat{H}^{\hat \Pi^{\ket{\psi}}_{j}} \right]\ket{\psi(t)}
\end{align}
and further commutators at arbitrary high orders.
Due to the dependence of the reduced problem currently solved onto the previous solution in \cref{eq:lie_trotter_approximation} the projectors $\hat{P}^{L,\ket{\psi}}_{j-1}$  in our integration scheme are in fact time-dependent because they need to be constructed from site tensors $A_{j^\prime<j}(t+\delta)$ that have already been evolved. In contrast, $\hat{P}^{R,\ket{\psi}}_{j+1}$ is constructed from unevolved site tensors $B_{j^\prime>j}(t)$.
We hence have to actually solve the $L$ initial value problems
\begin{align}
  -\mathrm i \frac{d}{dt} \ket{\psi^{\hat \Pi^{\ket{\psi}}_{j}}(t)} &= \left[ \hat{\Pi}^{\ket{\psi}}_{j}(t,t+\delta)\hat{H}\hat{\Pi}^{\ket{\psi}}_{j}(t,t+\delta)\right]\ket{\psi^{\hat \Pi^{\ket{\psi}}_{j}}(t)} \nonumber \\
\textrm{and} \quad  \ket{\psi^{\hat \Pi^{\ket{\psi}}_{j}}(t)} & = \ket{\psi^{\hat \Pi^{\ket{\psi}}_{j-1}}(t+\delta)}
\end{align}
with $\hat{\Pi}^{\ket{\psi}}_{j}(t,t+\delta) \equiv \hat{P}^{L,\ket{\psi^{\hat \Pi^{\ket{\psi}}_{j-1}}(t+\delta)}}_{j-1}\otimes\mathbf{\hat 1}_{j}\otimes\hat{P}^{R,\ket{\psi(t)}}_{j+1}$.
Note that while these problems appear to evolve over $L$ time steps with step size $\delta$, they only do so using different local Hamiltonians.
Those local problems are then connected via their initial conditions (c.f. \cref{eq:lie_trotter_approximation}) to generate a global time evolution by $\delta$ under (an approximation of) the global Hamiltonian.

Having decoupled the global Schr\"odinger equation into $L$ local problems, we can take advantage of the local representation of the state by multiplying each local differential equation at site $j$ with the bond maps $\overbar{\psi}^{L}_{j-1}\otimes \mathbf{\hat 1}_{j} \otimes \overbar{\psi}^{R}_{j+1}$
\begin{align}
        -\mathrm i \frac{d}{dt} \left[ \overbar{\psi}^{L}_{j-1}\otimes \mathbf{\hat 1}_{j} \otimes \overbar{\psi}^{R}_{j+1} \right] \ket{\psi^{\hat \Pi^{\ket{\psi}}_{j}}}
        &=
   \left[ \overbar{\psi}^{L}_{j-1}\otimes \mathbf{\hat 1}_{j} \otimes \overbar{\psi}^{R}_{j+1} \right] \hat{H}^{\hat \Pi^{\ket{\psi}}_{j}} \ket{\psi^{\hat \Pi^{\ket{\psi}}_{j}}} \notag \\
        \Rightarrow
        -\mathrm i \frac{d}{dt} M_{j} &= H^{\mathrm{eff}}_{j} M_{j} \label{eq:local-krylov:local-tdse}
\end{align}
where we defined effective reduced-site state and operator representations $M_{j}$ and $H^{\mathrm{eff}}_{j}$, respectively (see also \cref{fig:krylov_effective_state,fig:krylov_effective_operator}).
It can be checked easily that the projection of the effective problems onto their reduced site spaces at site $j$ leaves the solution invariant. 
Carrying out the differentiation on the left-hand site explicitly yields a sum of partial differentiations. 
However, on the right hand site $\hat{H}^{\hat \Pi^{\ket{\psi}}_{j}}$ acts only on the reduced site spaces and hence all differentials must vanish except the one acting on the tensor in the reduced site space.
Unfortunately, a direct adoption of the recursive solution strategy proposed above is not possible because the current problem at site $j$ requires the projectors to be constructed from left-/right-canonical site-tensors $A_{1},\ldots, A_{j-1},B_{j+1},\ldots, B_{L}$.
However, the solution $M_{j-1}$ of the previous reduced problem is not in general in a canonical representation so that in order to construct the next projector $\hat \Pi^{\ket{\psi}}_{j}$ we need to perform a basis transformation first.
There is no prescription in the derived decomposition scheme which corresponds to such a basis transformation keeping the evolved site tensors in canonical form.
On the other hand we can of course rewrite the updated site tensor as $M_{j-1}\rightarrow A_{j-1} R_{\underline{j-1}}$ in a left-canonical representation.
But, as already pointed out, we have to discard the transformation matrix $R_{\underline{j-1}}$ to obey the decomposition scheme.
To proceed with the next reduced problem, we need to resolve the situation of having $2$ different basis representations between the sites $(j-1,j)$.
Hence, we introduce a transformation between the unevolved and evolved site tensors $A_{j-1},M_{j}$.
The simplest guess is to consider a direct mapping between the unevolved and evolved site tensor basis sets.
This transformation can be readily obtained from the iterative solution strategy and is performed by projecting the evolved onto the unevolved bond space.
To see this we write the partially evolved state explicitely in the mixed canonical form
\begin{align}
	\ket{\psi(t,t+\delta)} 
		&= 
			\sum_{
					\substack
					{
						\sigma^{\noprime}_{1}, \ldots, \sigma^{\noprime}_{L},\\
						m^{\noprime}_{j-1}, m^{\noprime}_{j}
					}
				 } 
				\underbrace{
					A^{\sigma^{\noprime}_{1}}_{1}(t+\delta)
					\cdots 
					A^{\sigma^{\noprime}_{j-1}}_{j-1}(t+\delta)
				}_{
					\equiv \psi^{L; \sigma_1, \ldots, \sigma_{j-1}}_{j-1; m^{\noprime}_{j-1}}(t+\delta)
				}
				M^{\sigma^{\noprime}_{j}}_{j;m_{j-1}, m_j}(t)
				\underbrace{
					B^{\sigma^{\noprime}_{j+1}}_{j+1}(t)
					\cdots 
					B^{\sigma^{\noprime}_{L}}_{L}(t)
				}_{
					\equiv \psi^{R; \sigma_{j+1}, \ldots, \sigma_{L}}_{j+1; m^{\noprime}_{j}}(t)
				}
				\ket{\sigma^{\noprime}_{1} \cdots \sigma^{\noprime}_{L}}
\end{align}
% \vspace{-1.5em}
and introduce a compact notation for the left/right partition's basis states
\begin{align}
	\ket{\psi^L_{j}(t)}_{m_{j}} &= \sum_{\sigma_1, \ldots, \sigma_{j}} \psi^{L; \sigma_1, \ldots, \sigma_{j}}_{j; m^{\noprime}_{j}}(t) \ket{\sigma_1 \cdots \sigma_{j}}	&
	\ket{\psi^R_{j}(t)}_{m_{j-1}} &= \sum_{\sigma_j, \ldots, \sigma_L} \psi^{R; \sigma_{j}, \ldots, \sigma_{L}}_{j; m^{\noprime}_{j-1}}(t) \ket{\sigma_{j} \cdots \sigma_{L}}
\;.
\end{align}
Without truncation $\ket{\psi^L_{j}(t)}_{m_{j}}$ and $\ket{\psi^R_{j}(t)}_{m_{j-1}}$ would constitute complete maps from the physical degrees of freedom in the left/right partition and we could perform an exact basis transformation mapping the evolved into the unevolved left basis
\begin{align}
	\ket{\psi(t)}
		&=
			\sum_{m_{j-1}}
				\ket{\psi^L_{j-1}(t)}_{m_{j-1}}
				{\vphantom{\ket{\psi^L_{j-1}(t)}}}_{m_{j-1}}\hspace{-0.4em}\braket{\psi^L_{j-1}(t) \vert \psi(t,t+\delta)} 
\\
		&=	\sum_{
				\substack
				{
					m^{\noprime}_{j-1}, m^{\prime}_{j-1}, \\
					\sigma^{\noprime}_{j}, m^{\noprime}_{j}
				}
			}
				\ket{\psi^L_{j-1}(t)}_{m^{\noprime}_{j-1}}
				\underbrace{
					{\vphantom{\ket{\psi^L_{j-1}(t)}}}_{m_{j-1}}\hspace{-0.4em} \braket{\psi^L_{j-1}(t) \lvert \psi^{L}_{j-1}(t+\delta)}_{m^{\prime}_{j-1}}
				}_{
					Q_{\underline{j-1};m^{\noprime}_{j-1},m^{\prime}_{j-1}}(t,t+\delta)
				}
				M^{\sigma^{\noprime}_{j}}_{j;m^{\prime}_{j-1}, m^{\noprime}_j}(t)\ket{\sigma^{\noprime}_{j}}\ket{\psi^{R}_{j}(t)}_{m^{\noprime}_{j}} 
\\
		&= 
			\sum_{
				\substack
				{
					\sigma^{\noprime}_{1}, \ldots, \sigma^{\noprime}_{L},\\
					m^{\noprime}_{j-1}, m^{\noprime}_{j}\\
					m^{\prime}_{j-1}
				}
			}
				\psi^{L; \sigma^\noprime_1, \ldots, \sigma^\noprime_{j-1}}_{j-1; m^{\noprime}_{j-1}}(t)
				Q_{\underline{j-1};m^{\noprime}_{j-1},m^{\prime}_{j-1}}(t,t+\delta)
				M^{\sigma^{\noprime}_{j}}_{j;m^{\prime}_{j-1}, m^{\noprime}_j}(t)
				\psi^{R}_{j+1; m^{\noprime}_{j}}(t)
				\ket{\sigma^{\noprime}_{1} \cdots \sigma^{\noprime}_{L}}
\; .
\end{align}
and the matrix elements $Q_{\underline{j-1};m^{\noprime}_{j-1},m^{\prime}_{j-1}}(t,t+\delta)$ of the basis transformation $Q_{\underline{j-1}}(t,t+\delta)$ are constructed from
\begin{align}
	Q_{\underline{j-1};m^{\noprime}_{j-1},m^{\prime}_{j-1}}(t,t+\delta) 
		&=
			\sum_{
				\sigma^{\noprime}_{1}, \ldots, \sigma^{\noprime}_{j-1}
			}
				\overbar{\psi}^{L;\sigma^{\noprime}_{1},\ldots,\sigma^{\noprime}_{j-1}}_{j-1;m^{\noprime}_{j-1}}(t)
				\psi^{L;\sigma^{\noprime}_{1},\ldots,\sigma^{\noprime}_{j-1}}_{j-1;m^{\prime}_{j-1}}(t+\delta) \;.
\end{align}
The transformation $Q_{\underline{j}}(t,t+\delta)$ maps bond basis states $\ket{\psi^{L}_{j}(t+\delta)}_{m_j}$ which are optimized to represent the evolved state $\ket{\psi(t+\delta)}$ into bond basis states $\ket{\psi^{L}_{j}(t)}_{m_j}$ which are optimized to represent the unevolved state $\ket{\psi(t)}$.
Now, if we would let act $Q_{\underline{j-1}}(t,t+\delta)$ to the already optimized canonical site tensor $A^{\sigma_{j-1}}_{j-1}(t+\delta)$ then the effect would be to undo the previous site optimization.
Hence, the inverse transformation $\overbar Q_{\underline{j-1}}(t,t+\delta)$ is employed to transform the unevolved bond basis labeled by $m_{j-1}$ of the current site tensor $M_{j}(t)$.
Observing that $\overbar Q_{\underline{j-1}}$ is constructed from the above introduced bond maps $\overbar{\psi}^{L}_{j-1}(t)$ and  $\psi^{L}_{j-1}(t+\delta)$ which themselves should be build recursively, the transformation can be updated and applied before solving the $j$-th problem via
\begin{align}
	Q_{\underline{j-1}}(t, t+\delta) &= \overbar A_{j-1}(t) Q_{\underline{j-2}}(t, t+\delta) A_{j-1}(t+\delta), \quad M_{j}(t) \longrightarrow \overbar Q_{\underline{j-1}}(t, t+\delta) M_{j}(t) \; .
\end{align}
If we allow for truncation the error incurred by this mapping depends on the overlap $\braket{\psi(t+\delta)|\psi(t)}$ as well as the discarded weight.
This basis transformation is mostly motivated by its straightforward  availability during the sweeping procedure.
However, to the best of our knowledge there is no mathematical justification and we can only give the physical motivation that for small time steps $\delta$ the time-evolved state is expected to be relatively close to the unevolved state (deviation $\propto L\delta^2$ as follows from the consideration in \cref{sec:local-krylov:errors}).

Instead of mapping onto the space of a single site, in practice we map
onto the space of two sites. The two-site local TDSE is solved using
the time-dependent Lanczos approach to obtain $A_j(t+\delta)$.  The
original orthogonality center MPS tensor $M_{j+1}(t)$ is then
projected onto the new left basis as described above. This allows for
a flexible adaptation of not only the tensor $A_j$ itself but also
of the bond basis and -- if necessary -- MPS bond dimension between
sites $j$ and $j+1$.

Historically, only this two-site variant was used; but in analogy to
the 2TDVP method presented later, it may well make sense to initially
use the two-site local Krylov method until the desired maximal bond
dimension has been obtained and then switch to the single-site
integrator to save computational effort.

\subsubsection{\label{sec:local-krylov:errors}Errors}

Four errors are present in the local Krylov method when used in its (standard) two-site variant. The smallest of those stems from the inexact solution of the local TDSE \cref{eq:local-krylov:local-tdse}. This error can be made very small using a precise solver; in practice, a Krylov exponential as described in \cref{sec:te:krylov} with very few (4-5) vectors is sufficient. The second error is the standard truncation error incurred during the SVD to split the merged two-site tensors again while truncating to the desired bond dimension. This error can be measured and observed throughout the calculation and is much the same as in the other methods.

The third error is due to the approximation in
\cref{eq:local-krylov:approx-lie-decomp}. This projection error is
difficult to measure and strongly depends on the initial state. If the
initial state has a reasonably large bond dimension and the
Hamiltonian has reasonably short-range interactions, this error will
be very small. The longer the interactions in the Hamiltonian, the
larger the state has to be. In the two-site method, nearest-neighbor
interactions can be handled at all bond dimensions, in the single-site
variant, only on-site interactions are error-free at small bond
dimensions. The projection error is in particular problematic when
globally quenching from a product state with a very non-local
Hamiltonian (e.g.~resulting from a 2D $\to$ 1D map). When calculating
equilibrium Green's functions for short-range Hamiltonians, this error
is quite negligible.

Finally, there is an error due to the sequential solution of the local
TDSE as resulting from the Lie-Trotter decomposition. This error
\emph{can} be quantified, but doing so requires some additional work
which will follow now: We continue from the Taylor expansion
\cref{eq:local-krylov:taylor-comparison}. We emphasize that the action of the commutators
$\left[\hat{H}^{\hat \Pi^{\ket{\psi}}_{i}}, \hat{H}^{\hat \Pi^{\ket{\psi}}_{j}} \right]\ket{\psi(t_{0})}$
need to be evaluated with respect to the iteration the commutators are generated
from.
\begin{figure}[t!]
  \centering
  \tikzsetnextfilename{fig_local_krylov_projected_commutator}
  \begin{tikzpicture}
    [
    siteA/.style={scale=0.7,regular polygon, regular polygon sides=3, shape border rotate= 30, draw=red!50,fill=red!20,thick,inner sep=0.5pt,minimum width=4em,minimum height=4em},
    siteB/.style={scale=0.7,regular polygon, regular polygon sides=3, shape border rotate= -30, draw=green!50,fill=green!20,thick,inner sep=0.5pt,minimum width=4em,minimum height=4em},
    siteAFlipped/.style={scale=0.7,regular polygon, regular polygon sides=3, shape border rotate= -30, draw=red!50,fill=red!20,thick,inner sep=0.5pt,minimum width=4em,minimum height=4em},
    siteBFlipped/.style={scale=0.7,regular polygon, regular polygon sides=3, shape border rotate= 30, draw=green!50,fill=green!20,thick,inner sep=0.5pt,minimum width=4em,minimum height=4em},
    siteC/.style={scale=0.7,circle,thick,inner sep=0pt,minimum height=3em,draw=blue!50,fill=blue!20},
    ghostL/.style={scale=0.7,regular polygon, regular polygon sides=3, shape border rotate= 30, inner sep=0.5pt,minimum width=4em,minimum height=4em},
    ghostR/.style={scale=0.7,regular polygon, regular polygon sides=3, shape border rotate= -30, inner sep=0.5pt,minimum width=4em,minimum height=4em},
    ghostC/.style={scale=0.7,regular polygon, regular polygon sides=4, shape border rotate= 0, inner sep=0.5pt,minimum height=3.5em},
    Op/.style={scale=0.7,regular polygon, regular polygon sides=4, draw=orange!50, fill=orange!20, thick, inner sep=0.5pt, minimum width=4em, minimum height=4em},
    Intersite/.style={regular polygon, regular polygon sides=4, shape border rotate= 45, draw=black!50,fill=black!20,thick,inner sep=-2.5pt,minimum width=2em}
    ]
    \begin{scope}[node distance=0.7]
        \node[siteA] (site1) {$A_{1}$};
        \node[siteA] (site2) [right=of site1] {$A_{2}$};
        \node[siteC] (site3) [right=of site2]{$M_{3}$};
        \node[siteB] (site4) [right=of site3] {$B_{4}$};
        \node[ghostC] (site4topbaseline) [right=of site4] {};
        \node[Op] (W1) [below=of site1] {$W_{1}$};
        \node[Op] (W2) at (site2|-W1) {$W_{2}$};
        \node[Op] (W3) at (site3|-W1) {$W_{3}$};
        \node[Op] (W4) at (site4|-W1) {$W_{4}$};
        \node[ghostC] (W4topbaseline) [right=of W4] {};
        \node[siteA] (sitePdag1) [below=of W1] {$\overbar{A}_{1}$};
        \node[siteA] (sitePdag2) at (W2|-sitePdag1) {$\overbar{A}_{2}$};
        \node[ghostC] (sitePdag3) at (W3|-sitePdag1) {};
        \node[siteB] (sitePdag4) at (W4|-sitePdag1) {$\overbar{B}_{4}$};
        \node[ghostC] (sitePdag4topbaseline) [right=of sitePdag4] {};
        \node[siteA] (siteP1) [below=of sitePdag1] {$A_{1}$};
        \node[siteA] (siteP2) at (sitePdag2|-siteP1) {$A_{2}$};
        \node[ghostC] (siteP3) at (sitePdag3|-sitePdag1) {};
        \node[siteB] (siteP4) at (sitePdag4|-siteP1) {$B_{4}$};
        \node[ghostC] (siteP4topbaseline) [right=of siteP4] {};
        \node[ghostC] (sitedag1) [below=of siteP1] {$\sigma_{1}$};
        \node[ghostC] (sitedag2) at (sitedag1-|siteP2) {$\sigma_{2}$};
        \node[ghostC] (sitedag3) at (sitedag1-|siteP3) {$\sigma_{3}$};
        \node[ghostC] (sitedag4) at (sitedag1-|siteP4) {$\sigma_{4}$};

        \draw [] (site1) -- (site2);
        \draw [] (site2) -- (site3);
        \draw [] (site3) -- (site4);
        \draw [] (W1) -- (W2);
        \draw [] (W2) -- (W3);
        \draw [] (W3) -- (W4);
        \draw [] (site1.south) -| (W1.north);
        \draw [] (site2.south) -| (W2.north);
        \draw [] (site3.south) -| (W3.north);
        \draw [] (site4.south) -| (W4.north);
        \draw [] (W1.south) -| (sitePdag1.north);
        \draw [] (W2.south) -| (sitePdag2.north);
        \draw [] (W4.south) -| (sitePdag4.north);
        \draw [] (sitePdag1) -- (sitePdag2);
        \draw [] (siteP1) -- (siteP2);
        \draw [] (sitePdag2.east) |- ($(sitePdag2.east)!0.3!(sitePdag3)$) |- (siteP2.east);
        \draw [] (sitePdag4.west) -| ($(sitePdag4.west)!0.3!(sitePdag3)$) |- (siteP4.west);

        \node[Op] (W1) [below=of siteP1] {$W_{1}$};
        \node[Op] (W2) at (W1-|siteP2) {$W_{2}$};
        \node[Op] (W32) at (W1-|siteP3) {$W_{3}$};
        \node[Op] (W4) at (W1-|siteP4) {$W_{4}$};
        \node[ghostC] (W4botbaseline) [right=of W4] {};

        \node[ghostC] (eq1_baseline) at ($(siteP4)!0.5!(W4)$) {};

        \draw [] (W1) -- (W2);
        \draw [] (W2) -- (W32);
        \draw [] (W32) -- (W4);
        \draw [] (siteP1.south) -- (W1.north);
        \draw [] (siteP2.south) -- (W2.north);
        \draw [] (W3.south) -- (W32.north);
        \draw [] (siteP4.south) -- (W4.north);

        \node[siteA] (sitePdag1) [below=of W1] {$\overbar{A}_{1}$};
        \node[ghostC] (sitePdag2) at (sitePdag1-|W2) {};
        \node[siteB] (sitePdag3) at (sitePdag1-|W3) {$\overbar{B}_{3}$};
        \node[siteB] (sitePdag4) at (sitePdag1-|W4) {$\overbar{B}_{4}$};
        \node[ghostC] (sitePdag4botbaseline) [right=of sitePdag4] {};
        \node[siteA] (siteP1) [below=of sitePdag1] {$A_{1}$};
        \node[siteB] (siteP3) at (siteP1-|sitePdag3) {$B_{3}$};
        \node[siteB] (siteP4) at (siteP1-|sitePdag4) {$B_{4}$};
        \node[ghostC] (siteP4botbaseline) [right=of siteP4] {};

        \draw [] (W1) -- (sitePdag1);
        \draw [] (W32) -- (sitePdag3);
        \draw [] (W4) -- (sitePdag4);
        \draw [] (sitePdag3) -- (sitePdag4);
        \draw [] (siteP3) -- (siteP4);
        \draw [] (sitePdag1.east) |- ($(sitePdag1.east)!0.3!(sitePdag2)$) |- (siteP1.east);
        \draw [] (sitePdag3.west) -| ($(sitePdag3.west)!0.3!(sitePdag2)$) |- (siteP3.west);

        \node[ghostC] (sitedag1) [below=of siteP1] {$\sigma_{1}$};
        \node[ghostC] (sitedag2) at (sitedag1-|siteP2) {$\sigma_{2}$};
        \node[ghostC] (sitedag3) at (sitedag1-|siteP3) {$\sigma_{3}$};
        \node[ghostC] (sitedag4) at (sitedag1-|siteP4) {$\sigma_{4}$};

        \draw [] (siteP1) -- (sitedag1);
        \draw [] (W2) -- (sitedag2);
        \draw [] (siteP3) -- (sitedag3);
        \draw [] (siteP4) -- (sitedag4);

        \node[ghostC] (eq1) [right=of eq1_baseline] {$=$};
        \node[ghostC] (eq1_baseline) [right=of eq1] {};

        \node[ghostC] (WEffL21) at (eq1_baseline|-site4topbaseline) {};
        \node[ghostC] (WEffL22) at (eq1_baseline|-W4topbaseline) {};
        \node[ghostC] (WEffL23) at (eq1_baseline|-sitePdag4topbaseline) {};
        \node[draw=black!50,fill=black!20,thick, fit=(WEffL21) (WEffL22) (WEffL23),minimum width=4.5em,inner sep=0em] (WEffL2) {$L_{1}$};

        \node[ghostC] (WEffL11) at (eq1_baseline|-siteP4topbaseline) {};
        \node[ghostC] (WEffL12) at (eq1_baseline|-W4botbaseline) {};
        \node[ghostC] (WEffL13) at (eq1_baseline|-sitePdag4botbaseline) {};
        \node[draw=black!50,fill=black!20,thick, fit=(WEffL11) (WEffL12) (WEffL13),minimum width=4.5em,inner sep=0em] (WEffL1) {$L_{1}$};

    \end{scope}
    \begin{scope}[node distance=1.0]
        \node[siteA] (site2) [right=of WEffL21] {$A_{2}$};
        \node[Op] (W21) at (WEffL22-|site2) {$W_{2}$};
        \node[siteA] (sitePdag21) at (WEffL23-|W21) {$\overbar{A}_{2}$};
        \node[siteC] (site3) [right=of site2] {$A_{3}$};
        \node[Op] (W31) at (site3|-W21) {$W_{3}$};
        \node[ghostC] (sitePdag31) at (W31|-sitePdag21) {};

        \node[ghostC] (WEffR21) [right=of site3] {};
        \node[ghostC] (WEffR22) at (W31-|WEffR21) {};
        \node[ghostC] (WEffR23) at (sitePdag31-|WEffR21) {};
        \node[draw=black!50,fill=black!20,thick, fit=(WEffR21) (WEffR22) (WEffR23),minimum width=4.5em,inner sep=0em] (WEffR2) {$R_{4}$};

        \draw [] (site2.east) -- (site3.west);
        \draw [] (site2.south) -- (W21.north);
        \draw [] (site3.south) -- (W31.north);
        \draw [] (W21.east) -- (W31.west);
        \draw [] (W21.south) -- (sitePdag21.north);

        \node[siteA] (siteP21) at (WEffL11-|sitePdag21) {$A_{2}$};
        \node[Op] (W22) at (WEffL12-|siteP21) {$W_{2}$};
        \node[ghostC] (sitePdag22) at (WEffL13-|W22) {};
        \node[ghostC] (siteP32) at (siteP21-|sitePdag31) {};
        \node[Op] (W32) at (W22-|siteP32) {$W_{3}$};
        \node[siteB] (sitePdag32) at (sitePdag22-|W32) {$\overbar{B}_{3}$};

        \node[ghostC] (WEffR11) at (siteP32-|WEffR23) {};
        \node[ghostC] (WEffR12) at (W32-|WEffR11) {};
        \node[ghostC] (WEffR13) at (sitePdag32-|WEffR12) {};
        \node[draw=black!50,fill=black!20,thick, fit=(WEffR11) (WEffR12) (WEffR13),minimum width=4.5em] (WEffR1) {$R_{4}$};

        \draw [] (siteP21) -- (W22);
        \draw [] (W31) -- (W32);
        \draw [] (W22) -- (W32);
        \draw [] (W32) -- (sitePdag32);

        \node[siteA] (siteP12) at (siteP4botbaseline-|WEffL13) {$A_{1}$};
        \node[ghostC] (siteP22) at (siteP12-|sitePdag22) {};
        \node[siteB] (siteP32) at (siteP12-|sitePdag32) {$B_{3}$};
        \node[siteB] (siteP42) at (siteP12-|WEffR13) {$B_{4}$};

        \draw [] (siteP32) -- (siteP42);

        \node[ghostC] (sitedag1) at (sitedag4-|siteP12) {$\sigma_{1}$};
        \node[ghostC] (sitedag2) at (sitedag1-|siteP22) {$\sigma_{2}$};
        \node[ghostC] (sitedag3) at (sitedag1-|siteP32) {$\sigma_{3}$};
        \node[ghostC] (sitedag4) at (sitedag1-|siteP42) {$\sigma_{4}$};

        \draw [] (siteP12) -- (sitedag1);
        \draw [] (W22) -- (sitedag2);
        \draw [] (siteP32) -- (sitedag3);
        \draw [] (siteP42) -- (sitedag4);
    \end{scope}
    \begin{scope}[on background layer]
            \begin{scope}[node distance=1.0]
                \node[draw=purple!50, fill=purple!20,thick, fit=(site2) (site3)] {};
%                 \draw[draw=orange!70, fill=orange!40,thick]
%                         ($(WEffL2.north west)+(-0.15,0.15)$) -- ($(WEffL2.north east)+(0.15,0.15)$) |-
%                         ($(W21.north east)+(0.75,0.15)$) -- ($(W22.south east)+(0.75,-0.15)$) -| ($(WEffL1.south east)+(0.15,-0.15)$) -|
%                         ($(WEffL1.south west)+(-0.15,-0.15)$) -- ($(WEffL2.north west)+(-0.15,0.15)$) {};
%                 \draw[draw=orange!70, fill=orange!40,thick]
%                         ($(WEffR2.north west)+(-0.15,0.15)$) -- ($(WEffR2.north east)+(0.15,0.15)$) |-
%                         ($(WEffR1.south east)+(0.15,-0.15)$) -- ($(WEffR1.south west)+(-0.15,-0.15)$) |-
%                         ($(W32.south east)+(0.15,-0.15)$) -| ($(W31.north west)+(-0.15,0.15)$) -| ($(WEffR2.north west)+(-0.15,0.15)$) -- ($(WEffR2.north west)+(-0.15,0.15)$) {};
                \draw[draw=black!50, fill=black!20, fill opacity=0.5, thick]
                        ($(W21.north west)+(-0.1,0.15)$) |- ($(W31.north east)+(0.25,0.2)$) |-
                        ($(siteP32.south east)+(0.25,-0.25)$) -| ($(W21.north west)+(-0.1,0.15)$) {};
                \draw [] (WEffL21.east) -- (site2.west);
                \draw [] (site3.east) -- (WEffR21.west);
                \draw [] (WEffL22.east) -- (W21.west);
                \draw [] (W31.east) -- (WEffR22.east);
                \draw [] (W31.east) -- (WEffR22.west);
                \draw [] (WEffL23.east) -- (sitePdag21);
                \draw [] (sitePdag21.east) |- ($(sitePdag21.east)!0.3!(sitePdag31)$) |- (siteP21.east);
                \draw [] (sitePdag32.east) -- (WEffR13.west);
                \draw [] (WEffL11.east) -- (siteP21.west);
                \draw [] (WEffL12.east) -- (W22.west);
                \draw [] (W32.east) -- (WEffR12.west);
                \draw [] (WEffR23.west) -| ($(WEffR23)!0.7!(sitePdag31)$) |- (WEffR11.west);
                \draw [] (WEffL13.east) -| ($(WEffL13.east)!0.7!(sitePdag22)$) |- (siteP12.east);
                \draw [] (sitePdag32.west) -| ($(sitePdag32.west)!0.3!(sitePdag22)$) |- (siteP32);
            \end{scope}
    \end{scope}
  \end{tikzpicture}
  \caption{\label{fig:local-krylov:projected_commutator}
  Evaluation of $\hat{H}^{\hat \Pi^{\ket{\psi}}_{2}} \hat{H}^{\hat \Pi^{\ket{\psi}}_{3}}\ket{\psi^{\hat \Pi^{\ket{\psi}}_{3}}(t)}$ at the example of a four-site system. The tensors $L_{i}/R_{i}$ correspond to partially contracted MPS-MPO-MPS-networks.
  The burgundy-shaded rectangular area at the top encloses the reduced site tensor $\psi^{C;\sigma_{i},\ldots,\sigma_{j}}_{i,\ldots,j;m_{i-1},m_{j}}$.
%   and the dark orange t-shaped blocks define the tensors $J^{l,\hat{H}}_{j,k}$ (left, $j=2$ and $k=1$) and $J^{r,\hat{H}}_{j,k}$ (right, $j=3$ and $k=1$) from main text (compare \cref{eq:local-krylov:effective_left_operatorproduct} and \cref{eq:local-krylov:effective_right_operatorproduct}).
  %
  The commutator $\left[\hat{H}^{\hat \Pi^{\ket{\psi}}_{2}} \hat{H}^{\hat \Pi^{\ket{\psi}}_{3}}\right]\ket{\psi^{\Pi}_{3}(t)}$ requires also the calculation of $\hat{H}^{\hat \Pi^{\ket{\psi}}_{3}} \hat{H}^{\hat \Pi^{\ket{\psi}}_{2}}\ket{\psi^{\hat \Pi^{\ket{\psi}}_{3}}(t)}$ which is obtained from vertically flipping the tensors covered by the central rectangular grey area between the sites $(i,j)$, i.e., $(2,3)$ in the presented example.}
\end{figure}
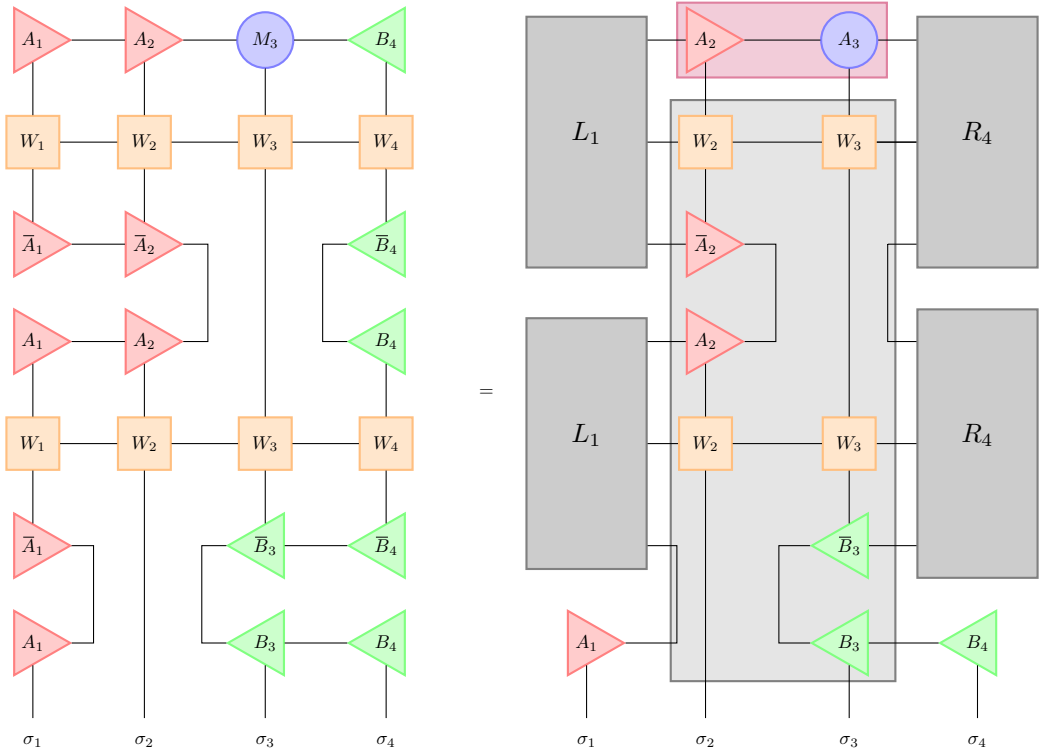
Consider, for instance, the action of $\hat{H}^{\hat \Pi^{\ket{\psi}}_{2}}\hat{H}^{\hat \Pi^{\ket{\psi}}_{3}}\ket{\psi(t)}$, which is generated from the first-order contribution $\hat{H}^{\hat \Pi^{\ket{\psi}}_{3}} \ket{\psi^{\hat \Pi^{\ket{\psi}}_{3}}}$ and subsequent application of $\hat{H}^{\hat \Pi^{\ket{\psi}}_{2}}$. Thus the commutator also needs to be evaluated considering the partial solution $\ket{\psi^{\hat \Pi^{\ket{\psi}}_{j}}}$, so that in general we have for $i<j$
\begin{align}
        \left[\hat{H}^{\hat \Pi^{\ket{\psi}}_{i}}, \hat{H}^{\hat \Pi^{\ket{\psi}}_{j}} \right]\ket{\psi(t)} &= \left[\hat{H}^{\hat \Pi^{\ket{\psi}}_{i}}, \hat{H}^{\hat \Pi^{\ket{\psi}}_{j}} \right]\ket{\psi^{\hat \Pi^{\ket{\psi}}_{j}}(t)}\; .
\end{align}
In \cref{fig:local-krylov:projected_commutator} the action of $\hat{H}^{\hat \Pi^{\ket{\psi}}_{i}} \hat{H}^{\hat \Pi^{\ket{\psi}}_{j}}\ket{\psi^{\hat \Pi^{\ket{\psi}}_{j}}(t)}$ is demonstrated in case of a four-site system with $i=2$, $j=3$ by performing most of the contractions graphically.
In order to obtain the matrix element with an arbitrary state $\ket{\phi}$ we will introduce a compact notation for contractions of MPS and MPO site-tensors with the boundary tensors (partially contracted MPS-MPO-MPS-networks $L_j/R_j$)
\begin{align}
	L^{\noprime}_{j-1 }E^{\noprime}_{j}
	&\equiv
	L^{\noprime}_{j-1} \overbar{M}^{\noprime}_{j} W^{\noprime}_{j} M^{\noprime}_{j} 
	= 
	\sum_{m^{\noprime}_{j-1}, \sigma^{\prime}_{j}} 
	\left(
		\sum_{w^{\noprime}_{j-1}, \sigma^{\noprime}_{j}}
		\left(
			\sum_{\overbar{m}^{\noprime}_{j-1}} 	
				L^{\overbar{m}^{\noprime}_{j-1},w^{\noprime}_{j-1}}_{j-1; m^{\noprime}_{j-1}} 
				\overbar{M}^{\sigma^{\noprime}_{j}}_{j; \overbar{m}^{\noprime}_{j-1},\overbar{m}^{\noprime}_{j}} 
		\right)
		W^{\sigma^{\noprime}_{j},\sigma^{\prime}_{j}}_{j,w^{\noprime}_{j-1},w^{\noprime}_{j}} 
	\right)	
	M^{\sigma^{\prime}_{j}}_{j,m^{\noprime}_{j-1},m^{\noprime}_{j}} 
\\
	E_{j} R_{j+1}
	&\equiv
	\overbar{M}^{\noprime}_{j} W^{\noprime}_{j} M^{\noprime}_{j} R^{\noprime}_{j+1} 
	=
	\sum_{\overbar{m}^\noprime_j, \sigma^\noprime_j}
		\overbar{M}^{\sigma^{\noprime}_{j}}_{j; \overbar{m}^{\noprime}_{j-1},\overbar{m}^{\noprime}_{j}}
		\left(
			\sum_{w^\noprime_j, \sigma^\prime_j}
				W^{\sigma^{\noprime}_{j},\sigma^{\prime}_{j}}_{j; w^{\noprime}_{j-1},w^{\noprime}_{j}}
				\left(
					\sum_{m^\noprime_j}
						M^{\sigma^{\prime}_{j}}_{j; m^{\noprime}_{j-1},m^{\noprime}_{j}} 
						R^{\overbar{m}^{\noprime}_{j},w^{\noprime}_{j}}_{j+1; m^{\noprime}_{j}}
				\right)
		\right)
\end{align}
with the transfer tensors $E^{\noprime}_{j} = \overbar{M}^{\noprime}_{j} W^{\noprime}_{j} M^{\noprime}_{j}$.
We also need transfer tensors with the target state which we define by $E^{\phi}_{j} = \overbar{M}^{\phi}_{j} W^{\noprime}_{j} M^{\noprime}_{j} $.
Finally, there will be \textit{open} bonds at sites $i$ and $j$ that correspond to the contractions originating from the ``brace''-contractions in the projectors $\hat{P}^{L, \ket{\psi}}_{i}$ and $\hat{P}^{R,\ket{\psi}}_{i}$ as well as $\hat{P}^{L,\ket{\psi}}_{j+1}$ and $\hat{P}^{R,\ket{\psi}}_{j+1}$, and we will label these bonds explicitly.
Considering for instance the first summand of the commutator at $i=2,j=3$, that is $\braket{\phi| \hat{H}^{\hat \Pi^{\ket{\psi}}_{2}}\hat{H}^{\hat \Pi^{\ket{\psi}}_{3}} |\psi^{\hat \Pi^{\ket{\psi}}_{3}}(t)}$ (c.f. \cref{fig:local-krylov:projected_commutator}) the left part of the contractions can be written as
\begin{align}
	\sum_{m^{\noprime}_{2},\overbar{m}^{\noprime}_{2}} \left(L^{\noprime}_{1} E^{\noprime}_{2} \right)_{\overbar{m}^{\noprime}_2} \otimes \left(L^{\noprime}_{1} E^{\phi}_{2} \right)_{m^{\noprime}_{2}} \delta_{\overbar{m}^{\noprime}_{2},m^{\noprime}_{2}} 
	&\equiv
	\sum_{m^{\noprime}_{2},\overbar{m}^{\noprime}_{2}}\left(L^{\noprime}_1 \otimes L^{\noprime}_{1} \right) \left( E^{\noprime}_{2} \otimes E^{\phi}_{2} \right)_{\overbar{m}^{\noprime}_{2},m^{\noprime}_{2}} \delta_{\overbar{m}^{\noprime}_{2},m^{\noprime}_{2}}\; .
\end{align}
To obtain a compact notation for the right part we introduce the ``dangling'' transfer tensors $D^{\noprime}_{j} = W^{\noprime}_{j}M^{\noprime}_{j}$ if only the ket site-tensor of the state is included and $\overbar{D}^{\noprime}_{j} = \overbar{M}^{\noprime}_{j} W^{\noprime}_{j}$ if only the bra site-tensor is considered.
In the same manner as for the left part we can now write for the right contractions compactly
\begin{align}
	\sum_{m^{\noprime}_{3},\overbar{m}^{\noprime}_{3}} \left(D_{3}R^{\noprime}_{4} \right)_{\overbar{m}^{\noprime}_{3}} \otimes \left( \overbar{D}^{\phi}_{3} R^{\noprime}_{4} \right)_{m^{\noprime}_{3}}\delta_{\overbar{m}^{\noprime}_{3},m^{\noprime}_{3}}
	&\equiv
	\sum_{m^{\noprime}_{3},\overbar{m}^{\noprime}_{3}} \left( D^{\noprime}_{3} \otimes \overbar{D}^{\phi}_{3} \right) \left( R^{\noprime}_{4} \otimes R^{\noprime}_{4} \right)_{\overbar{m}^{\noprime}_{3},m^{\noprime}_{3}}\delta_{\overbar{m}_{3},m^{\noprime}_{3}} \;.
\end{align}
In general we obtain for the matrix element of the commutator with the target state $\ket{\phi}$ (suppressing the Kronecker-$\delta$ for brevity)
\begin{align}
	&\braket{\phi|\left[\hat{H}^{\hat \Pi^{\ket{\psi}}_{i}}, \hat{H}^{\hat \Pi^{\ket{\psi}}_{j}} \right]|\psi^{\hat \Pi^{\ket{\psi}}_{j}}(t)} \notag \\
	&=
	\sum_{\substack{m^{\noprime}_{j-1},\overbar{m}^{\noprime}_{j-1} \\ m^{\noprime}_{j},\overbar{m}^{\noprime}_{j} }}
	\left(L^{\noprime}_{i-1} \otimes L^{\noprime}_{i-1} \right) \left(E^{\noprime}_{i}\cdots E_{j-1} \otimes E^{\phi}_{i}\cdots E^{\phi}_{j-1} \right)_{\overbar{m}^{\noprime}_{j-1},m^{\noprime}_{j-1}}\left(D_{j}\otimes \overbar{D}^{\phi}_{j} \right) \left( R^{\noprime}_{j+1} \otimes R^{\noprime}_{j+1} \right)_{\overbar{m}^{\noprime}_{j},m^{\noprime}_{j}} + \cdots \notag \\
	&- \sum_{\substack{m^{\noprime}_{i-1},\overbar{m}^{\noprime}_{i-1} \\ m^{\noprime}_{i},\overbar{m}^{\noprime}_{i} }}
	\left(L^{\noprime}_{i-1} \otimes L^{\noprime}_{i-1} \right)_{\overbar{m}^{\noprime}_{i-1},m^{\noprime}_{i-1}} \left(D^{\noprime}_{i}\otimes \overbar{D}^{\phi}_{i} \right)\left(E_{i+1}\cdots E^{\noprime}_{j}\otimes E^{\phi}_{i+1} \cdots E^{\phi}_{j} \right)_{\overbar{m}^{\noprime}_{i},m^{\noprime}_{i}} \left(R^{\noprime}_{j+1}\otimes R^{\noprime}_{j+1} \right)\; .
\end{align}
It becomes immediately clear that only those terms in the Hamiltonian contribute to the commutator which cross the bond $(j-1,j)$ or $(i,i+1)$.
For the current purpose it suffices to find a general estimate for the incurred error $\propto \delta^2$ in the Taylor expansion.
Thus, we will only consider the contributions from nearest-neighbor interactions and hence set $i=j-1$ so that
\begin{align}
	&\braket{\phi|\left[\hat{H}^{\hat \Pi^{\ket{\psi}}_{j-1}}, \hat{H}^{\hat \Pi^{\ket{\psi}}_{j}} \right]|\psi^{\hat \Pi^{\ket{\psi}}_{j}}(t)} \notag \\
	&=
	\sum_{\substack{m^{\noprime}_{j-1},\overbar{m}^{\noprime}_{j-1} \\ m^{\noprime}_{j},\overbar{m}^{\noprime}_{j} }}
	\left(L^{\noprime}_{j-2}\otimes L^{\noprime}_{j-2} \right)\left(E^{\noprime}_{j-1}\otimes E^{\phi}_{j-1} \right)_{\overbar{m}^{\noprime}_{j-1},m^{\noprime}_{j-1}} \left(D^{\noprime}_{j} \otimes \overbar{D}^{\phi}_{j} \right)\left(R^{\noprime}_{j+1} \otimes R^{\noprime}_{j-1} \right)_{\overbar{m}^{\noprime}_{j},m^{\noprime}_{j}} + \cdots \notag \\
	&-
	\sum_{\substack{m^{\noprime}_{j-2},\overbar{m}^{\noprime}_{j-2} \\ m^{\noprime}_{j-1},\overbar{m}^{\noprime}_{j-1} }}
	\left(L^{\noprime}_{j-2} \otimes L^{\noprime}_{j-2} \right)_{\overbar{m}^{\noprime}_{j-2},m^{\noprime}_{j-2}} \left(D^{\noprime}_{j-1} \otimes \overbar{D}^{\phi}_{j-1} \right) \left(E^{\noprime}_{j} \otimes E^{\phi}_{j} \right)_{\overbar{m}^{\noprime}_{j-1},m^{\noprime}_{j-1}} \left(R^{\noprime}_{j+1} \otimes R^{\noprime}_{j+1} \right) \;.
\end{align}
The crucial observation here is that for each ``open'' bond index pair we can treat the combined MPS-MPS and MPO-MPO tensor-contractions over the bond $(j-1,j)$ as scalar product between the left and right part of the system.
We can thus write for the first summand for each open index pair
\begin{align}
	&\underbrace{\left(L^{\noprime}_{j-2}\otimes L^{\noprime}_{j-2} \right)\left(E^{\noprime}_{j-1}\otimes E^{\phi}_{j-1} \right)_{\overbar{m}^{\noprime}_{j-1},m^{\noprime}_{j-1}}}_{\equiv \bra{e^{L}_{j-1}\otimes e^{L,\phi}_{j-1}}} \underbrace{\vphantom{\left(L^{\noprime}_{j-2}\otimes L^{\noprime}_{j-2} \right)\left(E^{\noprime}_{j-1}\otimes E^{\phi}_{j-1} \right)_{\overbar{m}^{\noprime}_{j-1},m^{\noprime}_{j-1}}}\left(D^{\noprime}_{j} \otimes \overbar{D}^{\phi}_{j} \right)\left(R^{\noprime}_{j+1} \otimes R^{\noprime}_{j-1} \right)_{\overbar{m}^{\noprime}_{j},m^{\noprime}_{j}}}_{\equiv \ket{d^{R}_{j}\otimes \overbar{d}^{R,\phi}_{j}}} \notag \\
	&=
	\frac{1}{2}\left( \left\lvert \ket{e^{L}_{j-1}\otimes e^{L,\phi}_{j-1}} + \ket{d^{R}_{j}\otimes \overbar{d}^{R,\phi}_{j}} \right\rvert^{2}  - \braket{e^{L}_{j-1}\otimes e^{L,\phi}_{j-1}|e^{L}_{j-1} \otimes e^{L,\phi}_{j-1}} - \braket{d^{R}_{j}\otimes \overbar{d}^{R,\phi}_{j}|d^{R}_{j}\otimes \overbar{d}^{R,\phi}_{j}} \right)
\end{align}
and for the second summand
\begin{align}
	&\underbrace{\vphantom{\left(E^{\noprime}_{j} \otimes E^{\phi}_{j} \right)_{\overbar{m}^{\noprime}_{j-1},m^{\noprime}_{j-1}} \left(R^{\noprime}_{j+1} \otimes R^{\noprime}_{j+1} \right)}\left(L^{\noprime}_{j-2} \otimes L^{\noprime}_{j-2} \right)_{\overbar{m}^{\noprime}_{j-2},m^{\noprime}_{j-2}} \left(D^{\noprime}_{j-1} \otimes \overbar{D}^{\phi}_{j-1} \right)}_{\equiv \bra{d^{L}_{j-1}\otimes \overbar{d}^{L,\phi}_{j-1}}} \underbrace{\left(E^{\noprime}_{j} \otimes E^{\phi}_{j} \right)_{\overbar{m}^{\noprime}_{j-1},m^{\noprime}_{j-1}} \left(R^{\noprime}_{j+1} \otimes R^{\noprime}_{j+1} \right)}_{\equiv \ket{e^{R}_{j}\otimes e^{R,\phi}_{j}}} \notag \\
	&= 
	-\frac{1}{2}\left( \left\lvert \ket{d^{L}_{j-1}\otimes \overbar{d}^{L,\phi}_{j-1}} -  \ket{e^{R}_{j}\otimes e^{R,\phi}_{j}}\right\rvert^{2} - \braket{d^{L}_{j-1}\otimes \overbar{d}^{L,\phi}_{j-1}|d^{L}_{j-1}\otimes \overbar{d}^{L,\phi}_{j-1}} - \braket{e^{R}_{j}\otimes e^{R,\phi}_{j}|e^{R}_{j}\otimes e^{R,\phi}_{j}} \right) \;.
\end{align}
We can bound the sums over the open bonds for the scalar products in terms of absolute values of expectation values of effective Hamiltonians $\hat{h}^{L/R,\mathrm{eff}}_{j}$ at site $j$
\begin{align}
	\sum_{\substack{ m^{\noprime}_{j-1} \\ \overbar{m}^{\noprime}_{j-1}}}\braket{e^{L}_{j-1}\otimes e^{L,\phi}_{j-1}|e^{L}_{j-1} \otimes e^{L,\phi}_{j-1}}
	&\leq \sum_{\substack{ m^{\noprime}_{j-1} \\ \overbar{m}^{\noprime}_{j-1}}} \left\lvert \ket{e^{L}_{j-1}\otimes e^{L,\phi}_{j-1}} \right\rvert^{2} 
	= \left\lvert \overbar{\psi}^{\mathrm{eff}}_{j-1}\hat{h}^{L,\mathrm{eff}}_{j-1}\psi^{\mathrm{eff}}_{j-1} \cdot \overbar{\phi}^{\mathrm{eff}}_{j-1}\hat{h}^{L,\mathrm{eff}}_{j-1}\psi^{\mathrm{eff}}_{j-1} \right\rvert \\
	\sum_{\substack{ m^{\noprime}_{j} \\ \overbar{m}^{\noprime}_{j} }} \braket{d^{R}_{j}\otimes \overbar{d}^{R,\phi}_{j}|d^{R}_{j}\otimes \overbar{d}^{R,\phi}_{j}}
	&\leq \sum_{\substack{ m^{\noprime}_{j} \\ \overbar{m}^{\noprime}_{j} }} \left\lvert \ket{d^{R}_{j}\otimes \overbar{d}^{R,\phi}_{j}} \right\rvert^{2}
	= \left\lvert \overbar{\phi}^{\mathrm{eff}}_{j}\left[\hat{h}^{R,\mathrm{eff}}_{j}\right]^2 \psi^{\mathrm{eff}}_{j} \right\rvert \\
	\sum_{\substack{ m^{\noprime}_{j-2} \\ \overbar{m}^{\noprime}_{j-2} }} \braket{d^{L}_{j-1}\otimes \overbar{d}^{L,\phi}_{j-1}|d^{L}_{j-1}\otimes \overbar{d}^{L,\phi}_{j-1}} 
	&\leq \sum_{\substack{ m^{\noprime}_{j-2} \\ \overbar{m}^{\noprime}_{j-2} }} \left\lvert \ket{d^{L}_{j-1}\otimes \overbar{d}^{L,\phi}_{j-1}}\right\rvert^{2} 
	= \left\lvert \overbar{\phi}^{\mathrm{eff}}_{j-1} \left[\hat{h}^{L,\mathrm{eff}}_{j-1}\right]^{2}\psi^{\mathrm{eff}}_{j-1} \right\rvert \\
	\sum_{\substack{ m^{\noprime}_{j-1} \\ \overbar{m}^{\noprime}_{j-1} }} \braket{e^{R}_{j}\otimes e^{R,\phi}_{j}|e^{R}_{j}\otimes e^{R,\phi}_{j}}
	&\leq \sum_{\substack{ m^{\noprime}_{j-1} \\ \overbar{m}^{\noprime}_{j-1} }} \left\lvert \ket{e^{R}_{j}\otimes e^{R,\phi}_{j}} \right\rvert^{2}
	= \left\lvert \overbar{\psi}^{\mathrm{eff}}_{j}\hat{h}^{R,\mathrm{eff}}_{j}\psi^{\mathrm{eff}}_{j} \cdot \overbar{\phi}^{\mathrm{eff}}_{j}\hat{h}^{R,\mathrm{eff}}_{j}\psi^{\mathrm{eff}}_{j} \right\rvert
\end{align}
where we defined
\begin{equation}
	\hat{h}^{L,\mathrm{eff}}_{j} = L^{\noprime}_{j-1} W^{\noprime} _{j}\quad \text{and} \quad \left[\hat{h}^{L,\mathrm{eff}}_{j}\right]^{2} = \left(L^{\noprime}_{j-1}\otimes L^{\noprime}_{j-1}\right) \left(W^{\noprime}_{j} \otimes W^{\noprime}_{j}\right)
\end{equation}
and in a similiar way $\hat{h}^{R,\mathrm{eff}}_{j}$ replacing $L_{j-1}\rightarrow R_{j+1}$.
The formal absolute values ``$\lvert \cdot \rvert$'' on the right side can be estimated by replacing the $L_{j}/R_{j}$ tensors with fractions of the overall energy expectation value.
This is a valid approximation as long as there are no interactions connecting the left/right contracted MPS-MPO-MPS tensor networks with sites to the right/left of them which was exactly the condition for non-vanishing contributions to the commutator.
We hence compare only squares of single-site expectation values with either one or two MPO site tensors sandwiched between the effective site tensors at sites $j-1,j$.
For a large system with smoothly varying site tensors the differences at neighboring sites are negligible so that the commutator can be estimated to
\begin{align}
	\braket{\phi|\left[\hat{H}^{\hat \Pi^{\ket{\psi}}_{j-1}}, \hat{H}^{\hat \Pi^{\ket{\psi}}_{j}} \right]|\psi^{\hat \Pi^{\ket{\psi}}_{j}}(t)}
	&\leq \frac{1}{2}\left( \left\lvert \ket{e^{L}_{j-1}\otimes e^{L,\phi}_{j-1}} + \ket{d^{R}_{j}\otimes \overbar{d}^{R,\phi}_{j}} \right\rvert^{2} + \left\lvert \ket{d^{L}_{j-1}\otimes \overbar{d}^{L,\phi}_{j-1}} - \ket{e^{R}_{j}\otimes e^{R,\phi}_{j}}\right\rvert^{2} \right) \notag \\
	&\leq \frac{1}{2}\left(\left\lvert \ket{e^{L}_{j-1}\otimes e^{L,\phi}_{j-1}}\right\rvert^{2} + \left\lvert \ket{d^{R}_{j}\otimes \overbar{d}^{R,\phi}_{j}} \right\rvert^{2} \right) \;.
\end{align}
The last expression can be estimated easily by the coupling strength of the nearest-neighbor interaction term which we denote by $\Gamma$.
The contributions from the boundary tensors $L_{j}/R_{j}$ are bounded by the square of the system size and cancel with the prefactors $\nicefrac{1}{L}$ of each projected Hamiltonian.
We hence conclude this analysis by the somewhat straightforward statement that in general the error at second order in $\delta$ scales with $L\cdot \Gamma^2$.
However, there is one special situation which simplifies the preceeding arguments drastically namely if we set $\ket{\phi}=\ket{\psi^{\hat \Pi^{\ket{\psi}}_{j}}(t)}$.
Then the commutator compares only local overlaps between site tensors sandwiched between either one or two MPO tensors on neighboring sites which scales as $\nicefrac{1}{L}$.
Therefore equal-time observables are evolved with very high precision $\sim \nicefrac{1}{L^2}$ and the contribution due to nearest-neighbor interactions is strongly suppressed.

\subsubsection{Algorithm}

\begin{algorithm}
  \caption{\label{alg:local-krylov:common-helpers}Common helper
    functions for the local Krylov method and the TDVP algorithms.}
  \begin{algorithmic}[1]
    \Procedure{Contract-Left}{$L_{j-1}$, $W_{j}$, $A_{j}$} \Comment{left-contraction as in DMRG}
    \State \textbf{return } 
		$
			L_{j; m^\noprime_j}^{\overbar{m}^\noprime_j, w^\noprime_j}
				= 
					\sum_{
						\sigma^\noprime_{j}, 
						\sigma^\prime_{j}, 
						w^\noprime_{j-1}, 
						m^\noprime_{j-1}, 
						\overbar{m}^\noprime_{j-1}
					} 
						L_{j-1; m^\noprime_{j-1}}^{\overbar{m}^\noprime_{j-1}, w^\noprime_{j-1}}
						\overbar{A}^{\sigma^\noprime_{j}}_{j;\overbar{m}^\noprime_{j-1}, \overbar{m}^\noprime_{j}} 
						W_{j; w^\noprime_{j-1}, w^\noprime_{j}}^{\sigma^\noprime_{j}, \sigma^\prime_{j}} 
						A^{\sigma^\prime_{j}}_{j; m^\noprime_{j-1}, m^\noprime_{j}}
		$
    \EndProcedure
    \Procedure{Contract-Right}{$R_{j+1}$, $W_{j}$, $B_{j}$} \Comment{right contraction as in DMRG}
    \State \textbf{return } 
		$
			R_{j; m^\noprime_{j-1}}^{\overbar{m}^\noprime_{j-1}, w^\noprime_{j-1}} 
				= 
					\sum_{
						\sigma^\noprime_{j}, 
						\sigma^\prime_{j}, 
						w^\noprime_j, 
						m^\noprime_j, 
						\overbar{m}^\noprime_j
					}
						\overbar{B}^{\sigma^\noprime_{j}}_{\overbar{m}^\noprime_{j-1}, \overbar{m}^\noprime_{j}} 
						W_{j; w^\noprime_{j-1}, w^\noprime_{j}}^{\sigma^\noprime_{j}, \sigma^\prime_{j}}
						B^{\sigma^\prime_{j}}_{j; m^\noprime_{j-1}, m^\noprime_{j}} R_{j+1; m^\noprime_j}^{\overbar{m}^\noprime_j, w^\noprime_j}$
    \EndProcedure
    \Procedure{Initialize}{MPO $\{ W_j \}_{j=1}^L$, MPS $\{ M_j \}_{j=1}^L$}
    \State $L_{0;m_0}^{\overbar{m}_0, w_0} \gets 1$ and $R_{L+1;m_L}^{\overbar{m}_L,w_L} \gets 1$
    \State Right-normalize $\{ M_j \}_{j=1}^L \to \{ B_j \}_{j=1}^L$ from right to left
    \For{$j \in [L, 2]$}
      \State $R_j \gets \Call{Contract-Right}{R_{j+1}, W_j, B_j}$
    \EndFor
    \State \textbf{return } $L_0$, $\{ R_j \}_{j=2}^{L+1}$, $\{ B_j \}_{j=1}^L$
    \EndProcedure
    \Procedure{Timestep}{$\delta$, $L_0$, $\{ R_j \}_{j=2}^{L+1}$, $\{ W_j \}$, $\{ M_1, B_j \}_{j=2}^L$}
    \State $\{ L_j \}_{j=0}^{L-1}$, $\{ A_j, M_L \}_{j=1}^{L-1}$ $\gets$ \Call{Sweep-Right}{$\nicefrac{\delta}{2}$, $L_0$, $\{ R_j \}_{j=2}^L$, $\{ W_j \}_{j=1}^L$, $\{ M_1, B_j \}_{j=2}^L$}
    \State $\{ R_j \}_{j=2}^{L+1}$, $\{ M_1, B_j \}_{j=2}^L$ $\gets$ \Call{Sweep-Left}{$\nicefrac{\delta}{2}$, $R_{L+1}$, $\{ L_j \}_{j=0}^{L-1}$, $\{ W_j \}_{j=1}^L$, $\{ A_j, M_L \}_{j=1}^{L-1}$}
    \EndProcedure
  \end{algorithmic}
\end{algorithm}

\begin{algorithm}
  \caption{\label{alg:local-krylov:method}The local 2-site Krylov
    method. Also cf.~\cref{alg:local-krylov:common-helpers} for
    definitions of \textsc{Initialize}, \textsc{Contract-Left},
    \textsc{Contract-Right} and the needed overall \textsc{Timestep}
    method.}
  \begin{algorithmic}[1]
    \Procedure{Sweep-Right}{$\delta$, $L_0$, $\{ R_j \}_{j=3}^{L+1}$, $\{ W_j \}_{j=1}^L$, $\{M_1, B_j \}_{j=2}^L$}
    \State $A_1, C_{\underline{1}} \gets M_1$ via QR; \quad $M_2 \gets C_{\underline{1}} \cdot B_2$
    \State $P^{\sigma_1}_{\overbar{m}_{0}, m_1} \gets A^{\sigma_1}_{1; m_0, m_1}$ \Comment{entry-for-entry copy, only label of $m_0$ adapted}
    \For{$j \in [1, L-1]$}
    \State $T_{j,j+1} \gets \sum_{m_j} A_{j;m_{j-1}, m_j}^{\sigma_j} M_{j+1;m_j, m_{j+1}}^{\sigma_{j+1}}$
    \State $T_{j,j+1} \gets \mathrm{exp}(-\mathrm{i}\delta \hat H^{\mathrm{eff}}_{(j,j+1)}) T_{j,j+1}$ using $\hat H^{\mathrm{eff}}_{(j,j+1)} \equiv L_{j-1} \cdot W_j \cdot W_{j+1} \cdot R_{j+2}$
    \If{$j \neq L-1$}
    \State $A^\prime_j \gets T_{j,j+1}$ via SVD and truncation, $S$ and $V$ discarded
    \State $L_j \gets \Call{Contract-Left}{L_{j-1}, W_j, A^\prime_j}$; \quad delete $R_{j+2}$
    \State $P^{\prime \overbar{m}_j}_{m_j} \gets \sum_{\sigma_j \overbar{m}_{j-1}} P^{\sigma_j}_{\overbar{m}_{j-1}, m_j} \cdot \overbar{A}^{\prime \sigma_j}_{j; \overbar{m}_{j-1}, \overbar{m}_j}$ \Comment{complete projector}
    \State $M^{\sigma_{j+1}}_{j+1; \overbar{m}_j, m_{j+1}} \gets \sum_{m_j} P^{\prime \overbar{m}_j}_{m_j} \cdot M^{\sigma_{j+1}}_{j+1; m_j, m_{j+1}}$ \Comment{project $M_{j+1}$ into new left basis}
    \State $A_{j+1}, C_{\underline{j+1}} \gets M_{j+1}$ via QR; \quad $M_{j+2} \gets C_{\underline{j+1}} \cdot B_{j+2}$
    \State $P^{\sigma_{j+1}}_{\overbar{m}_{j}, m_{j+1}} \gets \sum_{m_j} P^{\prime \overbar{m}_{j}}_{m_j} \cdot A^{\sigma_j}_{j+1; m_j,m_{j+1}}$
    \Else
    \State $A^\prime_j, C_{\underline{j}}, B_{j+1} \gets T_{j,j+1}$ via SVD and truncation
    \State $M_{j+1} \gets C_{\underline{j}} \cdot B_{j+1}$
    \EndIf
    \State $A_j \gets A^\prime_j$
    \EndFor
    \State \textbf{return } $\{ L_j \}_{j=0}^{L-2}$, $\{ A_j, M_L \}_{j=1}^{L-1}$
    \EndProcedure
    \Procedure{Sweep-Left}{$\delta$, $R_{L+1}$, $\{ L_j \}_{j=0}^{L-2}$, $\{ W_j \}_{j=1}^L$, $\{ A_j, M_L \}_{j=1}^{L-1}$}
    \State $B_L, C_{\underline{L-1}} \gets M_L$ via QR; \quad $M_{L-1} \gets A_{L-1} \cdot C_{\underline{L-1}}$
    \State $P^{\sigma_L}_{m_{L-1}, \overbar{m}_L} \gets B^{\sigma_L}_{L; m_{L-1}, m_L}$ \Comment{entry-for-entry copy, only label of $m_L$ adapted}
    \For{$j \in [L, 2]$}
    \State $T_{j-1,j} \gets \sum_{m_{j-1}} A_{j-1;m_{j-2}, m_{j-1}}^{\sigma_{j-1}} M_{j;m_{j-1}, m_{j}}^{\sigma_{j}}$
    \State $T_{j-1,j} \gets \mathrm{exp}(-\mathrm{i} \delta \hat H^{\mathrm{eff}}_{(j-1,j)}) T_{j-1,j}$ using $\hat H^{\mathrm{eff}}_{(j-1,j)} \equiv L_{j-2} \cdot W_{j-1} \cdot W_{j} \cdot R_{j+1}$
    \If{$j \neq 2$}
    \State $B^\prime_{j} \gets T_{j-1,j}$ via SVD and truncation, $U$ and $S$ discarded
    \State $R_j \gets \Call{Contract-Right}{R_{j+1}, W_j, B_j}$; \quad delete $L_{j-2}$
    \State $P^{\prime \overbar{m}_{j-1}}_{m_{j-1}} \gets \sum_{\sigma_j \overbar{m_j}} P^{\sigma_j}_{m_{j-1}, \overbar{m}_j} \overbar{B}^{\prime \sigma_j}_{j; \overbar{m}_{j-1}, \overbar{m}_j}$ \Comment{complete projector}
    \State $M^{\sigma_{j-1}}_{j-1; m_{j-2}, \overbar{m}_{j-1}} \gets \sum_{m_{j-1}} P^{\prime \overbar{m}_{j-1}}_{m_{j-1}} M^{\sigma_{j-1}}_{j-1; m_{j-2}, m_{j-1}}$
    \State $B_{j-1}, C_{\underline{j-2}} \gets M_{j-1}$ via QR; \quad $M_{j-2} \gets M_{j-2} \cdot C_{\underline{j-2}}$
    \State $P^{\sigma_{j-1}}_{m_{j-2}, \overbar{m}_{j-1}} \gets \sum_{m_{j-1}} P^{\prime \overbar{m}_{j-1}}_{m_{j-1}} B^{\sigma_{j-1}}_{j-1; m_{j-2}, m_{j-1}}$
    \Else
    \State $A_{j-1}, C_{\underline{j-1}}, B_j \gets T_{j-1,j}$ via SVD and truncation
    \State $M_{j-1} \gets A_{j-1} \cdot C_{\underline{j-1}}$
    \EndIf
    \State $B_j \gets B^\prime_j$
    \EndFor
    \State \textbf{return } $\{ R_j \}_{j=2}^{L+1}$, $\{ M_1, B_j \}_{j=2}^{L}$
    \EndProcedure
  \end{algorithmic}
\end{algorithm}

The two-site local Krylov method is described in detail in
\cref{alg:local-krylov:method}. It relies on some basic initialization
functions familiar from the DMRG algorithm which are summarized in
\cref{alg:local-krylov:common-helpers}.

%%% Local Variables:
%%% mode: latex
%%% TeX-master: "../time_evolution_review"
%%% End:

%% file: content/tdvp.tex
\clearpage

\subsection{\label{sec:tdvp}The time-dependent variational principle (TDVP)}

There is an alternative to the Lie-Trotter decomposition introduced in
the previous \cref{sec:local-krylov} which also results in a series of
local problems: the time-dependent variational
principle\cite{haegeman11:_time_depen_variat_princ_quant_lattic,
  haegeman16:_unify}. The motivation of this approach is quite
different: its primary aim is to constrain the time evolution to a
specific manifold of matrix-product states of a given initial bond
dimension. To do so, it projects the action of the Hamiltonian into
the tangent space to this manifold and then solves the TDSE solely
within the manifold. While ideally used in its single-site variant,
the two-site variant allows for flexibility in the bond dimension.

\input{content/tdvp_effrhs.tex}

\subsubsection{Derivation}

The main difference between the TDVP and the local Krylov method is in
the derivation of the series of these local time-dependent Schr\"odinger
equations and the recovery of the original-time step after each local
forward update: Instead of simply projecting the original site tensor
onto the new basis as done in the local Krylov approach, the TDVP
explicitly solves a backwards-evolution equation.
To embark on the derivation for the TDVP, we need to introduce a few
additional ingredients: First, we define the \emph{single-site tangent
  space} $T_{|\psi\rangle}$ of a given MPS $|\psi\rangle$ as the space
spanned by variations of single MPS tensors.  One may e.g.~change the
first site tensor of the MPS keeping all others fixed to obtain a new
state and combine the result with another MPS where only the second
site tensor was changed, but one may not change two (or more) site
tensors in the same basis state.  The projector
$\hat P_{T_{\ket{\psi}}}$ which projects onto this tangent space is
given by\cite{haegeman11:_time_depen_variat_princ_quant_lattic,
  haegeman16:_unify}
\begin{equation}
  \hat P_{T_{|\psi\rangle}} = \sum_{j=1}^L \hat P_{j-1}^{L, |\psi\rangle} \otimes \mathbf{\hat 1}^{\vphantom{L\ket{\psi}}}_j \otimes \hat P_{j+1}^{R,|\psi\rangle} - \sum_{j=1}^{L-1} \hat P_{j}^{L,|\psi\rangle} \otimes \hat P_{j+1}^{R,|\psi\rangle} \quad, \label{eq:tdvp:projector}
\end{equation}
where $\hat{P}^{L,\ket{\psi}}_j$ ($\hat{P}^{R,\ket{\psi}}_j$) projects on the sites left (right)
of and including site $n$ (cf.~\cref{fig:tdvp_projectors}) and is
exactly the same as the projectors used in the local Krylov method.
As before, these projectors use the gauge-fixed left- and
right-normalised MPS tensors, i.e.~they depend on the MPS
$|\psi\rangle$ and can be written as
\begin{align}
  \hat P^{L, \ket{\psi}}_{j; \overbar{\sigma}_1, \ldots, \overbar{\sigma}_j, \sigma_1, \ldots, \sigma_j} & = \sum_{m_j} \overbar{\psi}^{L; \overbar{\sigma}_1, \ldots, \overbar{\sigma}_j}_{j; m_j} \otimes \psi^{L; \sigma_1,\ldots, \sigma_j}_{j; m_j} \\
  \hat P^{R, \ket{\psi}}_{j; \overbar{\sigma}_j, \ldots, \overbar{\sigma}_L, \sigma_j, \ldots, \sigma_L} & = \sum_{m_{j-1}} \overbar{\psi}^{R; \overbar{\sigma}_j, \ldots, \overbar{\sigma}_L}_{j; m_{j-1}} \otimes \psi^{R; \sigma_j,\ldots, \sigma_L}_{j; m_{j-1}} 
\end{align}
where $\psi^{L(R)}_{j}$ is the collection of left- (right-) normalised
MPS tensors on site $j$ and to its left (right) as in the local Krylov
approach.  The first contributing sum in \cref{eq:tdvp:projector}
filters for all MPS which differ at most on one site from
$|\psi\rangle$, whereas the second contributing sum removes all those
states which coincide with $|\psi\rangle$.  Put differently,
individual tangent vectors are constructed by replacing any
orthogonality center tensor $M_j$ of the MPS by another tensor $N_j$
which is orthogonal to $M_j$, i.e.,~$M_j \cdot \bar{N}_{j} = 0$. In
contrast to the projectors $\hat \Pi_j^{\ket{\psi}}$ of the local
Krylov method, the total projector $\hat P_{T_{\ket{\psi}}}$ projects
onto some subspace of the Hilbert space and is only coincidentally
written as a sum of local terms.

Second, when inserting the projector $\hat P_{T_{|\psi\rangle}}$ into
the TDSE, we obtain
\begin{align}
  \frac{\partial}{\partial t}|\psi\rangle = & - \mathrm{i} \hat P_{T_{|\psi\rangle}} \hat H |\psi\rangle \\
                                          = & - \mathrm{i} \sum_{j=1}^L \hat P_{j-1}^{L,|\psi\rangle} \otimes \mathbf{\hat 1}^{\vphantom{L\ket{\psi}}}_j \otimes \hat P_{j+1}^{R,|\psi\rangle} \hat H |\psi\rangle + \mathrm{i} \sum_{j=1}^{L-1} \hat P_{j}^{L,|\psi\rangle} \otimes \hat P_{j+1}^{R,|\psi\rangle} \hat H |\psi\rangle \label{eq:tdvp:tdse-projected}\;.
\end{align}
While an exact solution of \cref{eq:tdvp:tdse-projected} is still not
possible, we can approximate it by solving each term individually and
sequentially, i.e.,~solve $L$ forward-evolving equations of the form
\begin{alignat}{2}
  \frac{\partial}{\partial t}|\psi\rangle &= - \mathrm{i} \hat P_{j-1}^{L,\ket{\psi}} \otimes \mathbf{\hat 1}^{\vphantom{L\ket{\psi}}}_j \otimes \hat P_{j+1}^{R,\ket{\psi}} \hat H |\psi\rangle\label{eq:tdvp:fw}
  \intertext{and $L-1$ backward-evolving equations of the form}
  \frac{\partial}{\partial t}|\psi\rangle &= + \mathrm{i} \hat P_{j}^{L,\ket{\psi}} \otimes \hat P_{j+1}^{R,\ket{\psi}} \hat H |\psi\rangle \;.\label{eq:tdvp:bw}
\end{alignat}
We then multiply each individual equation above by the single-site map
$\overbar{\psi}^L_{j-1} \otimes \overbar{\psi}^R_{j+1}$ or the
center-bond map $\overbar{\psi}^L_{j} \otimes \overbar{\psi}^R_{j+1}$,
respectively.  As a result, instead of having to work with the full
MPS $|\psi\rangle$, we can work with the effective single-site and
effective center matrix tensors and associated local Schr\"odinger
equations directly:
\begin{align}
  \frac{\partial}{\partial t}M^{\vphantom{\mathrm{eff}}}_j &= - \mathrm{i} \hat H^{\mathrm{eff}}_{j} M^{\vphantom{\mathrm{eff}}}_j\label{eq:tdvp:fw:projected} \\
  \frac{\partial}{\partial t}C^{\vphantom{\mathrm{eff}}}_{\underline{j}} &= + \mathrm{i} \hat H^\mathrm{eff}_{\underline{j}} C^{\vphantom{\mathrm{eff}}}_{\underline{j}} \quad.\label{eq:tdvp:bw:projected}
\end{align}
The tensor contraction in the RHS of \cref{eq:tdvp:fw:projected} is
graphically represented in \cref{fig:tdvp_efftdse}, while the RHS of
\cref{eq:tdvp:bw:projected} is shown in \cref{fig:tdvp_efftdse_bond}.
Each of these equations can be solved with a local application of the
Krylov method much as in DMRG or the local Krylov method of the
previous section.

Sweeping right-to-left (rather than left-to-right) through the system
results in solving the equations in reverse order.  This turns the
initial first-order integrator into a second-order integrator,
reducing the time step error (as described in \cref{sec:tdvp:errors})
from $O(\delta)$ to $O(\delta^2)$ if both sweeps are done with halved
time steps $\delta/2$.

An interesting property of the single-site TDVP variant (1TDVP) is
that the projection of the Hamiltonian onto the MPS manifold occurs
before the time evolution, the projection is the \emph{only} step
necessary to obtain the Lie-Trotter decomposition of the Hamiltonian,
and no truncation has to happen after the evolution.  As such, both
the norm and energy of the state are conserved under real-time
evolution. This is in contrast to the local Krylov method, where the
basis transformation generated by the $Q$ tensors (as described at the
end of \cref{sec:local-krylov}) is \emph{not} part of the projector
and hence introduces an additional error.
Alternatively, it is
straightforward to extend the mechanism to a two-site variant. This
2TDVP forward-evolves a local tensor $M_{(j,j+1)}$ which needs to be
split into two separate site tensors again following the evolution.
The advantage is that the bond dimension of the state can be adapted
on the fly.  However, norm and energy are now no longer conserved
exactly if a truncation of the evolved bond is necessary.

\subsubsection{\label{sec:tdvp:errors}Errors}

The TDVP has four sources of errors:
firstly, there is a \emph{projection error} due to the projection of
the full time-dependent Schr\"odinger equation (TDSE) onto the MPS
manifold of limited bond dimension.  This error is particularly severe
if the MPS in question has a small bond dimension, but it is exactly
zero if the MPS has maximal (exponentially growing) bond dimension.
However, the projection error occurs during the projection of the TDSE
onto the relevant subspace, i.e.,~\emph{before} the time evolution.  As
such, it cannot lead to a violation of energy conservation or change
the norm of the time-evolved state (during real-time evolution).
Using a two- or multi-site variance\cite{hubig18:_error} it is
possible to estimate this projection error.  If the $n$-site variance
of the state is large, the $(n-1)$TDVP will provide inadequate
results.  Vice versa, if the up-to-$n$-site variance of a state is
small, the $n$TDVP will consider this state an eigenstate of the
Hamiltonian and the time evolution will only add a global phase to the
state.  As a corollary, the 2TDVP can evolve Hamiltonians with only
nearest-neighbor interactions without incurring a projection error.

Second, the chain of forwards and backwards evolutions can be
considered a sequential solution of a series of coupled TDSE (which
are the result of the projection above), each describing the evolution
of any particular site tensor.  Except in the special case that all
these evolutions describe exactly the same dynamics (due to the state
having maximal bond dimension), there is a finite time-step error of
order $O(\delta^3)$ per time step and order $O(\delta^2)$ per unit
time.  In practice, the prefactor of this error is often much smaller
than, e.g., in a TEBD calculation, in particular if the bond dimension
of the input state is reasonably large. If the bond dimension is very
small, the time-step error will be relatively large.

Third, the 2TDVP contains a SVD to split the evolved two-site tensor
into two separate tensors again.  During this SVD, a truncation is
typically unavoidable, leading to a measurable truncation error.
Careful analysis of this truncation error is necessary as always, but
also proceeds in much the same way as always. In 1TDVP, this error is
exactly zero.

The fourth source of error lies in the inexact solution of the local
equations.  Using sufficiently many Krylov vectors locally, it is very
easy to make this error small. Therefore, one should always use
sufficiently many vectors such that the obtained error is at least
smaller than the truncation error in the previous step.

Note that changing the time-step size $\delta$ in the TDVP affects the
four errors differently: the projection and truncation error affect
each time step relatively independently of the size of that step.
Hence, increasing the number of time steps during a fixed total time
evolution increases the projection and truncation errors.  The finite
time-step error and the error from the inexact local solution, on the
other hand, decrease when increasing the number of time steps and the
total time is kept fixed.  As such, choosing a smaller $\delta$
decreases the time-step error but increases the projection and
truncation error.  It is hence typically useful to take some care when
choosing, e.g., the truncation threshold and the time-step size such as
to approximately balance the induced errors.

Additionally, the energy and norm of the state are conserved exactly
within the 1TDVP and only affected by the truncation error in the
2TDVP. This exact conservation may extend to those quantities which
are contained within the Hamiltonian\cite{leviatan17:_quant_matrix_produc_states, goto18:_perfor}.
While such energy conservation is certainly very helpful to obtain
long-time hydrodynamic observables such as diffusion constants, care
has to be taken when using only 1TDVP during the calculation as shown
in Ref.~\cite{kloss18:_time}. Specifically, one has to take great care
to ensure that the obtained data is completely converged in the bond
dimension of the state at all times.

\begin{algorithm}[tb]
  \caption{\label{alg:tdvp:1tdvp}The 1TDVP method. For a more
    detailled description of the Lanczos solver used in lines
    \ref{alg:tdvp:1tdvp:srsm}, \ref{alg:tdvp:1tdvp:srsc},
    \ref{alg:tdvp:1tdvp:slsm} and \ref{alg:tdvp:1tdvp:slsc} see
    \cref{sec:te:krylov}. Also
    cf.~\cref{alg:local-krylov:common-helpers} for definitions of
    \textsc{Initialize}, \textsc{Contract-Left},
    \textsc{Contract-Right} and the needed overall \textsc{Timestep}
    method. To run, one first initializes the worker object using
    \textsc{Initialize} and then does incremental time steps using
    the \textsc{Timestep} function.}
  \begin{algorithmic}[1]
    \Procedure{Sweep-Right}{$\delta$, $L_0$, $\{ R_j \}_{j=2}^{L+1}$, $\{ W_j \}_{j=1}^L$, $\{ M_1, B_j \}_{j=2}^L$}
    \For{$j \in [1, L]$}
      \State $M^{\vphantom{\mathrm{eff}}}_j \gets \mathrm{exp}(-\mathrm{i}\nicefrac{\delta}{2} \hat H^{\mathrm{eff}}_j) M^{\vphantom{\mathrm{eff}}}_j$ using $\hat H_j^{\mathrm{eff}} \equiv L^{\vphantom{\mathrm{eff}}}_{j-1} \cdot W^{\vphantom{\mathrm{eff}}}_j \cdot R^{\vphantom{\mathrm{eff}}}_{j+1}$ \Comment{e.g.~with Lanczos} \label{alg:tdvp:1tdvp:srsm}
      \State $A_j, C_{\underline{j}} \gets M_j$ via QR
      \State $L_j \gets \Call{Contract-Left}{L_{j-1}, W_j, A_j}$
      \If{$j \neq L$}
        \State $C^{\vphantom{\mathrm{eff}}}_{\underline{j}} \gets \mathrm{exp}(\mathrm{i}\nicefrac{\delta}{2} \hat H_{\underline{j}}^{\mathrm{eff}}) C^{\vphantom{\mathrm{eff}}}_{\underline{j}}$ using $\hat H_{\underline{j}}^{\mathrm{eff}} \equiv L^{\vphantom{\mathrm{eff}}}_{j} \cdot R^{\vphantom{\mathrm{eff}}}_{j+1}$ \Comment{e.g. with Lanczos} \label{alg:tdvp:1tdvp:srsc}
        \State $M_{j+1} \gets C_{\underline{j}} \cdot B_{j+1}$
        \State Delete $R_{i+1}$
      \EndIf
    \EndFor
    \State \textbf{return } $\{ L_j \}_{j=0}^{L-1}$, $\{ A_j, M_L \}_{j=1}^{L-1}$
    \EndProcedure
    \Procedure{Sweep-Left}{$\delta$, $R_{L+1}$, $\{ L_j \}_{j=0}^{L-1}$, $\{ W_j \}_{j=1}^L$, $\{ A_j, M_L \}_{j=1}^{L-1}$}
    \For{$j \in [L, 1]$}
      \State $M^{\vphantom{\mathrm{eff}}}_j \gets \mathrm{exp}(-\mathrm{i}\nicefrac{\delta}{2} \hat H^{\mathrm{eff}}_j) M^{\vphantom{\mathrm{eff}}}_j$ using $\hat H^{\mathrm{eff}}_j \equiv L^{\vphantom{\mathrm{eff}}}_{j-1} \cdot W^{\vphantom{\mathrm{eff}}}_j \cdot R^{\vphantom{\mathrm{eff}}}_{j+1}$ \Comment{e.g.~with Lanczos}  \label{alg:tdvp:1tdvp:slsm}
      \State $B_j, C_{\underline{j-1}} \gets M_j$ via QR
      \State $R_j \gets \Call{Contract-Right}{R_{j+1}, W_j, B_j}$
      \If{$i \neq 1$}
        \State $C^{\vphantom{\mathrm{eff}}}_{\underline{j-1}} \gets \mathrm{exp}(\mathrm{i}\nicefrac{\delta}{2} \hat H^{\mathrm{eff}}_{\underline{j-1}}) C^{\vphantom{\mathrm{eff}}}_{\underline{j-1}}$ using $\hat H^{\mathrm{eff}}_{\underline{j-1}} \equiv L^{\vphantom{\mathrm{eff}}}_{j-1} \cdot R^{\vphantom{\mathrm{eff}}}_{j}$ \Comment{e.g. with Lanczos}  \label{alg:tdvp:1tdvp:slsc}
        \State $A_{j-1} \gets A_{j-1} \cdot C_{\underline{j-1}}$
        \State Delete $L_{j-1}$
      \EndIf
    \EndFor
    \State \textbf{return } $\{ R_j \}_{j=2}^{L+1}$, $\{ M_1, B_j \}_{j=2}^L$
    \EndProcedure
  \end{algorithmic}
\end{algorithm}

\begin{algorithm}[tb]
  \caption{\label{alg:tdvp:2tdvp}The 2TDVP method. Also cf.~\cref{alg:local-krylov:common-helpers} for definitions of \textsc{Initialize}, \textsc{Contract-Left}, \textsc{Contract-Right} and the needed overall \textsc{Timestep} method.}
  \begin{algorithmic}[1]
    \Procedure{Sweep-Right}{$\delta$, $L_0$, $\{ R_j \}_{j=3}^{L+1}$, $\{ W_j \}_{j=1}^L$, $\{ M_1, B_j \}_{j=2}^L$}
    \For{$j \in [1, L-1]$}
    \State $T_{j,j+1} \gets \sum_{m_j} M_{j;m_{j-1}, m_j}^{\sigma_i} B_{j+1;m_j, m_{j+1}}^{\sigma_{j+1}}$
    \State $T^{\vphantom{\mathrm{eff}}}_{j,j+1} \gets \mathrm{exp}(-\mathrm{i}\nicefrac{\delta}{2} \hat H^{\mathrm{eff}}_{(j,j+1)}) T^{\vphantom{\mathrm{eff}}}_{j,j+1}$ using $\hat H^{\mathrm{eff}}_{(j,j+1)} \equiv L^{\vphantom{\mathrm{eff}}}_{j-1} \cdot W^{\vphantom{\mathrm{eff}}}_j \cdot W^{\vphantom{\mathrm{eff}}}_{j+1} \cdot R^{\vphantom{\mathrm{eff}}}_{j+2}$
    \State $A_j, C_{\underline{j}}, B_{j+1} \gets T_{j,j+1}$ via singular value decomposition and truncation
    \State $M_{j+1} \gets C_{\underline{j}} \cdot B_{j+1}$
    \If{$j \neq L-1$}
    \State $L_j \gets \Call{Contract-Left}{L_{j-1}, W_j, A_j}$
    \State $M^{\vphantom{\mathrm{eff}}}_{j+1} \gets \mathrm{exp}(\mathrm{i}\nicefrac{\delta}{2} \hat H^{\mathrm{eff}}_{j+1}) M^{\vphantom{\mathrm{eff}}}_{j+1}$ using $\hat H_{j+1}^{\mathrm{eff}} \equiv L^{\vphantom{\mathrm{eff}}}_{j} \cdot W^{\vphantom{\mathrm{eff}}}_{j+1} \cdot R^{\vphantom{\mathrm{eff}}}_{j+2}$
    \State Delete $R_{j+2}$
    \EndIf
    \EndFor
    \State \textbf{return } $\{ L_j \}_{j=0}^{L-2}$, $\{ A_j, M_L \}_{j=1}^{L-1}$
    \EndProcedure
    \Procedure{Sweep-Left}{$\delta$, $R_{L+1}$, $\{ L_j \}_{j=0}^{L-2}$, $\{ W_j \}_{j=1}^L$, $\{ A_j, M_L \}_{j=1}^{L-1}$}
    \For{$j \in [L, 2]$}
    \State $T_{j-1,j} \gets \sum_{m_{j-1}} A_{j-1;m_{j-2}, m_{j-1}}^{\sigma_{j-1}} M_{j;m_{j-1}, m_{i}}^{\sigma_{i}}$
    \State $T^{\vphantom{\mathrm{eff}}}_{j-1,j} \gets \mathrm{exp}(-\mathrm{i}\nicefrac{\delta}{2} \hat H^{\mathrm{eff}}_{(j-1,j)}) T^{\vphantom{\mathrm{eff}}}_{j-1,j}$ using $\hat H^{\mathrm{eff}}_{(j-1,j)} \equiv L^{\vphantom{\mathrm{eff}}}_{j-2} \cdot W^{\vphantom{\mathrm{eff}}}_{j-1} \cdot W^{\vphantom{\mathrm{eff}}}_{j} \cdot R^{\vphantom{\mathrm{eff}}}_{j+1}$
    \State $A_{j-1}, C_{\underline{j-1}}, B_{j} \gets T_{j-1,j}$ via singular value decomposition and truncation
    \State $M_{j-1} \gets A_{j-1} \cdot C_{\underline{j-1}}$
    \If{$j \neq 2$}
    \State $R_j \gets \Call{Contract-Right}{R_{j+1}, W_j, B_j}$
    \State $M^{\vphantom{\mathrm{eff}}}_{j-1} \gets \mathrm{exp}(\mathrm{i}\nicefrac{\delta}{2} \hat H^{\mathrm{eff}}_{j-1}) M^{\vphantom{\mathrm{eff}}}_{j-1}$ using $\hat H_{j-1}^{\mathrm{eff}} \equiv L^{\vphantom{\mathrm{eff}}}_{j-2} \cdot W^{\vphantom{\mathrm{eff}}}_{j-1} \cdot R^{\vphantom{\mathrm{eff}}}_{j}$
    \State Delete $L_{j-2}$
    \EndIf
    \EndFor
    \State \textbf{return } $\{ R_j \}_{j=2}^{L+1}$, $\{ M_1, B_j \}_{j=2}^{L}$
    \EndProcedure
  \end{algorithmic}
\end{algorithm}

\subsubsection{Algorithm}

In practice, the 1/2TDVP method is quite similar to the 1/2DMRG method
without subspace expansion or density matrix perturbation and nearly
identical to the local Krylov method. Compared to the DMRG method, one
of course has to replace the local eigensolver by a local
exponentiation. Compared to both the local Krylov and the DMRG
methods, we also need an additional backwards evolution step either on
each bond (1TDVP) or the second site of a two-site evolution
(2TDVP). This replacement of the ad-hoc basis transformation done by
the local Krylov method with a properly motivated backwards evolution
will result in smaller errors in each step.

The 1TDVP algorithm is described in detail in \cref{alg:tdvp:1tdvp},
the 2TDVP algorithm in \cref{alg:tdvp:2tdvp}.

%%% Local Variables: 
%%% mode: latex
%%% TeX-master: "../time_evolution_review"
%%% End: 

%% file: content/tdvp_effrhs.tex
\begin{figure}[p]
  \centering
  \tikzsetnextfilename{tdvp_efftdse}
  \begin{tikzpicture}
    [
    siteA/.style={scale=0.7,regular polygon, regular polygon sides=3, shape border rotate= 30, draw=red!50,fill=red!20,thick,inner sep=0.5pt,minimum width=4em,minimum height=4em},
    siteB/.style={scale=0.7,regular polygon, regular polygon sides=3, shape border rotate= -30, draw=green!50,fill=green!20,thick,inner sep=0.5pt,minimum width=4em,minimum height=4em},
    siteC/.style={scale=0.7,circle,thick,inner sep=0pt,minimum height=3em,draw=blue!50,fill=blue!20},
    ghostL/.style={scale=0.7,regular polygon, regular polygon sides=3, shape border rotate= 30, inner sep=0.5pt,minimum width=4em,minimum height=4em},
    ghostR/.style={scale=0.7,regular polygon, regular polygon sides=3, shape border rotate= -30, inner sep=0.5pt,minimum width=4em,minimum height=4em},
    ghostC/.style={scale=0.7, circle, inner sep=0.5pt,minimum height=3em},
    op/.style={scale=0.7,regular polygon, regular polygon sides=4, draw=orange!50, fill=orange!20, thick, inner sep=0.5pt, minimum width=4em, minimum height=4em}
    ]
    \begin{scope}[node distance=0.7]
      % top row, this is the original MPS |\psi>
      \node[siteA] (site1) {$A_1$};
      \node[siteA] (site2) [right=of site1] {$A_2$};
      \node[siteA] (site3) [right=of site2] {$A_3$};

      \node[siteC] (site4) [right=of site3] {$M_4$};

      \node[siteB] (site5) [right=of site4] {$B_5$};
      \node[siteB] (site6) [right=of site5] {$B_6$};
      
      \draw [decorate,decoration={brace,amplitude=3pt}] ($(site6.east)+(0.2,+0.5)$) -- +(0,-1) node [black,midway,right,xshift=3pt] {\footnotesize $\partial_{M_4} \Ket{\psi}$ and $M_4$};

      % now comes the MPO
      \node[op] (op1) [below=of site1] {$W_1$};
      \node[op] (op2) [below=of site2] {$W_2$};
      \node[op] (op3) [below=of site3] {$W_3$};
      \node[op] (op5) [below=of site5] {$W_5$};
      \node[op] (op4) at ($(op3)!0.5!(op5)$) {$W_4$};
      \node[op] (op6) [below=of site6] {$W_6$};

      \draw [] (site1) -- (site2);
      \draw [] (site2) -- (site3);
      \draw [dashed] (site3) -- (site4);
      \draw [dashed] (site4) -- (site5);
      \draw [] (site5) -- (site6);

      \draw [] (op1) -- (site1);
      \draw [] (op2) -- (site2);
      \draw [] (op3) -- (site3);
      \draw [dashed] (op4) -- (site4);
      \draw [] (op5) -- (site5);
      \draw [] (op6) -- (site6);

      \draw [] (op1) -- (op2);
      \draw [] (op2) -- (op3);
      \draw [] (op3) -- (op4);
      \draw [] (op4) -- (op5);
      \draw [] (op5) -- (op6);

      \draw [decorate,decoration={brace,amplitude=3pt}] ($(op6.east)+(0.13,+0.5)$) -- +(0,-1) node [black,midway,right,xshift=3pt] {\footnotesize $\hat H$};

      % and now the top row of the projector
      \node[siteA] (site1dag) [below=of op1] {$\overbar{A}_1$};
      \node[siteA] (site2dag) [below=of op2] {$\overbar{A}_2$};
      \node[siteA] (site3dag) [below=of op3] {$\overbar{A}_3$};

      \node[siteB] (site5dag) [below=of op5] {$\overbar{B}_5$};
      \node[siteB] (site6dag) [below=of op6] {$\overbar{B}_6$};

      \node[ghostC] (site4dag) at ($(site3dag)!0.5!(site5dag)$) {};

      \draw [decorate,decoration={brace,amplitude=5pt}] ($(site1dag.west)+(-0.2,-3.5)$) -- +(0,+7) node[black,midway,left,xshift=-5pt] {\footnotesize $\hat H^{\mathrm{eff}}_{4} \cdot M_{4}$};

      \draw [] (op1) -- (site1dag);
      \draw [] (op2) -- (site2dag);
      \draw [] (op3) -- (site3dag);
      \draw [] (op5) -- (site5dag);
      \draw [] (op6) -- (site6dag);
      
      \draw [] (site1dag) -- (site2dag);
      \draw [] (site2dag) -- (site3dag);
      \draw [] (site5dag) -- (site6dag);

      % now the bottom row of the projector
      \node[siteA] (site1b) [below=of site1dag] {$A_1$};
      \node[siteA] (site2b) [right=of site1b] {$A_2$};
      \node[siteA] (site3b) [right=of site2b] {$A_3$};
      \node[siteB] (site5b) [below=of site5dag] {$B_5$};
      \node[siteB] (site6b) [right=of site5b] {$B_6$};
      \node[ghostC] (site4b) at ($(site3b)!0.5!(site5b)$) {};

      \draw [] (site3b.east) -| ($(site3b)!0.5!(site4b)$) |- (site3dag.east);
      \draw [] (site5b.west) -| ($(site5b)!0.5!(site4b)$) |- (site5dag.west);

      \draw [] (site1b) -- (site2b);
      \draw [] (site2b) -- (site3b);

      \draw [] (site5b) -- (site6b);

      \draw [decorate,decoration={brace,amplitude=3pt}] ($(site6dag.east)+(0.2,+0.5)$) -- +(0,-2.4) node [black,midway,right,xshift=3pt] {\footnotesize $\hat P_{3}^{L, |\psi\rangle} \otimes \mathbf{\hat 1}_4 \otimes
    \hat P_{5}^{R,|\psi\rangle}$};

      % and now the isometry which turns the full MPS into an
      % effective single-site tensor
      \node[siteA] (site1dagb) [below=of site1b] {$\overbar{A}_1$};
      \node[siteA] (site2dagb) [below=of site2b] {$\overbar{A}_2$};
      \node[siteA] (site3dagb) [below=of site3b] {$\overbar{A}_3$};
      \node[siteB] (site5dagb) [below=of site5b] {$\overbar{B}_5$};
      \node[siteB] (site6dagb) [below=of site6b] {$\overbar{B}_6$};
      \node[ghostC] (site4dagb) at ($(site3dagb)!0.5!(site5dagb)$) {};

      \draw [decorate,decoration={brace,amplitude=3pt}] ($(site6dagb.east)+(0.2,+0.5)$) -- +(0,-1) node [black,midway,right,xshift=3pt] {\footnotesize $\overbar{\psi}^L_3, \overbar{\psi}^R_5$};

      \draw [] (site1dagb) -- (site2dagb);
      \draw [] (site2dagb) -- (site3dagb);
      \draw [dashed] (site3dagb) -- (site4dagb);
      \draw [dashed] (site4dagb) -- (site5dagb);
      \draw [] (site5dagb) -- (site6dagb);
      \draw [dashed] (op4) -- (site4dagb);

      \draw [] (site1dagb) -- (site1b);
      \draw [] (site2dagb) -- (site2b);
      \draw [] (site3dagb) -- (site3b);

      \draw [] (site5dagb) -- (site5b);
      \draw [] (site6dagb) -- (site6b);
    \end{scope}
  \end{tikzpicture}
  \caption{\label{fig:tdvp_efftdse} Right-hand side of the effective
    single-site forwards-evolving Schr\"odinger equation (with $j = 4$). The effective
    Hamiltonian $\hat H^{\mathrm{eff}}_j$ is given by the cornered
    green, orange and red tensors. The effective state is given by the
    blue circled tensor $M_j$. During the calculation, the connected
    dashed lines are contracted, resulting in a new tensor with three
    legs (the three open dashed lines).}
\end{figure}
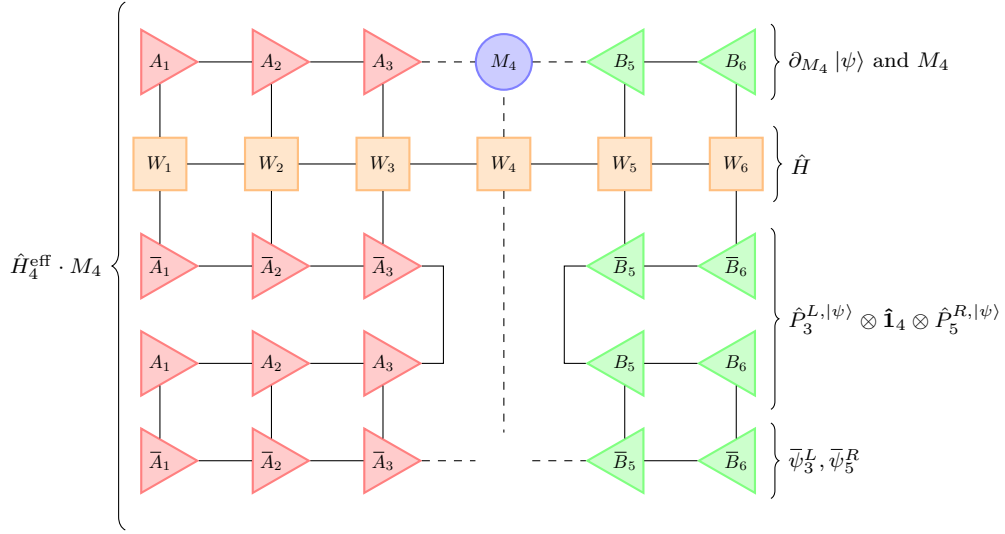

\begin{figure}[p]
  \centering
  \tikzsetnextfilename{tdvp_efftdse_bond}
  \begin{tikzpicture}
    [
    siteA/.style={scale=0.7,regular polygon, regular polygon sides=3, shape border rotate= 30, draw=red!50,fill=red!20,thick,inner sep=0.5pt,minimum width=4em,minimum height=4em},
    siteB/.style={scale=0.7,regular polygon, regular polygon sides=3, shape border rotate= -30, draw=green!50,fill=green!20,thick,inner sep=0.5pt,minimum width=4em,minimum height=4em},
    siteC/.style={scale=0.7,regular polygon, regular polygon sides=4, shape border rotate=45, draw=black!50,fill=black!20, thick,inner sep=0pt,minimum height=3em},
    ghostL/.style={scale=0.7,regular polygon, regular polygon sides=3, shape border rotate= 30, inner sep=0.5pt,minimum width=4em,minimum height=4em},
    ghostR/.style={scale=0.7,regular polygon, regular polygon sides=3, shape border rotate= -30, inner sep=0.5pt,minimum width=4em,minimum height=4em},
    ghostC/.style={scale=0.7,regular polygon, regular polygon sides=4, shape border rotate=60, inner sep=0pt,minimum height=3em},
    op/.style={scale=0.7,regular polygon, regular polygon sides=4, draw=orange!50, fill=orange!20, thick, inner sep=0.5pt, minimum width=4em, minimum height=4em}
    ]
    \begin{scope}[node distance=0.7]
      % top row, this is the original MPS |\psi>
      \node[siteA] (site1) {$A_1$};
      \node[siteA] (site2) [right=of site1] {$A_2$};
      \node[siteA] (site3) [right=of site2] {$A_3$};
      \node[siteA] (site4) [right=of site3] {$A_4$};
      \node[siteC] (siteC) [right=of site4] {$C_{\underline{4}}$};
      \node[siteB] (site5) [right=of siteC] {$B_5$};
      \node[siteB] (site6) [right=of site5] {$B_6$};
      
      \draw [decorate,decoration={brace,amplitude=3pt}] ($(site6.east)+(0.2,+0.5)$) -- +(0,-1) node [black,midway,right,xshift=3pt] {\footnotesize $\partial_{C_{\underline{4}}} \Ket{\psi}$ and $C_{\underline{4}}$};

      % now comes the MPO
      \node[op] (op1) [below=of site1] {$W_1$};
      \node[op] (op2) [below=of site2] {$W_2$};
      \node[op] (op3) [below=of site3] {$W_3$};
      \node[op] (op4) [below=of site4] {$W_4$};
      \node[op] (op5) [below=of site5] {$W_5$};
      \node[op] (op6) [below=of site6] {$W_6$};

      \draw [] (site1) -- (site2);
      \draw [] (site2) -- (site3);
      \draw [] (site3) -- (site4);
      \draw [dashed] (site4) -- (siteC);
      \draw [dashed] (siteC) -- (site5);
      \draw [] (site5) -- (site6);

      \draw [] (op1) -- (site1);
      \draw [] (op2) -- (site2);
      \draw [] (op3) -- (site3);
      \draw [] (op4) -- (site4);
      \draw [] (op5) -- (site5);
      \draw [] (op6) -- (site6);

      \draw [] (op1) -- (op2);
      \draw [] (op2) -- (op3);
      \draw [] (op3) -- (op4);
      \draw [] (op4) -- (op5);
      \draw [] (op5) -- (op6);

      \draw [decorate,decoration={brace,amplitude=3pt}] ($(op6.east)+(0.13,+0.5)$) -- +(0,-1) node [black,midway,right,xshift=3pt] {\footnotesize $\hat H$};

      % and now the top row of the projector
      \node[siteA] (site1dag) [below=of op1] {$\overbar{A}_1$};
      \node[siteA] (site2dag) [below=of op2] {$\overbar{A}_2$};
      \node[siteA] (site3dag) [below=of op3] {$\overbar{A}_3$};
      \node[siteA] (site4dag) [below=of op4] {$\overbar{A}_4$};
      \node[siteB] (site5dag) [below=of op5] {$\overbar{B}_5$};
      \node[siteB] (site6dag) [below=of op6] {$\overbar{B}_6$};

      \draw [decorate,decoration={brace,amplitude=5pt}] ($(site1dag.west)+(-0.2,-3.5)$) -- +(0,+7) node[black,midway,left,xshift=-5pt] {\footnotesize $\hat H^{\mathrm{eff}}_{\underline{4}} \cdot C_{\underline{4}}$};

      \draw [] (op1) -- (site1dag);
      \draw [] (op2) -- (site2dag);
      \draw [] (op3) -- (site3dag);
      \draw [] (op4) -- (site4dag);
      \draw [] (op5) -- (site5dag);
      \draw [] (op6) -- (site6dag);
      
      \draw [] (site1dag) -- (site2dag);
      \draw [] (site2dag) -- (site3dag);
      \draw [] (site3dag) -- (site4dag);
      \draw [] (site5dag) -- (site6dag);

      % now the bottom row of the projector
      \node[siteA] (site1b) [below=of site1dag] {$A_1$};
      \node[siteA] (site2b) [right=of site1b] {$A_2$};
      \node[siteA] (site3b) [right=of site2b] {$A_3$};
      \node[siteA] (site4b) [right=of site3b] {$A_4$};
      \node[siteB] (site5b) [below=of site5dag] {$B_5$};
      \node[siteB] (site6b) [right=of site5b] {$B_6$};

      \node[ghostC] (siteCb) at ($(site4b)!0.5!(site5b)$) {};

      \draw [] (site4b.east) -| ($(site4b)!0.5!(siteCb)$) |- (site4dag.east);
      \draw [] (site5b.west) -| ($(site5b)!0.5!(siteCb)$) |- (site5dag.west);

      \draw [] (site1b) -- (site2b);
      \draw [] (site2b) -- (site3b);
      \draw [] (site3b) -- (site4b);

      \draw [] (site5b) -- (site6b);

      \draw [decorate,decoration={brace,amplitude=3pt}] ($(site6dag.east)+(0.2,+0.5)$) -- +(0,-2.4) node [black,midway,right,xshift=3pt] {\footnotesize $\hat P_{4}^{L, |\psi\rangle} \otimes \hat P_{5}^{R,|\psi\rangle}$};

      % and now the isometry which turns the full MPS into an
      % effective single-site tensor
      \node[siteA] (site1dagb) [below=of site1b] {$\overbar{A}_1$};
      \node[siteA] (site2dagb) [below=of site2b] {$\overbar{A}_2$};
      \node[siteA] (site3dagb) [below=of site3b] {$\overbar{A}_3$};
      \node[siteA] (site4dagb) [below=of site4b] {$\overbar{A}_4$};
      \node[siteB] (site5dagb) [below=of site5b] {$\overbar{B}_5$};
      \node[siteB] (site6dagb) [below=of site6b] {$\overbar{B}_6$};
      \node[ghostC] (siteCdagb) at ($(site4dagb)!0.5!(site5dagb)$) {};

      \draw [] (site1dagb) -- (site2dagb);
      \draw [] (site2dagb) -- (site3dagb);
      \draw [] (site3dagb) -- (site4dagb);
      \draw [dashed] (site4dagb) -- (siteCdagb);
      \draw [dashed] (site5dagb) -- (siteCdagb);
      \draw [] (site5dagb) -- (site6dagb);

      \draw [] (site1dagb) -- (site1b);
      \draw [] (site2dagb) -- (site2b);
      \draw [] (site3dagb) -- (site3b);
      \draw [] (site4dagb) -- (site4b);

      \draw [] (site5dagb) -- (site5b);
      \draw [] (site6dagb) -- (site6b);

      \draw [decorate,decoration={brace,amplitude=3pt}] ($(site6dagb.east)+(0.2,+0.5)$) -- +(0,-1) node [black,midway,right,xshift=3pt] {\footnotesize $\overbar{\psi}^L_4, \overbar{\psi}^R_5$};

    \end{scope}
  \end{tikzpicture}
  \caption{\label{fig:tdvp_efftdse_bond} Right-hand side of the
    effective center matrix backward-evolving Schr\"odinger
    equation with $\underline{j} = 4$. The effective state over the bond between sites $j$ and
    $j+1$ is given by the grey diamond $C_{\underline{j}}$. During the
    calculation, the connected dashed lines are contracted, resulting
    in a new tensor with two legs (the two open dashed
    lines).}
\end{figure}
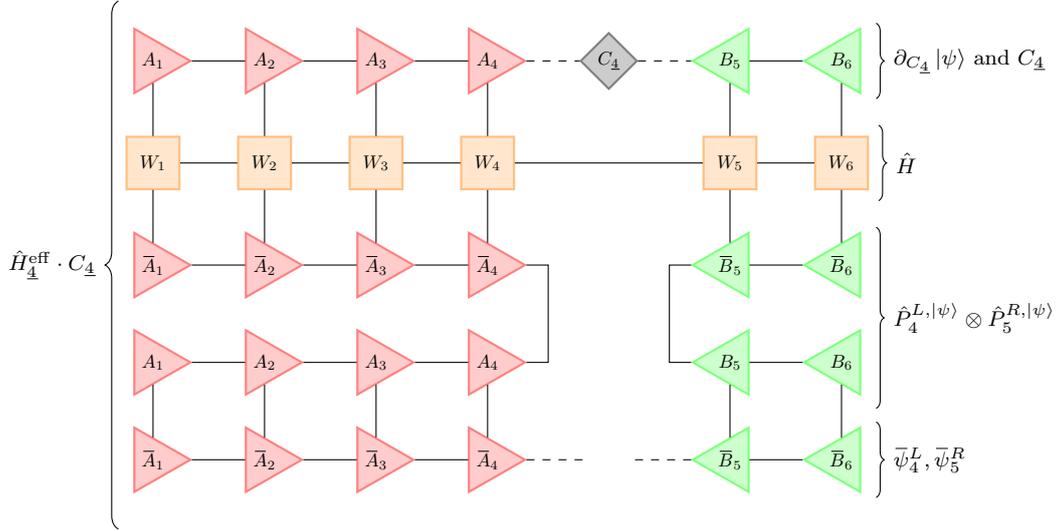

%% file: content/tricks.tex
\section{\label{sec:te:tricks}Additional tricks}

Having reviewed the most commonly used methods in detail, we will now
collect and summarize some generic improvements without any claim to
completeness. These are essentially tricks which are relatively
independent of the actual time-evolution method and can mostly be
implemented in all of the methods we just reviewed. They are not
strictly necessary to implement in conjunction with any method and as
such will not be benchmarked in detail later, but they are useful to
keep in mind in case of particularly hard or challenging problems.

\subsection{\label{sec:te:tricks:heisenberg}Combining Heisenberg
  and Schr\"odinger picture time evolution}

In the context of MPS methods, combining Schr\"odinger and Heisenberg
picture\cite{hartmann09:_densit_matrix_renor_group_heisen_pictur, prosen09:_matrix} time evolution was first proposed in
Ref.~\cite{kennes16:_exten}. Considering a time-dependent observable
\begin{equation}
  \langle \phi | \hat O(t) | \psi \rangle
\end{equation}
between two arbitrary states, in the Schr\"odinger picture we would evaluate
\begin{equation}
  \langle \phi | \hat O(t) | \psi \rangle = \left( \langle \phi | e^{\mathrm{i} t \hat H} \right) \hat O \left( e^{-\mathrm{i} t \hat H} | \psi\rangle \right) \quad.
\end{equation}
That is, we apply time-evolution operators to the states
$|\phi\rangle$ and $|\psi\rangle$ to obtain time-evolved states
$|\phi(t)\rangle$ and $|\psi(t)\rangle$ and then evaluate the
time-independent observable between them. The maximum time $t$
obtainable is then typically limited by the entanglement growth in the
states, resulting in larger and larger bond dimensions or errors.

In comparison, the Heisenberg picture would see us time evolve the
operator $\hat O$ as
\begin{equation}
  \langle \phi | \hat O(t) | \psi \rangle = \langle \phi | \left( e^{\mathrm{i} t \hat H} \hat O e^{-\mathrm{i}t \hat H} \right) | \psi\rangle
\end{equation}
while keeping the states $|\phi\rangle$, $|\psi\rangle$ static. Again,
the maximal obtainable time t is limited by the entanglement growth in
the operator $\hat O$ and the maximal bond dimension we can use to
represent it.\footnote{Note that it is difficult to compare errors
  between MPS and MPO truncations, as the error of the MPO truncation
  is given by the operator norm whereas during an MPO compression, we
  only control the 2-norm of the operator\cite{barthel16:_matrix}.} If we now combine the two
evolutions as
\begin{equation}
  \langle \phi | \hat O(t_1+t_2) | \psi \rangle = \left( \langle \phi | e^{\mathrm{i} t_1 \hat H} \right) \left( e^{\mathrm{i} t_2 \hat H} \hat O e^{-\mathrm{i}t_2 \hat H} \right) \left( e^{-\mathrm{i} t_1 \hat H} | \psi\rangle \right)
\end{equation}
we can obtain times $t_1 + t_2$ while only requiring MPO and MPS bond
dimensions typical of times $t_1$ and $t_2$ respectively. Note that in
this case, the computationally limiting operation is no longer the
time evolution itself but the evaluation of observables given as the
tensor network of a large-$w$ MPO between two large-$m$
MPS\cite{kennes16:_exten}.

\subsection{\label{sec:te:tricks:timestepchoice}Complex time steps}

Let us assume that we have a time-evolution operator
$\hat U^\prime(\delta) = \mathbf{\hat 1} - \mathrm{i} \delta \hat H$ which is exact
to first order. Applying this operator to a state will result in an
error $O(\delta^2)$ compared to the exact evolution with the operator
$\hat U(\delta) = e^{-\mathrm{i} \delta \hat H}$. Repeating the
process $T / \delta$ times to obtain a state at final time $T$, we
incur an error $O(\delta^2) \nicefrac{T}{\delta} = O(\delta)$.
However, if we allow complex intermediate steps $\delta_1$ and
$\delta_2$, we can solve
\begin{equation}
  \hat U^\prime(\delta_1) \hat U^\prime(\delta_2) = \mathbf{\hat 1} - \mathrm{i} \delta \hat H - \nicefrac{\delta}{2} \hat H^2
\end{equation}
for $\delta_1$ and $\delta_2$ by expanding the left-hand side:
\begin{align}
  & \quad \hat U^\prime(\delta_1) \hat U^\prime(\delta_2) \\
= & \quad \left( \mathbf{\hat 1} - \mathrm{i} \delta_1 \hat H \right) \left( \mathbf{\hat 1} - \mathrm{i} \delta_2 \hat H \right) \\
= & \quad \mathbf{\hat 1} - \mathrm{i} ( \delta_1 + \delta_2 ) \hat H - \delta_1 \delta_2 \hat H^2 \\
\Rightarrow & \quad \delta_1 + \delta_2 = \delta \quad \land \quad \delta_1 \delta_2 = \nicefrac{\delta}{2}\;.
\end{align}
Two solutions are possible, one of them is
\begin{align}
  \delta_1 = \frac{1 - \mathrm{i}}{2} \delta \quad \land \quad \delta_2 = \frac{1 + \mathrm{i}}{2} \delta \quad.
\end{align}
Choosing these values for $\delta_{1,2}$ then results in a third-order
error per time step and a second-order error overall. This choice of time steps is
suggested in particular in combination with the MPO \wiii method to obtain a better error per time step. The cost of the
method only grows linearly with the number of evolution operators and, 
e.g.,~four operators $\hat U^\prime(\delta_{1,2,3,4})$ are required for
a third-order error \cite{zaletel15:_time}.

The drawback is the loss of unitarity at each individual time step
which may be disadvantageous. Furthermore, if the time evolution is
purely imaginary (e.g., for finite-temperature calculations) and the
Hamiltonian does not contain complex coefficients, one may avoid
complex arithmetic entirely and only use real floating-point scalars
for 50\% less memory usage and an approximately four-fold speed-up on
matrix multiplications. Unfortunately, it is then impossible to use this
trick to reduce the time-step error.

\subsection{\label{sec:te:tricks:lowlying}Green's functions 1: Removal of low-lying states}

This trick was first proposed in Ref.~\cite{schmitteckert04:_noneq}
and is relatively straightforward to implement. Assume a ground state
$|0\rangle$ as an MPS obtained via DMRG and let us shift the
Hamiltonian such that this state has energy $E_0 = 0$,
$\hat H |0\rangle = 0$. We are then interested in the time-dependent
observable
\begin{equation}
  x(t) = \langle 0 | \hat A(t) \hat B | 0 \rangle
\end{equation}
where $\hat A$ and $\hat B$ are typically local operators such as
creators or annihilators. The evolution of $\hat B | 0 \rangle$ is
generically non-trivial and if we want to capture all frequency
contributions in $x(t)$, we need to evolve until at least times
$t^\prime = \nicefrac{1}{E_1}$ where $E_1$ is the energy of the lowest
eigenstate with non-zero energy $|1\rangle$ contained in
$\hat B |0 \rangle$. In contrast, to capture contributions of
higher-energy states $|n\rangle$ with energies $E_n > E_1$, we only
need to evolve to shorter times
$t^{\prime\prime} = \nicefrac{1}{E_n} < t^\prime$.

However, a few low-lying eigenstates can often be calculated also with
DMRG by orthogonalizing against previously-found eigenstates. Hence if
we run DMRG multiple times, we can obtain not just the ground state
$|0\rangle$ but also further eigenstates $|1\rangle$, $|2\rangle$
etc. If we use quantum numbers and $\hat B$ changes the quantum number
of the state, these additional eigenstates should be calculated in the
quantum number sector of $\hat B |0\rangle$. If we then orthogonalize
$\hat B |0\rangle$ against $|1\rangle$, $|2\rangle$ etc., we remove
the contributions which rotate (in real-time) or decay (in
imaginary-time evolutions) the slowest and hence require the longest
time evolutions. The evolution of the removed states can then be done
exactly as we know both their energy and initial weight in
$\hat B |0\rangle$. Even missing one of the eigenstates due to
convergence problems with DMRG does not introduce an error but merely
decreases the effectivity of the method.

\subsection{Green's functions 2: Linear prediction}

Calculating dynamical structure factors or more generally spectral
functions from time-dependent data requires two Fourier
transformations: first, one needs to transform from real-space to
momentum-space and second from real-time to frequency. The former
transformation is typically unproblematic, but the latter
transformation suffers either from overdamping or strong spectral
leakage if the available maximal time $t$ is insufficient. Linear
prediction\cite{vaidyanathan08:_theor_linear_predic,
  white08:_spect_heisen, barthel09:_spect} assumes that the real-time momentum-space
Green's function $G(k,t)$ is composed of multiple distinct exponentially
decaying and oscillating contributions arising from a distinct pole
structure of $G(k,\omega)$. If this is true for a time series $x_1, x_2, x_3, \ldots$,
an additional data point $\tilde{x}_n$ can be approximated well by the
form
\begin{equation}
  \tilde{x}_n = - \sum_{i=1}^p a_i x_{n-i-1}
\end{equation}
with suitably chosen coefficients $a_i$ independent of $n$. We hence
first compute a finite time series which is still as long as we can
manage with time-dependent MPS methods. Subsequently, we need to find
coefficients $a_i$ such that the above holds for the data we computed
exactly. Using those $a_i$, we can then extend the time series to
arbitrarily long times to generate a sufficiently long time series
that a subsequent Fourier transform only requires minimal damping and
hence provides for clear features.

It is useful to divide the available calculated data into three
segments: first, one should discard an interval $[x_0, \ldots, x_{D-1}]$
at the beginning which captures short-time physics irrelevant and
untypical of the longer-time behavior. Second, a fitting interval
$[x_{D}, \ldots, x_{D+N-1}]$ of N points is selected over which the
coefficients $a_i$ are minimized. Third, trust in the prediction is
increased if it coincides with additional calculated data
$[x_{D+N}, \ldots, x_{\mathrm{max}}]$ outside the fitting interval.

To select the $a_i$, we want to minimize the error
\begin{equation}
  \epsilon = \sum_{k=D}^{D+N-1} | \tilde{x}_k - x_k |^2 \quad.
\end{equation}
Note that to evaluate $\tilde{x}_k$, $D$ must be larger than the
number of coefficients $p$. The coefficient vector $\underline{a}$ is obtained as
\begin{equation}
  \underline{a} = - \underline{R}^{-1} \underline{r}
\end{equation}
where the matrix $\underline{R}$ and vector $\underline{r}$ have entries
\begin{align}
  R_{i,j} = & \sum_{k=D}^{D+N-1} x_{k-i}^\star x_{k-j} \\
  r_i    = & \sum_{k=D}^{D+N-1} x_{k-i}^\star x_k
\end{align}
respectively. Once the $a_i$ are obtained, data ideally can be
generated initially for the interval
$[x_{D+N}, \ldots, x_{\mathrm{max}}]$ and, once verified to coincide
with the calculated data, extended to arbitrary times.

Several numerical pitfalls need to be considered here: First, the
matrix $\underline{R}$ may be singular. Two possible remedies include
addition of a small shift $\varepsilon$ or reduction of the number of
parameters $p$. Ideally the latter should be considered, but may lead
to problems finding the optimal non-singular $p$. Second, if we
construct the vector
\begin{equation}
  \underline{x}_n = [x_{n-1}, \ldots, x_{n-p}]^T
\end{equation}
we can move it forward one step as
\begin{equation}
  \underline{\tilde{x}}_{n+1} = \underline{A} {\hspace{0.1cm}} \underline{x}_n
\end{equation}
where the matrix $\underline{A}$ is of the form
\begin{equation}
  \underline{A} = \begin{pmatrix} -a_1 & -a_2 & -a_3 & \cdots & -a_p \\
 1 & 0 & 0 & \cdots & 0 \\
 0 & 1 & 0 & \cdots & 0 \\
 \vdots & \ddots & \ddots & \ddots & \vdots \\
 0 & \cdots & 0 & 1 & 0
\end{pmatrix}
\end{equation}
and its eigenvalues $\alpha_i$ contain the frequencies and dampings of
the aforementioned oscillations and exponential decays. As such,
$\alpha_i > 1$ are unphysical and need to be dealt with, it appears
\cite{barthel09:_spect} that setting those contributions to zero works
best.

\subsection{\label{sec:tricks:pip}Purification insertion point (PIP)}

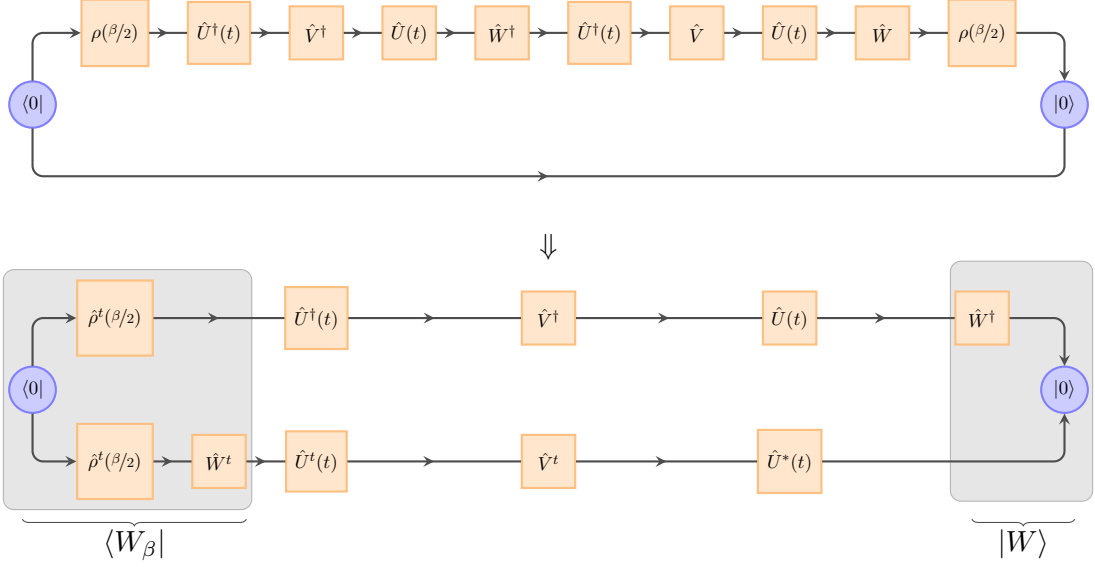
\begin{figure}[t!]
        \centering
        \tikzsetnextfilename{OTOCPurification1}
        \begin{tikzpicture}
                [
                        node distance = 0.5,
                        site/.style={circle,draw=blue!50,fill=blue!20,thick,minimum height=2em, scale=0.65},
                        op/.style={regular polygon, regular polygon sides=4, draw=orange!50, fill=orange!20, thick, minimum width=4em, minimum height=4em, inner sep = 0pt, scale=0.65}
                ]

                \node[site] (bra1) {$\bra{0}$};

                \node[] (topleftAnchor1) [above = of bra1] {};
                \node[] (botleftAnchor1) [below = of bra1] {};

                \node[op] (rhoLeft1) [right = of topleftAnchor1] {$\rho(\nicefrac{\beta}{2})$};
                \node[op] (Udagger11) [right = of rhoLeft1] {$\hat{U}^{\dagger}(t)$};
                \node[op] (Vdagger1) [right = of Udagger11] {$\hat{V}^{\dagger}$};
                \node[op] (U11) [right = of Vdagger1] {$\hat{U}(t)$};
                \node[op] (Wdagger1) [right = of U11] {$\hat{W}^{\dagger}$};
                \node[op] (Udagger21) [right = of Wdagger1] {$\hat{U}^{\dagger}(t)$};
                \node[op] (V1) [right = of Udagger21] {$\hat{V}$};
                \node[op] (U21) [right = of V1] {$\hat{U}(t)$};
                \node[op] (W1) [right = of U21] {$\hat{W}$};
                \node[op] (rhoRight1) [right = of W1] {$\rho(\nicefrac{\beta}{2})$};

                \node[] (toprightAnchor1) [right = of rhoRight1] {};
                \node[] (botrightAnchor1) at (botleftAnchor1-|toprightAnchor1) {};

                \node[site] (ket1) at (bra1-|toprightAnchor1) {$\ket{0}$};

                \draw[->,thick,draw=black!70,rounded corners] (bra1) -- ++($(topleftAnchor1) - (bra1)$) -- (rhoLeft1);
                \draw[->-,thick,draw=black!70] (rhoLeft1) -- (Udagger11);
                \draw[->-,thick,draw=black!70] (Udagger11) -- (Vdagger1);
                \draw[->-,thick,draw=black!70] (Vdagger1) -- (U11);
                \draw[->-,thick,draw=black!70] (U11) -- (Wdagger1);
                \draw[->-,thick,draw=black!70] (Wdagger1) -- (Udagger21);
                \draw[->-,thick,draw=black!70] (Udagger21) -- (V1);
                \draw[->-,thick,draw=black!70] (V1) -- (U21);
                \draw[->-,thick,draw=black!70] (U21) -- (W1);
                \draw[->-,thick,draw=black!70] (W1) -- (rhoRight1);
                \draw[->,thick,draw=black!70,rounded corners] (rhoRight1) -- ++($(toprightAnchor1) - (rhoRight1)$) -- (ket1);
                \draw[->-,thick,draw=black!70,rounded corners] (bra1) -- ++($(botleftAnchor1) - (bra1)$) -- ++($(botrightAnchor1) - (botleftAnchor1)$) -- (ket1);

                \node[] (centerAnchor1) at ($(botleftAnchor1)!0.5!(botrightAnchor1)$) {};
                \node[] (downArrow) [below = of centerAnchor1] {$\Downarrow$};
                \node[] (centerAnchor2) [below = of downArrow] {};

                \node[] (topleftAnchor2) at (centerAnchor2-|topleftAnchor1) {};
                \node[] (toprightAnchor2) at (centerAnchor2-|toprightAnchor1) {};

                \node[site] (bra2) [below = of topleftAnchor2] {$\bra{0}$};
                \node[site] (ket2) [below = of toprightAnchor2] {$\ket{0}$};

                \node[] (botleftAnchor2) [below = of bra2] {};
                \node[] (botrightAnchor2) [below = of ket2] {};

                \node[op] (rhoLeft2) at (botleftAnchor2-|rhoLeft1) {$\hat{\rho}^{t}(\nicefrac{\beta}{2})$};
                \node[op] (W2) at (botleftAnchor2-|Udagger11) {$\hat{W}^{t}$};
                \node[op] (U12) at (botleftAnchor2-|Vdagger1) {$\hat{U}^{t}(t)$};
                \node[op] (V2) at (botleftAnchor2-|downArrow) {$\hat{V}^{t}$};
                \node[op] (Udagger12) at (botleftAnchor2-|U21) {$\hat{U}^{*}(t)$};
                \node[] (rightbraceAnchor) at (botleftAnchor2-|rhoRight1) {};

                \node[op] (rhoRight2) at (topleftAnchor2-|rhoLeft2) {$\hat{\rho}^{t}(\nicefrac{\beta}{2})$};
                \node[op] (Udagger22) at (topleftAnchor2-|U12) {$\hat{U}^{\dagger}(t)$};
                \node[op] (Vdagger2) at (topleftAnchor2-|V2) {$\hat{V}^{\dagger}$};
                \node[op] (U22) at (topleftAnchor2-|Udagger12) {$\hat{U}(t)$};
                \node[op] (Wdagger2) at (topleftAnchor2-|rhoRight1) {$\hat{W}^{\dagger}$};

                \draw[->,thick,draw=black!70,rounded corners] (bra2) --++($(topleftAnchor2) - (bra2)$) -- (rhoRight2);
                \draw[->-,thick,draw=black!70] (rhoRight2) -- (Udagger22);
                \draw[->-,thick,draw=black!70] (Udagger22) -- (Vdagger2);
                \draw[->-,thick,draw=black!70] (Vdagger2) -- (U22);
                \draw[->-,thick,draw=black!70] (U22) -- (Wdagger2);
                \draw[->,thick,draw=black!70,rounded corners] (Wdagger2) --++($(toprightAnchor2) - (Wdagger2)$) -- (ket2);
                \draw[->,thick,draw=black!70,rounded corners] (bra2) --++($(botleftAnchor2) - (bra2)$) -- (rhoLeft2);
                \draw[->-,thick,draw=black!70] (rhoLeft2) -- (W2);
                \draw[->-,thick,draw=black!70] (W2) -- (U12);
                \draw[->-,thick,draw=black!70] (U12) -- (V2);
                \draw[->-,thick,draw=black!70] (V2) -- (Udagger12);
                \draw[->,thick,draw=black!70,rounded corners] (Udagger12) --++($(botrightAnchor2) - (Udagger12)$) -- (ket2);

                \draw[draw=black!70,decorate,decoration={brace,mirror,raise=2em}] (botleftAnchor2.west) -- node[below,yshift=-2em] {$\bra{W_{\beta}}$} (W2.east);
                \draw[draw=black!70,decorate,decoration={brace,mirror,raise=2em}] (rightbraceAnchor.west) -- node[below,yshift=-2em] {$\ket{W}$} (botrightAnchor2.east);

                \begin{scope}[on background layer]
                        \node[] (toprightboxAnchor) at (topleftAnchor2-|W2) {};
                        \node [draw=black!30, fill=black!10, scale = 1.2, xscale = 0.85, fit=(topleftAnchor2)(toprightboxAnchor)(botleftAnchor2)(W2), yshift=0.25em, xshift=-2.5pt, rounded corners] {};
                        \node [draw=black!30, fill=black!10, scale = 1.2, xscale = 0.85, fit=(Wdagger2)(toprightAnchor2)(botrightAnchor2)(rightbraceAnchor), yshift=0.0em, xshift=2.5pt, rounded corners] {};
                \end{scope}
        \end{tikzpicture}
        \caption{Shifting the purification insertion point to reduce computational complexity of the time evolution in OTOCs.}
        \label{fig:PIPShift}
\end{figure}

Calculating out-of-time-ordered correlators (OTOC) allows us to
measure the scrambling of quantum information and finds many
interesting and current applications. In general an OTOC of operators
$\hat{W},\hat{V}$ is given as an ensemble average
\begin{align}
        C^{\hat V, \hat W}_{\beta}(t) &= \frac{1}{2} \Tr \left\{\hat{\rho}(\beta) \left[\hat{V}(t),\hat{W}\right]^{\dagger} \left[\hat{V}(t),\hat{W}\right] \right\} \notag \\
        &= \underbrace{\Real \left[ \Tr \left\{\hat{\rho}(\beta) \hat{V}^{\dagger}(t)\hat{W}^{\dagger} \hat{V}(t)\hat{W} \right\} \right]}_{\equiv F^{\hat V, \hat W}_{\beta}(t)} + \text{time ordered}
\end{align}
wherein we have suppressed the time-ordered terms and define the OTOC as the out-of-time ordered part
$F^{\hat V, \hat W}_{\beta}(t)$. At finite temperature, we have to use a
purification to evaluate this quantity.  If we would calculate the
time evolutions in $F^{\hat V, \hat W}_{\beta}(t)$ naively by direct evolution
only in the physical degrees of freedom we would require
$\mathcal{O}(N^{2})$ time steps to obtain the OTOC at time
$t=N\delta$.
Clearly, the growing numerical expenses forbid to reach both large
system sizes and long time scales $t$. The graphical notion (see
\cref{fig:PIPShift}) immediately suggests to transform the operators
in the OTOC in some way as to evenly distribute the required time
evolutions leading to only linear scaling of effort in time $t$. In
the following, we will explain how to transform these operators in the
purification picture and alter the purification insertion point
(PIP). For related work in the framework of matrix-product operators,
cf.~Ref.~\cite{bohrdt17:_scram}.

Consider the ensemble average
$F_{\hat X, \hat Y, \hat Z,\beta} \equiv \Tr
\left\{\hat{\rho}(\beta)\hat{Z}\hat{Y}\hat{X}\right\}$ for some global
operators $\hat{X},\hat{Y},\hat{Z}$ at inverse temperature $\beta$.
Using the cyclic property of the trace the ensemble average can now be written as expectation value in the enlarged Hilbert space
\begin{equation}
        F_{\hat X, \hat Y, \hat Z,\beta} = \Tr \left\{\hat{\rho}(\beta)\hat{Z}\hat{Y}\hat{X}\right\} = \bra{0}\hat{\rho}(\nicefrac{\beta}{2})\hat{Z}\hat{Y}\hat{X}\hat{\rho}(\nicefrac{\beta}{2})\ket{0} \equiv \bra{\nicefrac{\beta}{2}}\hat{Z}\hat{Y}\hat{X}\ket{\nicefrac{\beta}{2}} \;,
\end{equation}
where we have introduced the purified finite temperature state $\ket{\nicefrac{\beta}{2}} \equiv \hat{\rho}(\nicefrac{\beta}{2})\ket{0}$ based on the infinite temperature state $\ket{0}$ (cf.~\cref{sec:mps:finitetemp}). 
A graphical representation of recasting the trace into an expectation value is given by the two networks \cref{fig:PurificationEqualities1} and \cref{fig:PurificationEqualities2} with out-going indices representing row vectors and in-going indices column vectors.
\begin{figure}
  \captionsetup[subfigure]{justification=centering}
        \begin{subfigure}{0.3\textwidth}
                \centering
                \caption{Ensemble average as trace}
                \label{fig:PurificationEqualities1}
                \tikzsetnextfilename{OPTracePurification}
                \begin{tikzpicture}
                        [
                                scale = 0.6,
                                node distance = 0.65,
                                site/.style={circle,draw=blue!50,fill=blue!20,thick,minimum height=2em},
                                op/.style={regular polygon, regular polygon sides=4, draw=orange!50, fill=orange!20, thick, minimum width=3em, minimum height=3em, inner sep = 0pt}
                        ]
                        \node[op] (rhoTop) {$\hat{\rho}(\nicefrac{\beta}{2})$};
                        \node[op] (opX) [below = of rhoTop] {$\hat{X}$};
                        \node[op] (opY) [below = of opX] {$\hat{Y}$};
                        \node[op] (opZ) [below = of opY] {$\hat{Z}$};
                        \node[op] (rhoBot) [below = of opZ] {$\hat{\rho}(\nicefrac{\beta}{2})$};
                        \node (separator) [right=of opY] {};

                        \draw[->-,thick,draw=black!70] (rhoBot) -- (opZ);
                        \draw[->-,thick,draw=black!70] (opZ) -- (opY);
                        \draw[->-,thick,draw=black!70] (opY) -- (opX);
                        \draw[->-,thick,draw=black!70] (opX) -- (rhoTop);
                        \draw[->-,thick,draw=black!70,rounded corners] (rhoTop) -- ++(0,2.0) -- ++(2.0,0) -- ++($(rhoBot) - (rhoTop)$) -- ++(0,-4.0) -- ++(-2.0,0) --(rhoBot);
                \end{tikzpicture}
        \end{subfigure}
        \hfill
        \begin{subfigure}{0.3\textwidth}
                \centering
                \caption{Canonically purified\\ensemble average}
                \label{fig:PurificationEqualities2}
                \tikzsetnextfilename{OPEPurification2}
                \begin{tikzpicture}
                        [
                                scale = 0.6,
                                node distance = 0.7,
                                site/.style={circle,draw=blue!50,fill=blue!20,thick,minimum height=2em},
                                op/.style={regular polygon, regular polygon sides=4, draw=orange!50, fill=orange!20, thick, minimum width=3em, minimum height=3em, inner sep = 0pt}
                        ]
                        \node[op] (rhoTop) {$\hat{\rho}(\nicefrac{\beta}{2})$};
                        \node[op] (opX) [below = of rhoTop] {$\hat{X}$};
                        \node[op] (opY) [below = of opX] {$\hat{Y}$};
                        \node[op] (opZ) [below = of opY] {$\hat{Z}$};
                        \node[op] (rhoBot) [below = of opZ] {$\hat{\rho}(\nicefrac{\beta}{2})$};
                        \node[] (rightAnchor) [right=of opY] {};
                        \node (separator) [left=of opY] {};

                        \node[site] (ket) [above=of rightAnchor] {$\ket{0}$};
                        \node[site] (bra) [below=of rightAnchor] {$\bra{0}$};
                        \node[] (ketHorzRef) at (ket-|opY) {};
                        \node[] (braHorzRef) at (bra-|opY) {};

                        \draw[->-,thick,draw=black!70] (rhoBot) -- (opZ);
                        \draw[->-,thick,draw=black!70] (opZ) -- (opY);
                        \draw[->-,thick,draw=black!70] (opY) -- (opX);
                        \draw[->-,thick,draw=black!70] (opX) -- (rhoTop);
                        \draw[->-,thick,draw=black!70,rounded corners] (rhoTop) -- ++(0,2.0) -- ++($(ket)-(ketHorzRef)$) -- ++(0,-2.0) -- (ket);
                        \draw[->-,thick,draw=black!70] (bra) -- (ket);
                        \draw[->-,thick,draw=black!70,rounded corners] (bra) -- ++($(rhoBot) - (braHorzRef)$) -- ++(0,-2.0) -- ++($(braHorzRef) - (bra)$) -- (rhoBot);
                \end{tikzpicture}
        \end{subfigure}~
        \hfill
        \begin{subfigure}{0.3\textwidth}
                \centering
                \caption{Ensemble average with shifted PIP}
                \label{fig:PurificationEqualities3}
                \tikzsetnextfilename{OPEPurification3}
                \begin{tikzpicture}
                        [
                                scale = 0.6,
                                node distance = 1.0 and 0.5,
                                site/.style={circle,draw=blue!50,fill=blue!20,thick,minimum height=2em},
                                op/.style={regular polygon, regular polygon sides=4, draw=orange!50, fill=orange!20, thick, minimum width=3em, minimum height=3em, inner sep = 0pt}
                        ]
                        \node[site] (ket) {$\ket{0}$};

                        \node[] (topleftAnchor) [left=of ket] {};
                        \node[op] (opY) [below = of topleftAnchor] {$\hat{Y}$};
                        \node[op] (opZ) [below = of opY] {$\hat{Z}$};
                        \node[op] (rhoLeft) [below = of opZ] {$\hat{\rho}(\nicefrac{\beta}{2})$};
                        \node[] (botleftAnchor) [below = of rhoLeft] {};

                        \node[] (toprightAnchor) [right = of ket] {};
                        \node[op] (opX) at (opZ-|toprightAnchor) {$\hat{X}^{t}$};
                        \node[op] (rhoRight) at (rhoLeft-|toprightAnchor) {$\hat{\rho}^{t}(\nicefrac{\beta}{2})$};
                        \node[] (botrightAnchor) at (botleftAnchor-|rhoRight) {};

                        \node[site] (bra) at (botleftAnchor-|ket) {$\bra{0}$};
        %               \node[] (ketHorzRef) at (ket-|opY) {};
        %               \node[] (braHorzRef) at (bra-|opY) {};

                        \draw[->,thick,draw=black!70,rounded corners] (bra) -- ++($(botleftAnchor) - (bra)$) -- (rhoLeft);
                        \draw[->-,thick,draw=black!70] (rhoLeft) -- (opZ);
                        \draw[->-,thick,draw=black!70] (opZ) -- (opY);
                        \draw[->,thick,draw=black!70,rounded corners] (opY) -- ++($(topleftAnchor) - (opY)$) -- (ket);
                        \draw[->,thick,draw=black!70,rounded corners] (bra) -- ++($(botrightAnchor) - (bra)$) -- (rhoRight);
                        \draw[->-,thick,draw=black!70] (rhoRight) -- (opX);
                        \draw[->,thick,draw=black!70,rounded corners] (opX) -- ++($(toprightAnchor) - (opX)$) -- (ket);
                \end{tikzpicture}
        \end{subfigure}
        \caption{Different choices of operator purifications for ensemble average $F_{\hat X, \hat Y, \hat Z,\beta} = \Tr \left\{\hat{\rho}(\beta)\hat{Z}\hat{Y}\hat{X}\right\}$.}
\end{figure}
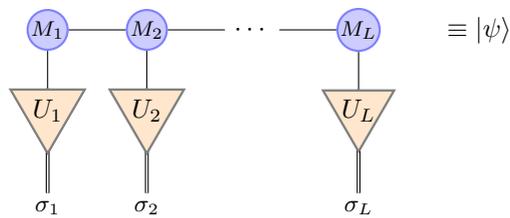
\begin{figure}
  \centering
  \tikzsetnextfilename{mps_lbo}
%  \tikzset{external/export next=false}
  \begin{tikzpicture}
    \begin{scope}[node distance = 0.5 and 0.75]
      \node[site, minimum width=1.5em] (site1) {$M_1$};
      \node[site, minimum width=1.5em] (site2) [right=of site1] {$M_2$};
      \node[ghost] (dots) [right=of site2] {$\cdots$};
      \node[site, minimum width=1.5em] (siteL) [right=of dots] {$M_L$};

      \node (psi) [right=of siteL] {$\equiv \left|\psi\right\rangle$};

      \node [draw, regular polygon, regular polygon sides=3, shape border rotate=180, draw=black!50, fill=orange!20, thick, minimum width=2em, inner sep=0pt] (lbo1) [below=of site1] {$U_1$};
      \node [draw, regular polygon, regular polygon sides=3, shape border rotate=180, draw=black!50, fill=orange!20, thick, minimum width=2em, inner sep=0pt] (lbo2) at (lbo1 -| site2) {$U_2$};
      \node [draw, regular polygon, regular polygon sides=3, shape border rotate=180, draw=black!50, fill=orange!20, thick, minimum width=2em, inner sep=-1pt] (lboL) at (lbo1 -| siteL) {$U_L$};

      \node[ld, inner sep=3pt] (sigma1) [below=of lbo1] {$\sigma_{1}$};
      \node[ld, inner sep=3pt] (sigma2) at (sigma1 -| site2) {$\sigma_{2}$};
      \node[ld, inner sep=3pt] (sigmaL) at (sigma1 -| siteL) {$\sigma_{L}$};

      \draw[] (lbo1) -- (site1);
      \draw[] (site1) -- (site2);
      \draw[] (lbo2) -- (site2);
      \draw[] (site2) -- (dots);
      \draw[] (dots) -- (siteL);
      \draw[] (lboL) -- (siteL);

      % lines go through the triangles, so we need to draw them again
      % drawing just the lines had them stop before the triangle,
      % which thomas thinks is ugly.
      \draw[double] (lbo1.north) -- (sigma1);
      \node [draw, regular polygon, regular polygon sides=3, shape border rotate=180, draw=black!50, fill=orange!20, thick, minimum width=2em, inner sep=0pt] (lbo1) [below=of site1] {$U_1$};
      \draw[double] (lbo2.north) -- (sigma2);
      \node [draw, regular polygon, regular polygon sides=3, shape border rotate=180, draw=black!50, fill=orange!20, thick, minimum width=2em, inner sep=0pt] (lbo2) at (lbo1 -| site2) {$U_2$};
      \draw[double] (lboL.north) -- (sigmaL);
      \node [draw, regular polygon, regular polygon sides=3, shape border rotate=180, draw=black!50, fill=orange!20, thick, minimum width=2em, inner sep=-1pt] (lboL) at (lbo1 -| siteL) {$U_L$};
      
    \end{scope}
  \end{tikzpicture}
  \caption
  {
    \label{fig:mps_lbo} Local basis optimization matrices $U_j$ are
    inserted on the physical legs of the MPS to transform a large
    physical basis (indicated by doubled lines) into a smaller
    effective basis of the MPS tensors $M_j$.}
\end{figure}
From the pictographical representation we motivate the infinite temperature state $\ket{0}$ to be represented by a rank $(2,0)$ tensor $\ket{0} \equiv \sum_{a,\bar{b}} D^{a,\bar{b}}\ket{a}\ket{\bar{b}}$ and correspondingly $\bra{0}$ by a rank $(0,2)$ tensor $\bra{0} \equiv \sum_{a,\bar{b}}D_{a,\bar{b}}\bra{a}\bra{\bar{b}}$, where we have placed a bar over those indices labeling ancilla degrees of freedom.

These tensors have to fulfill the orthogonality conditions
\begin{equation}
        \sum_{\bar{b}} D^{a,\bar{b}}D_{c,\bar{b}} = \delta^{a}_{c}, \quad \sum_{a} D^{a,\bar{b}}D_{a,\bar{c}} = \delta^{\bar{b}}_{\bar{c}}
\end{equation}
so that the tensor elements can be choosen to be $D^{a,\bar{b}} \equiv D^{a,\bar{b}}\delta^{\bar{a}}_{\bar{b}}$ and $D_{a,\bar{b}} \equiv D_{a,\bar{b}}\delta^{\bar{b}}_{\bar{a}}$.
When contracted over physical degrees of freedom, the action of these tensors is to convert row vectors into column vectors and vice versa
\begin{equation}
        D\hat O D^{\dagger} = \sum_{a,c}\sum_{\bar{b},\bar{d}}D^{a,\bar{b}}\hat O^{c}_{a}D_{c,\bar{d}}\ket{\bar{b}}\bra{\bar{d}} = \sum_{\bar{b},\bar{d}}\hat O^{\bar{a}}_{\bar{c}}\ket{\bar{a}}\bra{\bar{c}} = \hat O^{t} \;.
\end{equation}
If we now interpret indices carrying a bar as maps between ancilla degrees of freedom we can reformulate the purification in terms of the $D$ tensors
\begin{equation}
        F_{\hat X, \hat Y, \hat Z,\beta} = \sum_{a,c,\ldots,g,\bar{b}} D_{a,\bar{b}}\hat \rho^{a}_{c}(\nicefrac{\beta}{2})\hat Z^{c}_{d} \hat Y^{d}_{e} \hat X^{e}_{f}\hat \rho^{f}_{g}(\nicefrac{\beta}{2})D^{g,\bar{b}}\;.
\end{equation}
Inserting identities on the physical Hilbert space between $\hat{\rho}$ and $\hat{X}$ as well as $\hat{X}$ and $\hat{Y}$ and making explicit use of the representation of $\hat{D}$ we obtain
\begin{align}
        F_{\hat X, \hat Y, \hat Z,\beta} &= \sum_{\substack{a,c,d,g,\bar{b}, \\ \bar{e},\bar{f},e_l,e_r,f_l,f_r}} D_{a,\bar{b}}\rho^{a}_{c}(\nicefrac{\beta}{2}) \hat Z^{c}_{d} \hat Y^{d}_{e_{l}} \underbrace{D^{e_{l},\bar{e}}D_{e_{r},\bar{e}}}_{\delta^{e_{l}}_{e_{r}}} \hat X^{e_{r}}_{f_{l}} \underbrace{D^{f_{l},\bar{f}}D_{f_{r},\bar{f}}}_{\delta^{f_{l}}_{f_{r}}} \rho^{f_{r}}_{g}(\nicefrac{\beta}{2})D^{g,\bar{b}} \notag \\
                      &= \sum_{\substack{a,c,d,\bar{b}, \\ \bar{e},\bar{f},e_l}} D_{a,\bar{b}} \rho^{a}_{c}(\nicefrac{\beta}{2}) \hat Z^{c}_{d} \hat Y^{d}_{e_{l}} D^{e_{l},\bar{e}} \underbrace{\hat X^{\bar{f}}_{\bar{e}} \rho^{\bar{b}}_{\bar{f}}(\nicefrac{\beta}{2})}_{\text{act on $\mathcal{H}_{A}$}}\\
                      & = \bra{0}\left(\hat{\rho}^{t}(\nicefrac{\beta}{2}) \hat{X}^{t} \right)_{A}\otimes \left(\hat{\rho}(\nicefrac{\beta}{2}) \hat{Z} \hat{Y} \right)_{P} \ket{0}
\end{align}
so that now $\sum_{\bar{f}} \hat X^{\bar{f}}_{\bar{e}} \hat \rho^{\bar{b}}_{\bar{f}}(\nicefrac{\beta}{2}) \equiv \hat{\rho}^{t}(\nicefrac{\beta}{2})\hat{X}^{t}$ are acting on the ancilla space $\mathcal{H}_{A}$, i.e., we have shifted the purification insertion point.
Again these manipulations can be represented efficiently in a graphical notation and are given in \cref{fig:PurificationEqualities3}.

Using this procedure, we can rewrite the OTOC $F^{\hat V, \hat W}_{\beta}(t)$ as
\begin{align}
        F^{\hat V, \hat W}_{\beta}(t) &= \Real\left[ \Tr \left\{\hat{\rho}(\nicefrac{\beta}{2}) \hat{U}^{\dagger}(t) \hat{V}^{\dagger} \hat{U}(t) \hat{W}^{\dagger} \hat{U}^{\dagger}(t) \hat{V} \hat{U}(t) \hat{W} \hat{\rho}(\nicefrac{\beta}{2})\right\} \right] \notag \\
        &= \Real \left[ \bra{0} \left( \hat{U}^{\dagger}(t) \hat{V} \hat{U}(t) \hat{W} \hat{\rho}(\nicefrac{\beta}{2}) \right)^{t}_{A} \otimes \left(\hat{\rho}(\nicefrac{\beta}{2}) \hat{U}^{\dagger}(t) \hat{V}^{\dagger} \hat{U}(t) \hat{W}^{\dagger} \right)_{P} \ket{0} \right] \;.
\end{align}
Defining the initial states
\begin{align}
        \ket{W} &\equiv \hat{W}^{\dagger}_{P}\otimes \mathbf{\hat{1}}^\nodagger_{A}\ket{0}\;, \\
        \ket{W_{\beta}} &\equiv \hat{\rho}_{P}(\nicefrac{\beta}{2}) \otimes \left(\hat{W}^{*}\hat{\rho}_{A}(\nicefrac{\beta}{2})\right) \ket{0} \;,
        \intertext{and their purified time evolutions}
        \ket{W(t)} &\equiv \hat{U}^{\vphantom{*}}_{P}(t) \otimes \hat{U}^{*}_{A}(t)\ket{W} \;,\\
        \ket{W_{\beta}(t)} &\equiv \hat{U}^{\vphantom{*}}_{P}(t) \otimes \hat{U}^{*}_{A}(t)\ket{W_{\beta}}
        \intertext{the OTOC can be obtained by calculating the expectation value}
        F^{\hat V, \hat W}_{\beta}(t) &= \Real \left[ \bra{W_{\beta}(t)}\hat{V}^{\dagger}_{P} \otimes \hat{V}^{t}_{A}\ket{W(t)} \right]\;.
\end{align}
We hence only need $N$ steps to evaluate all expectation values for times $t = N \delta$.

From a more general point of view shifting the purification insertion
point in the OTOCs reformulates the multiple Schr\"odinger time
evolutions of the physical system in the canonical choice of the PIP
into a Heisenberg time evolution on both the physical and ancilla
system of a generalized observable
$\hat{V}^{\dagger}_{P}\otimes \hat{V}^{\vphantom{\dagger}t}_{A}$.

\subsection{\label{sec:tricks:lbo}Local basis optimization}

While the dimension of local Hilbert spaces is typically very limited
in spin and electronic systems, bosonic systems potentially require a
large local dimension $\sigma = \mathcal{O}(100)$. As this local
dimension typically enters at least quadratically in some operations
on matrix-product states, some way to dynamically select the most
relevant subspace of the local Hilbert space is potentially extremely
helpful. The local basis
transformation\cite{zhang98:_densit_matrix_approac_local_hilber_space_reduc,
  guo12:_critic_stron_coupl_phases_one, brockt15:_matrix} method
provides for just this: by inserting an additional matrix $U_j$ on
each physical leg of the MPS (cf.~\cref{fig:mps_lbo}), the large
physical dimension is transformed into a smaller effective basis. The
rank-3 MPS tensor then only has to work with the smaller effective
basis. The method was adapted for TEBD time evolution in
Ref.~\cite{brockt15:_matrix} but it is also straightforward to use in
the other time-evolution methods presented here. For the MPO \wiii and
global Krylov methods, only the MPO-MPS product has to be adapted to
generate additionally a new optimal local basis after each step. The DMRG, TDVP and
local Krylov method translate\cite{schroeder16:_simul, dorfner17:_numer,
  hubig17:_symmet_protec_tensor_networ} directly in much the same way
as DMRG.

%%% Local Variables:
%%% mode: latex
%%% TeX-master: "../time_evolution_review"
%%% End:

%% file: content/examples/intro.tex
\section{\label{sec:examples}Examples}	

The following four subsections serve as exemplary applications of the different time-evolution methods to demonstrate, justify and verify the theoretical remarks of the earlier reviews of each method.
To some extent it is our hope that the examples here are sufficiently general to be used as benchmarks for \emph{future} time-evolution methods as a way to increase the comparability with previous work.
A pairwise comparison of a new method with one of the methods tested here would hence also serve as a comparison with all other methods tested here.
To this end, we focus on a clear description of the problem setting and analysis of the runtime behaviour of the methods instead of analyzing the physical results in great detail.

In \cref{sec:examples:dsf}, we will calculate the dynamical structure factor of a XXZ spin chain in a staggered magnetic field at zero temperature.
This is a standard application of one-dimensional time-dependent MPS techniques which is straightforward to reproduce in any implementation.
\cref{sec:examples:hubbard} studies the cooling of a Hubbard chain in the canonical ensemble as an application of imaginary-time evolution and a study of the stability of each evolution method against the build-up of errors.
\cref{sec:example:melting} considers the melting of N\'eel order on a two-dimensional lattice in real-time, a very challenging problem due to very long-range interactions coming into play when projecting the system onto a chain geometry, which is a necessary step for treating higher-dimensional systems with MPS. 
Finally, in \cref{sec:examples:otoc} we simulate the evolution of out-of-time-order correlators on an interacting spin chain at infinite temperature.
This problem setting combines the need for a purification ansatz with real-time evolution and the calculation of different-times correlators, which, like the dynamical structure factor, requires the correct treatment of phases during the evolution.
%

%%% Local Variables: 
%%% mode: latex
%%% TeX-master: "../time_evolution_review"
%%% End: 

%% file: content/examples/dsf.tex
\subsection{\label{sec:examples:dsf}Dynamical spin structure factor (DSF)}

In this section we examine the longitudinal dynamical spin structure factor (DSF) $S^{zz}(q,\omega)$. This quantity can be measured directly in e.g.~neutron scattering experiments.
Our system of choice here is an anisotropic $\text{S}=1/2$ Heisenberg chain in a staggered magnetic field described by the Hamiltonian
\begin{align}
	\hat{H} 
	&= J\sum_{j=1}^{L}\left[\hat{s}^{x}_{j}\hat{s}^{x}_{j+1} + \hat{s}^{y}_{j}\hat{s}^{y}_{j+1} + \Delta\hat{s}^{z}_{j}\hat{s}^{z}_{j+1} - h^{s}_{j}\hat{s}^{z}_{j} \right] \notag\\
	&= J\sum_{j=1}^{L}\left[\frac{1}{2}\left(\hat{s}^{+}_{j}\hat{s}^{-}_{j+1} + \hat{s}^{-}_{j}\hat{s}^{+}_{j+1}\right) + \Delta\hat{s}^{z}_{j}\hat{s}^{z}_{j+1} - h^{s}_{j}\hat{s}^{z}_{j} \right],
	\quad h^{s}_{j} = (-1)^{j}h \; . \label{dsf:ham}
\end{align}
We set $\Delta \equiv J^{-1}$ and vary $J = 0.1, 0.2, 0.3$ and $J\cdot h = 0.01, 0.05, 0.1$.
The calculation is carried out on systems of sizes $L=100, 150, 200$ with open boundary conditions.
$S^{zz}(q,w)$ can then be obtained from the double Fourier transformation of the time-dependent spin-spin correlation function
\begin{align}
	\braket{\hat{s}^{z}_{j}(t)\hat{s}^{z}_{\nicefrac{L}{2}}(0)}_{cc} &= \braket{0|\hat{U}^{\dagger}(t)\hat{s}^{z}_{j}\hat{U}(t)\hat{s}^{z}_{\nicefrac{L}{2}}|0} - \braket{0|\hat{s}^{z}_{j}\hat{s}^{z}_{\nicefrac{L}{2}}|0}
\end{align}
with $\ket{0}$ the ground state of \cref{dsf:ham}. We then have
\begin{align}
	S^{zz}(q,\omega) 
	&= \frac{1}{L}\sum_{j=1}^{L}e^{-\mathrm i q (j - L/2)}\int_{-\infty}^{\infty}dt\; e^{i\omega t}\braket{\hat{s}^{z}_{j}(t)\hat{s}^{z}_{\nicefrac{L}{2}}(0)}_{cc} \\
	&\stackrel{\sim}{=} \frac{2\pi}{L T}\delta \sum_{j=1}^{L}e^{-\mathrm i q (j - L/2)}\sum_{n=0}^{N} e^{i(\omega + i\eta) t_{n}} 2\Real \braket{\hat{s}^{z}_{j}(t_{n})\hat{s}^{z}_{\nicefrac{L}{2}}(0)}_{cc}
\end{align}
where we discretize the time coordinate $t_{n}=n \delta$ and truncate the integral at a finite, maximum evolution time $T = N\delta$.
It is of course important to chose the system size $L$ sufficiently large that the initial central excitation does not reach the edges of the open chain during the finite simulation time $T$.
Due to this finite simulation time $T$, it is also necessary to introduce a damping factor $\eta>0$.
Without the damping factor, we would observe large spectral leakage due to the finite interval.
With $\eta > 0$, we obtain an artificial broadening of spectral lines instead.
The overall simulation procedure was to calculate the ground state $\ket{0}$ of \cref{dsf:ham} with DMRG at a maximum discarded weight of $10^{-14}$ and energy convergence better than $10^{-12}$.
The excited state $\ket{1}=\hat{s}^{z}_{\nicefrac{L}{2}}\ket{0}$ was then evolved in time until the final time $T=200$ in units of $\Delta$ was reached.
During the evolution, we need to evaluate the time-dependent correlator
\begin{align}
  S^{zz}(j,t_n) &= \braket{0|\hat{s}^{z}_{j}(t_{n})\hat{s}^{z}_{\nicefrac{L}{2}}(0)|0} - \braket{0|\hat{s}^{z}_{j}\hat{s}^{z}_{\nicefrac{L}{2}}|0}  \\
                &= \braket{0|\hat U^\dagger(t_n) \hat s^z_j \hat U(t_n) \hat s^z_{\nicefrac{L}{2}}|0} - \braket{0|\hat{s}^{z}_{j}\hat{s}^{z}_{\nicefrac{L}{2}}|0}\\
                &= e^{\I E_0 t_n} \braket{0|\hat s^z_j|1(t)} - \braket{0|\hat{s}^{z}_{j}\hat{s}^{z}_{\nicefrac{L}{2}}|0}
\end{align}
where the term $\braket{0|\hat{s}^{z}_{j}\hat{s}^{z}_{\nicefrac{L}{2}}|0}$ is of course only evaluated once.
The last equality only holds for representations of the time stepper $\hat{U}(\delta)$ which act on the ground state by multiplying a phase, i.e. $\hat{U} (\delta)\ket{0} = e^{-\mathrm i E_{0}\delta}\ket{0}$.
This should of course be the case ideally, but in practice is not achieved by all methods.
In general, $S^{zz}(q,\omega)$ is very sensitive to the build-up of global phases by the particular time stepper rendering this quantity an important testcase.

\subsubsection{\label{sec:examples:dsf:real-space}Real-space evolution}
\begin{figure}[t!]
	\tikzsetnextfilename{dsf_real_space_tdvp}
	\begin{tikzpicture}
		[node distance=0.05]\pgfplotsset
		{
			/pgfplots/colormap={temp}{
				rgb255=(36,0,217) 		% 0 "#2400d9"
				rgb255=(25,29,247) 		% 1 "#191df7"
				rgb255=(41,87,255) 		% 2 "#2957ff"
				rgb255=(61,135,255) 	% 3 "#3d87ff"
				rgb255=(87,176,255) 	% 4 "#57b0ff"
				rgb255=(117,211,255) 	% 5 "#75d3ff"
				rgb255=(153,235,255) 	% 6 "#99ebff"
				rgb255=(189,249,255) 	% 7 "#bdf9ff"
				rgb255=(211,255,255) 	% 8 "#d3ffff"
				rgb255=(255,255,255)
				rgb255=(255,255,211) 	% 9 "#ffffd3"
				rgb255=(255,242,189) 	% 10 "#fff2bd"
				rgb255=(255,214,153) 	% 11 "#ffd699"
				rgb255=(255,172,117) 	% 12 "#ffac75"
				rgb255=(255,120,87) 	% 13 "#ff7857"
				rgb255=(255,61,61) 		% 14 "#ff3d3d"
				rgb255=(247,40,54) 		% 15 "#f72836"
				rgb255=(217,22,48) 		% 16 "#d91630"
				rgb255=(166,0,33)		% 17 "#a60021"
			}
		}
% 		\pgfplotsset
% 		{
% 			/pgfplots/colormap={gnuplot_7_5_15}{
% 				rgb255=(0,0,0)
% 				rgb255=(57,0,79)
% 				rgb255=(81,0,150)
% 				rgb255=(99,1,206)
% 				rgb255=(114,2,243)
% 				rgb255=(128,4,255)
% 				rgb255=(140,7,243)
% 				rgb255=(151,11,206)
% 				rgb255=(161,16,150)
% 				rgb255=(171,23,79)
% 				rgb255=(180,32,0)
% 				rgb255=(189,42,-79)
% 				rgb255=(198,55,-150)
% 				rgb255=(206,70,-206)
% 				rgb255=(213,87,-243)
% 				rgb255=(221,108,-255)
% 				rgb255=(228,131,-243)
% 				rgb255=(235,157,-206)
% 				rgb255=(242,186,-150)
% 				rgb255=(249,219,-79)
% 				rgb255=(255,255,0)
% 			}
% 		}
		\pgfplotsset
		{
			/pgfplots/colormap={gnuplot}{rgb255=(0,0,0) rgb255=(46,0,53) rgb255=(65,0,103) rgb255=(80,0,149) rgb255=(93,0,189) rgb255=(104,1,220) rgb255=(114,2,242) rgb255=(123,3,253) rgb255=(131,4,253) rgb255=(139,6,242) rgb255=(147,9,220) rgb255=(154,12,189) rgb255=(161,16,149) rgb255=(167,20,103) rgb255=(174,25,53) rgb255=(180,31,0) rgb255=(186,38,0) rgb255=(191,46,0) rgb255=(197,55,0) rgb255=(202,64,0) rgb255=(208,75,0) rgb255=(213,87,0) rgb255=(218,100,0) rgb255=(223,114,0) rgb255=(228,130,0) rgb255=(232,147,0) rgb255=(237,165,0) rgb255=(241,185,0) rgb255=(246,207,0) rgb255=(250,230,0) rgb255=(255,255,0) }
		}
		\begin{groupplot}
		[
			group style = 
			{
				group size 		=	2 by 2,
				vertical sep		=	5em,
				horizontal sep		=	0.0mm,
% 				x descriptions at	=	edge bottom,
				y descriptions at	=	edge left,
			},
% 			height 	= 	0.35\textheight,
			width 	= 	1.0\textwidth,
% 			point meta min=-0.0025,
% 			point meta max=0.0025,
% 			xmin	= 0,
% 			xmax	= 100,
		]
		\nextgroupplot
			[
				axis on top,
				enlargelimits	= false,
				height		= 0.35\textheight,
				width		= 0.5\textwidth,
				xlabel		= Site $j$,
				ylabel		= {Time $t [\nicefrac{1}{\Delta}]$},
				ytick		= {0,50,...,200},
				xtick		= {25,50,...,100},
				title		= {$S^{zz}(j,t_{n})$ (2TDVP)},
			]
			\addplot graphics
			[
				xmin = 0, 
				xmax = 100, 
				ymin = 0, 
				ymax = 200,
			]
			{../ex1-spinchain/precompiled/syten_tdvp_L_100_t_0p1_J_1p0_h_0p01_dt_0p1.rt-1.real.pdf};
		\nextgroupplot
			[
				axis on top,
				enlargelimits	= false,
				height		= 0.35\textheight,
				width		= 0.5\textwidth,
				xlabel		= Site $j$,
% 				ylabel		= {time $[t]$},
				ytick		= {0,50,...,200},
				xtick		= {0,25,50,...,100},
				title		= {$S^{zz}(j,t_{n})$ (TEBD2)},
				colorbar right,
				colormap name	= temp,
				point meta min=-0.0025,
				point meta max=0.0025,
				every colorbar/.append style =
					{
% 						height			=	2*\pgfkeysvalueof{/pgfplots/parent axis height} + 1*\pgfkeysvalueof{/pgfplots/group/vertical sep},
% 						ylabel			=	{$\langle \hat N_j \rangle$},
						width			=	3mm,
						ytick			= 	{-0.0025,0.0,0.0025},
% 						yticklabels		=	{$-0.0025$,$0.0$,$0.0025$},
% 						ylabel shift 		=	-4pt,
						y tick scale label style=	{xshift=2.0em,yshift=0.5em},
					},
			]
			\addplot graphics
			[
				xmin = 0, 
				xmax = 100, 
				ymin = 0, 
				ymax = 200,
			]
			{../ex1-spinchain/precompiled/syten_tebd_L_100_t_0p1_J_1p0_h_0p01_dt_0p1.rt-1.real.pdf};
		\nextgroupplot
			[
				axis on top,
				enlargelimits	= false,
				height		= 0.2\textheight,
				width		= 0.5\textwidth,
				xlabel		= {Wave vector $k$},
				ylabel		= {Energy $\omega[\Delta]$},
				ytick		= {0,0.75,1.5},
				xtick		= {-50,-25,0,25,50},
				xticklabels	= {$0$, $\nicefrac{\pi}{2}$,$\pi$, $\nicefrac{3\pi}{2}$, $2\pi$},
				title		= {$S^{zz}(k,\omega)$ (2TDVP)},
			]
			\addplot graphics
			[
				xmin = -50, 
				xmax = 50, 
				ymin = 0, 
				ymax = 1.5,
			]
			{../ex1-spinchain/precompiled/syten_tdvp_L_100_t_0p1_J_1p0_h_0p01_dt_0p1.kw-1e-1.real.pdf};
		\nextgroupplot
			[
				axis on top,
				enlargelimits	= false,
				height		= 0.2\textheight,
				width		= 0.5\textwidth,
				xlabel		= {Wave vector $k$},
				ytick		= {0,0.75,1.5},
				xtick		= {-25,0,25,50},
				xticklabels	= {$\nicefrac{\pi}{2}$,$\pi$, $\nicefrac{3\pi}{2}$, $2\pi$},
				title		= {$S^{zz}(k,\omega)$ (TEBD2)},
				colorbar right,
				colormap name	= temp,
				point meta min=-4.0,
				point meta max=4.0,
				every colorbar/.append style =
					{
% 						height			=	2*\pgfkeysvalueof{/pgfplots/parent axis height} + 1*\pgfkeysvalueof{/pgfplots/group/vertical sep},
% 						ylabel			=	{$\langle \hat N_j \rangle$},
						width			=	3mm,
						ytick			= 	{-4.0,0.0,4.0},
% 						yticklabels		=	{$-0.0025$,$0.0$,$0.0025$},
% 						ylabel shift 		=	-4pt,
						y tick scale label style=	{xshift=2.0em,yshift=0.5em},
					},
			]
			\addplot graphics
			[
				xmin = -50, 
				xmax = 50, 
				ymin = 0, 
				ymax = 1.5,
			]
			{../ex1-spinchain/precompiled/syten_tebd_L_100_t_0p1_J_1p0_h_0p01_dt_0p1.kw-1e-1.real.pdf};
		\end{groupplot}
	\end{tikzpicture}
	\caption{\label{fig:dsf:compare_rt_kw}(DSF) Real-time evolution of the time-dependent correlator $S^{zz}(j, t_{n})$ with parameters $J=0.1$, $J\cdot h=0.01$ and time step $\delta=0.1$. Top panels show the evolution of the initial perturbation using \syten{} 2TDVP (left) and \syten{} TEBD2 (right) without damping. Bottom panels show the low-energy part of the dynamical spin structure factor using \syten{} 2TDVP (left) and \syten{} TEBD2 (right) stepper and damping $\eta$ such that $e^{200\eta}=0.1$.}
\end{figure}
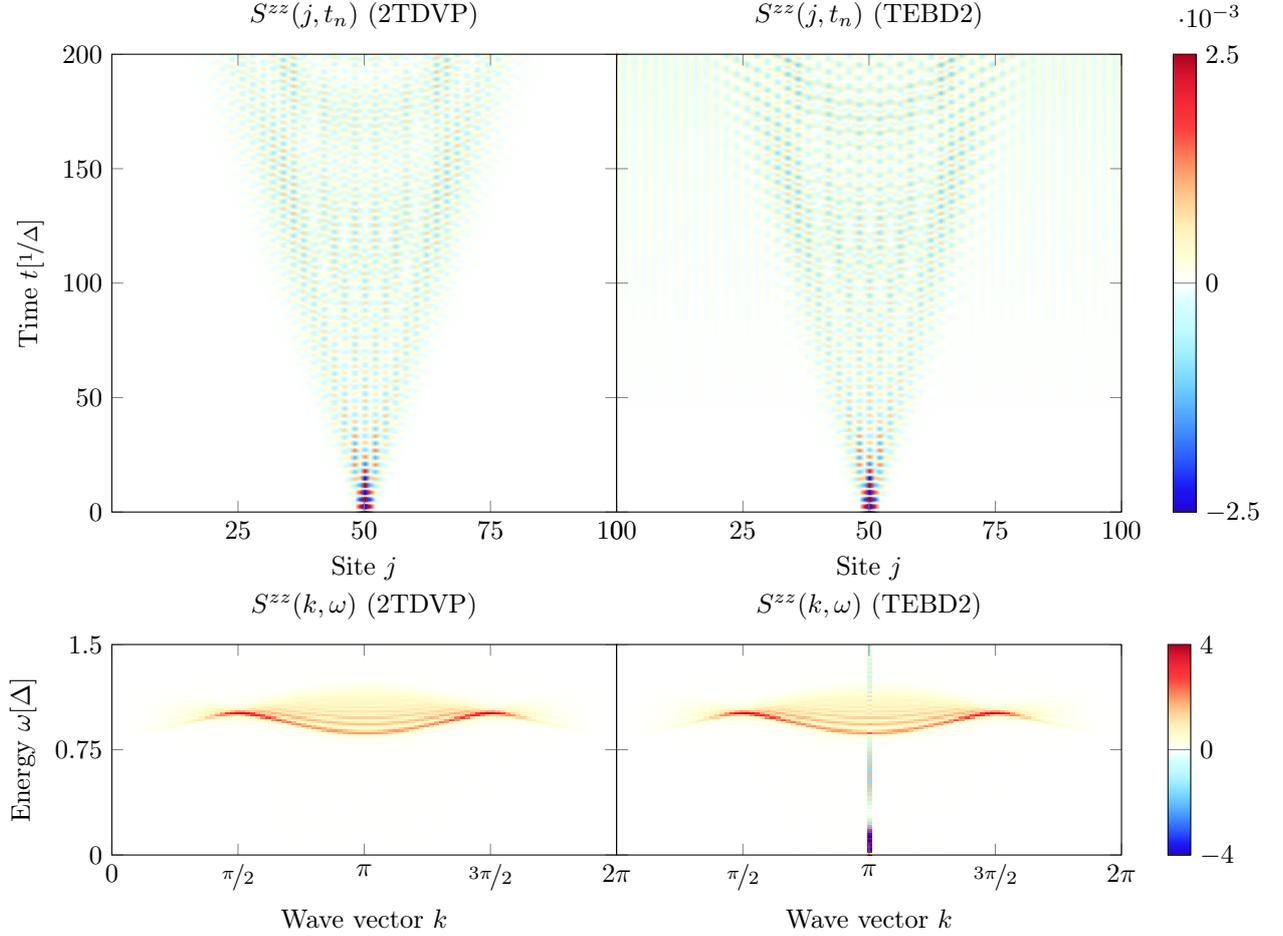
In \cref{fig:dsf:compare_rt_kw}, $S^{zz}(j,t_n)$ as obtained from \syten{} 2TDVP and TEBD2 is shown in the top panels.
Here we already see the main error source when using the raw data
to calculate the spin structure factor: a global phase shift in
the correlator which can be identified by a uniformly evolving
stripe pattern in the TEBD2 data imposed over the perturbation light cone.
When calculating the Fourier transformations this uniformly evolving phase will cause a $k=\pi$ signal which ultimatively may overlay the relevant spectral information.
In the bottom panels of \cref{fig:dsf:compare_rt_kw} we plot the low-energy part of the DSF with damping $\eta$ applied such that $e^{200\eta}=0.1$.
This plot reveals a dominant $k=\pi$ feature visible in the TEBD2 data which is absent in the 2TDVP data.
At least to some extent, this phase shift can be reduced by calculating the explicit time evolution of the ground state $\hat{U}(t_{n})\ket{0} \equiv \ket{0(t_{n})}$ in the evaluation of $S^{zz}(j,t_n)$.
\input{content/examples/dsf_lightcone_comparison.tex}

\subsubsection{\label{sec:examples:dsf:spectral-functions}Spectral functions}
\input{content/examples/dsf_spectral.tex}
More generally, most methods do not evolve the area outside the light cone exactly.
That is, while in this region the Hamiltonian should lead to trivial dynamics, small errors are introduced.
Given enough time, these errors can accumulate with two major results:
(i)~The slightly erroneous state is now even less of an eigenstate of the (effective) Hamiltonian and hence acquires dynamics, leading to a runaway effect.
(ii)~Additional entanglement accumulates and makes the calculation more difficult.
The precise way in which this error occurs slightly depends on the method:
For the TEBD2 and the MPO \wii method, the error is simply a finite time-step error due to the Trotterization or otherwise $\delta$-dependent error terms.
The local Krylov method is primarily subject to the error created by the basis transformation after solving each local problem, which is proportional to the step size $\delta$.
This error affects every site and, after comparably short times, results in site tensors which have lost the property of being exact eigenstates of the effective Hamiltonian outside the light cone.
Altogether, these methods can produce results at time step $\delta = 0.1$, but the results only qualitatively reproduce the spectrum, require a large bond dimension $m$ and also carry a large spectral weight around $\omega = 0$.
For the global Krylov method, we had initially selected a Krylov error $< 10^{-6}$, believing that this error should be sufficiently small to provide good results.
However, it turns out that an extremely small Krylov error $< 10^{-10}$ is necessary to avoid the runaway effect and the accumulation of errors outside the light cone (observed best in \cref{fig:dsf:benchmarks-chimax,fig:dsf:benchmarks-cputime}).
For this method, the additional dynamics also make the problem more complicated, hence requiring more Krylov vectors for a precise solution.
As these Krylov vectors are typically more strongly entangled, their truncation will be more severe.
Conversely, the Krylov error at a fixed number of Krylov vectors will increase, leading to yet larger errors.
In effect, the method runs into an entirely artificial exponential wall, built from accumulated errors outside the light cone and stopping the calculation in just a few steps (cf.~\cref{fig:dsf:benchmarks-cputime}).
Selecting a smaller time step size $\delta = 0.01$ and disabling the extrapolation (cf.~\cref{sec:krylov_dynamic_step:extra}) -- hence forcing a very small Krylov error -- leads to a very precise calculation which is also competitive in runtime.

The 2TDVP, in comparison, does evolve the area outside of the light cone exactly -- the MPS tensors there are exact eigenstates of the local effective Hamiltonian, as they are the result of a DMRG ground-state search procedure.
As such, 2TDVP also produces good results at $\delta = 0.1$ and (like the global Krylov at $\delta = 0.01$) no spectral weight around $\omega = 0$.
\cref{fig:dsf:compare_rt_log} shows the light cones produced by 2TDVP, TEBD2, the global Krylov method and the local Krylov method.
2TDVP and the global Krylov method give small, homogenous correlations outside the light cone.
With sufficiently large frequency resolution, we could therefore find some spectral weight exactly at $\omega = 0$; in practice our resolution is too low and the prefactor too small for this effect to be visible.
TEBD2, in contrast, also results in slow dynamics in this region.
These dynamics can be seen in some remaining spectral weight close to $\omega = 0$, but the light cone itself is produced very well.
The local Krylov method, in contrast, gives immediately large contributions to the correlations outside the light cone.
These correlations (i) give large spectral weight around $\omega = 0$ even with $\delta = 0.01$ and (ii) result in a shift of the overall spectral function.

As an example of the effects discussed above, consider $S^{zz}(k=\pi,\omega)$ plotted in \cref{fig:dsf:kpi}.
Most importantly, all methods are able to resolve the location of the primary peak at $\omega \approx 0.5$ and also subsequent peaks at $\omega \approx 0.67$, $\omega \approx 0.8$ and $\omega \approx 0.88$.
However, only 2TDVP produces the ``correct'' function at $\delta = 0.1$ and is only joined in this by the global Krylov method at $\delta = 0.01$.
All other methods produce additional unphysical features such as an
overall shift (local Krylov method at both $\delta$ values; TEBD2 and
the MPO \wii method at $\delta = 0.1$) or spectral weight towards
$\omega = 0$ (TEBD2, MPO \wii and local Krylov methods at all
$\delta$).

\input{content/examples/dsf_chimaxruntime.tex}

\subsubsection{\label{sec:examples:dsf:benchmark}Benchmark}

Finally, we compare the runtime and growth of the bond dimension in the different methods in \cref{fig:dsf:benchmarks-chimax,fig:dsf:benchmarks-cputime}.
All calculations where performed single threaded on an Intel Xeon Gold 6150 CPU with 192 GB of RAM and no hard-disk caching.
The growth of the bond dimension is slower when the time step used is smaller (especially $\delta = 0.01$), suggesting that at least some part of the entanglement structure is wrong due to a finite time-step error.
Still at $\delta = 0.01$, the bond dimension grows slowest during the 2TDVP evolution resulting in a bond dimension at $t = 200$ roughly a factor of two smaller than for the local Krylov and MPO \wii methods.
The local Krylov method at $\delta = 0.1$ and $\delta = 0.05$ very quickly saturates the maximal bond dimension $m = 200$ which is consistent with the picture of the site tensors being pushed away from the ground state.
The strong dependence on the step size is related to the error induced by the basis transformation which in turn can be estimated by the overlap between the unevolved and evolved state which is reduced for smaller time steps.
Curiously, the situation is comparable for the global Krylov method even though the underlying reasoning is very different.
In this case the strong dependence of the bond dimension on the time step is due to the incurred Krylov error.
Once the site tensors are no longer the ground-state solutions of the reduced two site problem, the incurred Krylov error increases even faster as the overall state rapidly evolves away from the global ground state and the small Krylov subspace can not faithfully represent the action of the operator exponential.

The growth of the bond dimension also translates directly to the computational effort, plotted in \cref{fig:dsf:benchmarks-cputime}.
Consistently, we find the local Krylov method to be much slower than the comparable 2TDVP even with the smallest time step.
Compared to TEBD2, the third method which produces good data at least in the range $0.4 < \omega < 1$, 2TDVP performance is acceptable though not superb.
The global Krylov method also performes comparably well as long as the time step is small enough $\delta=0.01$ and is completely unsuitable if $\delta=0.1$.

\subsubsection{\label{sec:examples:dsf:conclusion}Conclusion}

We calculated the dynamical spin structure factor which can be boiled down to the evaluation of non-equal time correlation functions.
In our analysis, we find 2TDVP to provide both the best numerical data and very efficient calculations due to its numerical stability also at larger time steps.
In fact, it is the only method which generated a stable time evolution if we set $\delta=0.1$.
The global Krylov method provides equally good numerical results as long as its Krylov errors are sufficiently small, which unfortunately results in a very small time step and no recycling of Krylov subspaces (cf.~\cref{sec:krylov_dynamic_step:extra}) being possible.
Putting the focus on the runtime, we find that TEBD2 can be considered as a method providing a satisfying trade-off between quick MPO-MPS application also at smaller time steps and satisfying numerical results, even though there is a spectral loss at very small energies $\omega \approx 0$ due to the build-up of global phases during the time evolution.
Finally, the local Krylov and the MPO \wii method are apparently the most susceptible for phase errors with very large errors at $\delta=0.1$. 
Even at $\delta=0.01$, the local Krylov method gives large spectral leakage over the entire frequency range while the MPO \wii method gives relatively reasonable data though slower and of worse quality than TEBD2.

An interesting remark is that counterintuitively, higher-precision calculations can be \emph{faster} than lower-precision calculations.
That is, as long as the time-evolution method only induces very small errors, the area outside the light-cone stays nearly a (local) eigenstate of the Hamiltonian.
If instead large errors are made by the method, these nonphysical errors then need to be evolved also in subsequent steps, which may be much harder than ``simply'' solving the physical problem.
The most obvious example of this behaviour is the global Krylov method.
In fact, however, \emph{all} methods exhibit larger bond dimensions at larger time step sizes (related to larger Trotterization errors) and a faster growth of bond dimensions, cf.~\cref{fig:dsf:benchmarks-chimax}.
Without the artificial limit to $m = 200$, we would hence expect an eventual cross-over where the more precise calculation at $\delta = 0.01$ with ten times more steps per unit time is faster than the large-step calculation at $\delta = 0.1$.

%% file: content/examples/dsf_lightcone_comparison.tex
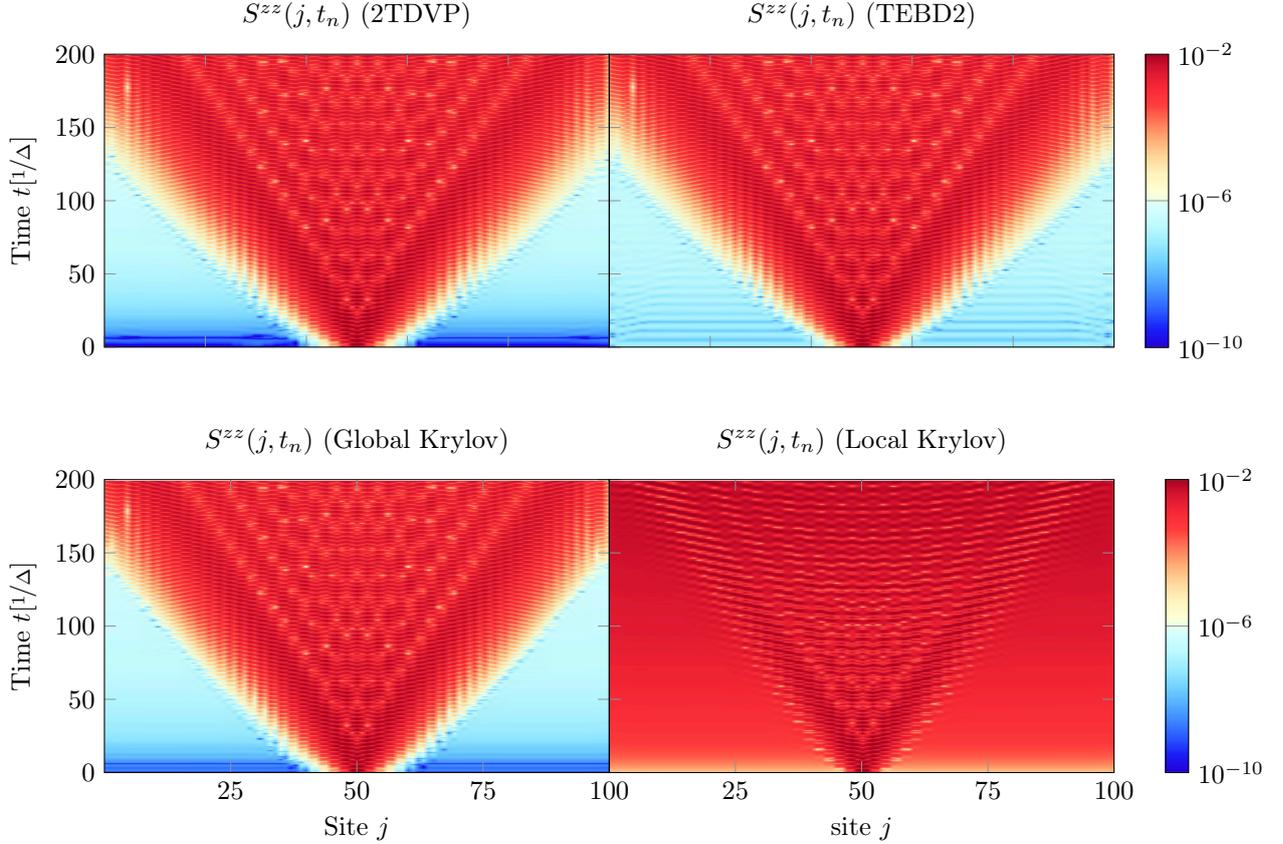
\begin{figure}[t!]
	\tikzsetnextfilename{dsf_real_space_compare_log}
	\begin{tikzpicture}
		[node distance=0.05]\pgfplotsset
		{
			/pgfplots/colormap={temp}{
				rgb255=(36,0,217) 		% 0 "#2400d9"
				rgb255=(25,29,247) 		% 1 "#191df7"
				rgb255=(41,87,255) 		% 2 "#2957ff"
				rgb255=(61,135,255) 	% 3 "#3d87ff"
				rgb255=(87,176,255) 	% 4 "#57b0ff"
				rgb255=(117,211,255) 	% 5 "#75d3ff"
				rgb255=(153,235,255) 	% 6 "#99ebff"
				rgb255=(189,249,255) 	% 7 "#bdf9ff"
				rgb255=(211,255,255) 	% 8 "#d3ffff"
				rgb255=(255,255,211) 	% 9 "#ffffd3"
				rgb255=(255,242,189) 	% 10 "#fff2bd"
				rgb255=(255,214,153) 	% 11 "#ffd699"
				rgb255=(255,172,117) 	% 12 "#ffac75"
				rgb255=(255,120,87) 	% 13 "#ff7857"
				rgb255=(255,61,61) 		% 14 "#ff3d3d"
				rgb255=(247,40,54) 		% 15 "#f72836"
				rgb255=(217,22,48) 		% 16 "#d91630"
				rgb255=(166,0,33)		% 17 "#a60021"
			}
		}
		\begin{groupplot}
		[
			group style = 
			{
				group size 		=	2 by 2,
				vertical sep		=	5em,
				horizontal sep		=	0.0mm,
% 				x descriptions at	=	edge bottom,
				y descriptions at	=	edge left,
			},
% 			height 	= 	0.35\textheight,
			width 	= 	1.0\textwidth,
% 			point meta min=-0.0025,
% 			point meta max=0.0025,
% 			xmin	= 0,
% 			xmax	= 100,
		]
		\nextgroupplot
			[
				axis on top,
				enlargelimits	= false,
				height		= 0.25\textheight,
				width		= 0.5\textwidth,
% 				xlabel		= site $j$,
				ylabel		= {Time $t [\nicefrac{1}{\Delta}]$},
				ytick		= {0,50,...,200},
				xticklabels	= {,,},
				title		= {$S^{zz}(j,t_{n})$ (2TDVP)},
			]
			\addplot graphics
			[
				xmin = 0, 
				xmax = 100, 
				ymin = 0, 
				ymax = 200,
			]
			{../ex1-spinchain/precompiled/scipal_tdvp_L_100_t_0p3_J_1p0_h_0p1_dt_0p01.rt-1.real.log.pdf};
		\nextgroupplot
			[
				axis on top,
				enlargelimits	= false,
				height		= 0.25\textheight,
				width		= 0.5\textwidth,
% 				xlabel		= Site $j$,
% 				ylabel		= {Time $[t]$},
				ytick		= {0,50,...,200},
				xticklabels	= {,,},
				title		= {$S^{zz}(j,t_{n})$ (TEBD2)},
				colorbar right,
				colormap name	= temp,
				point meta min=-10,
				point meta max=-2,
				every colorbar/.append style =
					{
% 						height			=	2*\pgfkeysvalueof{/pgfplots/parent axis height} + 1*\pgfkeysvalueof{/pgfplots/group/vertical sep},
% 						ylabel			=	{$\langle \hat N_j \rangle$},
						width			=	3mm,
						ytick			= 	{-10,-6,-2},
						yticklabels		=	{$10^{-10}$, $10^{-6}$, $10^{-2}$},
% 						ylabel shift 		=	-4pt,
						y tick scale label style=	{xshift=2.0em,yshift=0.5em},
					},
			]
			\addplot graphics
			[
				xmin = 0, 
				xmax = 100, 
				ymin = 0, 
				ymax = 200,
			]
			{../ex1-spinchain/precompiled/syten_tebd_L_100_t_0p3_J_1p0_h_0p1_dt_0p01.rt-1.real.log.pdf};
		\nextgroupplot
			[
				axis on top,
				enlargelimits	= false,
				height		= 0.25\textheight,
				width		= 0.5\textwidth,
				xlabel		= Site $j$,
				ylabel		= {Time $t [\nicefrac{1}{\Delta}]$},
				ytick		= {0,50,...,200},
				xtick		= {25,50,...,100},
				title		= {$S^{zz}(j,t_{n})$ (Global Krylov)},
			]
			\addplot graphics
			[
				xmin = 0, 
				xmax = 100, 
				ymin = 0, 
				ymax = 200,
			]
			{../ex1-spinchain/precompiled/syten_global_krylov_L_100_t_0p3_J_1p0_h_0p1_dt_0p01.rt-1.real.log.pdf};
		\nextgroupplot
			[
				axis on top,
				enlargelimits	= false,
				height		= 0.25\textheight,
				width		= 0.5\textwidth,
				xlabel		= site $j$,
% 				ylabel		= {time $t [\nicefrac{1}{\Delta}]$},
				ytick		= {0,50,...,200},
				xtick		= {25,50,...,100},
				title		= {$S^{zz}(j,t_{n})$ (Local Krylov)},
				colorbar right,
				colormap name	= temp,
				point meta min=-10,
				point meta max=-2.0,
				every colorbar/.append style =
					{
% 						height			=	2*\pgfkeysvalueof{/pgfplots/parent axis height} + 1*\pgfkeysvalueof{/pgfplots/group/vertical sep},
% 						ylabel			=	{$\langle \hat N_j \rangle$},
						width			=	3mm,
						ytick			= 	{-10,-6,-2},
						yticklabels		=	{$10^{-10}$, $10^{-6}$, $10^{-2}$},
% 						ylabel shift 		=	-4pt,
						y tick scale label style=	{xshift=2.0em,yshift=0.5em},
					},
			]
			\addplot graphics
			[
				xmin = 0, 
				xmax = 100, 
				ymin = 0, 
				ymax = 200,
			]
			{../ex1-spinchain/precompiled/scipal_local_krylov_L_100_t_0p3_J_1p0_h_0p1_dt_0p01.rt-1.real.log.pdf};
		\end{groupplot}
	\end{tikzpicture}
	\caption{\label{fig:dsf:compare_rt_log}(DSF) Real-time evolution of the time-dependent correlator $S^{zz}(j, t_{n})$ with parameters $J=0.3$, $J\cdot h=0.1$ and time step $\delta=0.01$. Panels show the evolution of the initial perturbation using \scipal{} 2TDVP (top left), \syten{} TEBD2 (top right), \syten{} global Krylov and \scipal{} local Krylov without damping.}
\end{figure}

%% file: content/examples/dsf_spectral.tex
\begin{figure}[t!]
% 	\begin{center}
% 		\tikzsetnextfilename{dsf_spectral_functions_compare_dt_legend}
% 		\ref{dsf-spectral-functions-compare-dt_legend}
% 	\end{center}
% 	\tikzset{external/export next=false}
	\tikzsetnextfilename{dsf_spectral_functions_compare_dt}
	\begin{tikzpicture}
		\pgfplotsset
		{
%			/pgfplots/colormap={temp}{
%				rgb255=(36,0,217) 		% 0 "#2400d9"
%				rgb255=(25,29,247) 		% 1 "#191df7"
%				rgb255=(41,87,255) 		% 2 "#2957ff"
%				rgb255=(61,135,255) 	% 3 "#3d87ff"
%				rgb255=(87,176,255) 	% 4 "#57b0ff"
%				rgb255=(117,211,255) 	% 5 "#75d3ff"
%				rgb255=(153,235,255) 	% 6 "#99ebff"
%				rgb255=(189,249,255) 	% 7 "#bdf9ff"
%				rgb255=(211,255,255) 	% 8 "#d3ffff"
%				rgb255=(255,255,211) 	% 9 "#ffffd3"
%				rgb255=(255,242,189) 	% 10 "#fff2bd"
%				rgb255=(255,214,153) 	% 11 "#ffd699"
%				rgb255=(255,172,117) 	% 12 "#ffac75"
%				rgb255=(255,120,87) 	% 13 "#ff7857"
%				rgb255=(255,61,61) 		% 14 "#ff3d3d"
%				rgb255=(247,40,54) 		% 15 "#f72836"
%				rgb255=(217,22,48) 		% 16 "#d91630"
%				rgb255=(166,0,33)		% 17 "#a60021"
%			}
			/pgfplots/colormap={temp}{
				rgb255=(255,255,255)	% 0 "#ffffff"
				rgb255=(85,85,255) 		% 1 "#5555ff"
				rgb255=(51,51,255) 		% 2 "#3333ff"
				rgb255=(12,12,255)	 	% 3 "#1111ff"
				rgb255=(0,0,187)	 	% 4 "#0000bb"
				rgb255=(0,0,153)	 	% 5 "#000099"
				rgb255=(0,0,102)	 	% 6 "#000066"
				rgb255=(0,0,51)		 	% 7 "#000033"
				rgb255=(0,0,0)		 	% 8 "#000000"
			}
		}
		\begin{axis}
		[
			width=\textwidth,
			axis lines=left,
			xlabel={Frequency $\omega [\Delta]$},
			font = {\footnotesize},
			ylabel={$S^{zz}(k=\pi,\omega)$},
			xmin=0,
			xmax=2,
			ymin=-1,
			ymax=3,
% 			legend style = {anchor=north west, at = {(0,1)}, draw = none, fill = none, font={\footnotesize}},
% 			legend cell align = left,
% 			legend to name = dsf-spectral-functions-compare-dt_legend,
% 			legend columns=4,
% 			clip mode = individual
		]	
			\addplot[thick,smooth,opacity=0.5,draw opacity=0.5, color=black!40!green, solid, mark=star, mark options=solid, restrict x to domain = 0:2, restrict y to domain = -10:200, mark phase=20, mark repeat=20]
			table[x expr = \coordindex*2*pi/409.6, y expr = \thisrowno{0}*0.01]{../ex1-spinchain/syten/GlobalKrylov/L_100_t_0p3_J_1p0_h_0p1_dt_0p01/results/k_eq_pi_w-1e-1.real};
 			\addplot[thick,smooth,opacity=0.5,draw opacity=0.5, color=blue, solid, mark=diamond, mark options=solid, restrict x to domain = 0:2, restrict y to domain = -10:200, mark phase=55, mark repeat=40]
 			table[x expr = \coordindex*2*pi/409.6, y expr = \thisrowno{0}*0.01]{../ex1-spinchain/scipal/TDVP/L_100_t_0p3_J_1p0_h_0p1_dt_0p01/results/k_eq_pi_w-1e-1.real};
 			\addplot[thick,smooth,opacity=0.5,draw opacity=0.5, color=red, solid, mark=o, mark options=solid, restrict x to domain = 0:2, restrict y to domain = -10:200, mark phase=50, mark repeat=40]
 			table[x expr = \coordindex*2*pi/409.6, y expr = \thisrowno{0}*0.01]{../ex1-spinchain/syten/TEBD/L_100_t_0p3_J_1p0_h_0p1_dt_0p01/results/k_eq_pi_w-1e-1.real};
 			\addplot[thick,smooth,opacity=0.5,draw opacity=0.5, color=orange, solid, mark=triangle, mark options=solid, restrict x to domain = 0:2, restrict y to domain = -10:200, mark phase=40, mark repeat=40]
 			table[x expr = \coordindex*2*pi/409.6, y expr = \thisrowno{0}*0.01]{../ex1-spinchain/scipal/LocalKrylov/L_100_t_0p3_J_1p0_h_0p1_dt_0p01/results/k_eq_pi_w-1e-1.real};
 			\addplot[thick,smooth,opacity=0.5,draw opacity=0.5, color=green!90!black, solid, mark=square, mark options=solid, restrict x to domain = 0:2, restrict y to domain = -10:200, mark phase=45, mark repeat=40]
 			table[x expr = \coordindex*2*pi/409.6, y expr = \thisrowno{0}*0.01]{../ex1-spinchain/syten/MPO/L_100_t_0p3_J_1p0_h_0p1_dt_0p01/results/k_eq_pi_w-1e-1.real};

 			\addplot[thick,smooth, color=blue, dotted, forget plot, mark=diamond, mark options=solid, restrict x to domain = 0:2, restrict y to domain = -10:200, mark phase=15, mark repeat=40]
 			table[x expr = \coordindex*2*pi/409.6, y expr = \thisrowno{0}*0.1]{../ex1-spinchain/scipal/TDVP/L_100_t_0p3_J_1p0_h_0p1_dt_0p1/results/k_eq_pi_w-1e-1.real};
 			\addplot[thick,smooth, color=red, dotted, forget plot, mark=o, mark options=solid, restrict x to domain = 0:2, restrict y to domain = -10:200, mark phase=10, mark repeat=40]
 			table[x expr = \coordindex*2*pi/409.6, y expr = \thisrowno{0}*0.1]{../ex1-spinchain/syten/TEBD/L_100_t_0p3_J_1p0_h_0p1_dt_0p1/results/k_eq_pi_w-1e-1.real};
			\addplot[thick,smooth, color=orange, dotted, forget plot, mark=triangle, mark options=solid, restrict x to domain = 0:2, restrict y to domain = -10:200, mark phase=0, mark repeat=40]
			table[x expr = \coordindex*2*pi/409.6, y expr = \thisrowno{0}*0.1]{../ex1-spinchain/scipal/LocalKrylov/L_100_t_0p3_J_1p0_h_0p1_dt_0p1/results/k_eq_pi_w-1e-1.real};
 			\addplot[thick,smooth, color=green!90!black, dotted, forget plot, mark=square, mark options=solid, restrict x to domain = 0:2, restrict y to domain = -10:200, mark phase=5, mark repeat=40]
 			table[x expr = \coordindex*2*pi/409.6, y expr = \thisrowno{0}*0.1]{../ex1-spinchain/syten/MPO/L_100_t_0p3_J_1p0_h_0p1_dt_0p1/results/k_eq_pi_w-1e-1.real};
		
 			\addlegendentry{Global Krylov};
 			\addlegendentry{2TDVP};
 			\addlegendentry{TEBD2};
 			\addlegendentry{Local Krylov};
 			\addlegendentry{MPO \wii};
 			\coordinate (insetPosition) at (axis cs:1.0,0.5);
		\end{axis}
		\begin{axis}
		[
			width=0.45*\textwidth,
			height=0.25*\textheight,
			at = {(insetPosition)},
			axis on top,
			enlargelimits	= false,
			xlabel		= {wave vector $k$},
			ytick		= {0,0.75,1.5},
			xtick		= {-50,-25,0,25,50},
			xticklabels	= {$0$, $\nicefrac{\pi}{2}$,$\pi$, $\nicefrac{3\pi}{2}$, $2\pi$},
			title		= {$S^{zz}(k,\omega)$ (2TDVP)},
			colorbar right,
			colormap name	= temp,
			point meta min=0,
			point meta max=3.5,
			every colorbar/.append style =
				{
% 						height			=	2*\pgfkeysvalueof{/pgfplots/parent axis height} + 1*\pgfkeysvalueof{/pgfplots/group/vertical sep},
% 						ylabel			=	{$\langle \hat N_j \rangle$},
					width			=	3mm,
					ytick			= 	{0.0,1.75,3.5},
% 						yticklabels		=	{$-0.0025$,$0.0$,$0.0025$},
% 						ylabel shift 		=	-4pt,
					y tick scale label style=	{xshift=2.0em,yshift=0.5em},
				},
		]
		\addplot graphics
		[
			xmin = -50, 
			xmax = 50, 
			ymin = 0, 
			ymax = 1.5,
		]
		{../ex1-spinchain/precompiled/syten_tdvp_L_100_t_0p3_J_1p0_h_0p1_dt_0p1.kw-1e-1.real.pdf};
		\end{axis}
	\end{tikzpicture}
	\caption{\label{fig:dsf:kpi} (DSF) $k = \pi$ cut through the
          band structure at $J=0.3$, $J \cdot h = 0.1$ for different
          methods and step sizes indicated as dotted ($\delta = 0.1$)
          and solid ($\delta = 0.01$) lines. Only 2TDVP (at
          $\delta = 0.1$ and $\delta = 0.01$) and the global Krylov
          method (with $\delta=0.01$) produce no spectral weight
          around $\omega = 0$ (their curves are atop of each other for
          all $\omega$). All other methods produce some spectral
          weight around $\omega = 0$; the errors of the local Krylov
          method also produce consistently negative
          intensities.\\Inset: Full spectrum as produced by 2TDVP for
          these parameters. The main graph x-axis corresponds to a
          vertical line at $k = \pi$ in the inset.}
\end{figure}
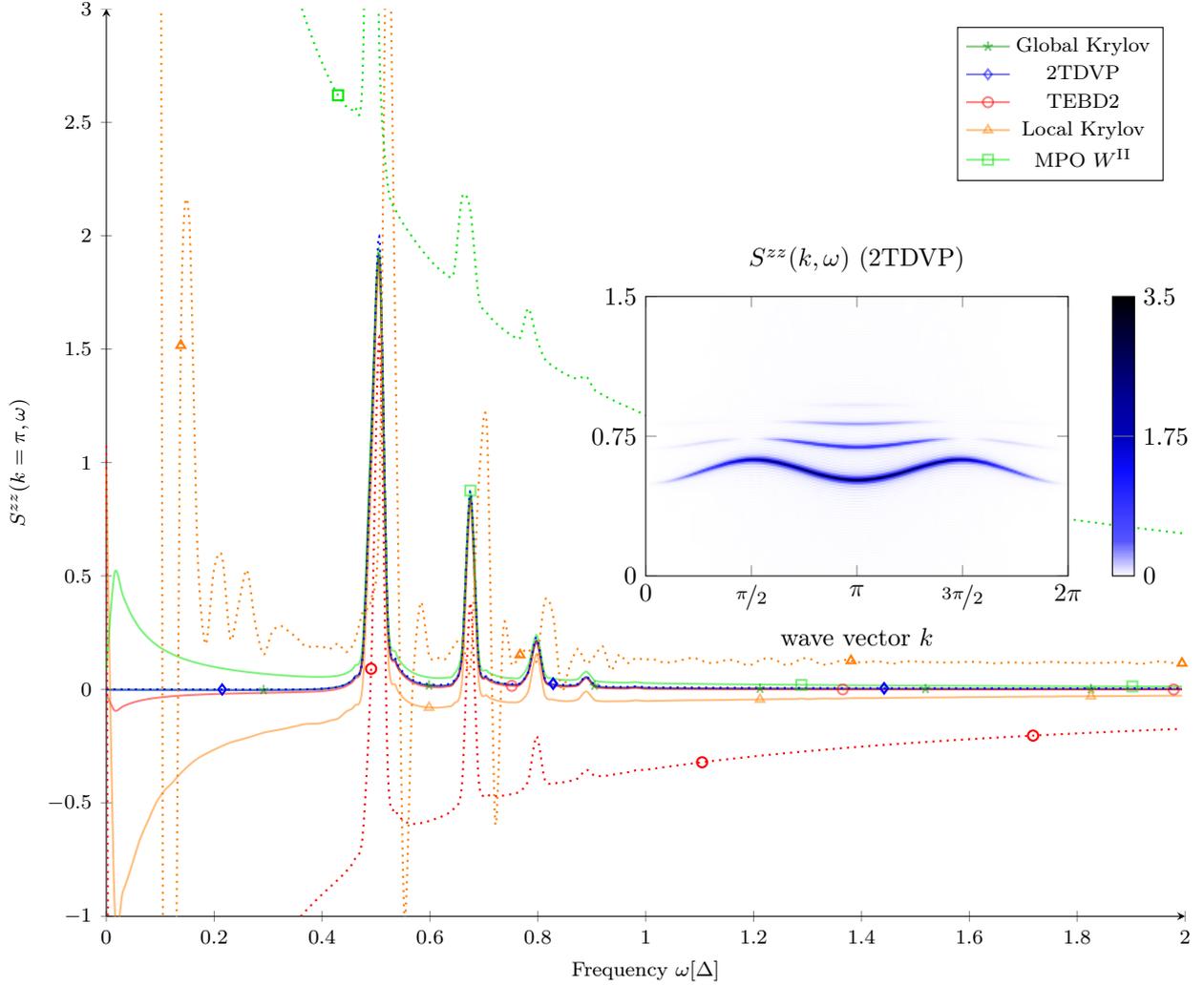

%% file: content/examples/dsf_chimaxruntime.tex
\begin{figure}[t!]
  \begin{center}
    \tikzsetnextfilename{dsf_legend}
    \ref{dsf_legend}
  \end{center}
	\begin{minipage}[t]{0.49\textwidth}
		\tikzsetnextfilename{dsf_benchmarks_chimax}
		\begin{tikzpicture}
			\begin{axis}
			[
				width=\textwidth,
				axis lines=left,
				xlabel={Time $t$},
				font = {\footnotesize},
				ylabel={Max. bond dimension $m$},
				xmin=0,
				xmax=210,
				ymax=210,
				legend style = {anchor=north east, at = {(1,0.8)}, draw = none, fill = none, font={\footnotesize}},
				legend to name = dsf_legend,
				legend columns=5,
				legend cell align = left,
			]
				\addplot[thick, smooth, color=black!40!green, solid, mark=star, mark options=solid, mark phase=4, mark repeat=40]
				table[x expr = \coordindex, y expr = \thisrowno{0}]{../ex1-spinchain/syten/GlobalKrylov/L_100_t_0p3_J_1p0_h_0p1_dt_0p01/chi_max.output};
				\addplot[thick, smooth, color=blue, solid, mark=diamond, mark options=solid, mark phase=3, mark repeat=40]
				table[x expr = \coordindex, y expr = \thisrowno{0}]{../ex1-spinchain/syten/TDVP/L_100_t_0p3_J_1p0_h_0p1_dt_0p01/chi_max.output};
				\addplot[thick, smooth, color=red, solid, mark=o, mark options=solid, mark phase=3, mark repeat=40]
				table[x expr = \coordindex, y expr = \thisrowno{0}]{../ex1-spinchain/syten/TEBD/L_100_t_0p3_J_1p0_h_0p1_dt_0p01/chi_max.output};
				\addplot[thick, smooth, color=orange, solid, mark=triangle, mark options=solid, mark phase=6, mark repeat=40]
				table[x expr = \coordindex, y expr = \thisrowno{0}]{../ex1-spinchain/scipal/LocalKrylov/L_100_t_0p3_J_1p0_h_0p1_dt_0p01/chi_max.output};
				\addplot[thick, smooth, color=green!90!black, solid, mark=square, mark options=solid, mark phase=6, mark repeat=40]
				table[x expr = \coordindex, y expr = \thisrowno{0}]{../ex1-spinchain/syten/MPO/L_100_t_0p3_J_1p0_h_0p1_dt_0p01/chi_max.output};

				\addplot[forget plot, thick, smooth, color=black!40!green, dotted, mark=star, mark options=solid, mark phase=6, mark repeat=40]
				table[x expr = \coordindex, y expr = \thisrowno{0}]{../ex1-spinchain/syten/GlobalKrylov/L_100_t_0p3_J_1p0_h_0p1_dt_0p1/chi_max.output};
				
				\addplot[forget plot, thick, smooth, color=blue, dotted, mark=diamond, mark options=solid, mark phase=2, mark repeat=40]
				table[x expr = \coordindex, y expr = \thisrowno{0}]{../ex1-spinchain/syten/TDVP/L_100_t_0p3_J_1p0_h_0p1_dt_0p1/chi_max.output};
				
                                \addplot[forget plot, thick, smooth, color=red, dotted, mark=o, mark options=solid, mark phase=2, mark repeat=40]
				table[x expr = \coordindex, y expr = \thisrowno{0}]{../ex1-spinchain/syten/TEBD/L_100_t_0p3_J_1p0_h_0p1_dt_0p1/chi_max.output};

				\addplot[forget plot, thick, smooth, color=orange, dotted, mark=triangle, mark options=solid, mark phase=5, mark repeat=40]
				table[x expr = \coordindex, y expr = \thisrowno{0}]{../ex1-spinchain/scipal/LocalKrylov/L_100_t_0p3_J_1p0_h_0p1_dt_0p1/chi_max.output};
				
				\addplot[forget plot, thick, smooth, color=green!90!black, dotted, mark=square, mark options=solid, mark phase=5, mark repeat=40]
				table[x expr = \coordindex, y expr = \thisrowno{0}]{../ex1-spinchain/syten/MPO/L_100_t_0p3_J_1p0_h_0p1_dt_0p1/chi_max.output};
				
				\addlegendentry{Global Krylov}
				\addlegendentry{2TDVP}
				\addlegendentry{TEBD2}
				\addlegendentry{Local Krylov}
				\addlegendentry{MPO \wii}
			\end{axis}
		\end{tikzpicture}
		\caption{\label{fig:dsf:benchmarks-chimax}(DSF) Maximal bond dimensions $m$ during time evolutions of systems with $L=100$ sites and different time steps $\delta=0.1$ (dotted) and $\delta=0.01$ (solid). All methods result in a smaller maximal bond dimension if the calculation is more precise and the step size is smaller.}
	\end{minipage}
	\hfill
	\begin{minipage}[t]{0.49\textwidth}
          \tikzsetnextfilename{dsf_benchmarks_cputime}
		\begin{tikzpicture}
			\begin{semilogyaxis}
			[
				width=\textwidth,
				axis lines=left,
				xlabel={Time $t$},
				font = {\footnotesize},
				ylabel={Avg. cpu time $t$ per unit time $\nicefrac{1}{\Delta}$},
				xmin=0,
				xmax=210,
				legend style={at={(1,0.05)}, anchor=south east},
% 				legend entries = {$t=2$,$t=14$,$t=26$,$t=38$, $t=50$, $t=62$, 2TDVP, TEBD2, Local Krylov, MPO\wii},
% 				ymin=1e-4,
% 				ymax=1050,
			]
				\addplot[thick, smooth, color=black!40!green, solid, mark=star, mark options=solid, mark phase=5, mark repeat=40]
				table[x expr = \coordindex, y expr = \thisrowno{0}]{../ex1-spinchain/syten/GlobalKrylov/L_100_t_0p3_J_1p0_h_0p1_dt_0p01/teo.output};
                                \addplot[thick, smooth, color=green!90!black, solid, mark=square, mark options=solid, mark phase=6, mark repeat=40]
				table[x expr = \coordindex, y expr = \thisrowno{0}]{../ex1-spinchain/syten/MPO/L_100_t_0p3_J_1p0_h_0p1_dt_0p01/teo.output};
				\addplot[thick, smooth, color=blue, solid, mark=diamond, mark options=solid, mark phase=3, mark repeat=40]
				table[x expr = \coordindex, y expr = \thisrowno{0}]{../ex1-spinchain/syten/TDVP/L_100_t_0p3_J_1p0_h_0p1_dt_0p01/teo.output};
				\addplot[thick, smooth, color=red, solid, mark=o, mark options=solid, mark phase=3, mark repeat=40]
				table[x expr = \coordindex, y expr = \thisrowno{0}]{../ex1-spinchain/syten/TEBD/L_100_t_0p3_J_1p0_h_0p1_dt_0p01/teo.output};
				\addplot[thick, smooth, color=orange, solid, mark=triangle, mark options=solid, mark phase=6, mark repeat=40]
				table[x expr = \coordindex, y expr = \thisrowno{0}*100]{../ex1-spinchain/scipal/LocalKrylov/L_100_t_0p3_J_1p0_h_0p1_dt_0p01/teo.output};

				\addplot[thick, smooth, color=black!40!green, dotted, mark=star, mark options=solid, mark phase=6, mark repeat=40]
				table[x expr = \coordindex, y expr = \thisrowno{0}]{../ex1-spinchain/syten/GlobalKrylov/L_100_t_0p3_J_1p0_h_0p1_dt_0p1/teo.output};
				\addplot[forget plot, thick, smooth, color=green!90!black, dotted, mark=square, mark options=solid, mark phase=5, mark repeat=40]
				table[x expr = \coordindex, y expr = \thisrowno{0}]{../ex1-spinchain/syten/MPO/L_100_t_0p3_J_1p0_h_0p1_dt_0p1/teo.output};
				\addplot[forget plot, thick, smooth, color=blue, dotted, mark=diamond, mark options=solid, mark phase=2, mark repeat=40]
				table[x expr = \coordindex, y expr = \thisrowno{0}]{../ex1-spinchain/syten/TDVP/L_100_t_0p3_J_1p0_h_0p1_dt_0p1/teo.output};
				\addplot[forget plot, thick, smooth, color=red, dotted, mark=o, mark options=solid, mark phase=2, mark repeat=40]
				table[x expr = \coordindex, y expr = \thisrowno{0}]{../ex1-spinchain/syten/TEBD/L_100_t_0p3_J_1p0_h_0p1_dt_0p1/teo.output};
				\addplot[forget plot, thick, smooth, color=orange, dotted, mark=triangle, mark options=solid, mark phase=5, mark repeat=40]
				table[x expr = \coordindex, y expr = \thisrowno{0}*10]{../ex1-spinchain/scipal/LocalKrylov/L_100_t_0p3_J_1p0_h_0p1_dt_0p1/teo.output};
			\end{semilogyaxis}
		\end{tikzpicture}
		\caption{\label{fig:dsf:benchmarks-cputime}(DSF) Average CPU time per unit time $\frac{1}{\Delta}$ evolving systems with $L=100$ sites and different time step sizes $\delta=0.1$ (dotted) and $\delta=0.01$ (solid). Note that physically reasonable results at $\delta=0.1$ are only obtained for 2TDVP.}
% Remark: The scaling of the local Krylov data was checked to be correct, the content of the file is simply different.
	\end{minipage}
\end{figure}

%% file: content/examples/hubbard.tex
\subsection{\label{sec:examples:hubbard}Cooling of a doped Hubbard chain}

Let us now consider the cooling of a Hubbard chain from infinite
temperature ($\beta = 0$) to a finite temperature $\beta \ge 0$. We
will work with a purified state in a canonical
ensemble\cite{barthel16:_matrix} conserving both particle number and
spin $S^z$ projection,
i.e. we implement the associated $\mathrm{U}(1)_N \times \mathrm{U}(1)_{S^z}$ symmetries. The local physical
dimension is four with each quantum number sector denoted as $(N,S^z)$  only
containing one single state (i.e., $(0,0)$,$(1,-\nicefrac{1}{2})$, $(1,+\nicefrac{1}{2})$ and $(2,0)$). 
The physical chain has a length of $L = 24$
sites resulting in an MPS with $L^\prime = 48$ tensors. Physical sites
will be labelled by indices $1, \ldots, 24$, the auxiliary sites as
$a(1), \ldots, a(24)$. The initial state is constructed
(cf.~\cref{sec:mps:finitetemp}) by applying the operator
\begin{equation}
  \hat C^\dagger_{\mathrm{tot}} = \sum_{j=1}^{L} \hat c^\dagger_{j,\uparrow} \hat c^\dagger_{a(j),\downarrow} + \hat c^\dagger_{j,\downarrow} \hat c^\dagger_{a(j),\uparrow}
\end{equation}
$20$ times to a vacuum state, resulting in $\nicefrac{5}{6}$
filling. The generated state has a maximal bond dimension $m = 80$ and
is used as the initial state for imaginary time evolution under the
Hamiltonian
\begin{equation}
  \hat H = -t \sum_{j=1}^{L-1} \left( \hat c^\dagger_{j,\uparrow} \hat c^{\mathstrut{}}_{j+1,\uparrow} + \hat c^\dagger_{j,\downarrow} \hat c^{\mathstrut{}}_{j+1,\downarrow} + \hat c^\dagger_{j+1,\uparrow} \hat c^{\mathstrut{}}_{j,\uparrow} + \hat c^\dagger_{j+1,\downarrow} \hat c^{\mathstrut{}}_{j,\downarrow} \right) + U \sum_{j=1}^{L} \hat n_{j,\uparrow} \hat n_{j,\downarrow}
\end{equation}
with $t=1$ and $U=4$. To simplify notation in this example, we will
time evolve using the evolution operator $e^{-\delta \hat H}$ and
refer to ``times'' via the real-valued $\beta$ variable. That is, we
drop the prefactor $-\I$ in $\delta$ (which would be necessary to
perform an imaginary time evolution using $e^{-\I\delta\hat H}$) and
we also drop the factor of $\nicefrac{1}{2}$ (which relates the target
``temperature'' $\beta$ as the result of the evolution to the actual
inverse temperature $\beta$ in the physical system,
cf.~\cref{sec:tricks:pip}). In
practice, all calculations are done using complex-valued arithmetic
and negative imaginary time steps.

In the following, we primarily consider the long-range spin-spin
correlator
\begin{align}
  Z_1 & = \sum_{j=2}^L \langle \hat s^z_1 \hat s^z_j \rangle \;.
\end{align}
The on-site term $\langle \hat s^z_1 \hat s^z_1 \rangle$ is about the
same order of magnitude as all the long-range terms which is why it is not included here. 
During the cooling simulation, the value of
this correlator evolves monotonously from 0 until it reaches a plateau,
ideally at the ground-state value (however, in practice, it may over- or undershoot, see below). 
In an error-free calculation, the simulated state would then be the
ground state of the Hamiltonian and only acquires a global prefactor upon
further imaginary time evolution. In practice, due to accumulated
errors, particles are able to leave the physical system and move into the
auxiliary system of equal size and the state can ``tunnel'' into the
global ground state in the combined physical and auxiliary degrees of freedom, which is not what is looked for. 
Measuring the particle number in the physical system is, therefore, a measure for the stability of the procedure. 
This process is monitored by evaluating $\hat N_p = \sum_{j=1}^L \hat n_j$ and comparing it to its true value, which is $\langle \hat N_p \rangle = 20$ here.

We run calculations at bond dimensions $m = 100, 200, 300, 400$ and
$500$ and time step sizes $\delta = 0.01, 0.05$ and $0.1$ up to
$\beta = 20$. In this problem, the global Krylov method has to be
directly disqualified as the highly entangled Krylov vectors generated
by it exceed the time and memory resources of the other methods by
more than a factor of 20 (RAM) and 100 (CPU time), respectively. The
1TDVP method is also unsuitable here
(cf.~\cref{sec:examples:hubbard:projection} below). We hence focus on
the 2TDVP and MPO \wii method (both with \syten{} and \scipal) as well
as the second-order TEBD and local Krylov methods (with \syten{}
only). \syten{} was configured to apply MPOs (in the MPO \wii method
and TEBD2) using the zip-up method, whereas the MPO \wii method in
\scipal{} is configured to use the zip-up method followed by a few
variational optimization sweeps. As always, for the plots we have
selected the more suitable of the two implementations.

\subsubsection{Loss of particles}

\begin{figure}
  \begin{center}
    \tikzsetnextfilename{hubbard_lost_particles_legend}
    \ref{hubbard-lost-particles_legend}
  \end{center}
  \begin{minipage}[t]{0.49\textwidth}
    \tikzsetnextfilename{hubbard_particles_lost_delta}
    \pgfplotsset{width=\textwidth}
    \pgfplotsset{every axis plot/.append style={
        line width=1pt,
        tick style={line width=0.6pt}}}
    \begin{tikzpicture}
      \begin{axis}[
        width=\textwidth,
        axis lines=left,
        xlabel={Time $\beta$},
        font = {\footnotesize},
        ylabel={$20 - \langle \hat N_p \rangle$},
        xmin=10,
        xmax=20,
        ymin=-0.1,
        ymax=3,
        legend style = {anchor=north west, at = {(0,1)}, draw = none, fill = none, font={\footnotesize}},
        legend cell align = left,
        legend to name = hubbard-lost-particles_legend,
        legend columns=4,
        ]

        \addplot[forget plot,color=green!90!black,dotted,mark=square,mark options=solid,mark phase=400,mark repeat=2000]table[y expr = 20 - \thisrowno{1}]{../ex2-hubbard/scipal-comp/N/mpo_d_0.01_m_200.n};
        \addplot[forget plot,color=green!90!black,dashed,mark=square,mark options=solid,mark phase=170,mark repeat=2000]table[y expr = 20 - \thisrowno{1}]{../ex2-hubbard/scipal-comp/N/mpo_d_0.05_m_200.n};
        \addplot[color=green!90!black,solid,mark=square,mark options=solid,mark phase=90,mark repeat=2000]table[y expr = 20 - \thisrowno{1}]{../ex2-hubbard/scipal-comp/N/mpo_d_0.1_m_200.n};

        \addplot[color=red,dotted,forget plot,mark=o,mark options=solid,mark phase=600,mark repeat=2000]table[y expr = 20 - \thisrowno{1}]{../ex2-hubbard/comparison/20180914-comparison_tebd2_d_0.01_m_200_data.N};
        \addplot[color=red,dashed,forget plot,mark=o,mark options=solid,mark phase=150,mark repeat=2000]table[y expr = 20 - \thisrowno{1}]{../ex2-hubbard/comparison/20180914-comparison_tebd2_d_0.05_m_200_data.N};
        \addplot[color=red,solid,mark=o,mark options=solid,mark repeat=2000,mark phase=95]table [y expr = 20 - \thisrowno{1}]{../ex2-hubbard/comparison/20180914-comparison_tebd2_d_0.1_m_200_data.N};

        \addplot[color=blue,dotted,forget plot,mark=diamond,mark options=solid,mark phase=750,mark repeat=2000]table[y expr = 20 - \thisrowno{1}]{../ex2-hubbard/comparison/20180914-comparison_2tdvp_d_0.01_m_200_data.N};
        \addplot[color=blue,dashed,forget plot,mark=diamond,mark options=solid,mark phase=195,mark repeat=2000]table[y expr = 20 - \thisrowno{1}]{../ex2-hubbard/comparison/20180914-comparison_2tdvp_d_0.05_m_200_data.N};
        \addplot[color=blue,solid,mark=diamond,mark options=solid,mark phase=98,mark repeat=20000]table [y expr = 20 - \thisrowno{1}]{../ex2-hubbard/comparison/20180914-comparison_2tdvp_d_0.1_m_200_data.N};

        \addplot[color=orange,dotted,forget plot,mark=triangle,mark options=solid,mark phase=500,mark repeat=2000]table[y expr = 20 - \thisrowno{1}]{../ex2-hubbard/local-krylov/d_0.01_m_200_data.N};
        \addplot[color=orange,dashed,forget plot,mark=triangle,mark options=solid,mark phase=190,mark repeat=2000]table[y expr = 20 - \thisrowno{1}]{../ex2-hubbard/local-krylov/d_0.05_m_200_data.N};
        \addplot[color=orange,solid,mark=triangle,mark options=solid,mark phase=95,mark repeat=2000]table[y expr = 20 - \thisrowno{1}]{../ex2-hubbard/local-krylov/d_0.1_m_200_data.N};

        \addlegendentry{MPO \wii};
        \addlegendentry{TEBD2};
        \addlegendentry{2TDVP};
        \addlegendentry{Local Krylov};
      \end{axis}
    \end{tikzpicture}
    \caption{\label{fig:ex:hubbard:particles-lost-delta}(Cooling) Loss of
      particles at $m = 200$ for step sizes $\delta = 0.1$ (solid),
      $\delta = 0.05$ (dashed) and $\delta = 0.01$ (dotted). At smaller
      $\delta$, truncation occurs more often and particle loss tends
      to set on earlier.}
  \end{minipage}
  \hfill
  \begin{minipage}[t]{0.49\textwidth}
    \tikzsetnextfilename{hubbard_particles_lost_methods}
    \pgfplotsset{width=\textwidth}
    \pgfplotsset{every axis plot/.append style={
        line width=1pt,
        tick style={line width=0.6pt}}}
    \begin{tikzpicture}
      \begin{axis}[
        width=\textwidth,
        axis lines=left,
        xlabel={Time $\beta$},
        font = {\footnotesize},
        ylabel={$20 - \langle \hat N_p \rangle$},
        xmin=15,
        xmax=20,
        ymin=-0.1,
        ymax=3,
        ]
        \addplot[forget plot,color=green!90!black,dotted,mark=square,mark options=solid,mark phase=90,mark repeat=2000]table[y expr = 20 - \thisrowno{1}]{../ex2-hubbard/scipal-comp/N/mpo_d_0.05_m_100.n};
        \addplot[forget plot,color=green!90!black,dashed,mark=square,mark options=solid,mark phase=95,mark repeat=2000]table[y expr = 20 - \thisrowno{1}]{../ex2-hubbard/scipal-comp/N/mpo_d_0.05_m_300.n};
        \addplot[color=green!90!black,solid,mark=square,mark options=solid,mark phase=95,mark repeat=2000]table[y expr = 20 - \thisrowno{1}]{../ex2-hubbard/scipal-comp/N/mpo_d_0.05_m_500.n};

        \addplot[color=red,dotted,forget plot,mark=o,mark options=solid,mark phase=80,mark repeat=100]table[y expr = 20 - \thisrowno{1}]{../ex2-hubbard/comparison/20180914-comparison_tebd2_d_0.05_m_100_data.N};
        \addplot[color=red,dashed,forget plot,mark=o,mark options=solid,mark phase=40,mark repeat=100]table[y expr = 20 - \thisrowno{1}]{../ex2-hubbard/comparison/20180914-comparison_tebd2_d_0.05_m_300_data.N};
        \addplot[color=red,solid,mark=o,mark options=solid,mark phase=65,mark repeat=100]table [y expr = 20 - \thisrowno{1}]{../ex2-hubbard/comparison/20180914-comparison_tebd2_d_0.05_m_500_data.N};

        \addplot[color=blue,dotted,forget plot,mark=diamond,mark options=solid,mark phase=99,mark repeat=100]table[y expr = 20 - \thisrowno{1}]{../ex2-hubbard/comparison/20180914-comparison_2tdvp_d_0.05_m_100_data.N};
        \addplot[color=blue,dashed,forget plot,mark=diamond,mark options=solid,mark phase=80,mark repeat=100]table[y expr = 20 - \thisrowno{1}]{../ex2-hubbard/comparison/20180914-comparison_2tdvp_d_0.05_m_300_data.N};
        \addplot[color=blue,solid,mark=diamond,mark options=solid,mark phase=80,mark repeat=100]table [y expr = 20 - \thisrowno{1}]{../ex2-hubbard/comparison/20180914-comparison_2tdvp_d_0.05_m_500_data.N};

        \addplot[color=orange,dotted,forget plot,mark=triangle,mark options=solid,mark phase=97,mark repeat=2000]table[y expr = 20 - \thisrowno{1}]{../ex2-hubbard/local-krylov/d_0.05_m_100_data.N};
        \addplot[color=orange,dashed,forget plot,mark=triangle,mark options=solid,mark phase=85,mark repeat=2000]table[y expr = 20 - \thisrowno{1}]{../ex2-hubbard/local-krylov/d_0.05_m_300_data.N};
        \addplot[color=orange,solid,mark=triangle,mark options=solid,mark phase=85,mark repeat=2000]table[y expr = 20 - \thisrowno{1}]{../ex2-hubbard/local-krylov/d_0.05_m_500_data.N};
      \end{axis}
    \end{tikzpicture}
    \caption{\label{fig:ex:hubbard:particles-lost-methods}(Cooling) Loss of
      particles at $\delta = 0.05$ for bond dimensions $m = 100$
      (dotted), $300$ (dashed) and $500$ (solid). No correlation
      between bond dimension and onset of particle loss can be
      observed.}
  \end{minipage}
\end{figure}

As we would like to simulate a canonical ensemble, it is quite
relevant that loss of particles to the auxiliary system does not occur
too early. Studying $\langle \hat N_p \rangle$, we find that, first,
increasing the step size usually reduces the loss of
particles (cf.~\cref{fig:ex:hubbard:particles-lost-delta}). Second,
there is no clear relation between the selected bond dimension and the
onset of loss of particles: E.g. 2TDVP at $\delta = 0.1$ starts losing
particles at $m=400$ around $\beta = 15$ but is stable until
$\beta = 17$ for $m = 100, 200, 300$ and $500$ (not shown) and while
the MPO \wii method is stable at $\delta =0.05$ and $m = 100$ and
$m=500$, it loses particles at $m=300$
(cf.~\cref{fig:ex:hubbard:particles-lost-methods}). As expected, the
behaviour of the local Krylov method and 2TDVP on this relatively
large-scale observable is very comparable.

The configuration difference in the MPO \wii routines to apply the
operator to the state between \syten{} and \scipal{} leads to
different results here: \syten{} was configured to only use the zip-up
method whereas \scipal{} also did a few variational sweeps afterwards
to optimize the state. This variational optimization leads to
considerably better stability and mostly no loss of particles at all,
but is computationally more expensive.

\subsubsection{\label{sec:examples:hubbard:spinspin}Spin-spin correlator}

\begin{figure}
  \begin{center}
    \tikzsetnextfilename{hubbard_z0_conv_m_legend}
    % note: You cannot compile this from within a standard pdflatex run because reasons.
    % compile with: pdflatex -shell-escape -halt-on-error -interaction=batchmode -jobname "figures/autogen/hubbard_z0_conv_m_legend" "\def\tikzexternalrealjob{time_evolution_review}\input{time_evolution_review}"
    \ref{hubbard-convergence-z1-legend}
  \end{center}
  \begin{minipage}[t]{0.49\textwidth}
    \tikzsetnextfilename{hubbard_z0_conv_m}
    \pgfplotsset{width=\textwidth}
    \pgfplotsset{every axis plot/.append style={
        line width=1pt,
        tick style={line width=0.6pt}}}
    \begin{tikzpicture}
      \begin{axis}[
        width=\textwidth-3.032pt,
        axis lines=left,
        xlabel={Bond dimension $m$},
        font = {\footnotesize},
        ylabel={$Z_1$},
        xmin=50,
        xmax=550,
        ymin=-0.0583,
        ymax=-0.0575,
        legend style = {anchor=north east, at = {(1,0.8)}, draw = none, fill = none, font={\footnotesize}},
        legend to name = hubbard-convergence-z1-legend,
        legend columns=4,
        legend cell align = left,
        ]
        \addplot[color=green!90!black,dotted,forget plot,mark=square,mark options=solid]table[y expr = \thisrowno{3}]{../ex2-hubbard/comparison/scipal-mpo-conv-at-beta-1};
        \addplot[color=green!90!black,dashed,forget plot,mark=square,mark options=solid]table[y expr = \thisrowno{2}]{../ex2-hubbard/comparison/scipal-mpo-conv-at-beta-1};
        \addplot[color=green!90!black,solid,mark=square,mark options=solid]table[y expr = \thisrowno{1}]{../ex2-hubbard/comparison/scipal-mpo-conv-at-beta-1};

        \addplot[color=red,dotted,forget plot,mark=o,mark options=solid]table[y expr = \thisrowno{6}]{../ex2-hubbard/comparison/convergence-at-beta-1};
        \addplot[color=red,dashed,forget plot,mark=o,mark options=solid]table[y expr = \thisrowno{5}]{../ex2-hubbard/comparison/convergence-at-beta-1};
        \addplot[color=red,solid,mark=o,mark options=solid]table[y expr = \thisrowno{4}]{../ex2-hubbard/comparison/convergence-at-beta-1};

        \addplot[color=blue,dotted,forget plot,mark=diamond,mark options=solid]table[y expr = \thisrowno{9}]{../ex2-hubbard/comparison/convergence-at-beta-1};
        \addplot[color=blue,dashed,forget plot,mark=diamond,mark options=solid]table[y expr = \thisrowno{8}]{../ex2-hubbard/comparison/convergence-at-beta-1};
        \addplot[color=blue,solid,mark=diamond,mark options=solid]table[y expr = \thisrowno{7}]{../ex2-hubbard/comparison/convergence-at-beta-1};

        \addplot[color=orange,dotted,forget plot,mark=triangle,mark options=solid]table[y expr = \thisrowno{2}]{../ex2-hubbard/local-krylov/Z0_0.01.m};
        \addplot[color=orange,dashed,forget plot,mark=triangle,mark options=solid]table[y expr = \thisrowno{2}]{../ex2-hubbard/local-krylov/Z0_0.05.m};
        \addplot[color=orange,solid,mark=triangle,mark options=solid]table[y expr = \thisrowno{2}]{../ex2-hubbard/local-krylov/Z0_0.1.m};

        \addlegendentry{MPO \wii};
        \addlegendentry{TEBD2};
        \addlegendentry{2TDVP};
        \addlegendentry{Local Krylov};
      \end{axis}
    \end{tikzpicture}
    \caption{\label{fig:ex:hubbard:z0-conv-m}(Cooling) Convergence of $Z_1$ with
      bond dimension at $\beta = 1$ for $\delta = 0.1$ (solid),
      $\delta = 0.05$ (dashed) and $\delta = 0.01$ (dotted). For
      $m \geq 300$ and small time steps, three methods converge to the
      same result, but the time-step error of the \wii method and the
      TEBD2 is very large at $\delta = 0.1$. Even at $\delta = 0.01$,
      the local Krylov method does not converge to the correct
      result.}
  \end{minipage}
  \hfill
  \begin{minipage}[t]{0.49\textwidth}
    \tikzsetnextfilename{hubbard_runtime}
    \pgfplotsset{width=\textwidth}
    \pgfplotsset{every axis plot/.append style={
        line width=1pt,
        tick style={line width=0.6pt}}}
    \begin{tikzpicture}
      \begin{axis}[
        width=\textwidth,
        axis lines=left,
        xlabel={Bond dimension $m$},
        font = {\footnotesize},
        ylabel={CPU time [s]},
        xmin=50,
        xmax=550,
        ymin=0,
        ymax=100,
        restrict y to domain=0:100,
        ]
        \addplot[color=green!90!black,dotted,forget plot,mark=square,mark options=solid]table[y expr = \thisrowno{3}]{../ex2-hubbard/runtime-scipal-mpo};
        \addplot[color=green!90!black,dashed,forget plot,mark=square,mark options=solid]table[y expr = \thisrowno{2}]{../ex2-hubbard/runtime-scipal-mpo};
        \addplot[color=green!90!black,solid,mark=square,mark options=solid]table[y expr = \thisrowno{1}]{../ex2-hubbard/runtime-scipal-mpo};

        \addplot[color=red,dotted,forget plot,mark=o,mark options=solid]table[y expr = \thisrowno{3}]{../ex2-hubbard/runtime-syten-tebd2};
        \addplot[color=red,dashed,forget plot,mark=o,mark options=solid]table[y expr = \thisrowno{2}]{../ex2-hubbard/runtime-syten-tebd2};
        \addplot[color=red,solid,mark=o,mark options=solid]table[y expr = \thisrowno{1}]{../ex2-hubbard/runtime-syten-tebd2};

        \addplot[color=blue,dotted,forget plot,mark=diamond,mark options=solid]table[y expr = \thisrowno{3}]{../ex2-hubbard/runtime-syten-2tdvp};
        \addplot[color=blue,dashed,forget plot,mark=diamond,mark options=solid]table[y expr = \thisrowno{2}]{../ex2-hubbard/runtime-syten-2tdvp};
        \addplot[color=blue,solid,mark=diamond,mark options=solid]table[y expr = \thisrowno{1}]{../ex2-hubbard/runtime-syten-2tdvp};

        \addplot[color=orange,dotted,forget plot,mark=triangle,mark options=solid]table[y expr = \thisrowno{3}]{../ex2-hubbard/local-krylov/runtimes};
        \addplot[color=orange,dashed,forget plot,mark=triangle,mark options=solid]table[y expr = \thisrowno{2}]{../ex2-hubbard/local-krylov/runtimes};
        \addplot[color=orange,solid,mark=triangle,mark options=solid]table[y expr = \thisrowno{1}]{../ex2-hubbard/local-krylov/runtimes};
      \end{axis}
    \end{tikzpicture}
    \caption{\label{fig:ex:hubbard:runtime}(Cooling) CPU time required by each
      method for the single step from $\beta = 1-\delta$ to
      $\beta = 1$. Solid lines are for $\delta = 0.1$, dashed for
      $\delta = 0.05$, dotted for $\delta = 0.01$. Only 2TDVP and the
      local Krylov method depend on the step size, as larger $\delta$
      requires more local Krylov vectors.}
  \end{minipage}
\end{figure}

Our proxy for a long-range observable, the sum of correlators $Z_1$,
obtains values comparable to the ground-state value around
$\beta = 15$ to $\beta = 18$. This is already deep in the regime where
particles tend to be lost to the auxiliary system. Additionally, it is
often unnecessary to consider this point when using a
finite-temperature method, as we could conceivably use DMRG directly to
obtain ground-state properties.

Hence, we will concentrate on a higher temperature at $\beta = 1$ and
consider the value of the correlator as a function of $m$ and
$\delta$, cf.~\cref{fig:ex:hubbard:z0-conv-m}. The first relevant
observation is that 2TDVP, TEBD2 and the MPO \wii method converge to
the same result once $m \geq 300$ and (for TEBD2 and MPO $W^\mathrm{II}$)
$\delta = 0.01$. Curiously, the local Krylov method appears to suffer
a relatively large time-step error here; only its results with
$\delta = 0.01$ are comparable to the other methods.  In contrast,
2TDVP provides very good results starting at $m = 200$ which are also
nearly independent of the time-step size. Curiously, the remaining
dependence of the 2TDVP error on the step size is such that the $\delta = 0.01$ calculation has the highest deviation from the other data points at $m \geq 300$
. That is, while the MPO \wii and TEBD2
methods suffer from a simple finite time-step error which becomes
consistently smaller with smaller time steps, the behavior of 2TDVP is
more complicated. It can be understood as the projection error being
2TDVP's primary error source. The projection error acts on each
individual time step, hence increasing the number of time steps also
increases the projection error. As the initial state is exact at
$m = 80$, we cannot increase the MPS manifold to make up for the
increased projection error.

Here again we observe that the variational optimization following the
zip-up considerably reduces the error in the MPO \wii method as used
by \scipal{} compared to the \syten{} implementation (not shown). In
contrast, 2TDVP provides exactly the same result in both
implementations.

\subsubsection{\label{sec:examples:hubbard:projection}Projection error
  in 1TDVP and 2TDVP}

We have excluded 1TDVP from the set of viable time-evolution
methods here as the nature of the problem -- global cooling and
redistribution of particles -- does not yield itself to the very
restricted manifold available to 1TDVP. Indeed, imaginary time evolution with
1TDVP results in a fixed state around $\beta = 3$ with energy zero. In
comparison, a 2TDVP calculation at $\beta = 3$ results in an energy of
$\approx -17.2$. Additionally, calculating the one-site variance and
the two-site variance\cite{hubig18:_error} (initially, they sum to
$\approx 45.9$) in both cases shows that 1TDVP continously minimizes
the one-site variance but leaves the two-site variance
invariant. 2TDVP minimizes the two-site variance \emph{at the first
  time step} and then slowly minimizes the one-site variance while the
two-site variance grows again (from $O(10^-5)$ to $O(10^{-2})$ at
$\beta = 20$). The large value of one- and two-site variance initially
suggests that also the three-site variance is large and 2TDVPs initial
abrupt minimization of the two-site variance also suggests that its
projection induces some relatively large error which cannot be reduced
further by using smaller step sizes.

\subsubsection{\label{sec:examples:hubbard:runtime}Runtime
  comparison}

In this experiment, each calculation was run on a single core of a
Xeon E5-2630 v4 with 64 GB of RAM and no hard-disk caching.
Comparing run times, the major advantage of the 2TDVP due to its very
limited time-step error is apparent. Judging from
\cref{fig:ex:hubbard:z0-conv-m}, the MPO \wii method and TEBD2 both
require a step size of $\delta = 0.01$ to provide data comparable to
2TDVP at $\delta = 0.1$. Supposedly, the local Krylov method would
need a yet smaller time step size to provide perfect results. Per
individual time step, we compare runtimes in
\cref{fig:ex:hubbard:runtime}. This runtime per step is lowest for the
TEBD2 method and highest for the MPO \wii method. The 2TDVP and local
Krylov method perform steps similarly quickly. The relatively long
runtime of the MPO \wii method is largely due to the variational
optimization necessary to obtain good results, the MPO \wii method in
\syten{} (using only the zip-up) is approximately a factor of two
faster than TEBD2 but also incurs large errors and loss of particles.

Taking into account the larger time step available with 2TDVP compared
to TEBD2 and the MPO \wii method (as seen in
\cref{fig:ex:hubbard:z0-conv-m}), the 2TDVP method is certainly the
fastest approach here. We also confirm that if the number of Krylov
vectors is chosen adaptively during the local solution of the 2TDVP
problem, the 2TDVP displays slightly longer CPU times per step for
larger time steps (as more local Krylov vectors are required). The MPO
\wii and the TEBD2 CPU times per step generically do not depend on the step size.
Only at $m = 500$
the bond dimension is not yet saturated at $\beta = 1$ if too few time
steps have been taken, resulting in slightly lower runtimes for larger
step sizes.

\subsubsection{\label{sec:examples:hubbard:conclusion}Conclusion}

To summarize the results of this example, we find that for small
$\beta$, 2TDVP provides adequate results very quickly, with an overall
speed-up of 5-10 compared to TEBD2 and the MPO \wii method. The local
Krylov method incurs large errors which -- while having comparable
runtime to 2TDVP -- make it unsuitable here. At large $\beta$, the MPO
\wii method with variational optimization tends to be the most stable,
displaying nearly no particle loss. We did not investigate a possible
combination of 2TDVP initially with a later switch to 1TDVP once the
bond dimension has grown sufficiently: this would then ensure
stability against particle loss at large $\beta$ combined with the
computational efficiency of the TDVP method here.

%%% Local Variables: 
%%% mode: latex
%%% TeX-master: "../../time_evolution_review"
%%% End: 

%% file: content/examples/2dmelting.tex
\subsection{\label{sec:example:melting}Melting of N\'eel order in the two-dimensional Heisenberg model}

\begin{figure}
  \centering
  \tikzsetnextfilename{2d_mapping}
  \begin{tikzpicture}
    \begin{scope}[node distance=0.5 and 0.5,minimum size=2em,inner sep=1]
      \node[site, draw=black, fill=blue!70!white, text=white] (s1) {$1$};
      \node[site, draw=black, fill=red!10!white] (s2) [right=of s1] {$2$};
      \node[site, draw=black, fill=blue!70!white, text=white] (s4) [right=of s2] {$4$};
      \node[site, draw=black, fill=red!10!white] (s7) [right=of s4] {$7$};
      
      \node[site, draw=black, fill=red!10!white] (s3) [below=of s1] {$3$};
      \node[site, draw=black, fill=blue!70!white, text=white] (s5) [below=of s2] {$5$};
      \node[site, draw=black, fill=red!10!white] (s8) [below=of s4] {$8$};
      \node[site, draw=black, fill=blue!70!white, text=white] (s11) [below=of s7] {$11$};
      
      \node[site, draw=black, fill=blue!70!white, text=white] (s6) [below=of s3] {$6$};
      \node[site, draw=black, fill=red!10!white] (s9) [below=of s5] {$9$};
      \node[site, draw=black, fill=blue!70!white, text=white] (s12) [below=of s8] {$12$};
      \node[site, draw=black, fill=red!10!white] (s14) [below=of s11] {$14$};
      
      \node[site, draw=black, fill=red!10!white] (s10) [below=of s6] {$10$};
      \node[site, draw=black, fill=blue!70!white, text=white] (s13) [below=of s9] {$13$};
      \node[site, draw=black, fill=red!10!white] (s15) [below=of s12] {$15$};
      \node[site, draw=black, fill=blue!70!white, text=white] (s16) [below=of s14] {$16$};
      
      \draw (s1) -- (s2);
      \draw (s2) -- (s3);
      \draw (s3) -- (s4);
      \draw (s4) -- (s5);
      \draw (s5) -- (s6);
      \draw (s6) -- (s7);
      \draw (s7) -- (s8);
      \draw (s8) -- (s9);
      \draw (s9) -- (s10);
      \draw (s10) -- (s11);
      \draw (s11) -- (s12);
      \draw (s12) -- (s13);
      \draw (s13) -- (s14);
      \draw (s14) -- (s15);
      \draw (s15) -- (s16);
    \end{scope}
  \end{tikzpicture}
  \caption{\label{fig:example:2dmelting:mapping}(2D) Exemplary $4\times 4$ lattice
    and the mapping of sites to a one-dimensional MPS chain.  It was
    suggested\cite{1705.05578} that this tilted z-mapping helps
    reduce the required bond dimension.  Encoded in pink/light and
    blue/dark are the two sub-lattices $A$ and $B$, respectively. In our test, we use an $8 \times 8$ lattice with open boundary conditions. }
\end{figure}
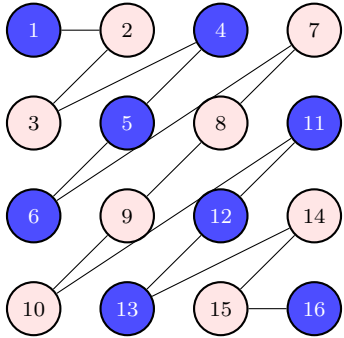

In this example we study the melting of N\'eel order in the
ferromagnetic Heisenberg model on the two-dimensional square lattice
of side length $L$.
The Hamiltonian reads
\begin{equation}
  \hat{H} = -\sum_{\braket{i, j}}\frac{1}{2}\left(\hat{s}^+_i \hat{s}^-_j +  \hat{s}^+_j \hat{s}^-_i \right) + \hat{s}^z_i \hat{s}^z_j\;.
\end{equation}
where $\braket{i,j}$ denotes nearest-neighbor site indices.
The initial state at $t=0$ is chosen to be the N\'eel state, i.e.~spins pointing up
and down in a checkerboard pattern
\begin{equation}
\ket{\psi(0)} =\bigotimes_{i\in A}\ket{\downarrow}_i\, \bigotimes_{j\in B}\ket{\uparrow}_j
\end{equation}
with sublattices $A$ and $B$ visualized in
\cref{fig:example:2dmelting:mapping}.
To map the two-dimensional system onto a one-dimensional MPS, we use a
tilted z-shaped mapping (cf.~\cref{fig:example:2dmelting:mapping}) as
suggested in Ref.~\cite{1705.05578}.
This mapping transports entanglement between the vertical and
horizontal bipartitions of the lattice through $O(L)$ bonds while
diagonal bipartitions only cut a single bond.
However, nearest-neighbor interactions now have to be carried through
$\approx \sqrt{2} L$ other sites.\footnote{It is not clear a priori, which mapping would be the optimal one for the dynamics of this system; for the sake of simplicity we have chosen this one.}
%Here, we do not investigate whether this mapping or another one is the most suitable for the square lattice N\'eel state, we simply have to pick one to continue.}
%
Under real-time evolution, the N\'eel order of the state is expected
to decay, and at the same time the entanglement of this initial product state will grow with time. 
To monitor this decay,
we consider the staggered magnetization per
site, which we define as 
\begin{equation}
\braket{\tilde{m}(t)}= \frac{1}{L^2}\left(\sum_{i\in B}\braket{\hat s^z_i(t)} - \sum_{j\in A}\braket{\hat s^z_j(t)}\right) \;.
\end{equation}
Initially, $\braket{\tilde{m}(0)} = \nicefrac{1}{2}$.
In addition to $\tilde{m}$, we also measure the energy of the state to
check the accuracy of energy conservation in the local Krylov and
TDVP approaches. Furthermore, we note that the system has an inversion symmetry:
Mirroring the initial N\'eel state on the MPS chain at the central bond results in the same
%$\braket{\hat s^z_i}$ 
local expectation values.  
Since the Hamiltonian does not break this symmetry, it has to be preserved under real-time evolution.
Hence, computing the deviation from this symmetry gives a measure for the error of the time evolution scheme. 
%measure the inversion symmetry error: 

%
In our test, we set $L = 8$, i.e.~work with $8 \times 8$ lattices of
$64$ sites with open boundary conditions in both directions.
We limit the CPU time for computing the time evolution to one hour on a
single core of a Xeon E5-2630 v4 clocked at
$2.20\mathrm{\;GHz}$.
After this hour, we then evaluate observables on the stored MPS
representing the individual time-evolved states.
The bond dimension of the states is limited to $m = 200$ and the
discarded weight varied between $10^{-8}$ and $10^{-12}$. It appears that the maximal bond dimension limit $m = 200$ is not the leading source of errors at least at short times $t < 1$ -- for such short times, calculations running for 24 CPU hours at $m = 2000$ produced essentially identical results. 
As this is a highly challenging problem for MPS methods, we prefer at this point not to investigate the limitations of our state-of-the-art approaches, but instead compare them for rather restricted bond dimensions focusing on the short time regime.  
It would be interesting to see, which time scales can be reached when further optimizing our procedure (in particular increase the number of kept states $m$, which quickly leads to a substantial increase of the needed computational resources, but also further aspects, like the mapping to the chain geometry), and we hope that our considerations here on the short time scales may help future developments of such optimized schemes for treating 2D or quasi-2D systems.   
%Due to the large number of parameters and methods and relatively constrained computational resources, we have opted not to run proper large-scale ($m = O(10^4)$, thousands of CPU hours) calculations here.
%

In the following, we choose time steps $\delta$ to be either $0.1$ or $0.01$ to verify approximate
convergence and to investigate the error caused by the finite time step.

\paragraph{TDVP and product initial states}
The state $\ket{\psi(0)}$ is a product state and can be represented at
bond dimension $m=1$.
Attempting to time-evolve this MPS with TDVP or the local Krylov
method is pointless, as many of the long-range interactions are lost
during the projection step and the resulting time evolution is simply
wrong.
To avoid this problem, we run an
initial DMRG calculation with the Hamiltonian
\begin{equation}
  \hat{H}_{\textrm{DMRG}} = -\alpha \left( \sum_{\braket{i, j}}\nicefrac{1}{2}\left(\hat{s}^+_i \hat{s}^-_j +  \hat{s}^+_j \hat{s}^-_i \right) + \hat{s}^z_i \hat{s}^z_j \right) + \sum_{i \in A} \hat s^z - \sum_{i \in B} \hat s^z_i\;.
\end{equation}
We initialize $\alpha = 1$ and then for ten successive sweeps reduce
it by a factor of ten each time.
Finally, we set $\alpha = 0$ and run five more sweeps.
Throughout, we use the subspace expansion to increase the bond
dimension of the state up to $m_{\mathrm{init}} = [50, 100, 200]$ and
set the discarded weight threshold to zero to keep the state at this
artificially large bond dimension.
Using the time-evolution Hamiltonian $\hat H$ together with the
subspace expansion guarantees that the generated additional states
have some relevance to the physical problem by being generated from
(partial) applications of $\hat H$ to $\ket{\psi(0)}$.
The DMRG calculation takes up to 120 seconds, the allowed runtime of
the TDVP and local Krylov methods is hence reduced accordingly.

\paragraph{Time-evolving block decimation} The TEBD2 requires
splitting the Hamiltonian into internally-commuting parts. To minimize
the bond dimensions of the MPOs, each column and each row of the
lattice is split into two Hamiltonians. These Hamiltonians then do not
contain overlapping gates and have a maximal bond dimension of
$4$. For optimal inversion symmetry, we apply gates on rows 1 and 8
first, then on rows 2 and 7, rows 3 and 6 and finally rows 4 and 5 (or
vice versa) and proceed in the same way with the column terms. These
MPOs are applied using the zip-up method which was tested to be as
precise as the variational optimization here, but much faster.

\paragraph{The global Krylov method} As can be expected, repeatedly
applying $\hat H$ to the N\'eel state generates large amounts of
entanglement. This is problematic for the global Krylov method, as its
Krylov vectors would ideally have a very large bond dimension. To
still proceed with the calculation, we force the truncation of Krylov
vectors also to a maximal bond dimension $m = 200$. This truncation
can lead to very wrong results when measuring long-range correlators
but should be acceptable for $\braket{\tilde{m}}$. Given the initial product
state, the variational orthogonalization has to work in the two-site
variant.

\subsubsection{The staggered magnetization $\braket{\tilde{m}(t)}$}

\input{content/examples/2dmelting_tdvpkrylov.tex}

\input{content/examples/2dmelting_all.tex}

We find that neither 1TDVP nor 2TDVP results change with the
step size, but some minimal changes can be seen in the local Krylov
results between $\delta = 0.1$ and $\delta = 0.01$
(cf.~\cref{fig:examples:melting:tdvpkrylov-dt}). Furthermore, we find that the 2TDVP and local Krylov method results for
$\braket{\tilde{m}(t)}$ do not change with the initial bond dimension
$m_{\mathrm{init}}=[50, 100, 200]$ as obtained by DMRG and converge to the same result at least for short times. The 1TDVP
results depend on this initial bond dimension
(cf.~\cref{fig:examples:melting:tdvpkrylov-m}) and even in the best
case of $m_{\mathrm{init}}=200$ only qualitatively reproduce the other
results. In the following, we
hence use $m_{\mathrm{init}} = 50$, $\delta = 0.1$ for 2TDVP,
$m_{\mathrm{init}} = 100, \delta = 0.1$ for 1TDVP and
$m_{\mathrm{init}} = 50, \delta = 0.1$ and $\delta = 0.01$ for the
local Krylov method.

The global Krylov results do not change with step size $\delta = 0.01$
or $\delta = 0.1$ nor with the discarded weight $10^{-8, -10, -12}$
and are in both cases in good agreement with the other methods until
$t \approx 0.7$. Beyond this point, it provides worse data than 1TDVP
due to the very strong truncation of the Krylov vectors. TEBD2
provides data within the range of the other methods at $\delta = 0.1$
with the zip-up algorithm and a discarded weight of $10^{-8}$. At
smaller $\delta$ or when using the variational optimization for
operator application, the evolution only obtains times $t \approx 0.5$
(but in agreement with all other methods).

When using the MPO \wii method with large time steps $\delta = 0.1$,
the results very quickly deviate from the other methods; in the
\syten{} implementation (using either zip-up or variational
application), additional instabilities occur around $t \approx 4$. At
$\delta = 0.01$, data is in line with the other methods mostly
independent of the discarded weight. The variational operator
application is much slower than the zip-up method here, so we use the
latter also for the MPO \wii method.

\cref{fig:examples:melting:all} shows the staggered magnetization for
the selected configurations of each method. The global Krylov method
is handicapped by its need to represent highly-entangled Krylov
vectors at $m = 200$, which leads to large errors. The MPO \wii method
suffers a large time-step error which in turn increases runtime by
approximately a factor of ten compared to those methods which can use
a larger step size. TEBD2 provides very reasonable data up to $t = 3$
and 2TDVP and the local Krylov method are able to go beyond up to
$t \approx 7$ and $t \approx 10$ respectively. 
%However, at reported truncation errors of order $10^{-2}$ per step size in both methods at larger $t$, this data at best can be used for qualitative arguments.
However, as the truncation error is of the order of $10^{-2}$ per step size in both methods at larger $t$, the results will have large quantitative errors at large times, as discussed next. 

\subsubsection{Energy conservation and inversion symmetry}

\input{content/examples/2dmelting_invsym.tex}

\input{content/examples/2dmelting_energy.tex}

Under the unitary time evolution, both the energy of the system and
the inversion symmetry over the central MPS bond should be
conserved. To measure the latter, we consider the maximal deviation
\begin{equation}
  \Delta I(t) = \max_{i = 1}^{L^2} \left| \braket{\hat s^z_i(t)} - \braket{\hat s^z_{L^2-i+1}(t)} \right| \;.
\end{equation}
Then considering the same configurations as before, we find in
\cref{fig:examples:melting:invsym} that the inversion error stays
relatively small for 2TDVP, the local Krylov and the TEBD2 methods, while it
becomes up to an order of magnitude larger for 1TDVP, the global Krylov
and the MPO \wii method at intermediate times. At very short times, we
find TEBD2 and (with $\delta = 0.01$) the local Krylov method to have the
smallest inversion symmetry error. The other methods (run at
$\delta = 0.01$) also provide similarly good results (not shown).

Considering
$\left| E(t) - E_0 \right|/L^2$, the error in the energy per site,
(cf.~\cref{fig:examples:melting:energy}) provides nearly the same
picture except for a much larger difference between the local Krylov
method and 2TDVP -- the latter conserves energy to a much higher
precision than the former. 1TDVP (not shown) provides accuracy in
energy up to $10^{-7}$ even at very late times while the global Krylov
method is very precise until $t \approx 0.7$, where its error quickly
increases.

\subsubsection{Conclusion}

%The most relevant take-away here is that by using a trick to increase
Obtaining the time evolution in 2D is, even for a non-entangled product initial state, a challenge for all methods. 
Even though the projection steps in the 2TDVP, the local Krylov, and the 1TDVP methods incur problems with such product initial states, this can be healed by simply increasing the initial bond dimension as discussed above, which is also often done in tDMRG in such situations. 
Apart from better energy conservation, 2TDVP and
the local Krylov method are nearly comparable, with the latter being
approximately 20-30\% faster (however, the maximal bond dimension $m = 200$ is here already
exhausted at $t < 1$). TEBD2 mostly suffers from slow MPO-MPS products
at large bond dimensions but actually has an acceptable error caused by 
the finite time step, allowing us to choose $\delta = 0.1$. In contrast, the
MPO method sports fast MPO-MPS products (due to the smaller
$\hat \wii$) but incurs a large time step error. It hence appears that
for longer time scales, the 2TDVP or local Krylov method are the most
promissing approaches, while for short times, any of the methods work reasonably
well.

%%% Local Variables: 
%%% mode: latex
%%% TeX-master: "../../time_evolution_review"
%%% End: 

%% file: content/examples/2dmelting_tdvpkrylov.tex
\begin{figure}
  \begin{center}
    \tikzsetnextfilename{tdvpkrylov_legend}
    \ref{tdvpkrylov_legend}
  \end{center}
  \begin{minipage}[t]{0.49\textwidth}
    \tikzsetnextfilename{melting_tdvplockrylov_dt}
    \pgfplotsset{width=\textwidth}
    \pgfplotsset{every axis plot/.append style={
        line width=1pt,
        tick style={line width=0.6pt}}}
    \begin{tikzpicture}
      \begin{axis}[
        width=\textwidth,
        axis lines=left,
        xlabel={Time $t$},
        font = {\footnotesize},
        ylabel={Staggered magnetization $\braket{\tilde{m}(t)}$},
        xmin=0,
        xmax=1,
        ymin=0,
        ymax=0.5,
        ]
        \addplot[color=brown,solid,mark=*,mark options=solid,mark phase=1,mark repeat=70]table[y expr = \thisrowno{3}]{../ex3-melting/syten/data/tdvp_1tdvp_dt=0.01_minit=100};
        \addplot[color=blue,solid,mark=diamond,mark options=solid,mark phase=11,mark repeat=70]table[y expr = \thisrowno{3}]{../ex3-melting/syten/data/tdvp_2tdvp_dt=0.01_minit=100};
        \addplot[color=orange,solid,mark=triangle,mark options=solid,mark phase=21,mark repeat=70]table[y expr = \thisrowno{3}]{../ex3-melting/syten/data/tdvp_lockrylov_dt=0.01_minit=100};
        \addplot[color=red,solid,mark=o,mark options=solid,mark phase=3,mark repeat=5]table[y expr = \thisrowno{3}]{../ex3-melting/syten/data/tebd_dt=0.1_w=1e-12_a=zipup_ft=1e-10};

        \addplot[color=blue,forget plot,dashed,mark=diamond,mark options=solid,mark phase=5,mark repeat=7]table[y expr = \thisrowno{3}]{../ex3-melting/syten/data/tdvp_2tdvp_dt=0.1_minit=100};

        \addplot[color=orange,forget plot,dashed,mark=triangle,mark options=solid,mark phase=6,mark repeat=7]table[y expr = \thisrowno{3}]{../ex3-melting/syten/data/tdvp_lockrylov_dt=0.1_minit=100};

        \addplot[color=brown,forget plot,dashed,mark=*,mark options=solid,mark phase=7,mark repeat=7]table[y expr = \thisrowno{3}]{../ex3-melting/syten/data/tdvp_1tdvp_dt=0.1_minit=100};
      \end{axis}
    \end{tikzpicture}
    \caption{\label{fig:examples:melting:tdvpkrylov-dt}(2D) Staggered magnetization $\braket{\tilde{m}(t)}$ during time evolution with maximal bond dimension $m_{\mathrm{max}}=200$. Comparison of the dependence on the step size of the 1TDVP, 2TDVP and the local Krylov method; we use $\delta = 0.01$ (solid) and $\delta = 0.1$ (dashed) at initial $m = 100$. 1TDVP and 2TDVP data is step-size independent but the
      local Krylov results differ slightly between $\delta = 0.1$ and
      $\delta = 0.01$. Due to the fixed bond dimension, 1TDVP at best
      provides qualitative results. }
  \end{minipage}
  \hfill
  \begin{minipage}[t]{0.49\textwidth}
    \tikzsetnextfilename{melting_tdvplockrylov_m}
    \pgfplotsset{width=\textwidth}
    \pgfplotsset{every axis plot/.append style={
        line width=1pt,
        tick style={line width=0.6pt}}}
    \begin{tikzpicture}
      \begin{axis}[
        width=\textwidth,
        axis lines=left,
        xlabel={Time $t$},
        font = {\footnotesize},
        ylabel={Staggered magnetization $\braket{\tilde{m}(t)}$},
        xmin=0,
        xmax=10,
        ymin=-0.2,
        ymax=0.5,
        legend style = {anchor=north east, at = {(1,0.8)}, draw = none, fill = none, font={\footnotesize}},
        legend cell align = left,
        legend to name = tdvpkrylov_legend,
        legend columns=4,
        ]
        \addplot[color=blue,solid,mark=diamond,mark options=solid,mark phase=2,mark repeat=5]table[y expr = \thisrowno{3}]{../ex3-melting/syten/data/tdvp_2tdvp_dt=0.1_minit=50};
        \addlegendentry{2TDVP};
        \addplot[color=blue,forget plot,dashed,mark=diamond,mark options=solid,mark phase=2,mark repeat=5]table[y expr = \thisrowno{3}]{../ex3-melting/syten/data/tdvp_2tdvp_dt=0.1_minit=100};
        \addplot[color=blue,forget plot,dotted,mark=diamond,mark options=solid,mark phase=2,mark repeat=5]table[y expr = \thisrowno{3}]{../ex3-melting/syten/data/tdvp_2tdvp_dt=0.1_minit=200};

        \addplot[color=orange,solid,mark=triangle,mark options=solid,mark phase=4,mark repeat=5]table[y expr = \thisrowno{3}]{../ex3-melting/syten/data/tdvp_lockrylov_dt=0.1_minit=50};
        \addlegendentry{Local Krylov};
        \addplot[color=orange,forget plot,dashed,mark=triangle,mark options=solid,mark phase=4,mark repeat=5]table[y expr = \thisrowno{3}]{../ex3-melting/syten/data/tdvp_lockrylov_dt=0.1_minit=100};
        \addplot[color=orange,forget plot,dotted,mark=triangle,mark options=solid,mark phase=4,mark repeat=5]table[y expr = \thisrowno{3}]{../ex3-melting/syten/data/tdvp_lockrylov_dt=0.1_minit=200};

        \addplot[color=brown,solid,mark=*,mark options=solid,mark phase=5,mark repeat=5]table[y expr = \thisrowno{3}]{../ex3-melting/syten/data/tdvp_1tdvp_dt=0.1_minit=50};
        \addlegendentry{1TDVP};
        \addplot[color=brown,forget plot,dashed,mark=*,mark options=solid,mark phase=6,mark repeat=5]table[y expr = \thisrowno{3}]{../ex3-melting/syten/data/tdvp_1tdvp_dt=0.1_minit=100};
        \addplot[color=brown,forget plot,dotted,mark=*,mark options=solid,mark phase=7,mark repeat=5]table[y expr = \thisrowno{3}]{../ex3-melting/syten/data/tdvp_1tdvp_dt=0.1_minit=200};
        
        \addplot[color=red,solid,mark=o,mark options=solid,mark phase=0,mark repeat=5]table[y expr = \thisrowno{3}]{../ex3-melting/syten/data/tebd_dt=0.1_w=1e-12_a=zipup_ft=1e-10};
        \addlegendentry{TEBD2};
      \end{axis}
    \end{tikzpicture}
    \caption{\label{fig:examples:melting:tdvpkrylov-m}(2D) 
      Staggered magnetization $\braket{\tilde{m}(t)}$ during time evolution with maximal bond dimension $m_{\mathrm{max}}=200$.
      Comparison of the dependence of the 
      1TDVP, 2TDVP and the local Krylov method on the initial bond
      dimension; $m = 50$ (solid), $m = 100$ (dashed) and $m = 200$
      (dotted). Step size is $\delta = 0.1$. The best TEBD2 result is
      plotted as a reference. The 2TDVP and the local Krylov method do not strongly depend
      on the initial bond dimension.}
  \end{minipage}
\end{figure}

%% file: content/examples/2dmelting_all.tex
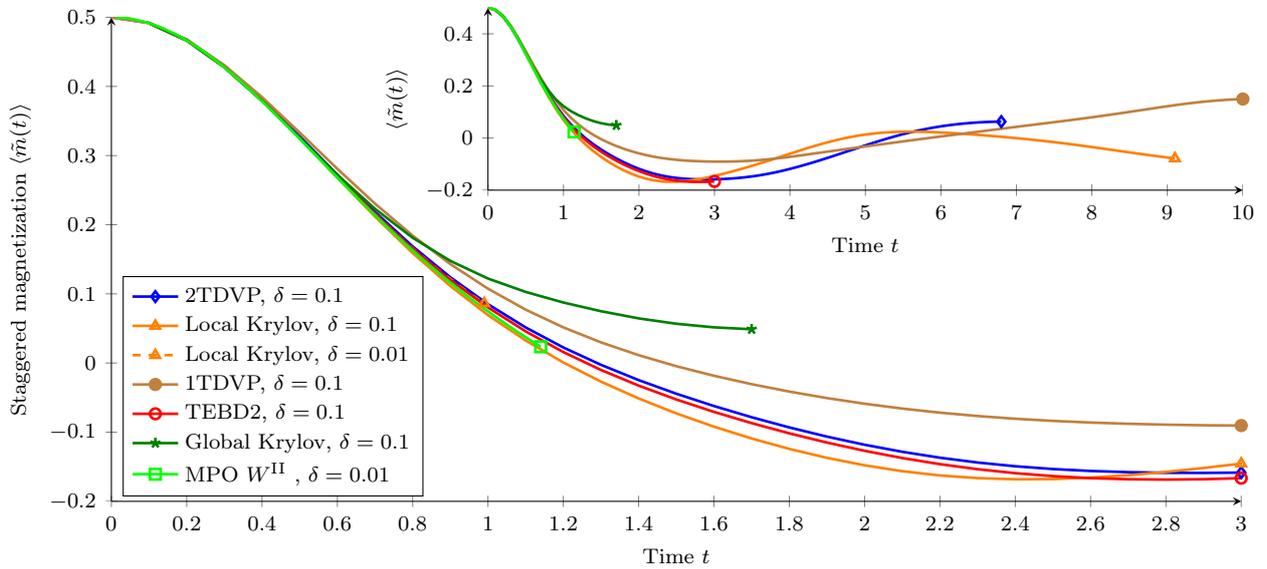
\begin{figure}
  \tikzsetnextfilename{melting_all}
  \pgfplotsset{width=\textwidth}
  \pgfplotsset{every axis plot/.append style={
      line width=1pt,
      tick style={line width=0.6pt}}}
  \begin{tikzpicture}
    \begin{axis}[
      width=\textwidth,
      height=8cm,
      axis lines=left,
      xlabel={Time $t$},
      font = {\footnotesize},
      ylabel={Staggered magnetization $\braket{\tilde{m}(t)}$},
      xmin=0,
      xmax=3,
      ymin=-0.2,
      ymax=0.5,
      legend style = {anchor=south west, at = {(0.01,0.01)}, font={\footnotesize}},
      legend cell align = left,
      ]
      \addplot[color=blue,solid,mark=diamond,mark options=solid,mark phase=31]table[y expr = \thisrowno{3}]{../ex3-melting/syten/data/tdvp_2tdvp_dt=0.1_minit=50};
      \addlegendentry{2TDVP, $\delta = 0.1$};

        \addplot[color=orange,solid,mark=triangle,mark options=solid,mark phase=31]table[y expr = \thisrowno{3}]{../ex3-melting/syten/data/tdvp_lockrylov_dt=0.1_minit=50};
        \addlegendentry{Local Krylov, $\delta = 0.1$};

        \addplot[color=orange,dashed,mark=triangle,mark options=solid,mark phase=100]table[y expr = \thisrowno{3}]{../ex3-melting/syten/data/tdvp_lockrylov_dt=0.01_minit=50};
        \addlegendentry{Local Krylov, $\delta = 0.01$};
        
        \addplot[color=brown,solid,mark=*,mark options=solid,mark phase=31]table[y expr = \thisrowno{3}]{../ex3-melting/syten/data/tdvp_1tdvp_dt=0.1_minit=100};
        \addlegendentry{1TDVP, $\delta = 0.1$};
        
        \addplot[color=red,solid,mark=o,mark options=solid,mark phase=30]table[y expr = \thisrowno{3}]{../ex3-melting/syten/data/tebd_dt=0.1_w=1e-8_a=zipup_ft=1e-10};
        \addlegendentry{TEBD2, $\delta = 0.1$};

        \addplot[color=green!50!black,solid,mark=star,mark options=solid,mark phase=17]table[y expr = \thisrowno{3}]{../ex3-melting/syten/data/krylov_krylov_dt=0.1_w=1e-8};
        \addlegendentry{Global Krylov, $\delta = 0.1$};

        \addplot[color=green,solid,mark=square,mark options=solid,mark phase=114]table[y expr = \thisrowno{3}]{../ex3-melting/syten/data/mpo_mpo_dt=0.01_a=zip_w=1e-8};
        \addlegendentry{MPO \wii, $\delta = 0.01$};
        
        \coordinate (insetPosition) at (axis cs:1.0,0.25);
      \end{axis}

      \begin{axis}[
        width=0.7*\textwidth,
        height=4cm,
        at = {(insetPosition)},
        axis lines=left,
        xlabel={Time $t$},
        font = {\footnotesize},
        ylabel={$\braket{\tilde{m}(t)}$},
        xmin=0,
        xmax=10,
        ymin=-0.2,
        ymax=0.5,
        ]
        \addplot[color=blue,solid,mark=diamond,mark options=solid,mark phase=69]table[y expr = \thisrowno{3}]{../ex3-melting/syten/data/tdvp_2tdvp_dt=0.1_minit=50};

        \addplot[color=orange,solid,mark=triangle,mark options=solid,mark phase=92]table[y expr = \thisrowno{3}]{../ex3-melting/syten/data/tdvp_lockrylov_dt=0.1_minit=50};

        \addplot[color=brown,solid,mark=*,mark options=solid,mark phase=101]table[y expr = \thisrowno{3}]{../ex3-melting/syten/data/tdvp_1tdvp_dt=0.1_minit=100};

        \addplot[color=red,solid,mark=o,mark options=solid,mark phase=30]table[y expr = \thisrowno{3}]{../ex3-melting/syten/data/tebd_dt=0.1_w=1e-8_a=zipup_ft=1e-10};

        \addplot[color=green!50!black,solid,mark=star,mark options=solid,mark phase=17]table[y expr = \thisrowno{3}]{../ex3-melting/syten/data/krylov_krylov_dt=0.1_w=1e-8};

        \addplot[color=green,solid,mark=square,mark options=solid,mark phase=114]table[y expr = \thisrowno{3}]{../ex3-melting/syten/data/mpo_mpo_dt=0.01_a=zip_w=1e-8};
      \end{axis}
      
    \end{tikzpicture}
    \caption{\label{fig:examples:melting:all}(2D) 
      Staggered magnetization $\braket{\tilde{m}(t)}$ during time evolution with maximal bond dimension $m_{\mathrm{max}}=200$ for all available methods.
      Only the local Krylov method shows
      dependence on the time step size. 1TDVP and the global Krylov method
      show large errors at $t \geq 0.5$. At intermediate times, 2TDVP
      seems to match the behaviour of TEBD2 better than the local
      Krylov method at $\delta = 0.1$. At long times, both the local Krylov 
      data and the 2TDVP data appear reasonable though not
      necessarily trustworthy. The 1TDVP, while able to obtain times
      $t > 100$, produces certainly wrong long-time results.}
\end{figure}

%% file: content/examples/2dmelting_invsym.tex
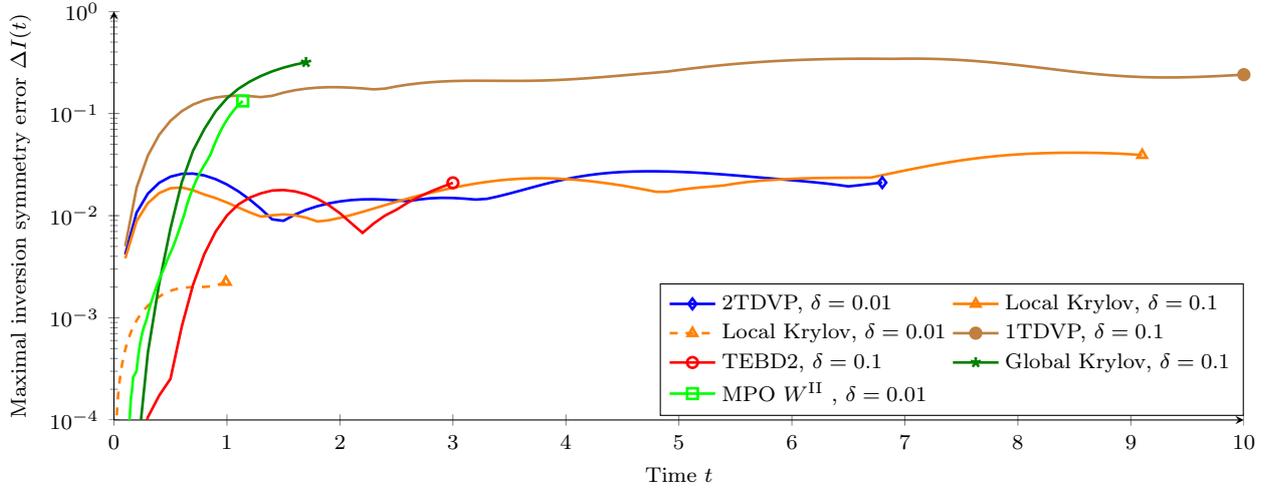
\begin{figure}
  \tikzsetnextfilename{melting_invsym}
  \pgfplotsset{width=\textwidth}
  \pgfplotsset{every axis plot/.append style={
      line width=1pt,
      tick style={line width=0.6pt}}}
  \begin{tikzpicture}
    \begin{axis}[
      width=\textwidth,
      height=7cm,
      axis lines=left,
      xlabel={Time $t$},
      font = {\footnotesize},
      ylabel={Maximal inversion symmetry error $\Delta I(t)$},
      xmin=0,
      xmax=10,
      ymin=1e-4,
      ymax=1,
      ymode=log,
      legend style = {anchor=south east, at = {(1,0.01)}, font={\footnotesize}},
      legend cell align = left,
      legend columns=2,
      ]
      \addplot[color=blue,solid,mark=diamond,mark options=solid,mark phase=68]table[y expr = \thisrowno{2}]{../ex3-melting/syten/data/tdvp_2tdvp_dt=0.1_minit=50};
      \addlegendentry{2TDVP, $\delta = 0.01$};

        \addplot[color=orange,solid,mark=triangle,mark options=solid,mark phase=91]table[y expr = \thisrowno{2}]{../ex3-melting/syten/data/tdvp_lockrylov_dt=0.1_minit=50};
        \addlegendentry{Local Krylov, $\delta = 0.1$};

        \addplot[color=orange,dashed,mark=triangle,mark options=solid,mark phase=97]table[y expr = \thisrowno{2}]{../ex3-melting/syten/data/tdvp_lockrylov_dt=0.01_minit=50};
        \addlegendentry{Local Krylov, $\delta = 0.01$};
        
        \addplot[color=brown,solid,mark=*,mark options=solid,mark phase=100]table[y expr = \thisrowno{2}]{../ex3-melting/syten/data/tdvp_1tdvp_dt=0.1_minit=100};
        \addlegendentry{1TDVP, $\delta = 0.1$};
        
        \addplot[color=red,solid,mark=o,mark options=solid,mark phase=28]table[y expr = \thisrowno{2}]{../ex3-melting/syten/data/tebd_dt=0.1_w=1e-8_a=zipup_ft=1e-10};
        \addlegendentry{TEBD2, $\delta = 0.1$};

        \addplot[color=green!50!black,solid,mark=star,mark options=solid,mark phase=15]table[y expr = \thisrowno{2}]{../ex3-melting/syten/data/krylov_krylov_dt=0.1_w=1e-8};
        \addlegendentry{Global Krylov, $\delta = 0.1$};

        \addplot[color=green,solid,mark=square,mark options=solid,mark phase=101]table[y expr = \thisrowno{2}]{../ex3-melting/syten/data/mpo_mpo_dt=0.01_a=zip_w=1e-8};
        \addlegendentry{MPO \wii, $\delta = 0.01$};
      \end{axis}
    \end{tikzpicture}
    \caption{\label{fig:examples:melting:invsym}(2D) Maximal inversion
      symmetry error $\Delta I(t)$. The simulations using TEBD2, 2TDVP and the local Krylov method 
      result in approximately the same errors at intermediate
      times. At short times, the local Krylov method with $\delta = 0.01$ and
      TEBD2 provide results with a relative error of a fraction of 1\%.}
\end{figure}

%% file: content/examples/2dmelting_energy.tex
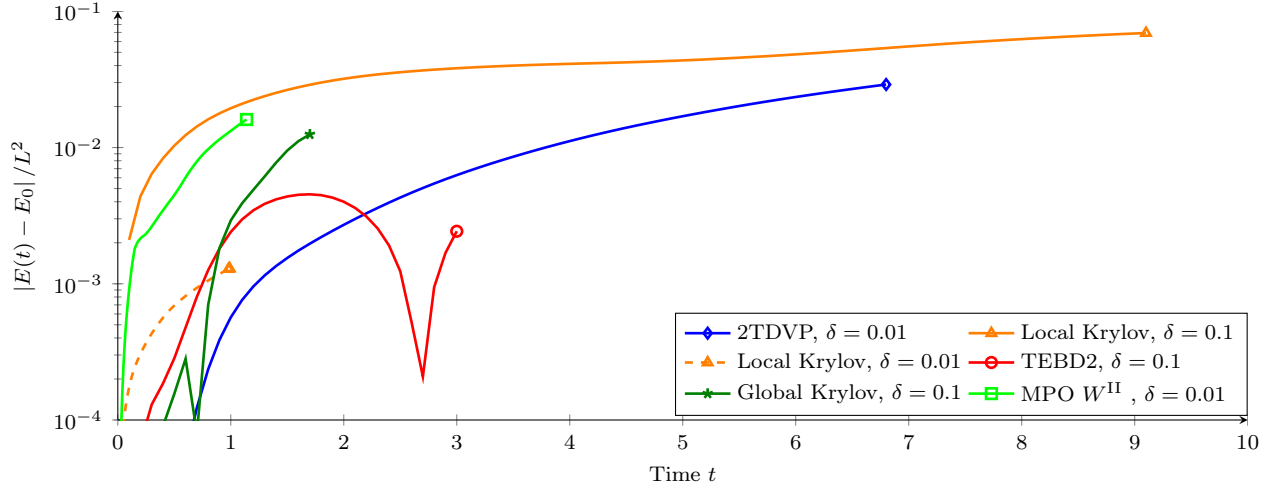
\begin{figure}
  \tikzsetnextfilename{melting_energy}
  \pgfplotsset{width=\textwidth}
  \pgfplotsset{every axis plot/.append style={
      line width=1pt,
      tick style={line width=0.6pt}}}
  \begin{tikzpicture}
    \begin{axis}[
      width=\textwidth,
      height=7cm,
      axis lines=left,
      xlabel={Time $t$},
      font = {\footnotesize},
      ylabel={$\left|E(t) - E_0\right|/L^2$},
      xmin=0,
      xmax=10,
      ymin=1e-4,
      ymax=1e-1,
      ymode=log,
      legend style = {anchor=south east, at = {(1,0.01)}, font={\footnotesize}},
      legend cell align = left,
      legend columns=2,
      ]
      \addplot[color=blue,solid,mark=diamond,mark options=solid,mark phase=62]table[y expr = abs(28 - \thisrowno{1})/64]{../ex3-melting/syten/data/tdvp_2tdvp_dt=0.1_minit=50.energy};
      \addlegendentry{2TDVP, $\delta = 0.01$};

      \addplot[color=orange,solid,mark=triangle,mark options=solid,mark phase=91]table[y expr = abs(28 - \thisrowno{1})/64]{../ex3-melting/syten/data/tdvp_lockrylov_dt=0.1_minit=50.energy};
      \addlegendentry{Local Krylov, $\delta = 0.1$};

      \addplot[color=orange,dashed,mark=triangle,mark options=solid,mark phase=93]table[y expr = abs(28 - \thisrowno{1})/64]{../ex3-melting/syten/data/tdvp_lockrylov_dt=0.01_minit=50.energy};
      \addlegendentry{Local Krylov, $\delta = 0.01$};
      
      % \addplot[color=brown,solid,mark=*,mark options=solid,mark phase=100]table[y expr = abs(28 - \thisrowno{1})/64]{../ex3-melting/syten/data/tdvp_1tdvp_dt=0.1_minit=50.energy};
      % \addlegendentry{1TDVP, $\delta = 0.1$};
      
      \addplot[color=red,solid,mark=o,mark options=solid,mark phase=28]table[y expr = abs(28 - \thisrowno{1})/64]{../ex3-melting/syten/data/tebd_dt=0.1_w=1e-8_a=zipup_ft=1e-10.energy};
      \addlegendentry{TEBD2, $\delta = 0.1$};

      \addplot[color=green!50!black,solid,mark=star,mark options=solid,mark phase=12]table[y expr = abs(28 - \thisrowno{1})/64]{../ex3-melting/syten/data/krylov_krylov_dt=0.1_w=1e-8.energy};
      \addlegendentry{Global Krylov, $\delta = 0.1$};

      \addplot[color=green,solid,mark=square,mark options=solid,mark phase=111]table[y expr = abs(28 - \thisrowno{1})/64]{../ex3-melting/syten/data/mpo_mpo_dt=0.01_a=zip_w=1e-8.energy};
      \addlegendentry{MPO \wii, $\delta = 0.01$};
    \end{axis}
  \end{tikzpicture}
  \caption{\label{fig:examples:melting:energy}(2D) Error in energy per site
    $\left|E(t) - E_0\right|/L^2$ during the real-time evolution. Not
    shown is single-site TDVP with error $< 10^{-7}$ at all
    times. Comparing 2TDVP and the local Krylov method, the better
    energy conservation of 2TDVP is obvious. The global Krylov method
    is good up to $t \approx 0.7$ when the truncation of Krylov
    vectors becomes problematic.}
\end{figure}

%% file: content/examples/otoc.tex
\subsection{\label{sec:examples:otoc}OTOCs for interacting spins}
The final example is motivated by the study of information spreading and scrambling in closed quantum many-body systems\cite{Chen:2016qpx,bohrdt17:_scram,Keyserlingk18:otocs}.
This can be done by studying the dynamics of out-of-time-order correlators (OTOCs) of local observables under an initial perturbation.
We consider a periodically-driven Ising chain with a conserved $\hat{S}^{z}$ quantum number which recently has been shown to exhibit characteristic long-time relaxation behavior: a power law decay in the OTOCs which can be related to a hydrodynamical picture of diffusion\cite{Rakovszky18:otocs}.
The time evolution is generated by a Floquet driving protocol $\hat{U}(kT) = \left[\hat{U}(T)\right]^{k}$ with driving period $T$ and
\begin{align}
	\hat{U}(T) &= e^{-\mathrm i \tau \hat{H}_{xy}}e^{-\mathrm i \tau \hat{H}_{zz}}e^{-\mathrm i \tau \hat{H}_{xy}}e^{-\mathrm i \tau \hat{H}_{z}} \\
	\hat{H}_{xy} &= J\sum_{j}\hat{s}^{x}_{j}\hat{s}^{x}_{j+1} + \hat{s}^{y}_{j}\hat{s}^{y}_{j+1} \\
	\hat{H}_{z}  &= J_{z}\sum_{j}\hat{s}^{z}_{j}\hat{s}^{z}_{j+1} \\
	\hat{H}_{zz} &= J_{zz}\sum_{j}\hat{s}^{z}_{j}\hat{s}^{z}_{j+2}\;.
\end{align}
The driving period is chosen so that $T = 1 = 4 \tau$ and the coupling constants are given by 
\begin{equation}
	J_{xy} = \frac{2\sqrt{3}+3}{7}, \quad J_{z} = \frac{\sqrt{3}+5}{6} \quad \text{and} \quad J_{zz} = \frac{\sqrt{5}}{2}\;.
\end{equation}
The OTOC we calculate describes correlations in the Heisenberg time evolution between operators $\hat{s}^{z}_{\nicefrac{L}{2}}(t\equiv 0)\equiv \hat{z}_{0}$ and $\hat{s}^{z}_{\nicefrac{L}{2}+r}(t) \equiv \hat{z}_{r}$ and is given by an ensemble average
\begin{align}
	F^{\hat{z}_{r}\hat{z}_{0}}_{\beta} &= \Real \left(\Tr \left\{\hat{\rho}(\beta)\hat{z}^{\dagger}_{r}(t)\hat{z}^{\dagger}_{0}\hat{z}_{r}(t)\hat{z}_{0} \right\}\right)
\end{align}
which can be interpreted physically as a measure for the causal relation between correlators during time evolution.
This is typically done at non-zero temperature so that we need to describe the system in an enlarged Hilbert space $\mathcal{H}=\mathcal{H}_{P}\otimes\mathcal{H}_{A}$ to be able to represent mixed states, cf.~\cref{sec:mps:finitetemp}. Here, we work in the infinite-temperature limit $\beta = 0$ of a canonical ensemble with $S^z = 0$.
As carried out in \cref{sec:tricks:pip} the time evolution of a purified, (in-)finite temperature state can be manipulated by shifting the purification insertion point so that we can evaluate the OTOC performing two time evolutions and calculate only local expectation values:
\begin{align}
	F^{\hat{z}_{r}\hat{z}_{0}}_{\beta} &= \braket{Z_{0}(\beta,t)|\hat{z}^{\dagger}_{P;r}\otimes\hat{z}^{t}_{A;r}|Z_{0}(t)} \\
	\ket{Z_{0}(t)} &= \left(\hat{U}_{P}(t)\otimes\hat{U}^{\star}_{A}(t) \right)\left(\hat{z}^{\dagger}_{P;0}\otimes \mathbf{\hat 1}\right)\ket{0} \\
	\ket{Z(\beta,t)} &= \left(\hat{U}_{P}(t)\otimes\hat{U}^{\star}_{A}(t) \right) \left(\hat{\rho}_{P}(\nicefrac{\beta/2}) \otimes \hat{z}^{\star}_{A;0}\hat{\rho}_{A}(\nicefrac{\beta}{2})\right)\ket{0} \;.
\end{align}
We calculated the time evolutions for $32,64$ and $128$ physical sites with step sizes $\delta=0.05,0.01$ and kept $m=1000$ as the maximum number of states.
We find the global Krylov method to be unsuitable for the current problem. Due to the infinite temperature initial state and time evolutions on both the physical and auxiliary system, the Krylov vectors are already at the beginning highly entangled, forbidding a calculation at reasonable time scales.
\subsubsection{\label{sec:examples:otoc:lightcone-spreading}Spreading of the light cone}
\begin{figure}
	\begin{minipage}[t]{0.49\textwidth}
		\tikzsetnextfilename{otoc_lightcone_tdvp}
		\begin{tikzpicture}
			\pgfplotsset
			{
				/pgfplots/colormap={gnuplot}{rgb255=(0,0,0) rgb255=(46,0,53) rgb255=(65,0,103) rgb255=(80,0,149) rgb255=(93,0,189) rgb255=(104,1,220) rgb255=(114,2,242) rgb255=(123,3,253) rgb255=(131,4,253) rgb255=(139,6,242) rgb255=(147,9,220) rgb255=(154,12,189) rgb255=(161,16,149) rgb255=(167,20,103) rgb255=(174,25,53) rgb255=(180,31,0) rgb255=(186,38,0) rgb255=(191,46,0) rgb255=(197,55,0) rgb255=(202,64,0) rgb255=(208,75,0) rgb255=(213,87,0) rgb255=(218,100,0) rgb255=(223,114,0) rgb255=(228,130,0) rgb255=(232,147,0) rgb255=(237,165,0) rgb255=(241,185,0) rgb255=(246,207,0) rgb255=(250,230,0) rgb255=(255,255,0) }
			}
			\begin{axis}
			[
				width	= 0.85\textwidth,
				height	= 0.3\textheight,
				axis on top,
				title	= {OTOC $F^{\hat{z}_{r}\hat{z}_{0}}_{\beta=0}(t)$ obtained with 2TDVP},
				xlabel	= {Relative position $r$},
				ylabel	= {Time $t$},
				font	= {\footnotesize},
				xmin	= -64,
				xmax	= 63,
				ymin	= 0,
				ymax	= 63.5,
				colorbar right,
				colormap name	= gnuplot,
				point meta min=-0.01,
				point meta max= 0.07,
				every colorbar/.append style =
					{
% 						height			=	2*\pgfkeysvalueof{/pgfplots/parent axis height} + 1*\pgfkeysvalueof{/pgfplots/group/vertical sep},
% 						ylabel			=	{$\langle \hat N_j \rangle$},
						width			=	3mm,
						ytick			= 	{-0.01,0.03,0.07},
% 						yticklabels		=	{$-0.0025$,$0.0$,$0.0025$},
% 						ylabel shift 		=	-4pt,
						y tick scale label style=	{xshift=2.0em,yshift=0.5em},
					},
			]
			\addplot graphics
			[
				xmin = -64, 
				xmax = 63, 
				ymin = 0, 
				ymax = 64,
			]
			{../ex4-otocs/precompiled/syten_tdvp_L_128_dt_0p05_beta_0p0.otoc.pdf};
			\draw[draw=black, thick, dashed] (axis cs:-2,2) -- (axis cs:-24,64);
			\draw[draw=black, thick, dashed] (axis cs:2,2) -- (axis cs:24,64);
			\draw[draw=white, thick, dotted] plot [smooth] coordinates { (axis cs: -1,1) (axis cs: -8,40) (axis cs: -10,64) };
			\draw[draw=white, thick, dotted] plot [smooth] coordinates { (axis cs:1,1) (axis cs: 8,40) (axis cs:10,64) };
			\end{axis}
		\end{tikzpicture}
		\caption{\label{fig:otoc:lightcone-tdvp}(OTOC) $F^{\hat{z}_{r}\hat{z}_{0}}_{\beta=0}(t)$ for a  periodically driven Ising chain with local operators $\hat{z}_{r} = \hat{s}^{z}_{\nicefrac{L}{2}+r}$ as obtained with 2TDVP and time step $\delta=0.01$. The broadening of the lightcone is marked 
		for the envelope of the lightcone front (dashed lines) and for the bulk (dotted lines).}
		%by the spreading of the envelope of the lightcone front (dashed) and the bulk (dotted).}
	\end{minipage}
	\hfill
	\begin{minipage}[t]{0.49\textwidth}
		\tikzsetnextfilename{otoc_lightcone_local_krylov}
		\begin{tikzpicture}
			\pgfplotsset
			{
				/pgfplots/colormap={gnuplot}{rgb255=(0,0,0) rgb255=(46,0,53) rgb255=(65,0,103) rgb255=(80,0,149) rgb255=(93,0,189) rgb255=(104,1,220) rgb255=(114,2,242) rgb255=(123,3,253) rgb255=(131,4,253) rgb255=(139,6,242) rgb255=(147,9,220) rgb255=(154,12,189) rgb255=(161,16,149) rgb255=(167,20,103) rgb255=(174,25,53) rgb255=(180,31,0) rgb255=(186,38,0) rgb255=(191,46,0) rgb255=(197,55,0) rgb255=(202,64,0) rgb255=(208,75,0) rgb255=(213,87,0) rgb255=(218,100,0) rgb255=(223,114,0) rgb255=(228,130,0) rgb255=(232,147,0) rgb255=(237,165,0) rgb255=(241,185,0) rgb255=(246,207,0) rgb255=(250,230,0) rgb255=(255,255,0) }
			}
			\begin{axis}
			[
				width	= 0.85\textwidth,
				height	= 0.3\textheight,
				axis on top,
				title	= {OTOC $F^{\hat{z}_{r}\hat{z}_{0}}_{\beta=0}(t)$ obtained with local Krylov},
				xlabel	= {Relative position $r$},
				ylabel	= {Time $t$},
				font	= {\footnotesize},
				xmin	= -64,
				xmax	= 63,
				ymin	= 0,
				ymax	= 63.5,
				colorbar right,
				colormap name	= gnuplot,
				point meta min=-0.01,
				point meta max= 0.07,
				every colorbar/.append style =
					{
% 						height			=	2*\pgfkeysvalueof{/pgfplots/parent axis height} + 1*\pgfkeysvalueof{/pgfplots/group/vertical sep},
% 						ylabel			=	{$\langle \hat N_j \rangle$},
						width			=	3mm,
						ytick			= 	{-0.01,0.03,0.07},
% 						yticklabels		=	{$-0.0025$,$0.0$,$0.0025$},
% 						ylabel shift 		=	-4pt,
						y tick scale label style=	{xshift=2.0em,yshift=0.5em},
					},
			]
			\addplot graphics
			[
				xmin = -64, 
				xmax = 63, 
				ymin = 0, 
				ymax = 64,
			]
			{../ex4-otocs/precompiled/scipal_local_krylov_L_128_dt_0p05_beta_0p0.otoc.pdf};
			\draw[draw=black, thick, dashed] (axis cs:-2,2) -- (axis cs:-24,64);
			\draw[draw=black, thick, dashed] (axis cs:2,2) -- (axis cs:24,64);
			\draw[draw=white, thick, dotted] plot [smooth] coordinates { (axis cs: -1,1) (axis cs: -8,40) (axis cs: -10,64) };
			\draw[draw=white, thick, dotted] plot [smooth] coordinates { (axis cs:1,1) (axis cs: 8,40) (axis cs:10,64) };
			\end{axis}
		\end{tikzpicture}
		\caption{\label{fig:otoc:lightcone-local-krylov}(OTOC) $F^{\hat{z}_{r}\hat{z}_{0}}_{\beta=0}(t)$ for a periodically driven Ising chain with local operators $\hat{z}_{r} = \hat{s}^{z}_{\nicefrac{L}{2}+r}$ as obtained with the local Krylov method and time step $\delta=0.05$. The lightcone broadening seen in the 2TDVP simulations is 
		%marked by the spreading 
		indicated for the envelope of the lightcone front (dashed lines) and for the bulk (dotted lines).}
	\end{minipage}
\end{figure}
In our setup we considered open boundary conditions so that we need to ensure that excitations from the boundary do not disturb the OTOC.
With the chosen set of coupling constants this is fulfilled by restricting the maximal computed time evolution to $kT \leq \nicefrac{L}{2}$ so that in order to study the hydrodynamic tail we will discuss only the results for the largest system with $L=128$ physical sites.
In \cref{fig:otoc:lightcone-tdvp} the OTOC $F^{\hat{z}_{r}\hat{z}_{0}}_{\beta=0}(t)$ obtained with 2TDVP and a step width $\delta=0.01$ is shown as an example for the time evolution at all relative lattice distances $r$ with respect to the central site of the system.
We have marked the envelope of the spreading lightcone front and the bulk for the case of 2TDVP which gives a quantitative picture of the broadening of the lightcone front.
%
% We can identify clearly the spreading and broadening of the lightcone.\todo{what is spreading, what is broadening?}
%
\cref{fig:otoc:lightcone-local-krylov} displays the same OTOC obtained with the local Krylov stepper and $\delta=0.05$.
Comparing to the time evolution obtained with 2TDVP we find that both the spreading of the lightcone front and the broadening of the lightcone itself is much less pronounced and disturbed by \textit{rays} in the lightcone which decay only slowly.
These rays are generated by structure inside the lightcone which are absent in the 2TDVP datasets.
Apart from even-odd effects, we would expect the dynamics inside the lightcone to generate a homogenous amplitude of the OTOC, strengthening the 2TDVP results.
Similar plots can be obtained for the MPO \wii method which suffers from the same effects as the local Krylov method.
\begin{figure}[t!]
  \begin{minipage}[t]{0.49\textwidth}
    \pgfplotsset{width=\textwidth}
    \tikzsetnextfilename{otoc_lightcone_spreading}
    \begin{tikzpicture}
      \begin{axis}
        [
        axis lines=left,
        xlabel={Relative position $r$},
        font = {\footnotesize},
        ylabel={OTOC $C^{\hat{Z}_{0}\hat{Z}_{1}}_{\beta=0}(t)$},
        width=0.92\textwidth,
        height=\textwidth,
        scaled y ticks = false,
        xmin=65,
        xmax=100,
        legend style={at={(1,0.05)}, anchor=south east},
        legend entries = {$t=2$,$t=14$,$t=26$,$t=38$, $t=50$, $t=62$, 2TDVP, TEBD2, Local Krylov, MPO \wii},
        % ymin=1e-4,
        % ymax=1e-1,
        ]
				\addlegendimage{no markers, thick, loosely dotted}
				\addlegendimage{no markers, thick, dotted}
				\addlegendimage{no markers, thick, densely dotted}
				\addlegendimage{no markers, thick, loosely dashed}
				\addlegendimage{no markers, thick, dashed}
				\addlegendimage{no markers, thick, densely dashed}
				\addlegendimage{mark=diamond, solid, thick, color=blue}
				\addlegendimage{mark=o, solid, thick, color=red}
				\addlegendimage{mark=triangle, solid, thick, color=orange}
				\addlegendimage{mark=square, solid, thick, color=green!90!black}
				
				\addplot[forget plot, thick, smooth, color=blue, loosely dotted, mark=diamond, mark options=solid, mark phase=0, mark repeat=40]
				table[x expr = \coordindex, y expr = \thisrowno{1}]{../ex4-otocs/syten/TDVP/L_128_dt_0p05_beta_0p0/results/otoc_t_slices};
				\addplot[forget plot, thick, smooth, color=blue, dotted, mark=diamond, mark options=solid, mark phase=3, mark repeat=40]
				table[x expr = \coordindex, y expr = \thisrowno{7}]{../ex4-otocs/syten/TDVP/L_128_dt_0p05_beta_0p0/results/otoc_t_slices};
				\addplot[forget plot, thick, smooth, color=blue, densely dotted, mark=diamond, mark options=solid, mark phase=10, mark repeat=40]
				table[x expr = \coordindex, y expr = \thisrowno{13}]{../ex4-otocs/syten/TDVP/L_128_dt_0p05_beta_0p0/results/otoc_t_slices};
				\addplot[forget plot, thick, smooth, color=blue, loosely dashed, mark=diamond, mark options=solid, mark phase=15, mark repeat=40]
				table[x expr = \coordindex, y expr = \thisrowno{19}]{../ex4-otocs/syten/TDVP/L_128_dt_0p05_beta_0p0/results/otoc_t_slices};
				\addplot[forget plot, thick, smooth, color=blue, dashed, mark=diamond, mark options=solid, mark phase=20, mark repeat=40]
				table[x expr = \coordindex, y expr = \thisrowno{25}]{../ex4-otocs/syten/TDVP/L_128_dt_0p05_beta_0p0/results/otoc_t_slices};
				\addplot[forget plot, thick, smooth, color=blue, densely dashed, mark=diamond, mark options=solid, mark phase=25, mark repeat=40]
				table[x expr = \coordindex, y expr = \thisrowno{31}]{../ex4-otocs/syten/TDVP/L_128_dt_0p05_beta_0p0/results/otoc_t_slices};
				
				\addplot[forget plot, thick, smooth, color=red, loosely dotted, mark=o, mark options=solid, mark phase=1, mark repeat=40]
				table[x expr = \coordindex, y expr = \thisrowno{1}]{../ex4-otocs/syten/TEBD/L_128_dt_0p05_beta_0p0/results/otoc_t_slices};
				\addplot[forget plot, thick, smooth, color=red, dotted, mark=o, mark options=solid, mark phase=3, mark repeat=40]
				table[x expr = \coordindex, y expr = \thisrowno{7}]{../ex4-otocs/syten/TEBD/L_128_dt_0p05_beta_0p0/results/otoc_t_slices};
				\addplot[forget plot, thick, smooth, color=red, densely dotted, mark=o, mark options=solid, mark phase=0, mark repeat=10]
				table[x expr = \coordindex, y expr = \thisrowno{13}]{../ex4-otocs/syten/TEBD/L_128_dt_0p05_beta_0p0/results/otoc_t_slices};
				\addplot[forget plot, thick, smooth, color=red, loosely dashed, mark=o, mark options=solid, mark phase=0, mark repeat=10]
				table[x expr = \coordindex, y expr = \thisrowno{19}]{../ex4-otocs/syten/TEBD/L_128_dt_0p05_beta_0p0/results/otoc_t_slices};
% 				\addplot[forget plot, thick, smooth, color=red, dashed, mark=diamond, mark options=solid, mark phase=0, mark repeat=10]
% 				table[x expr = \coordindex, y expr = \thisrowno{25}]{../ex4-otocs/syten/TEBD/L_128_dt_0p05_beta_0p0/results/otoc_t_slices};
% 				\addplot[forget plot, thick, smooth, color=red, densely dashed, mark=diamond, mark options=solid, mark phase=0, mark repeat=10]
% 				table[x expr = \coordindex, y expr = \thisrowno{31}]{../ex4-otocs/syten/TEBD/L_128_dt_0p05_beta_0p0/results/otoc_t_slices};
				
				\addplot[forget plot, thick, smooth, color=orange, loosely dotted, mark=triangle, mark options=solid, mark phase=2, mark repeat=40]
				table[skip first n=2,x expr = \coordindex, y expr = \thisrowno{1}]{../ex4-otocs/scipal/LocalKrylov/L_128_dt_0p05_beta_0p0/loc_observables/otoc_t_slices};
				\addplot[forget plot, thick, smooth, color=orange, dotted, mark=triangle, mark options=solid, mark phase=4, mark repeat=40]
				table[skip first n=2,x expr = \coordindex, y expr = \thisrowno{7}]{../ex4-otocs/scipal/LocalKrylov/L_128_dt_0p05_beta_0p0/loc_observables/otoc_t_slices};
				\addplot[forget plot, thick, smooth, color=orange, densely dotted, mark=triangle, mark options=solid, mark phase=10, mark repeat=40]
				table[skip first n=2,x expr = \coordindex, y expr = \thisrowno{13}]{../ex4-otocs/scipal/LocalKrylov/L_128_dt_0p05_beta_0p0/loc_observables/otoc_t_slices};
				\addplot[forget plot, thick, smooth, color=orange, loosely dashed, mark=triangle, mark options=solid, mark phase=12, mark repeat=10]
				table[skip first n=2,x expr = \coordindex, y expr = \thisrowno{19}]{../ex4-otocs/scipal/LocalKrylov/L_128_dt_0p05_beta_0p0/loc_observables/otoc_t_slices};
% 				\addplot[forget plot, thick, smooth, color=orange, dashed, mark=triangle, mark options=solid, mark phase=0, mark repeat=10]
% 				table[skip first n=2,x expr = \coordindex, y expr = \thisrowno{25}]{../ex4-otocs/scipal/LocalKrylov/L_128_dt_0p05_beta_0p0/loc_observables/otoc_t_slices};
% 				\addplot[forget plot, thick, smooth, color=orange, densely dashed, mark=triangle, mark options=solid, mark phase=0, mark repeat=10]
% 				table[skip first n = 2,x expr = \coordindex, y expr = \thisrowno{31}]{../ex4-otocs/scipal/LocalKrylov/L_128_dt_0p05_beta_0p0/loc_observables/otoc_t_slices};
				
				\addplot[forget plot, thick, smooth, color=green!90!black, loosely dotted, mark=square, mark options=solid, mark phase=3, mark repeat=40]
				table[skip first n=2,x expr = \coordindex, y expr = \thisrowno{1}/3]{../ex4-otocs/scipal/MPO/L_128_dt_0p05_beta_0p0/loc_observables/otoc_t_slices};
				\addplot[forget plot, thick, smooth, color=green!90!black, dotted, mark=square, mark options=solid, mark phase=4, mark repeat=40]
				table[skip first n=2,x expr = \coordindex, y expr = \thisrowno{7}/3]{../ex4-otocs/scipal/MPO/L_128_dt_0p05_beta_0p0/loc_observables/otoc_t_slices};
				\addplot[forget plot, thick, smooth, color=green!90!black, densely dotted, mark=square, mark options=solid, mark phase=9, mark repeat=40]
				table[skip first n=2,x expr = \coordindex, y expr = \thisrowno{13}/3]{../ex4-otocs/scipal/MPO/L_128_dt_0p05_beta_0p0/loc_observables/otoc_t_slices};
				\addplot[forget plot, thick, smooth, color=green!90!black, loosely dashed, mark=square, mark options=solid, mark phase=0, mark repeat=10]
				table[skip first n=2,x expr = \coordindex, y expr = \thisrowno{19}/3]{../ex4-otocs/scipal/MPO/L_128_dt_0p05_beta_0p0/loc_observables/otoc_t_slices};
% 				\addplot[forget plot, thick, smooth, color=green!90!black, dashed, mark=square, mark options=solid, mark phase=0, mark repeat=10]
% 				table[skip first n=2,x expr = \coordindex, y expr = \thisrowno{25}/3]{../ex4-otocs/scipal/MPO/L_128_dt_0p05_beta_0p0/loc_observables/otoc_t_slices};
% 				\addplot[forget plot, thick, smooth, color=orange, densely dashed, mark=square, mark options=solid, mark phase=0, mark repeat=10]
% 				table[skip first n=2,x expr = \coordindex, y expr = \thisrowno{31}/3]{../ex4-otocs/scipal/MPO/L_128_dt_0p05_beta_0p0/loc_observables/otoc_t_slices};
			\end{axis}
		\end{tikzpicture}
		\caption{\label{fig:otoc:lightcone-spreading}(OTOC) Spreading of the lightcone front for $F^{\hat{z}_{r}\hat{z}_{0}}_{\beta=0}(t)$ of a periodically driven Ising chain with local operators $\hat{z}_{r} = \hat{s}^{z}_{\nicefrac{L}{2}+r}$ at constant times and $\delta = 0.05$. 2TDVP appears to resemble the expected spreading, while TEBD2, the local Krylov and the MPO \wii methods exhibit only a reduced spreading. Instead, the latter methods create artificial offsets (results from these simulations for time slices $t=50,62$ are not shown here for the sake of clarity).}
	\end{minipage}
	\hfill
	\begin{minipage}[t]{0.49\textwidth}
          \pgfplotsset{width=\textwidth}
          \tikzsetnextfilename{otoc_compare_dt}
          \begin{tikzpicture}
            \begin{loglogaxis}
              [
              axis lines=left,
              xlabel={Time $t$},
              font = {\footnotesize},
              ylabel={OTOC $F^{\hat{z}_{1}\hat{z}_{0}}_{\beta=0}(t)$},
              width=0.96\textwidth,
              height=\textwidth,
              xmin=5,
              xmax=100,
              legend style={at={(1,0.05)}, anchor=south east},
              ymin=1e-4,
              ymax=1e-1,
              ]
				\addplot[thick,smooth, color=blue, dotted, forget plot, mark=diamond, mark options=solid, mark phase=0, mark repeat=200]
				table[x expr = \coordindex*0.05, y expr = \thisrowno{0}]{../ex4-otocs/syten/TDVP/L_128_dt_0p05_beta_0p0/results/otoc_x_65};
				\addplot[thick,opacity=0.5,draw opacity=0.5,smooth, color=blue, solid, mark=diamond, mark options=solid, mark phase=25, mark repeat=200]
				table[x expr = \coordindex*0.05, y expr = \thisrowno{0}]{../ex4-otocs/syten/TDVP/L_128_dt_0p01_beta_0p0/results/otoc_x_65};
				
				\addplot[thick,smooth, color=red, dotted, forget plot, mark=o, mark options=solid,  mark phase=5, mark repeat=200]
				table[x expr = \coordindex*0.05, y expr = \thisrowno{0}]{../ex4-otocs/syten/TEBD/L_128_dt_0p05_beta_0p0/results/otoc_x_65};
				\addplot[thick,opacity=0.5,draw opacity=0.5,smooth, color=red, solid, mark=o, mark options=solid, mark phase=35, mark repeat=200]
				table[x expr = \coordindex*0.05, y expr = \thisrowno{0}]{../ex4-otocs/syten/TEBD/L_128_dt_0p01_beta_0p0/results/otoc_x_65};
				
				\addplot[thick,smooth, color=orange, dotted, forget plot, mark=triangle, mark options=solid, mark phase=10, mark repeat=200]
				table[x expr = \coordindex*0.05, y expr = \thisrowno{0}]{../ex4-otocs/scipal/LocalKrylov/L_128_dt_0p05_beta_0p0/loc_observables/otoc_x_65};
				\addplot[thick,opacity=0.5,draw opacity=0.5,smooth, color=orange, solid, mark=triangle, mark options=solid, mark phase=40, mark repeat=200]
				table[x expr = \coordindex*0.05, y expr = \thisrowno{0}]{../ex4-otocs/scipal/LocalKrylov/L_128_dt_0p01_beta_0p0/loc_observables/otoc_x_65};
				
				\addplot[thick,smooth, color=green!90!black, dotted, forget plot, mark=square, mark options=solid, mark phase=10, mark repeat=200]
				table[x expr = \coordindex*0.05, y expr = \thisrowno{0}/3]{../ex4-otocs/scipal/MPO/L_128_dt_0p05_beta_0p0/loc_observables/otoc_x_65};
				\addplot[thick,opacity=0.5,draw opacity=0.5,smooth, color=green!90!black, solid, mark=square, mark options=solid, mark phase=40, mark repeat=200]
				table[x expr = \coordindex*0.05, y expr = \thisrowno{0}/3]{../ex4-otocs/scipal/MPO/L_128_dt_0p01_beta_0p0/loc_observables/otoc_x_65};
				
				\addplot[thick, smooth, color=black!60, solid, mark=none, domain=5:100] {1/(sqrt((10^(2*1.00883))*x-1))};
				
				\addlegendentry{2TDVP};
				\addlegendentry{TEBD2};
				\addlegendentry{Local Krylov}
				\addlegendentry{MPO \wii};
				\addlegendentry{$f(t) = \frac{a}{\sqrt{v_{B}\cdot t - 1}}$};
				
				\node[thick] at (axis cs:70,0.005) { \large $\propto \frac{1}{\sqrt{v_{b}\cdot t - 1}}$};
			\end{loglogaxis}
		\end{tikzpicture}
		\caption{\label{fig:otoc:compare-dt}(OTOC) $F^{\hat{z}_{1}\hat{z}_{0}}_{\beta=0}(t)$ of a periodically driven Ising chain with local operators $\hat{z}_{i} = \hat{s}^{z}_{\nicefrac{L}{2}+i}$. Equal colors and symbols refer to the same time evolution method, while the different line shapes indicate different step sizes. Dotted lines represent simulations with $\delta=0.05$ and solid lines represent $\delta=0.01$. The black solid line indicates the analytical form of the hydrodynamic tail, which is $f(t)=\frac{1}{\sqrt{v_{B}\cdot t - 1}}$.}
	\end{minipage}
\end{figure}
In order to investigate the lightcone spreading more carefully, we show in \cref{fig:otoc:lightcone-spreading} the decay of the lightcone at various slices throughout the right half of the system at constant times $t=2,14,26,38,50$ and $62$ at fixed time step $\delta=0.05$.
Only at the first two time slices all methods (nearly) agree with each other displaying the expected broading of the lightcone front.
The results for the 2TDVP method show an overall decay for the slope at the lightcone front at later times, as expected.
In turn the local Krylov, TEBD2 and MPO \wii methods produce a decay for the slope of the lightcone front under the time evolution which can be attributed to additional ray-like structures in the ligtcone, c.f. \cref{fig:otoc:lightcone-local-krylov}.
Furthermore, we find that the saturation value outside the lightcone for these three methods lowers during the time evolution.
This can be linked to additional phases built up during the time evolutions between the two states $\ket{Z_{0}(t)}$ and $\ket{Z(\beta=0,t)}$, as already observed in \cref{sec:examples:dsf}.

\subsubsection{\label{sec:examples:otoc:algebraic-tail}Hydrodynamic tail}
In the case of periodic boundary conditions the long time behavior behind the lightcone, which travels with the butterfly velocity $v_{B}$, is expected to exhibit an algebraic power law decay, i.e., for $v_{B}\cdot t \gg \lvert r \rvert$ the OTOC is dominated by $\frac{1}{\sqrt{v_{B}\cdot t - \lvert r \rvert}}$.
\cref{fig:otoc:compare-dt} compares the OTOCs at fixed relative position $r\equiv 1$ obtained with the methods under consideration (2TDVP, TEBD2, MPO \wii and local Krylov) and the time steps $\delta=0.01,0.05$.
2TDVP qualitatively reproduces the expected hydrodynamic tail as can be seen by comparing to an indicated reference function $f(t)=\frac{a}{\sqrt{v_{B}\cdot t -1}}$ where the parameters $a$ and $v_{B}$ are obtained by a suitable fit of the 2TDVP datasets.
Furthermore there are only little deviations when tuning the time steps, strengthening the validity of these results.
However, comparing against the local Krylov, MPO \wii and TEBD2 datasets there is only in agreement until $t\approx 10T$ after which the methods start to deviate and exhibit quite distinct oscillatoric behavior.
It seems that at least for the local Krylov (at $\delta=0.01$) and MPO \wii (at $\delta=0.05,0.01$) method one can identify the expected hydrodynamic tail but the data is far from being reliable.
Since the initial states describe a (perturbed) ensemble at infinite temperature with comparably large initial bond dimensions ($m\sim 150$) we could expect the projection errors in 2TDVP and the local Krylov method, respectively, to be very small because there will be a finite overlap with a large portion of the states in the many-body Hilbert space.
Subsequently, the effect of the perturbation $\hat{s}^{z}_{\nicefrac{L}{2}}$ on the infinite temperature state is then to create quasiparticles from the entire spectrum resulting in a complicated dynamics which is a tough challenge for all time evolution schemes.
Nevertheless, since TDVP (1TDVP exact, 2TDVP approximately) fulfills conservations laws for the energy, particles etc. it is not too surprising to find the proper long-time dynamics being captured by TDVP.
Even though the datasets created by 2TDVP are physically reasonable, the local Krylov as the other method approximating the action of the exponential suffers from similar phase accumulation as in \cref{sec:examples:dsf}.
Consistently we also find the strongest dependence on the step size $\delta$ for the local Krylov method.

\subsubsection{\label{sec:examples:otoc:benchmarks}Benchmark}
\begin{figure}[t!]
	\begin{center}
	  \tikzsetnextfilename{otoc_benchmarks_legend}
	  \ref{otoc-benchmarks-legend}
	\end{center}
	\begin{minipage}[t]{0.49\textwidth}
		\tikzsetnextfilename{otoc_benchmarks_chimax}
		\begin{tikzpicture}
			\begin{axis}
			[
				width=\textwidth,
				axis lines=left,
				xlabel={Time $t$},
				font = {\footnotesize},
				ylabel={max. bond dimension $m$},
				xmin=1,
				xmax=15,
				% legend style={at={(1,0.05)}, anchor=south east},
				% legend entries = {$t=2$,$t=14$,$t=26$,$t=38$, $t=50$, $t=62$, 2TDVP, TEBD2, Local Krylov, MPO\wii},
				% ymin=1e-4,
				ymax=1050,
				legend style = {anchor=north east, at = {(1,0.8)}, draw = none, fill = none, font={\footnotesize}},
				legend to name = otoc-benchmarks-legend,
				legend columns=4,
				legend cell align = left,
			]
				\addplot[forget plot, thick, smooth, color=blue, dotted, mark=diamond, mark options=solid, mark phase=2, mark repeat=40]
				table[x expr = \coordindex, y expr = \thisrowno{0}]{../ex4-otocs/syten/TDVP/L_128_dt_0p05_beta_0p0/chi_max_W1.output};
				\addplot[thick, smooth, color=blue, solid, mark=diamond, mark options=solid, mark phase=3, mark repeat=40]
				table[x expr = \coordindex, y expr = \thisrowno{0}]{../ex4-otocs/syten/TDVP/L_128_dt_0p01_beta_0p0/chi_max_W1.output};
				
				\addplot[forget plot, thick, smooth, color=red, dotted, mark=o, mark options=solid, mark phase=2, mark repeat=40]
				table[x expr = \coordindex, y expr = \thisrowno{0}]{../ex4-otocs/syten/TEBD/L_128_dt_0p05_beta_0p0/chi_max_W1.output};
				\addplot[thick, smooth, color=red, solid, mark=o, mark options=solid, mark phase=3, mark repeat=40]
				table[x expr = \coordindex, y expr = \thisrowno{0}]{../ex4-otocs/syten/TEBD/L_128_dt_0p01_beta_0p0/chi_max_W1.output};
				
				\addplot[forget plot, thick, smooth, color=orange, dotted, mark=triangle, mark options=solid, mark phase=5, mark repeat=40]
				table[x expr = \coordindex, y expr = \thisrowno{0}]{../ex4-otocs/scipal/LocalKrylov/L_128_dt_0p05_beta_0p0/chi_max_W1.output};
				\addplot[thick, smooth, color=orange, solid, mark=triangle, mark options=solid, mark phase=6, mark repeat=40]
				table[x expr = \coordindex, y expr = \thisrowno{0}]{../ex4-otocs/scipal/LocalKrylov/L_128_dt_0p01_beta_0p0/chi_max_W1.output};
				
				\addplot[forget plot, thick, smooth, color=green!90!black, dotted, mark=square, mark options=solid, mark phase=5, mark repeat=40]
				table[x expr = \coordindex, y expr = \thisrowno{0}]{../ex4-otocs/scipal/MPO/L_128_dt_0p05_beta_0p0/chi_max_W1.output};
				\addplot[thick, smooth, color=green!90!black, solid, mark=square, mark options=solid, mark phase=6, mark repeat=40]
				table[x expr = \coordindex, y expr = \thisrowno{0}]{../ex4-otocs/scipal/MPO/L_128_dt_0p01_beta_0p0/chi_max_W1.output};
				
				\addlegendentry{2TDVP}
				\addlegendentry{TEBD2}
				\addlegendentry{Local Krylov}
				\addlegendentry{MPO \wii}
			\end{axis}
		\end{tikzpicture}
		\caption{\label{fig:otoc:benchmarks-chimax}(OTOC) Maximal bond dimension $m$ during the evolution of $\ket{Z_{0}(t)}$ plotted over real time $t$. The datapoints are obtained after a complete driving cycle $t=kT$ and the line styles indicate the used time step: dotted lines correspond to $\delta=0.05$, solid lines to $\delta=0.01$. For all methods $m$ grows quickly and reaches the maximal permitted value at times $t< 10T$. 2TDVP exhibits the fastest increase with the maximum $m$ being reached at times $t<5T$ while MPO \wii shows the slowest growth.}
	\end{minipage}
	\hfill
	\begin{minipage}[t]{0.49\textwidth}
		\tikzsetnextfilename{otoc_benchmarks_cputime}
		\begin{tikzpicture}
			\begin{semilogyaxis}
			[
				width=\textwidth,
				axis lines=left,
				xlabel={Time $t$},
				font = {\footnotesize},
				ylabel={avg. cpu time per unit time step $T$},
				xmin=0,
				xmax=64,
% 				ymin=1e-4,
% 				ymax=1050,
			]
				\addplot[forget plot, thick, smooth, color=blue, dotted, mark=diamond, mark options=solid, mark phase=2, mark repeat=40]
				table[x expr = \coordindex, y expr = \thisrowno{0}]{../ex4-otocs/syten/TDVP/L_128_dt_0p05_beta_0p0/teo.output};
				\addplot[thick, smooth, color=blue, solid, mark=diamond, mark options=solid, mark phase=3, mark repeat=40]
				table[x expr = \coordindex, y expr = \thisrowno{0}]{../ex4-otocs/syten/TDVP/L_128_dt_0p01_beta_0p0/teo.output};
				
				\addplot[forget plot, thick, smooth, color=red, dotted, mark=o, mark options=solid, mark phase=2, mark repeat=40]
				table[x expr = \coordindex, y expr = \thisrowno{0}]{../ex4-otocs/syten/TEBD/L_128_dt_0p05_beta_0p0/teo.output};
				\addplot[thick, smooth, color=red, solid, mark=o, mark options=solid, mark phase=3, mark repeat=40]
				table[x expr = \coordindex, y expr = \thisrowno{0}]{../ex4-otocs/syten/TEBD/L_128_dt_0p01_beta_0p0/teo.output};
 				
				\addplot[forget plot, thick, smooth, color=orange, dotted, mark=triangle, mark options=solid, mark phase=5, mark repeat=40]
				table[x expr = \coordindex, y expr = \thisrowno{0}]{../ex4-otocs/scipal/LocalKrylov/L_128_dt_0p05_beta_0p0/teo.output};
				\addplot[thick, smooth, color=orange, solid, mark=triangle, mark options=solid, mark phase=6, mark repeat=40]
				table[x expr = \coordindex, y expr = \thisrowno{0}]{../ex4-otocs/scipal/LocalKrylov/L_128_dt_0p01_beta_0p0/teo.output};
				\addplot[forget plot, thick, smooth, color=green!90!black, dotted, mark=square, mark options=solid, mark phase=5, mark repeat=40]
				table[x expr = \coordindex, y expr = \thisrowno{0}]{../ex4-otocs/scipal/MPO/L_128_dt_0p05_beta_0p0/teo.output};
				\addplot[thick, smooth, color=green!90!black, solid, mark=square, mark options=solid, mark phase=6, mark repeat=40]
				table[x expr = \coordindex, y expr = \thisrowno{0}]{../ex4-otocs/scipal/MPO/L_128_dt_0p01_beta_0p0/teo.output};
			\end{semilogyaxis}
		\end{tikzpicture}
		\caption{\label{fig:otoc:benchmarks-cputime}(OTOC) Elapsed CPU time per unit time $T$ during real-time evolution of $\ket{Z_{0}(t)}$ plotted over real time $t$. The datapoints are obtained after a complete driving cycle $t=kT$ and the line styles indicate the used time step: dotted lines correspond to $\delta=0.05$, solid lines to $\delta=0.01$. By far, 2TDVP shows the best performance while the TEBD2, MPO \wii and local Krylov methods at least permit for reasonable calculation of the time-evolution at time steps $\delta=0.05$.}
	\end{minipage}
\end{figure}
Finally we take a look at the benchmark data for the calculated time evolutions at the example of the evolution of the state $\ket{Z_{0}(t)}$.
The calculations where performed single threaded on an Intel Xeon Gold 6150 CPU with 192 GB of RAM and no hard-disk caching.
All methods are subject to a rapid increase of the maximal bond dimension $m$ (cf.~\cref{fig:otoc:benchmarks-chimax}).
2TDVP exhibits the quickest increase and already reaches the maximal permitted $m$ at times $t<5T$ while the MPO \wii method has the slowest growth and saturates around $t\approx 10T$ which may be attributed to the variational optimization during the MPO application.
In \cref{fig:otoc:benchmarks-cputime} we plot the CPU time required per unit time $T$ averaged over one driving cycle.
We find the best performance for 2TDVP followed by TEBD2 and the MPO $W^\mathrm{II}$.
Interestingly, the performance of the MPO \wii method at $\delta=0.05$ is very competitive with TEBD2 at the same step size but when reducing the step size towards $\delta=0.01$ the variational update in the MPO application, which is absolutely necessary to ensure numerical stability in this testcase, renders the simulations extremly slow.
The large difference between the runtimes for 2TDVP and the local Krylov method can be explained by inspecting the number of local Krylov vectors which are required to achieve the target precision of $10^{-10}$.
TDVP typically requires only $5$ iterations of the local Krylov solver to converge while the local Krylov method reaches the maximum number of permitted iterations ($12$) already after a few time steps.
Consequently the local Krylov method is unsuitable for small time steps $\delta = 0.01$.

\subsubsection{\label{sec:examples:otoc:conclusion}Conclusion}
We have tested four time evolution methods: 2TDVP, TEBD2, local Krylov and MPO \wii for their ability to describe the time evolution of out-of-time-order correlators for a periodically-driven Ising chain.
The best results are provided by 2TDVP with respect to both physical accuracy and performance.
TEBD2 again yields a competitive runtime at least for larger step sizes but the data fails to reproduce the broadening of the lightcone and even the hydrodynamic tail of the OTOC is only qualitatively reproduced.
Both the local Krylov and MPO \wii method fail to adequately reproduce the broadening of the lightcone front employing the numerically feasible step size $\delta=0.05$ and, more severely, pick up global offsets in the OTOC at long runtimes.
The hydrodynamic tail may be identified in these datasets but the reliability should be strengthened by a very careful analysis with more time steps and different maximal bond dimensions $m$.
Since fulfilling conservation laws seems to be advantageous for capturing the complicated dynamics inside the lightcone the possibly best strategy would be to perform 2TDVP time evolution until the bulk bond dimensions have saturated to the permitted maximum value and then switch to 1TDVP.
In addition to the results presented here using the purification approach, it was recently shown that TDVP when used with METTS can also provide reasonable data for OTOCs\cite{hemery19}.

%% file: content/outlook.tex
\section{\label{sec:outlook}Future developments}

While one should of course not lose sight of new alternative methods
to evaluate excitation spectra\cite{haegeman12:_variat,
  haegeman13:_post,
  vanderstraeten17:_tensor_networ_states_effec_partic,
  vanderstraeten18:_simul, vanderstraeten19:_tangen} or Green's
functions\cite{xie18:_reort_cheby}, the direct evaluation of real-time
observables will always find interesting applications. For the future,
at least three distinct approaches for new developments of real-time
evolution methods should be considered:

\paragraph{Improved MPS-local methods} Following the spirit of
td-DMRG, the local Krylov and TDVP may provide the best avenue towards
hydrodynamics with the 1TDVP method already enforcing complete energy
conservation though still suffering from a limited bond dimension\cite{kloss18:_time}. Combining e.g.~the 2TDVP method with a truncation scheme
which \emph{also} ensures energy conservation\cite{white18:_quant} may
prove particularly fruitful.

\paragraph{Easier and more accurate construction of $\hat U(\delta)$}
At the moment one has to use either the TEBD2 or the MPO \wiii method
to construct a MPO representation of $\hat U(\delta)$. Both incur
relatively large time-step errors and require more than simply a set
of tensors describing the MPO as input. One potential alternative here is a construction as
$(1 - \I \delta^\prime \hat H)^N$ with $\delta^\prime N = \delta$ and
$N \gg 1$. The ingredients can all be represented within the MPO
framework and MPO compression may provide a very helpful tool to
conserving as much accuracy as possible during the
construction\cite{greene17:_tensor_train_split_operat_fourier}. Alternatively,
combining different approaches, the newest example being a
large-scale Trotter decomposition combined with a small-scale Krylov
method, may also lead to flexible time steppers which only suffer from
minimal time step-size
errors\cite{hashizume18:_dynam_phase_trans_two_dimen}.

\paragraph{``Better'' Krylov vectors} The major drawback of the
global Krylov method at the moment is the fast growth of entanglement
in the Krylov vectors. While it is used where absolutely
necessary -- due to its flexibility in the choice of time step sizes
-- alternative methods provide much faster long-time evolution. In
addition to this flexibility, it also only requires an MPO
representation as input, making it potentially useful in applications
where decompositions are not straightforward. Hence, an approach to
use it also in complex settings would be highly appreciated. This
could be done by using less entangled Krylov-like subspaces whose
basis vectors can be represented efficiently as matrix-product
states. Alternatively, one can envision a second basis transformation
finding minimally-entangled basis vectors for the standard Krylov
subspace.